\documentclass[oldversion]{aa}
\usepackage{graphicx}
\usepackage{amssymb}
\usepackage{amsmath}
\usepackage{natbib}
\usepackage{txfonts}

\def\void#1{{}}

\begin{document}
   \title{Multi-object spectroscopy of low-redshift EIS clusters}
   \subtitle{IV. Reliability of matched-filter results at $z\sim0.3-0.4$
	 \thanks{Based on observations made with the Danish1.5-m
	 telescope at ESO, La Silla,
	 Chile.}$^,$
	 \thanks{Table~\ref{tab:galredshifts} is only available
	 in electronic form at the CDS via anonymous ftp to
	 cdsarc.u-strasbg.fr (130.79.128.5) or via
	 http://cdsweb.u-strasbg.fr/cgi-bin/qcat?J/A+A/}}

   \author{L.F. Grove\inst{1}\thanks{L.F. Grove previously published under L.F. Olsen}
           \and
           L. da Costa \inst{2}
           \and
	   C. Benoist\inst{3}
%	  \and
%	   H.E. J{\o}rgensen\inst{1}
%	   \and
%	   L. Hansen\inst{1}
	   }

   \offprints{L.F. Grove, lisbeth@dark-cosmology.dk}

   \institute{Dark Cosmology Centre, Niels Bohr Institute, University of Copenhagen, Juliane Maries Vej 30, DK-2100 Copenhagen, Denmark
            \and Observat\'orio Nacional, Rua Gen. Jos\'e Cristino 77, Rio de
Janeiro, R.J., Brazil 
	    \and Laboratoire Cassiop\'ee, CNRS, UMR 6202, Observatoire de la C\^ote d'Azur, BP4229,  06304 Nice Cedex 4, France
}

   \date{Received .....; accepted .....}

   \abstract{ This paper is the last in a series investigating
low-redshift galaxy systems identified by the matched-filter technique
in a moderately deep $I-$band survey. In this paper we present new
redshifts for 747 galaxies in 23 ESO Imaging Survey (EIS) cluster
fields. We use the ``gap''-technique to search for significant
overdensities in redshift space for identifying groups/clusters of
galaxies corresponding to the original EIS matched-filter cluster
candidates. In this way we spectroscopically confirm systems in 10 of
the 23 cluster candidate fields with a matched-filter estimated
redshift $z_{MF}=0.3-0.4$ and with spectroscopic redshifts in the
range from $z=0.158$ to $z=0.534$, with the observations favouring the
confirmation of systems at the lower redshift end. After careful
analysis of the redshift distribution, one system was split into two
very close clumps in redshift space. We find that the systems
identified in the present paper span a broad range of velocity
dispersion and richness. The measured one-dimensional velocity
dispersion range from $175\mathrm{km/s}$ to $497\mathrm{km/s}$,
consistent with the values obtained in previous papers using much
larger samples for systems over the same redshift range. Both
undersampling and contamination by substructures contribute to the
uncertainty of these measurements.  The richness range corresponds to
clusters with an estimated total luminosity in the range
$12L^*<L<65L^*$, but these estimates are very uncertain as are their
relation to the velocity dispersion (mass) of the systems. From the
analysis of the colours of the galaxy populations we find that
$\sim$60\% of the spectroscopically confirmed systems have a
``significant'' red sequence.  We find that the colour of the red
sequence galaxies matches passive stellar evolution predictions. With
this paper we complete our spectroscopic survey of the fields of 58
EIS cluster candidates with estimated redshifts $z\leq0.4$. We have
measured a total of 1954 galaxy redshifts in the range $z=0.0065$ to
$z=0.6706$. Of the 58 systems we confirm 42 ($\sim$75\%) with
redshifts between $z=0.095$ and $z=0.534$.  } \keywords{ galaxies:
clusters: general -- cosmology: observations -- galaxies: distances
and redshifts -- galaxies: photometry }

   \maketitle
%
%__________________________________________________________________

\section{Introduction}
%__________________________________________________________________

Clusters of galaxies have long been recognised as important targets
for cosmology and astrophysics, both to constrain cosmological
parameters and for evolutionary studies of large-scale structure and
of cluster members. However, to carry out these studies large samples
of clusters of galaxies with well-defined selection criteria,
representative of the entire population of clusters and spanning a
wide redshift range are required.

Over the past decades a number of systematic efforts to assemble
catalogues of galaxy clusters
\cite[e.g. ][]{gunn86,postman96,scodeggio99, gladders01, gonzalez01,
bahcall03, olsen07, vanbreukelen2006, koester2007} have been
undertaken based on large optical and infrared imaging surveys in one
or more passbands, using different search techniques. This has greatly
expanded the available samples both at low and high redshifts, as
compared to those detected in X-rays
\citep[e.g. ][]{rosati98a,bohringer00}. Unfortunately, systematic
studies determining the yield of real bound systems from these image
based searches has not progressed as fast, mainly because
spectroscopic data covering large areas of the sky are difficult to
assemble especially at intermediate and high redshifts. Furthermore,
each detection technique is based on some different characteristic of
the cluster such as its luminosity function, galaxy population or gas
content. Therefore, it is important to understand the impact that
these underlying assumptions may have on the detection and on the
estimates of redshift and richness.  In order to be able to use
cluster samples in statistical studies, especially to constrain
cosmological parameters, it is important not only to detect the
systems but also be able to establish how reliable estimates for the
redshift and richness are, how well they relate to the underlying mass
of the system and the selection function of the sample. To determine
that requires both extensive spectroscopic follow-up and the use of
realistic numerical simulations from which mock galaxy catalogues can
be built.  Here we continue the effort of analysing the ESO Imaging
Survey (EIS) Cluster Candidate Catalogue constructed using a
matched-filter algorithm \citep{olsen99a,olsen99b,scodeggio99}. A
first estimate of the selection function for this catalogue based on
simplistic simulations was given in \cite{olsen00}.  In this paper we
concentrate on the results of a comprehensive spectroscopic follow-up
of these cluster candidates, covering a broad range of redshifts.
While firm conclusions cannot be drawn due to the small size of the
sample, the issues that may affect the use of galaxy clusters for
statistical studies can be explored and will serve as a guide for
further investigation based on on-going numerical simulations.

This paper is part of a series reporting the results obtained
from a spectroscopic follow-up of selected cluster candidate fields
drawn from the original EIS cluster candidate sample assembled by
\cite{olsen99a,olsen99b,scodeggio99}. This sample was constructed from
the EIS $I-$band survey over 17~square degrees of the sky using the
matched-filter technique. Out of 302 identified cluster candidates we
have so far studied more than 50 candidate cluster fields chosen to
contain candidates at different estimated redshift bins. So far we
have probed fields with candidates at low \citep[$z_{MF}\sim0.2$,
][hereafter Paper~I, II and III]{hansen02,olsen03,olsen05},
intermediate \citep[$0.5\leq z_{MF}\leq0.7$, ][]{ramella00} and high
redshifts \citep[$z_{MF}\geq0.6$, ][]{benoist02, olsen05b} using
different telescopes and spectrographs. In this paper we extend the
previous work at the low-redshift end by carrying out spectroscopic
observations of 23 fields around cluster candidates selected in the
estimated redshift bin $0.3\leq z_{MF}\leq0.4$.

The paper is structured as follows: Sect.~\ref{sec:obs} gives an
overview of the observations and data reduction.
Sect.~\ref{sec:results} reviews the identification of systems in
redshift space as well as the procedure adopted for associating the
redshift groups to the EIS detections. Sect.~\ref{sec:properties}
describes the photometric properties of the spectroscopically
confirmed systems including an analysis of the colour properties of
the galaxy populations.  Finally, in Sect.~\ref{sec:conclusions} a
brief discussion of the conclusions of the results obtained in the
present paper is presented, followed by a summary of the results of
the entire series in Sect.~\ref{sec:summary}.

%__________________________________________________________________

\section{Observations and data reduction}
%__________________________________________________________________

\label{sec:obs}

In the present work, all cluster candidates with estimated
redshifts in the range $0.3 \leq z_{MF} \leq 0.4$ within EIS patches~A
and B covering an area of $\sim3.5$~square degrees
\citep{olsen99a,olsen99b} were selected for investigation. A total of
23 fields were observed, one of which containing two cluster
candidates. Unfortunately, one of them (EISJ2243-4008) was not
properly sampled and is not included in the present analysis.  We
list the 23 targeted cluster fields in Table~\ref{tab:cl_targets},
giving in Col.~1 an identifier for the cluster field; in Col.~2 the
name of the field referring to the notation adopted by
\cite{olsen99a,olsen99b}; in Cols.~3 and 4 the matched filter position
in J2000; in Col.~5 the matched filter estimated redshift and in
Col.~6 the $\Lambda_{cl, org}$-richness, which is the original
richness estimate of the matched filter detection. The candidates
roughly cover a richness range corresponding to Abell richness classes
$\leq 1$ \citep[e.g.][]{postman96}.

\begin{table}
\caption{Cluster candidates selected for observations.}
\label{tab:cl_targets}
\begin{minipage}{\columnwidth}
\begin{center}
\begin{tabular}{rlllrr}
\hline\hline ID & Field\footnote{Here we have added a ``J'' in the
name to conform with international standards. The EIS identification
is the same except for this ``J''.} & $\alpha_{J2000}$ &
$\delta_{J2000}$ & $z_{MF}$ & $\Lambda_{cl,org}$ \\
\hline 
1 & EISJ0044-2950A & 00 44 58.6 & -29 50 49.5 & 0.3 & 23.1 \\ 
2 & EISJ0045-2944 & 00 45 00.8 & -29 44 57.7 & 0.4 & 35.5 \\ 
3 & EISJ0047-2942 & 00 47 23.0 & -29 42 59.0 & 0.4 & 30.4 \\ 
4 & EISJ0048-2928 & 00 48 25.8 & -29 28 50.1 & 0.4 & 36.6  \\ 
5 & EISJ0049-2920 & 00 49 31.3 & -29 20 34.1 & 0.3 & 35.7 \\ 
6 & EISJ2236-3935 & 22 36 02.9 & -39 35 33.7 & 0.3 & 44.5 \\ 
7 & EISJ2236-4026 & 22 36 47.6 & -40 26 17.4 & 0.4 & 44.0 \\ 
8 & EISJ2236-4014 & 22 36 52.6 & -40 14 53.0 & 0.4 & 37.4 \\ 
9 & EISJ2237-4000 & 22 37 11.4 & -40 00 16.1 & 0.3 & 31.3 \\ 
10 & EISJ2238-3934 & 22 38 03.4 & -39 34 50.4 & 0.3 & 41.8 \\ 
11 & EISJ2239-3954 & 22 39 18.4 & -39 54 34.9 & 0.3 & 62.5 \\ 
12 & EISJ2240-4021 & 22 40 07.8 & -40 21 08.0 & 0.3 & 41.2  \\ 
13 & EISJ2241-4006 & 22 41 26.7 & -40 06 24.7 & 0.3 & 32.6 \\ 
14 & EISJ2241-3932 & 22 41 31.3 & -39 32 10.4 & 0.4 & 44.5 \\ 
15 & EISJ2243-4010A & 22 43 01.9 & -40 10 24.8 & 0.3 & 39.1 \\ 
16 & EISJ2243-3952 & 22 43 19.4 & -39 52 41.2 & 0.3 & 50.9 \\ 
17 & EISJ2243-3959 & 22 43 29.4 & -39 59 33.5 & 0.3 & 45.0 \\ 
18 & EISJ2243-4010B & 22 43 42.7 & -40 10 30.4 & 0.3 & 32.4  \\ 
$-$\footnote{Overlaps with EISJ2243-4014B and separate observations were not carried out.}&
EISJ2243-4008 & 22 43 47.4 & -40 08 47.0 & 0.3 & 34.3 \\ 
19 & EISJ2243-3947 & 22 43 56.1 & -39 47 28.8 & 0.4 & 48.6 \\
20 & EISJ2244-4008 & 22 44 21.8 & -40 08 21.6 & 0.4 & 31.1 \\ 
21 & EISJ2244-4019 & 22 44 28.4 & -40 19 46.5 & 0.3 & 38.3 \\ 
22 & EISJ2246-4012B & 22 46 48.5 & -40 12 48.2 & 0.4 & 39.5\\ 
23 & EISJ2248-4015 & 22 48 54.8 & -40 15 18.8 & 0.3 & 36.2 \\ 
\hline
\end{tabular}
\end{center}
\end{minipage}
\end{table}

\begin{table*}
\begin{center}
\caption{Summary of spectroscopic coverage for each target cluster.}
\label{tab:spec_compl}
\begin{tabular}{rlccrrc}
\hline\hline ID & Field & Covered area & \#targets & \#z & Compl. & Efficiency \\
\hline 
1 & EISJ0044-2950A & 56.2  & 35 & 20 & 0.54 & 0.57 \\ 
2 & EISJ0045-2944 &  60.4  & 40 & 32  & 0.67 & 0.80 \\ 
3 & EISJ0047-2942 &  78.9  & 36 & 31 & 0.46 & 0.86  \\ 
4 & EISJ0048-2928 &  75.8  & 38 & 26 & 0.50 & 0.68 \\ 
5 & EISJ0049-2920 &  71.6  & 39 & 35 & 0.56 & 0.90 \\ 
6 & EISJ2236-3935 &  123.4 & 52 & 42 & 0.39 & 0.81 \\ 
7 & EISJ2236-4026 &  64.3  & 41 & 21 & 0.52 & 0.51 \\ 
8 & EISJ2236-4014 &  76.4  & 42 & 28 & 0.52 & 0.68 \\ 
9 & EISJ2237-4000 &  111.6 & 55 & 35 & 0.49 & 0.64 \\ 
10 & EISJ2238-3934 & 118.0 & 55 & 45 & 0.38 & 0.82 \\ 
11 & EISJ2239-3954 & 100.1 & 48 & 42 & 0.43 & 0.88 \\ 
12 & EISJ2240-4021 & 102.0 & 48 & 36 & 0.48 & 0.75 \\ 
13 & EISJ2241-4006 & 91.5  & 57 & 36 & 0.53 & 0.63 \\ 
14 & EISJ2241-3932 & 61.7  & 38 & 22 & 0.55 & 0.58 \\ 
15 & EISJ2243-4010A & 82.8 & 45 & 21 & 0.40 & 0.47 \\ 
16 & EISJ2243-3952 &  97.5 & 49 & 44 & 0.41 & 0.90 \\ 
17 & EISJ2243-3959 &  91.1 & 47 & 43 & 0.47 & 0.91 \\ 
18 & EISJ2243-4010B & 93.9 & 46 & 26 & 0.38 & 0.57 \\ 
19 & EISJ2243-3947 &  65.3 & 37 & 35 & 0.51 & 0.95 \\
20 & EISJ2244-4008 &  65.6 & 41 & 35 & 0.63 & 0.85 \\ 
21 & EISJ2244-4019 &  88.1 & 51 & 40 & 0.49 & 0.78 \\ 
22 & EISJ2246-4012B & 54.8 & 17 & 16 & 0.27 & 0.94 \\ 
23 & EISJ2248-4015 & 61.3 & 42 & 36 & 0.61 & 0.86 \\ 
\hline
\end{tabular}
\end{center}
\end{table*}

The observations were carried out at the Danish 1.54m telescope at La
Silla, Chile. We used the Danish Faint Object Spectrograph and Camera
(DFOSC) in the Multi-Object Spectroscopy (MOS)-mode. The field of view
of DFOSC corresponds to $3.4\mathrm{Mpc}$ and $4.1\mathrm{Mpc}$ at
$z=0.3$ and $z=0.4$ (assuming $\mathrm{H}_0=75\mathrm{km/s/Mpc}$,
$\Omega_\mathrm{m} = 0.3$ and $\Omega_\Lambda=0.7$). This is a
good match to the typical size of a cluster. The effective field that
could be covered with MOS slit masks was typically
$12\farcm4\times6\farcm5$, depending on the exact configuration of
galaxy positions in each field. The slit width was set to $2\arcsec$,
and the slit length varied according to the extent of each galaxy.  We
used grism \#4, giving a dispersion of $220\mathrm{{\AA}/mm}$, and
covering, on average, a wavelength range from $3800$ to
$7500\mathrm{{\AA}}$. The resolution, as determined from HeNe line
spectra, was found to be $16.6\mathrm{{\AA}\;FWHM}$.   For each
field we created two slit masks for an exposure time of 1.5~hours per
mask. We estimate the S/N of the spectra for which we could measure
the redshift to be in the range 5 to 15.

We preferentially targeted bright galaxies with $I-$magnitude
$I\leq20.0$ (Vega\footnote{All magnitudes are quoted in the EIS
magnitude system as provided by the EIS team, see
\protect\cite{nonino99,prandoni99,benoist99}}).  The Schechter
magnitude is estimated to be $I^*\sim18.5$ and $19.3$ at $z=0.3$ and
$0.4$, respectively, using an absolute Schechter magnitude in the
Cousin filter of $M^*_I=-22.33$ as commonly adopted
\citep[e.g.][]{postman96,olsen99a}. The corresponding apparent
magnitude was computed using the K-correction for an elliptical galaxy
template spectrum from the Kinney library \citep{kinney96}. We thus
estimate our survey to cover galaxies to between $\sim1.5$ and
$\sim0.7$~magnitudes fainter than the Schechter magnitude.  The
target galaxies were selected without applying any additional colour
selection. This allows us to investigate whether systems without red
sequences are present in our sample. The matched-filter sample does
not have any built-in bias against such systems and therefore provides
the necessary input for such an investigation.

The data reduction was performed using standard procedures available
in IRAF\footnote{ IRAF is distributed by the National Optical
Astronomy Observatories, which is operated by AURA Inc. under contract
with NSF.} packages as described in detail in previous papers of this
series (in particular Papers~I and III). The redshifts were measured
by Fourier cross-correlating our spectra with standard galaxy spectra
templates from \cite{kinney96}.  
However, the redshift estimated by the cross-correlation was only
accepted if a prominent spectral feature could be identified at a
redshift consistent with that estimated by the cross-correlation.
In Paper~III the accuracy of the measured redshifts was
estimated to be $\sigma_z=0.0005$. Since the typical magnitudes are
similar we adopt the same value here.

For all target clusters two slit masks were produced regardless of the
local galaxy density. Therefore, we do not reach the same level of
completeness (defined as the fraction of targeted galaxies in a field)
in all fields. Also the efficiency for obtaining redshifts (the
fraction of obtained spectra that yielded a redshift measurement)
varies, mainly due to varying observing conditions.  In
Table~\ref{tab:spec_compl} we summarise the results of the
spectroscopic campaign. The table lists in Col.~1 the field
identifier; Col.~2 the field name; in Col.~3 the area in square arcmin
containing the target galaxies; in Col.~4 the number of target
galaxies; in Col.~5 the number of derived redshifts; in Col.~6 the
overall completeness, as defined below; in Col.~7 the overall
efficiency of obtaining redshifts (the ratio between Col.~5 and
Col.~4). The overall completeness given in Col.~6 is defined as the
ratio of observed to all galaxies brighter than $I=20.0$ within a
rectangular region. The latter is defined as the smallest rectangle
covering all observed galaxies and is outlined by dashed lines in
Figs.~\ref{fig:conf} and \ref{fig:nonconf}.

%in Col.~8 the fraction of galaxies with redshifts among all
%galaxies with magnitudes within one magnitude of the expected $I^*$.

\begin{figure}
\begin{center}
\resizebox{0.9\columnwidth}{!}{\includegraphics{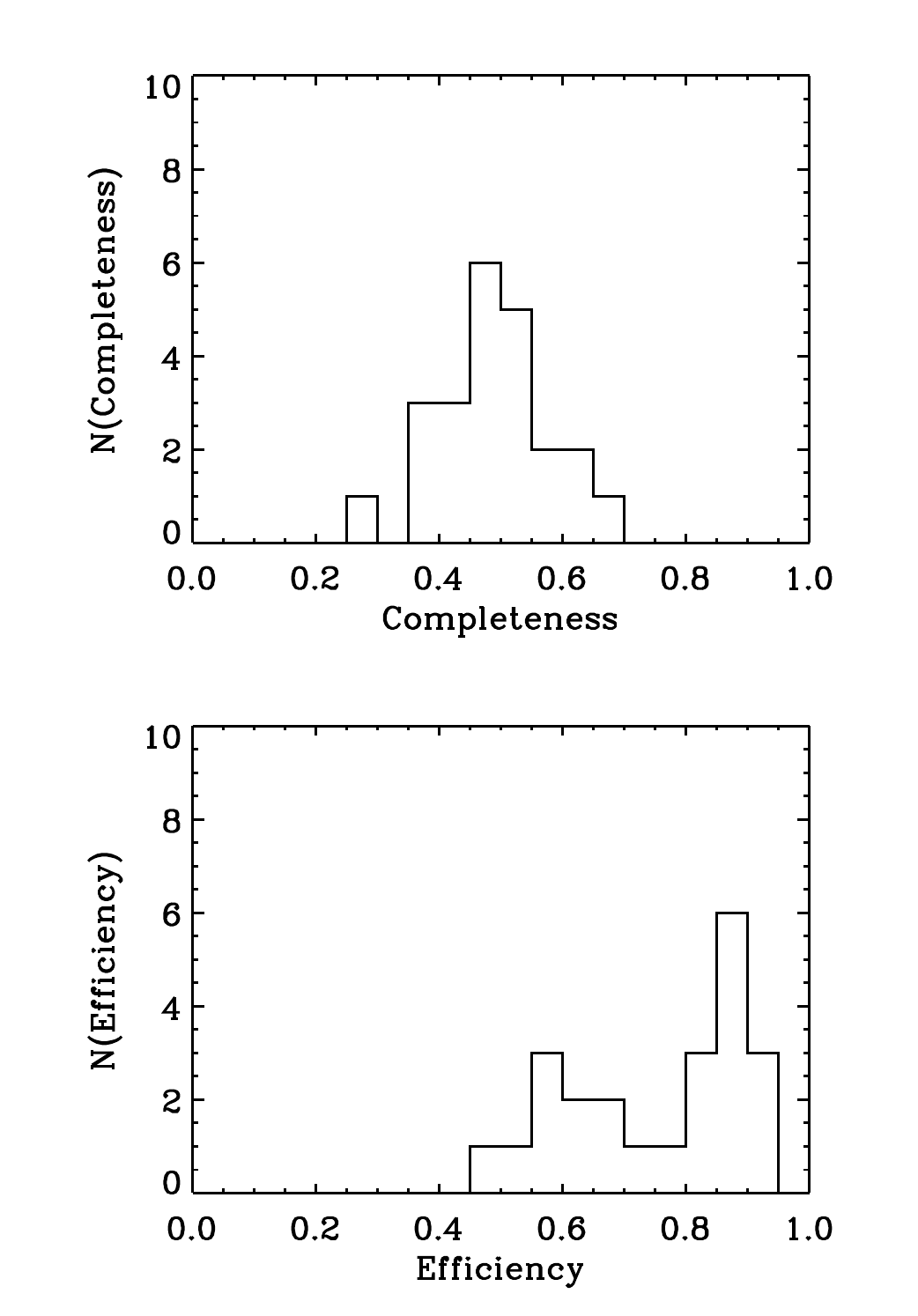}}
\end{center}
\caption{ The distribution of overall completeness, the fraction of
targeted galaxies to all galaxies (upper panel), and overall
efficiency, the fraction of spectra that yielded a redshift
determination (lower panel) per field. The overall completeness is
defined for galaxies brighter than $I=20$.}
\label{fig:avrg_compl}
\end{figure}

In Fig.~\ref{fig:avrg_compl} we show the distributions of the overall
completeness and overall efficiency. We find that in general the
overall completeness is $\sim50\%$. For one field (EISJ2246-4010B,
\#22) only one mask was observed yielding a low completeness of only
27\%. The distribution of overall efficiencies show two peaks,
corresponding to fields that were observed under good conditions and
fields observed in less than optimal conditions due to clouds
or the presence of the moon. The two peaks are centred around
$\sim60\%$ and $\sim85\%$.

\begin{figure}
\begin{center}
\resizebox{0.75\columnwidth}{!}{\includegraphics{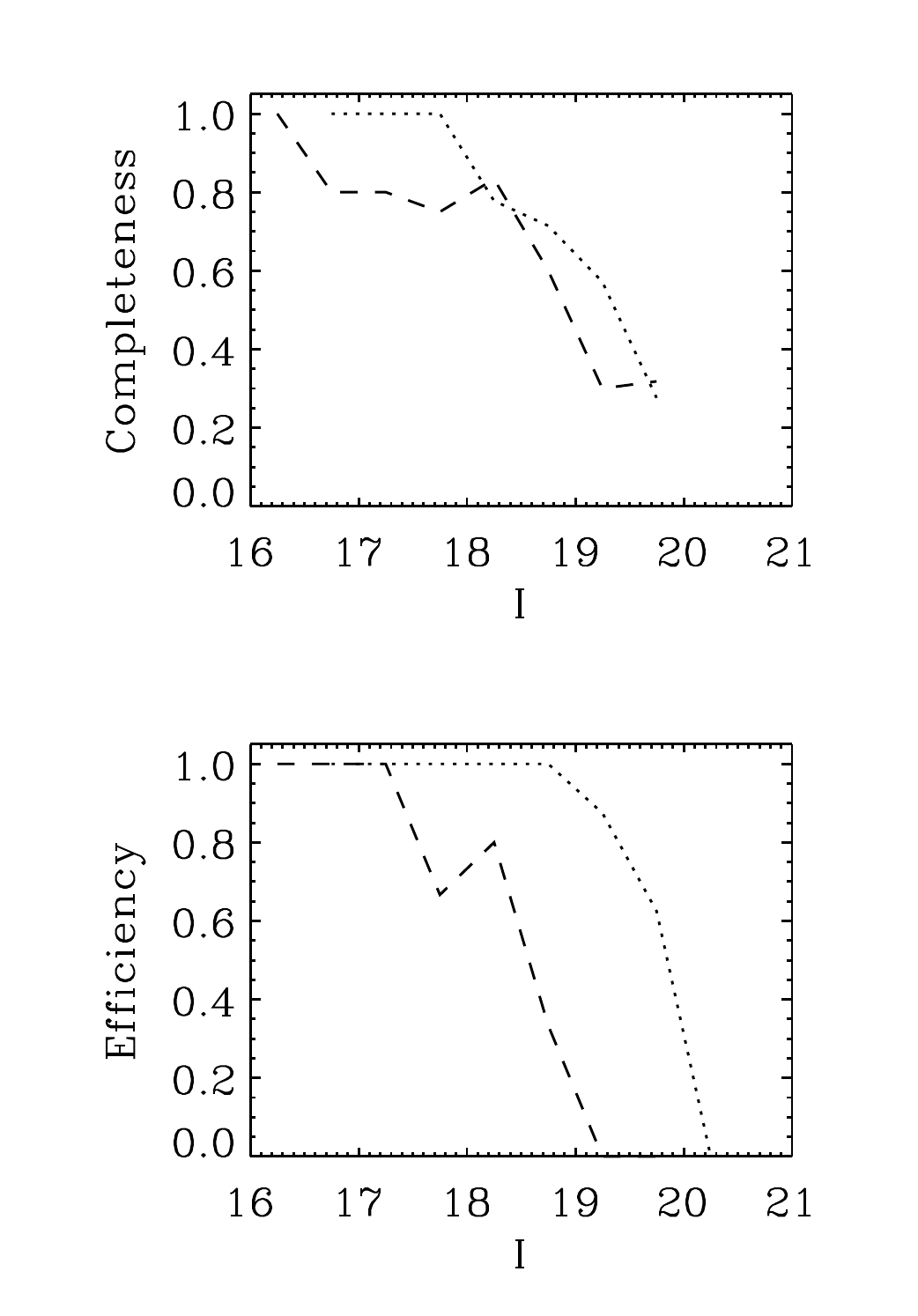}}
\end{center}
\caption{ The completeness (upper panel) and efficiency (lower panel)
as function of magnitude as found for the clusters EISJ0049-2920 (\#5,
dotted line), a typical case of high efficiency, and EISJ2236-4026
(\#7, dashed line), a typical case of low efficiency. }
\label{fig:completeness}
\end{figure}

In Fig.~\ref{fig:completeness} we show two examples of
completeness (upper panel) and efficiency (lower panel) as a function
of magnitude. The two examples are typical for systems with low
(EISJ2236-4026, \#7, dashed line) and high efficiency (EISJ0049-2920,
\#5, dotted line), respectively. It can be seen, from the upper panel
of the figure, that the completeness is very high at the brightest
magnitudes but decreases to $\sim25\%$ at about $I=19.75$.  The
efficiency in measuring redshifts, shown in the lower panel, is quite
high, reaching $\sim50\%$ at $I\sim20.0$ for the systems observed in
good conditions.  Unfortunately, the observing conditions were not
homogeneous and for some systems the 50\% efficiency is reached at
much brighter magnitudes ($I\sim18.5$). From this it is clear that for
some systems we are not able to obtain redshifts for galaxies with
luminosities close to $L^*$, corresponding to the magnitude range
18.5-19.3 in the $I-$band. Therefore, for these systems the
confirmation is questionable. 

The variation of the effective magnitude limit of our
observations from field-to-field can introduce a strong bias in our
results. This can happen in two ways: one by severely undersampling
higher $z$ clusters and not being able to confirm them, biasing the
sample to lower redshifts; the other, similar to the first but leading
to the misidentification of the more distant clusters associating them
to foreground structures, and biasing our conclusions regarding the
reliability of the redshift and richness estimates. We use the ratio
of faint to bright galaxies around $M^*$ as a measure of how we probe
the luminosity function at the redshift estimated for the cluster
candidate. The bright galaxies are defined to be in the magnitude
interval $[I^*-1:I^*]$ and faint in $[I^*:I^*+1]$.

%__________________________________________________________________

\section{Spectroscopic results}
\label{sec:results}
%__________________________________________________________________

\begin{table}
\begin{center}
\caption{Redshifts measured for the individual galaxies. Here we give
the first five lines of the table. The entire table is available at CDS.}
\label{tab:galredshifts}
\begin{tabular}{r c c c c}
\hline\hline
 & $\alpha$ (J2000) & $\delta$ (J2000) & I & $z$ \\
\hline
    1 & 11.126234 & -29.822523 & 17.89 & 0.2576\\
    2 & 11.136551 & -29.839476 & 19.90 & 0.5082:\\
    3 & 11.153917 & -29.852741 & 18.31 & 0.2572\\
    4 & 11.156263 & -29.831465 & 18.80 & 0.1738\\
    5 & 11.187512 & -29.898016 & 18.97 & 0.6706\\
\hline
\end{tabular}
\end{center}
\end{table}

We measured a total of 747 redshifts for galaxies in the 23 EIS
cluster fields considered, with the number of derived redshifts
ranging between 16 and 45 per field.  Table~\ref{tab:galredshifts},
available at the CDS, lists in Col.~1 an identifier for each galaxy;
in Cols.~2 and 3 the right ascension and declination in J2000 for the
galaxy; Col.~4 the $I-$magnitude from the EIS object catalogues
\citep{nonino99,prandoni99} and in Col.~5 the measured redshift. A
colon (``:'') indicates that the redshift was assigned exclusively
based on the cross-correlation without any recognisable spectroscopic
features. An ``e'' indicates that emission lines are present in the
spectrum.

In an attempt to increase the number of redshifts in each field
we searched the NASA Extragalactic Database (NED). For several fields
we found no new redshifts. In others we found no more than five but,
in general, these either were in common with our observations or lay
at the outskirts of the field having no bearing on the confirmation of
the candidate clusters. In only two cases we find one redshift for a
galaxy in the central part of the fields. These cases are
EISJ2246-4012B (\#22) and EISJ2248-4015 (\#23). None of these had an
impact on the conclusions drawn based on our own observations, and are
therefore not included in the analysis below.

\begin{figure*}
\begin{center}
\resizebox{0.22\textwidth}{!}{\includegraphics[bb=0 0 283 226,clip]{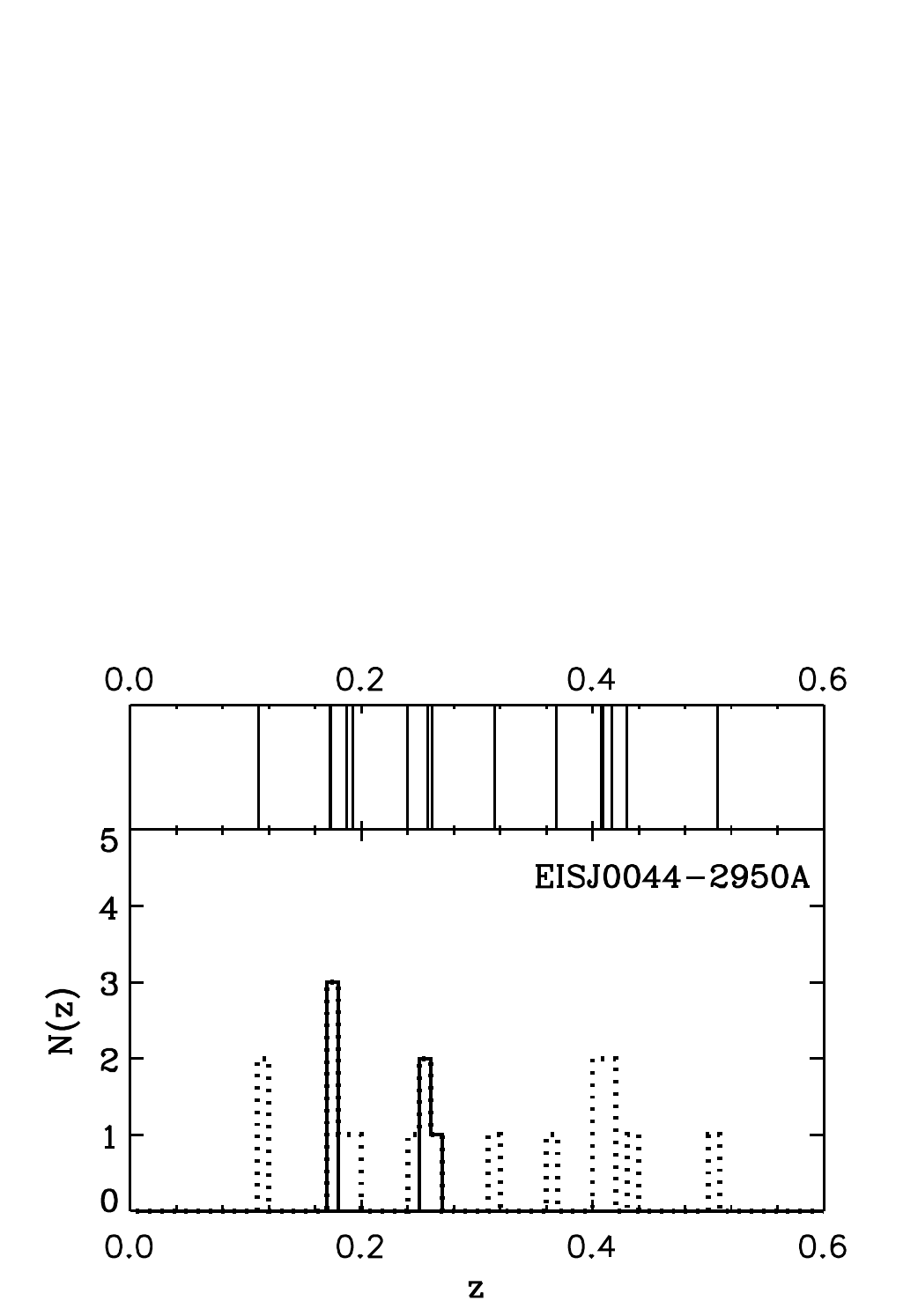}}
\resizebox{0.22\textwidth}{!}{\includegraphics[bb=0 0 283 226,clip]{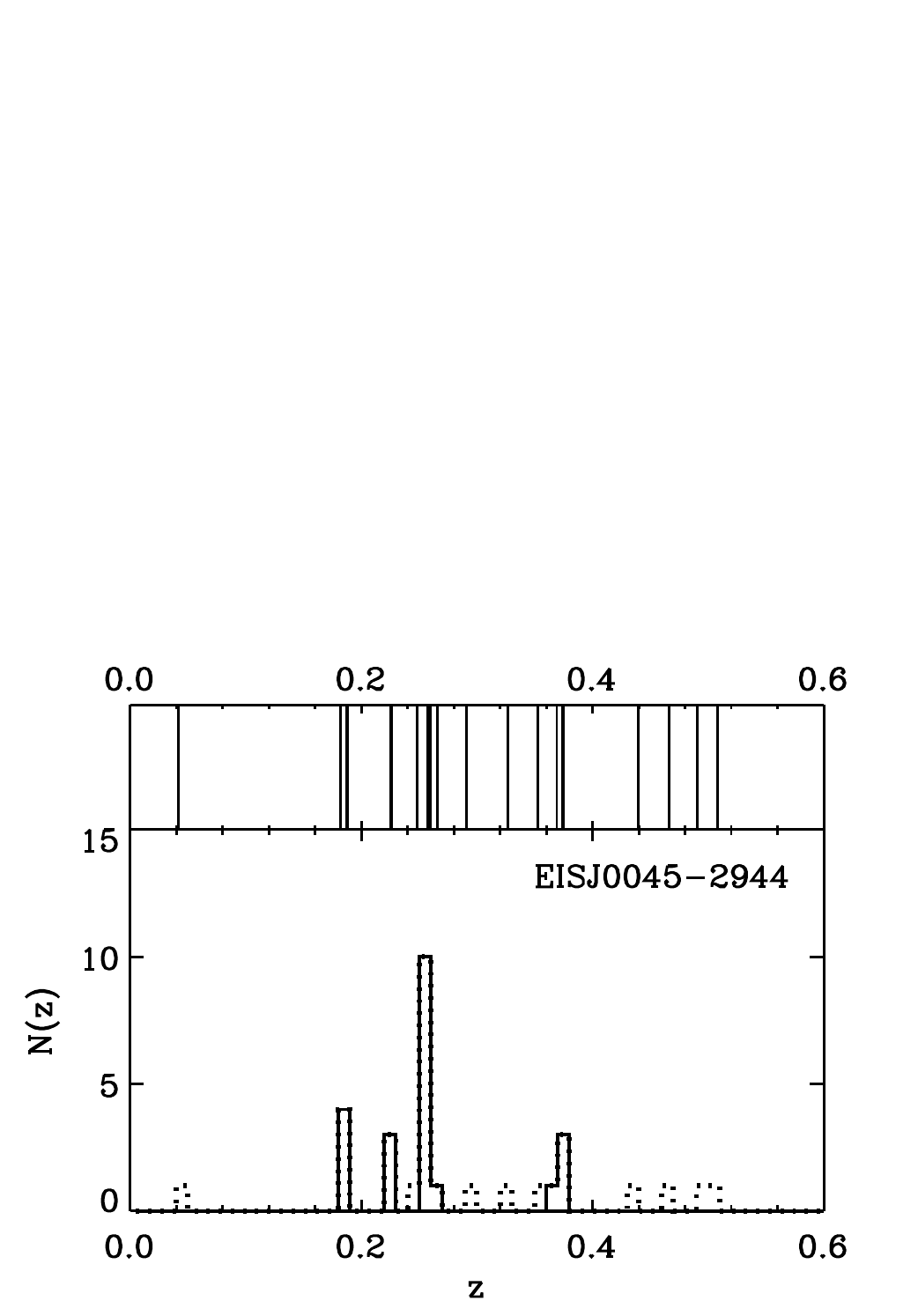}}
\resizebox{0.22\textwidth}{!}{\includegraphics[bb=0 0 283 226,clip]{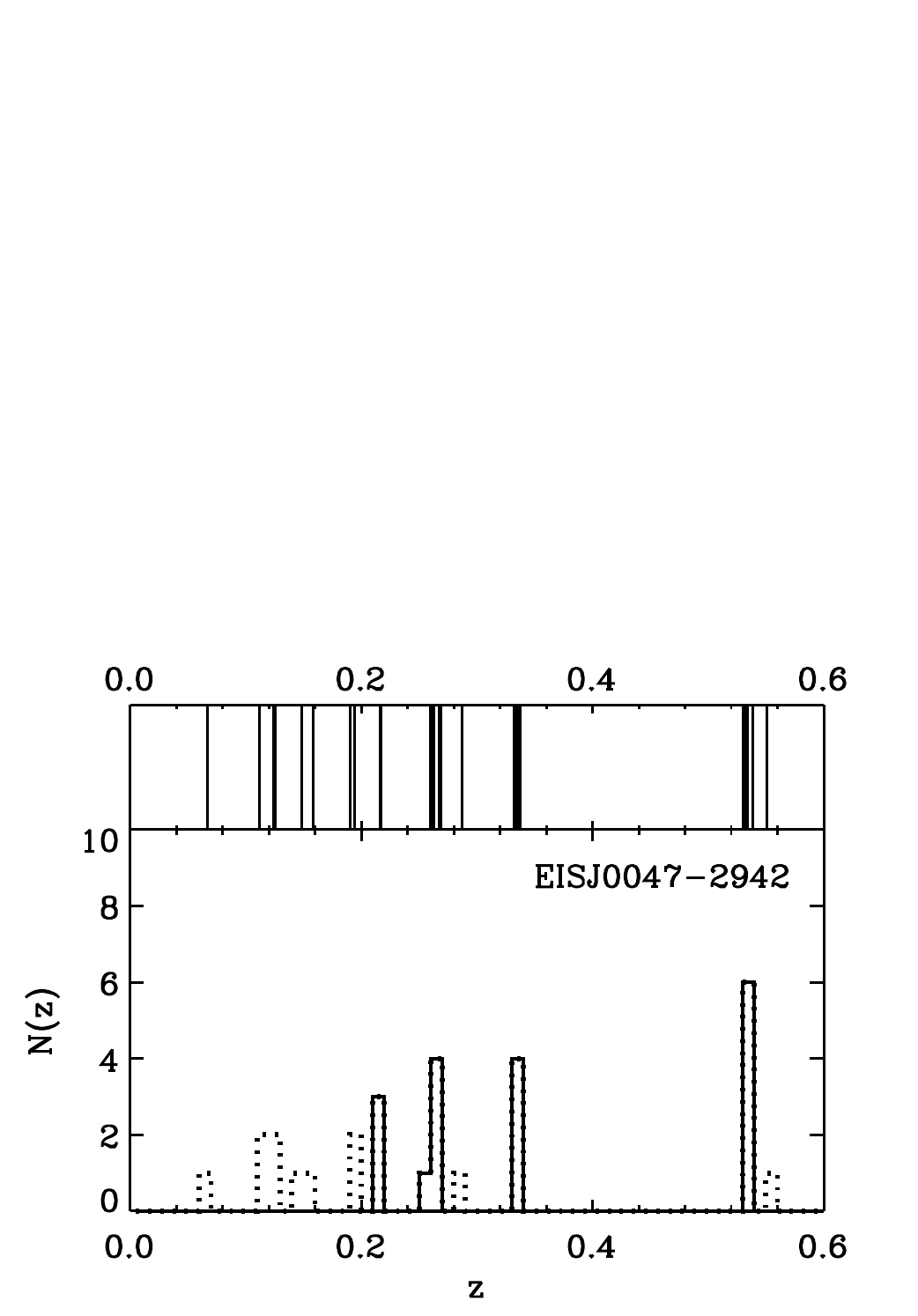}}
\resizebox{0.22\textwidth}{!}{\includegraphics[bb=0 0 283 226,clip]{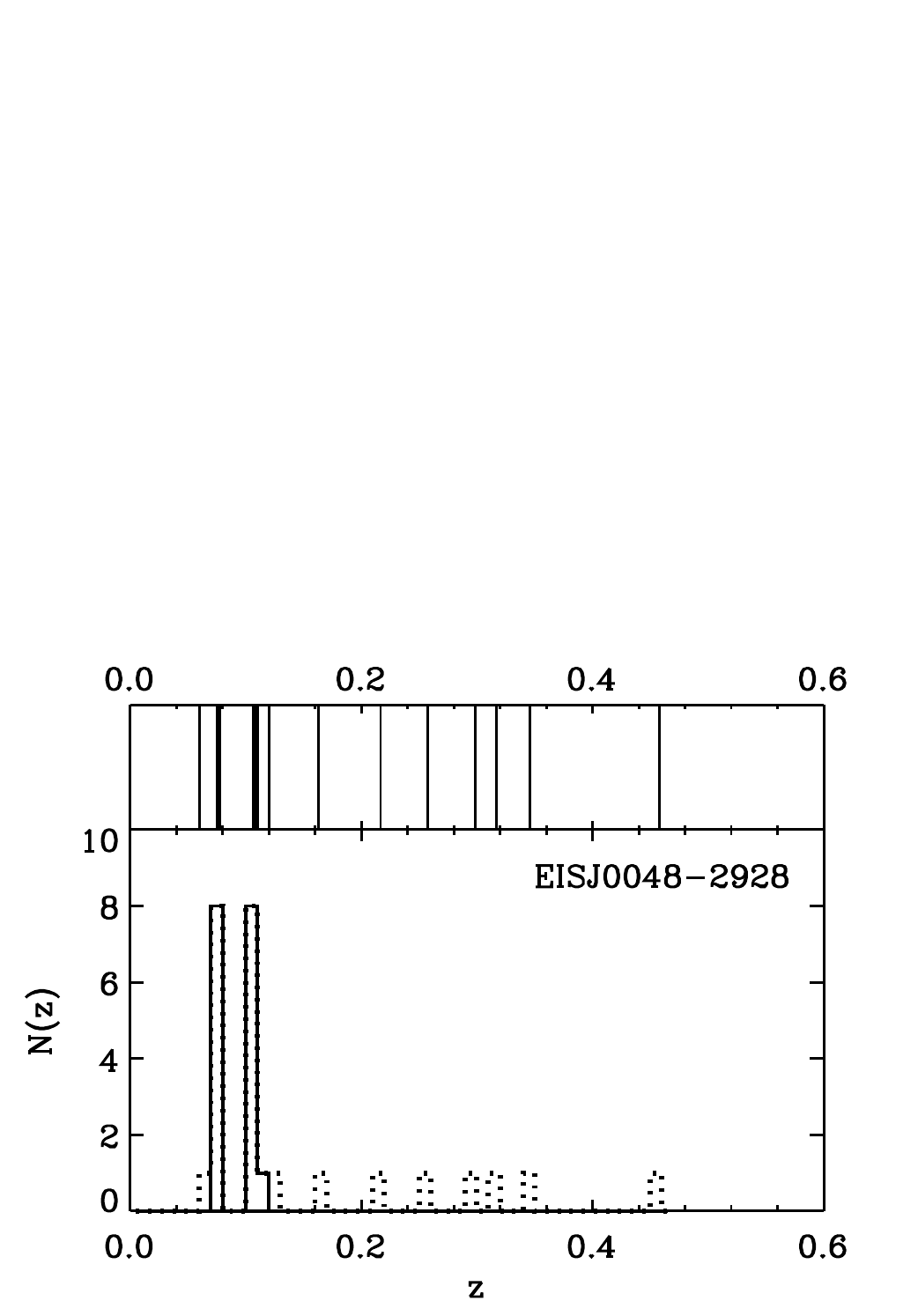}}
\resizebox{0.22\textwidth}{!}{\includegraphics[bb=0 0 283 226,clip]{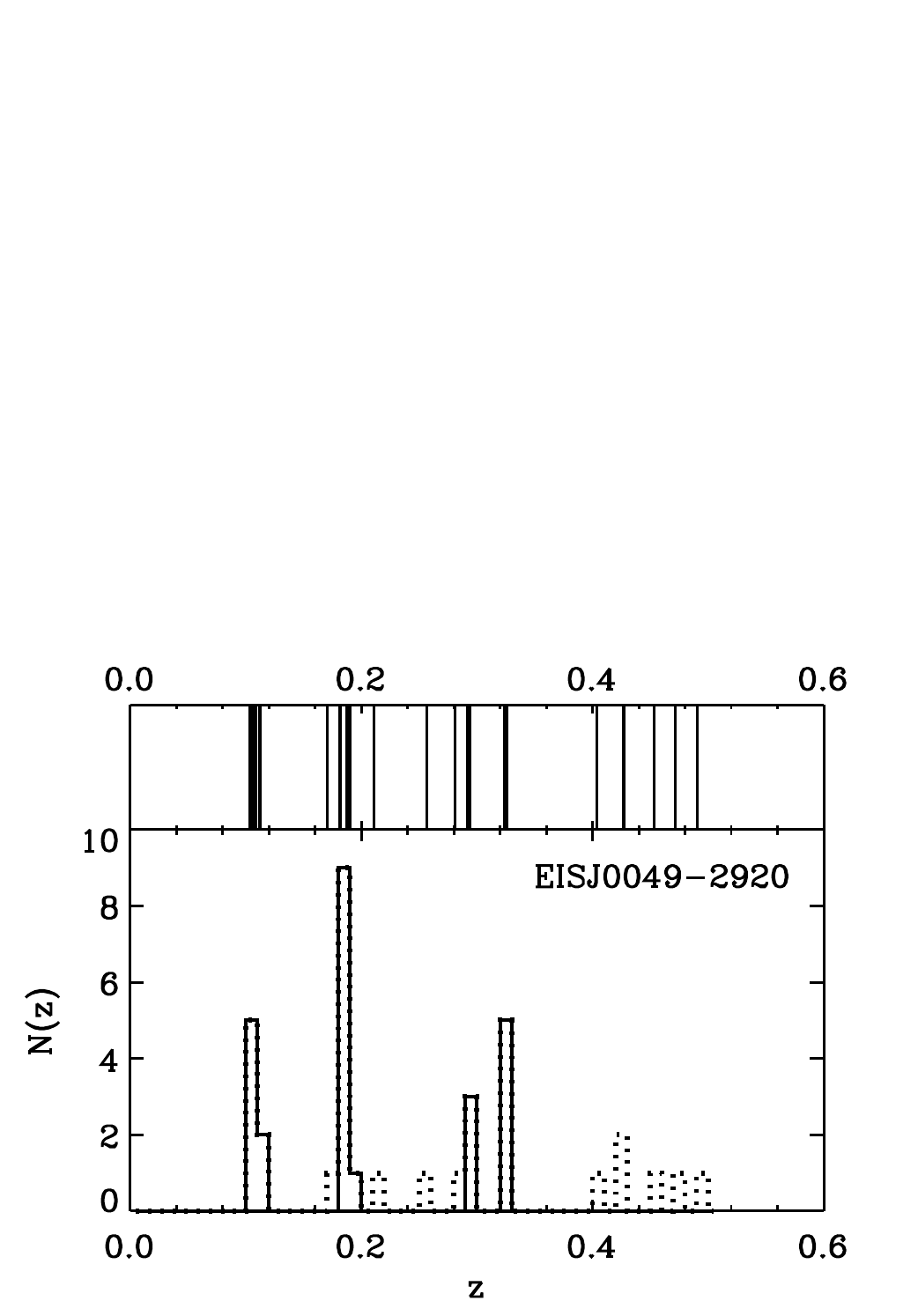}}
\resizebox{0.22\textwidth}{!}{\includegraphics[bb=0 0 283 226,clip]{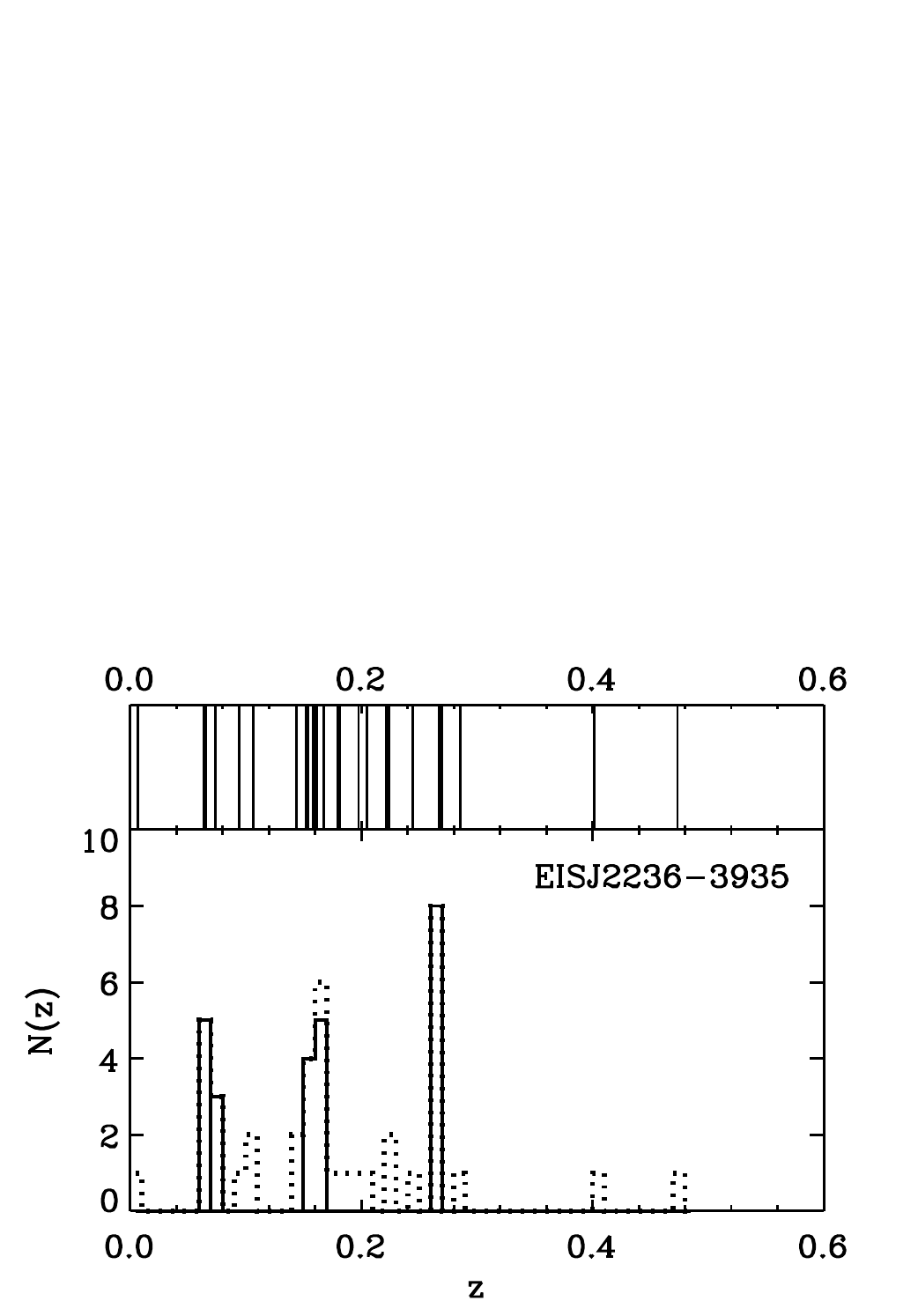}}
\resizebox{0.22\textwidth}{!}{\includegraphics[bb=0 0 283 226,clip]{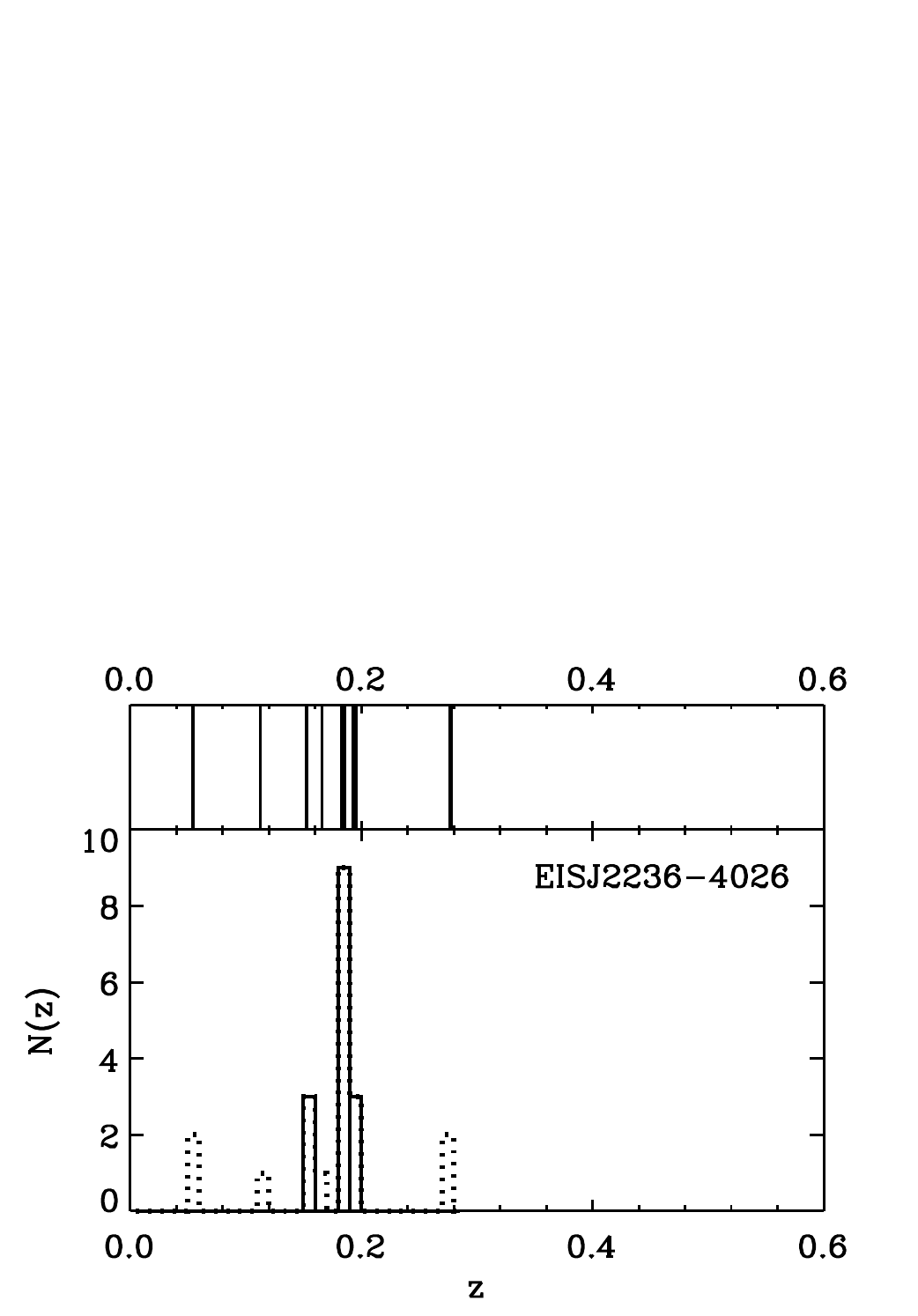}}
\resizebox{0.22\textwidth}{!}{\includegraphics[bb=0 0 283 226,clip]{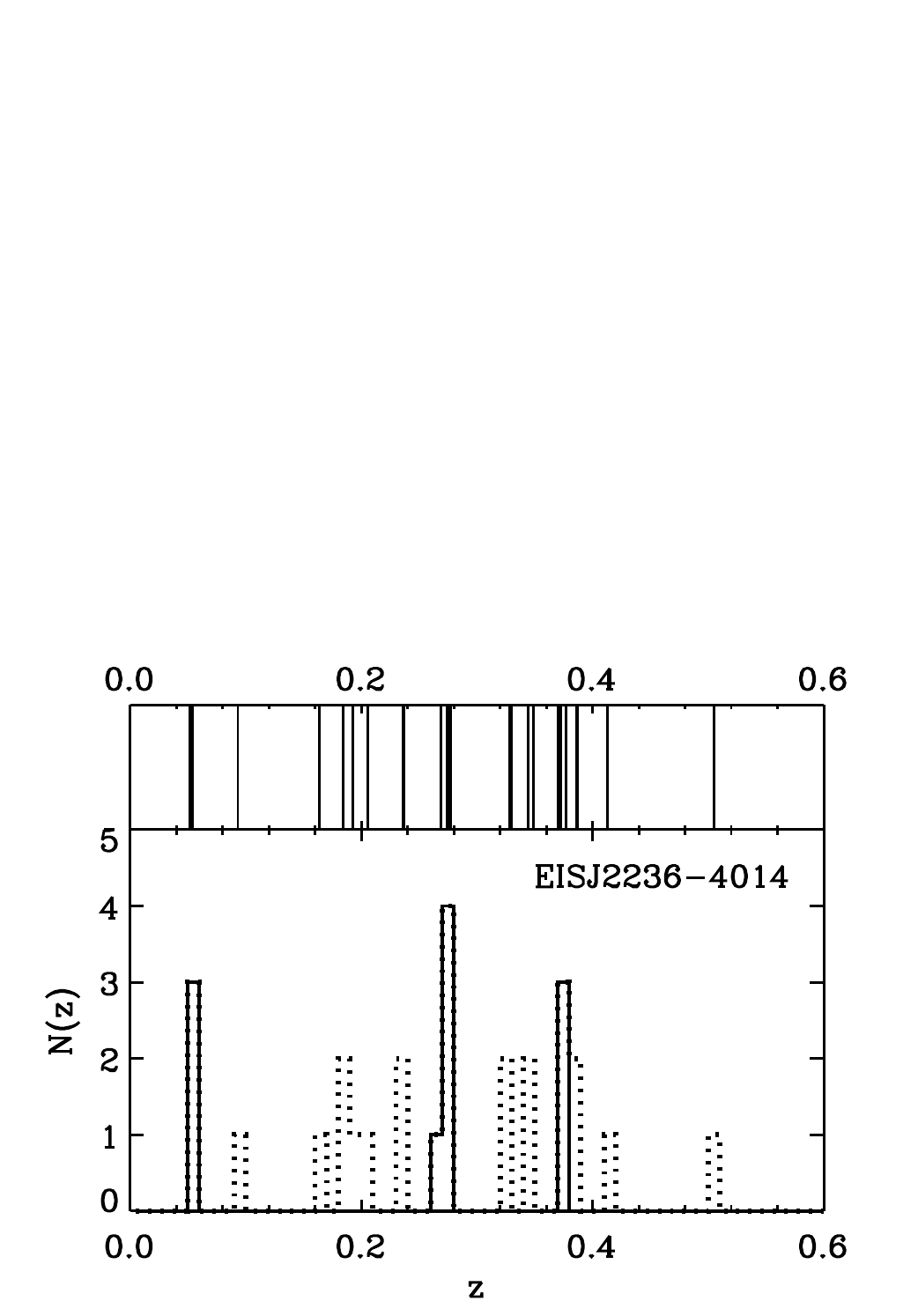}}
\resizebox{0.22\textwidth}{!}{\includegraphics[bb=0 0 283 226,clip]{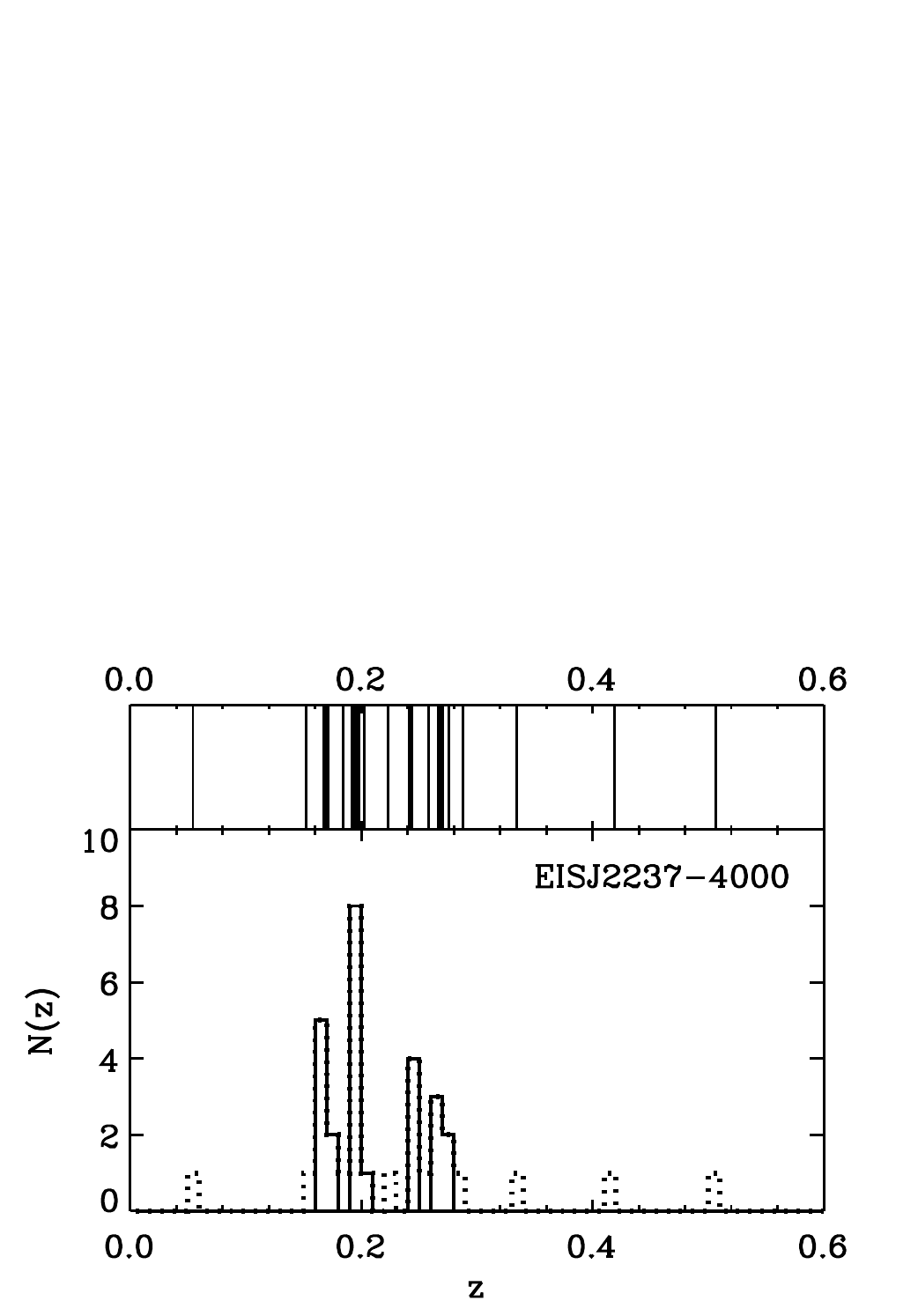}}
\resizebox{0.22\textwidth}{!}{\includegraphics[bb=0 0 283 226,clip]{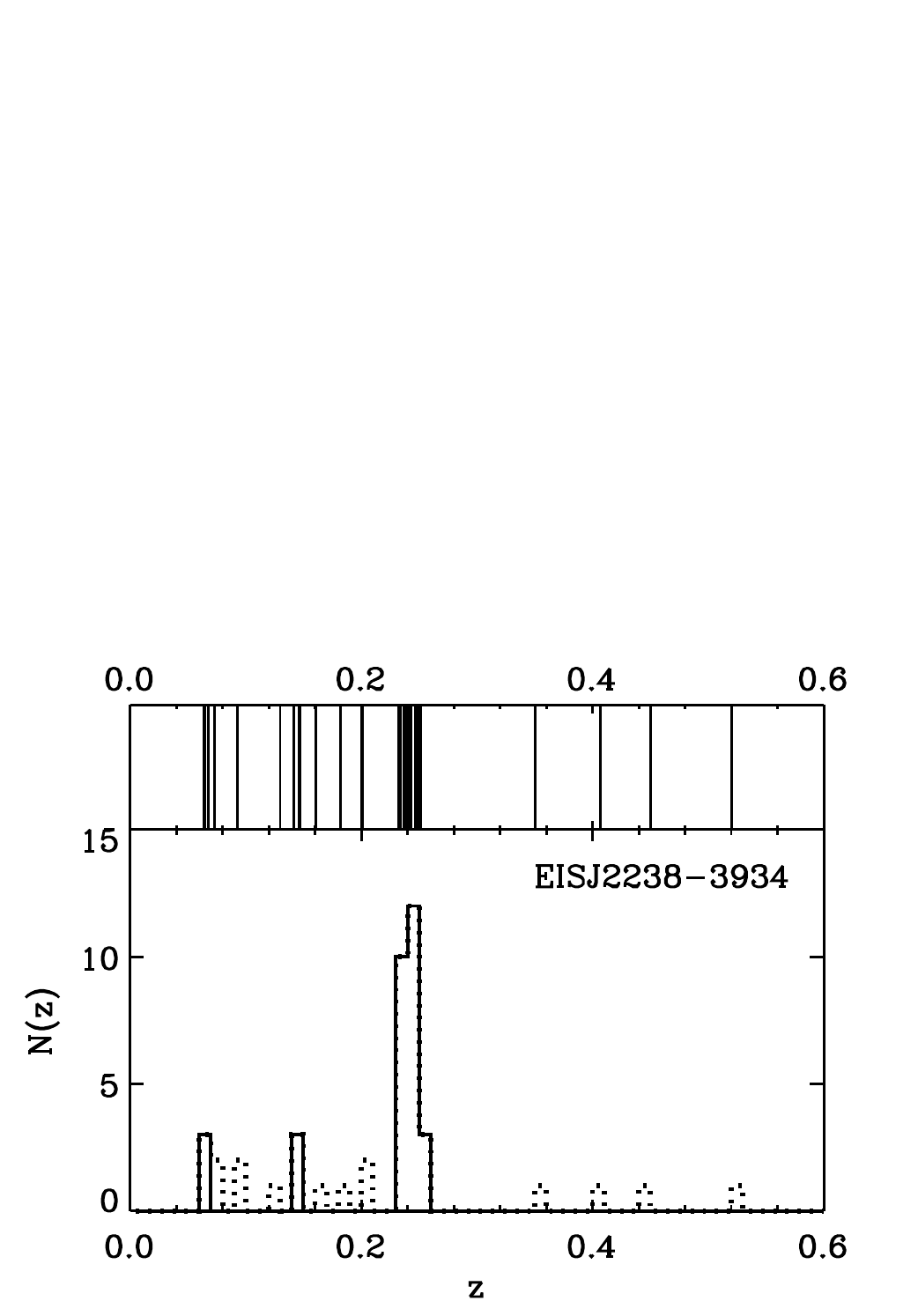}}
\resizebox{0.22\textwidth}{!}{\includegraphics[bb=0 0 283 226,clip]{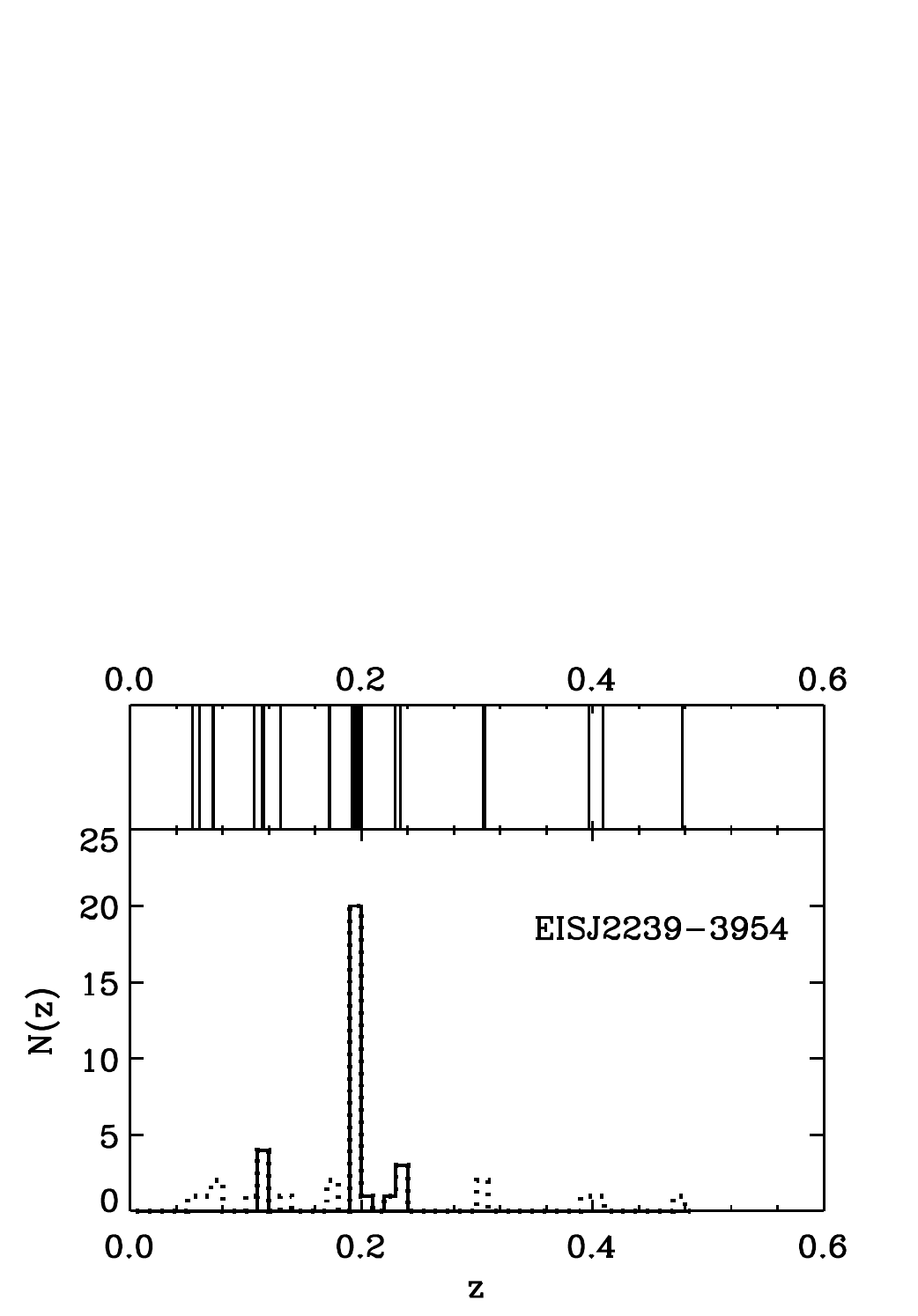}}
\resizebox{0.22\textwidth}{!}{\includegraphics[bb=0 0 283 226,clip]{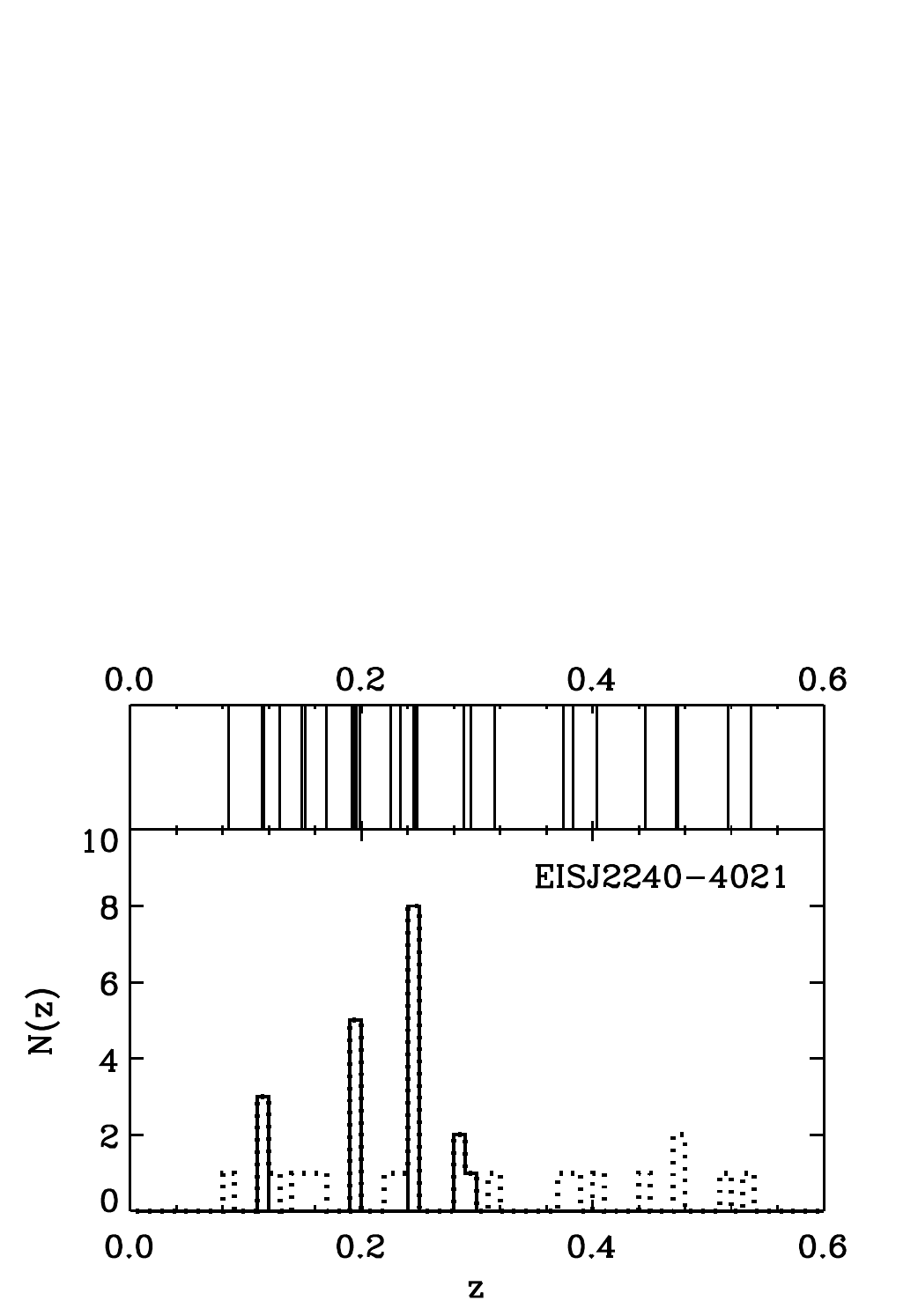}}
\resizebox{0.22\textwidth}{!}{\includegraphics[bb=0 0 283 226,clip]{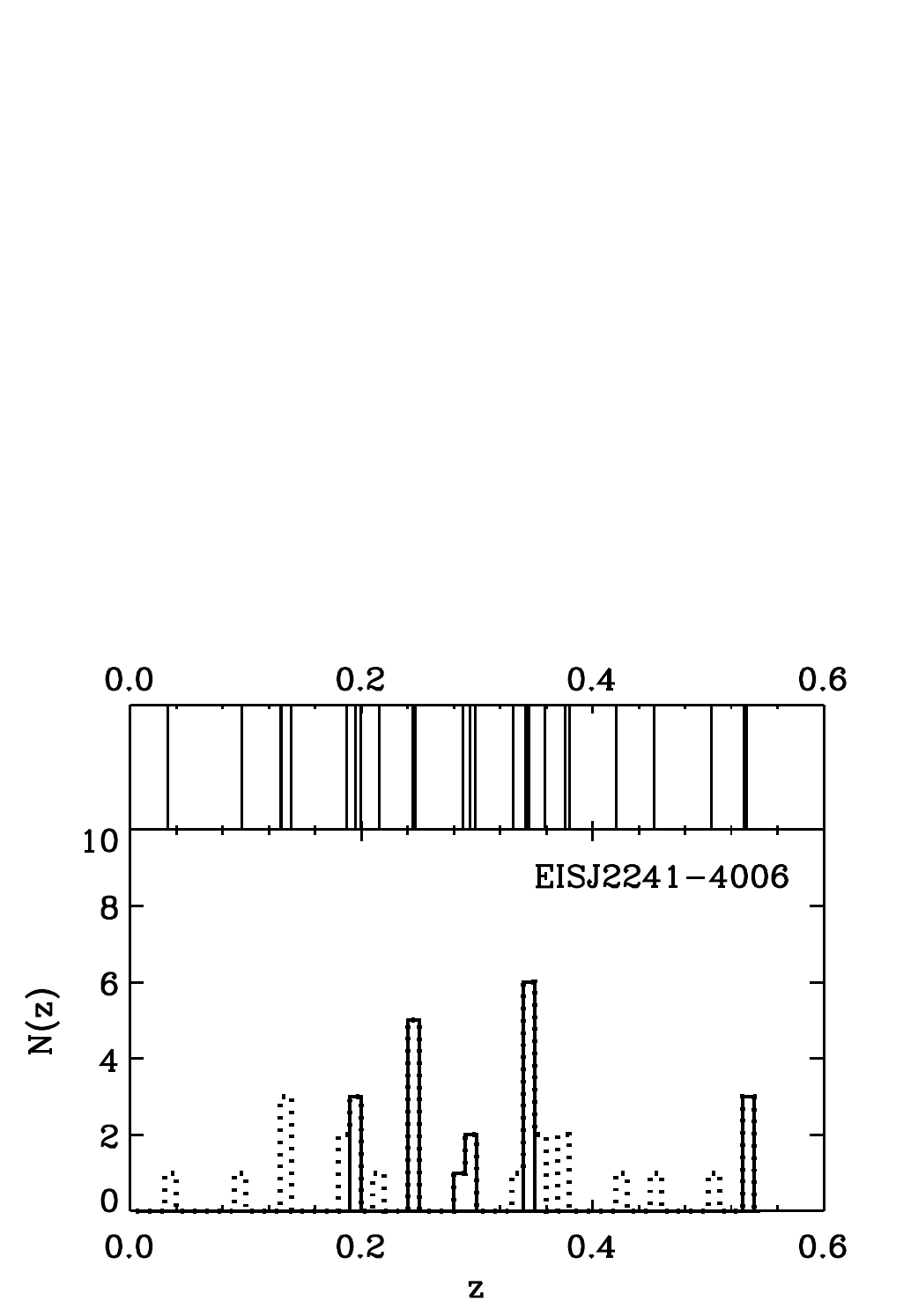}}
\resizebox{0.22\textwidth}{!}{\includegraphics[bb=0 0 283 226,clip]{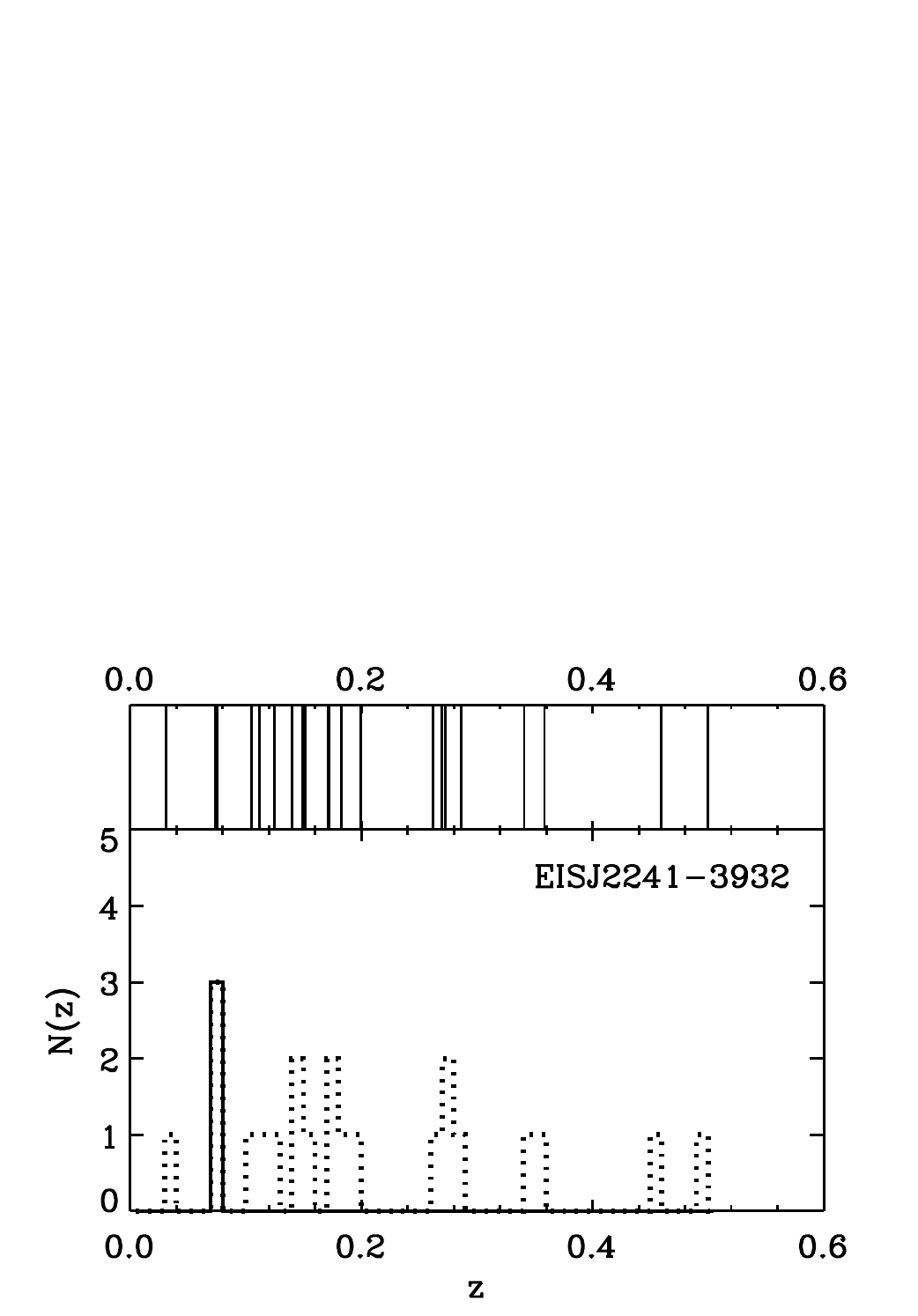}}
\resizebox{0.22\textwidth}{!}{\includegraphics[bb=0 0 283 226,clip]{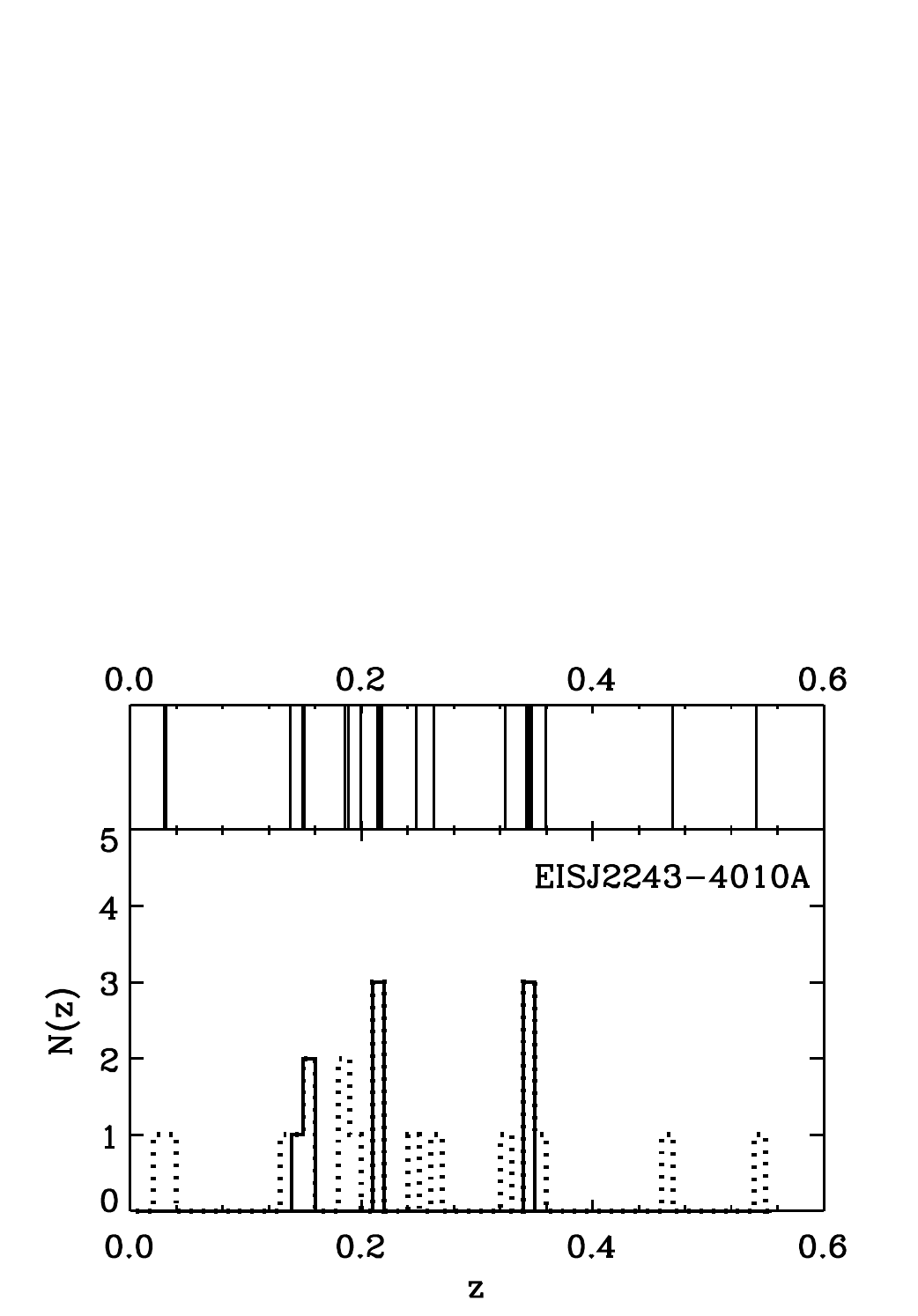}}
\resizebox{0.22\textwidth}{!}{\includegraphics[bb=0 0 283 226,clip]{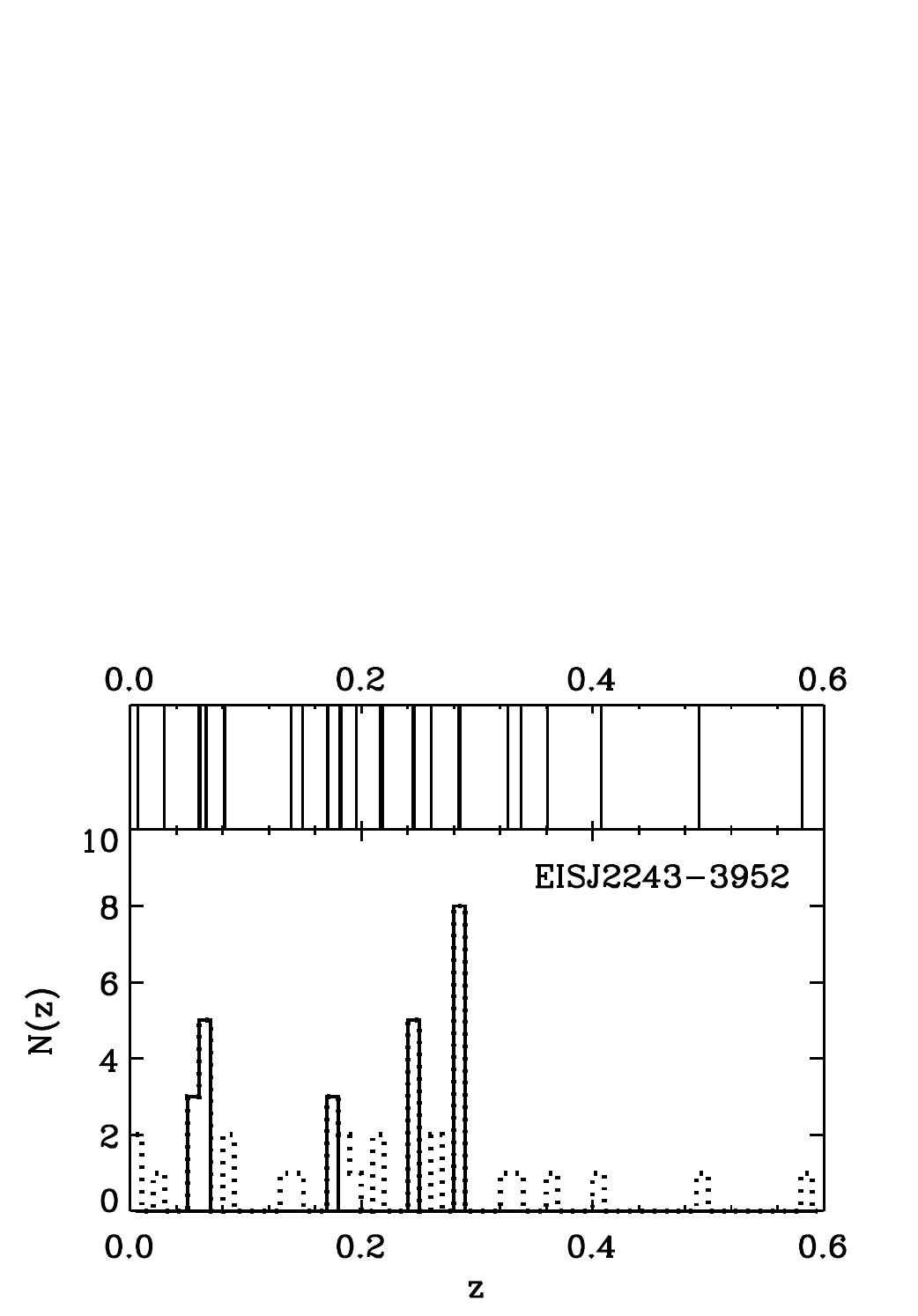}}
\resizebox{0.22\textwidth}{!}{\includegraphics[bb=0 0 283 226,clip]{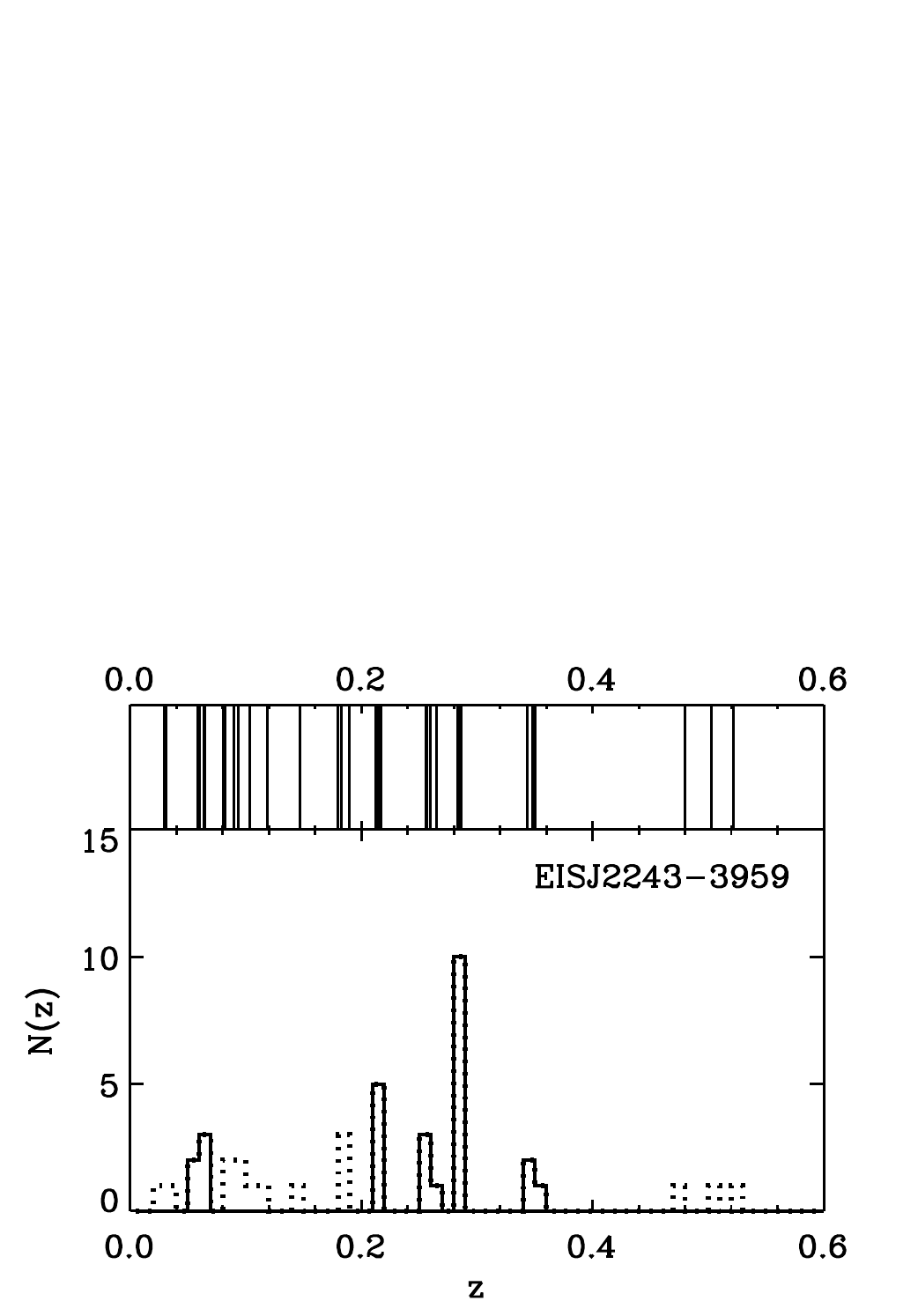}}
\resizebox{0.22\textwidth}{!}{\includegraphics[bb=0 0 283 226,clip]{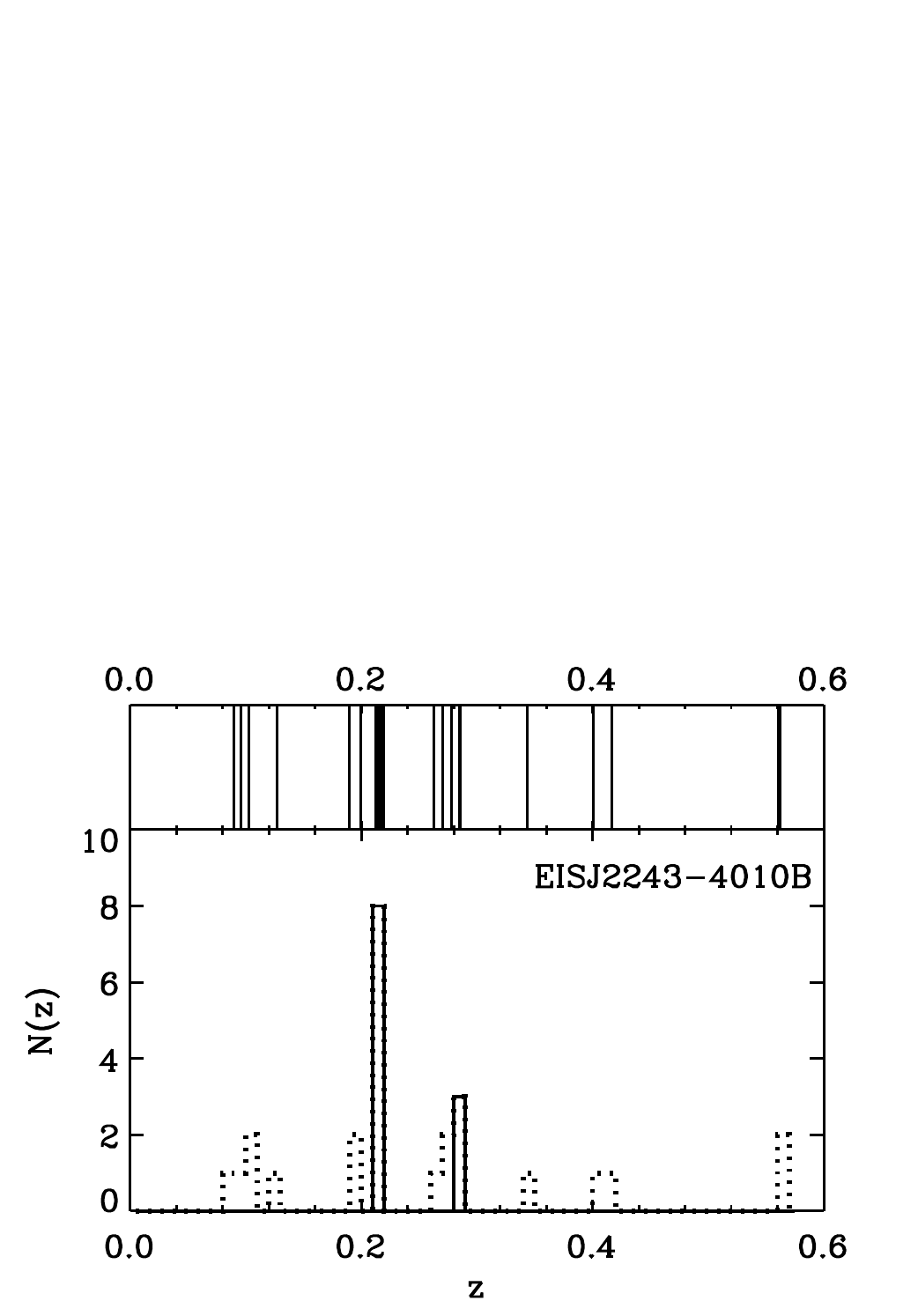}}
\resizebox{0.22\textwidth}{!}{\includegraphics[bb=0 0 283 226,clip]{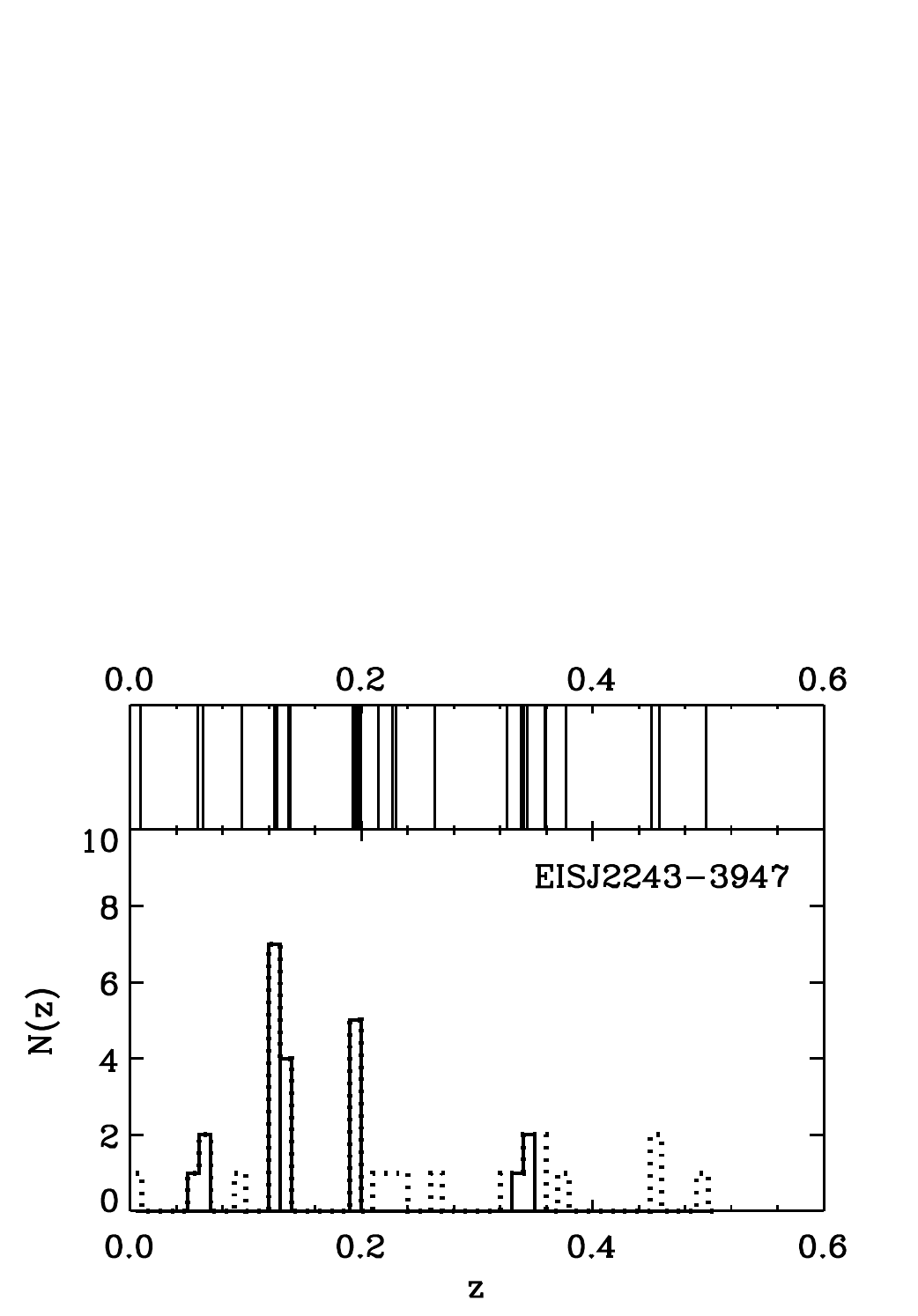}}
\resizebox{0.22\textwidth}{!}{\includegraphics[bb=0 0 283 226,clip]{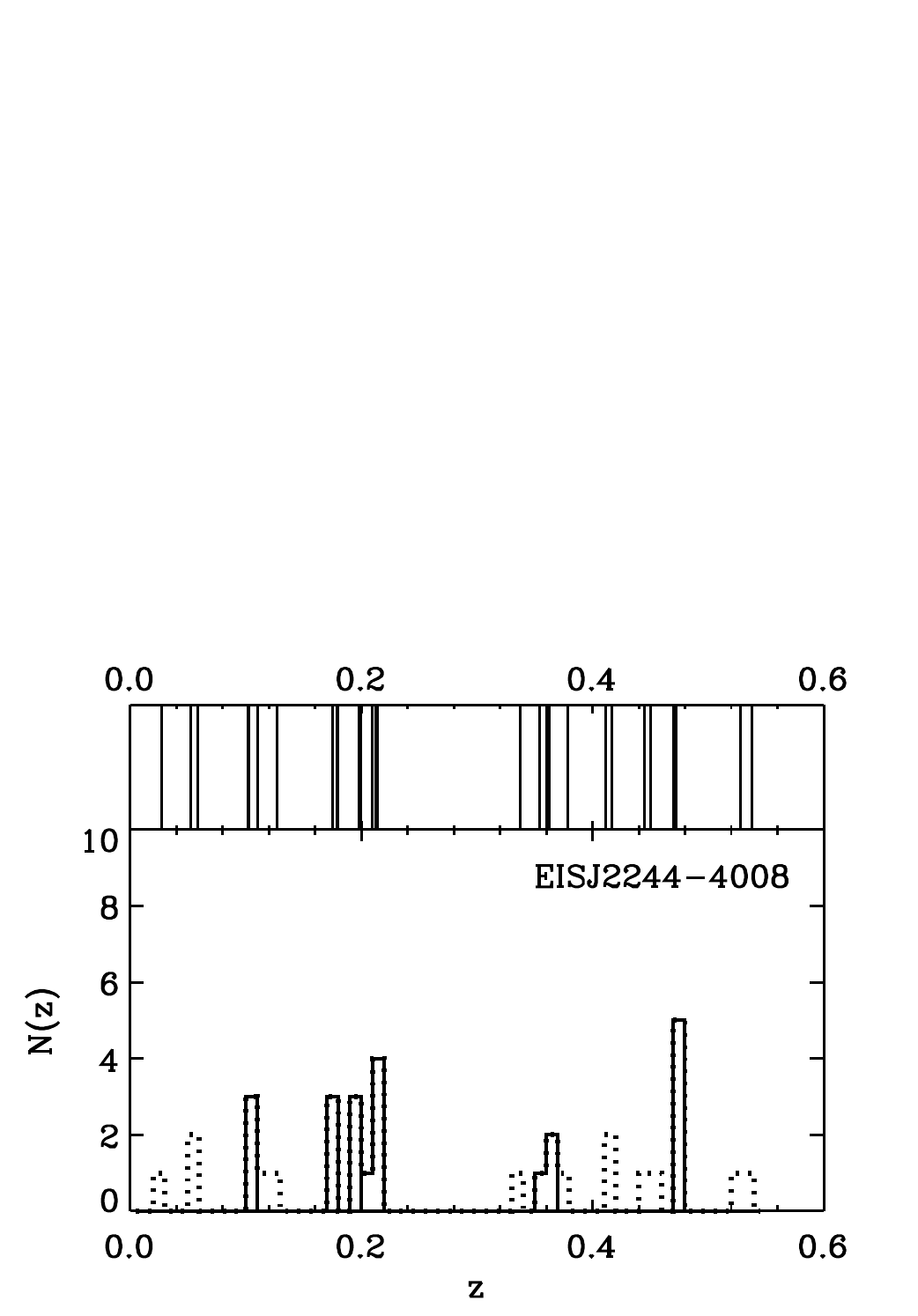}}
\resizebox{0.22\textwidth}{!}{\includegraphics[bb=0 0 283 226,clip]{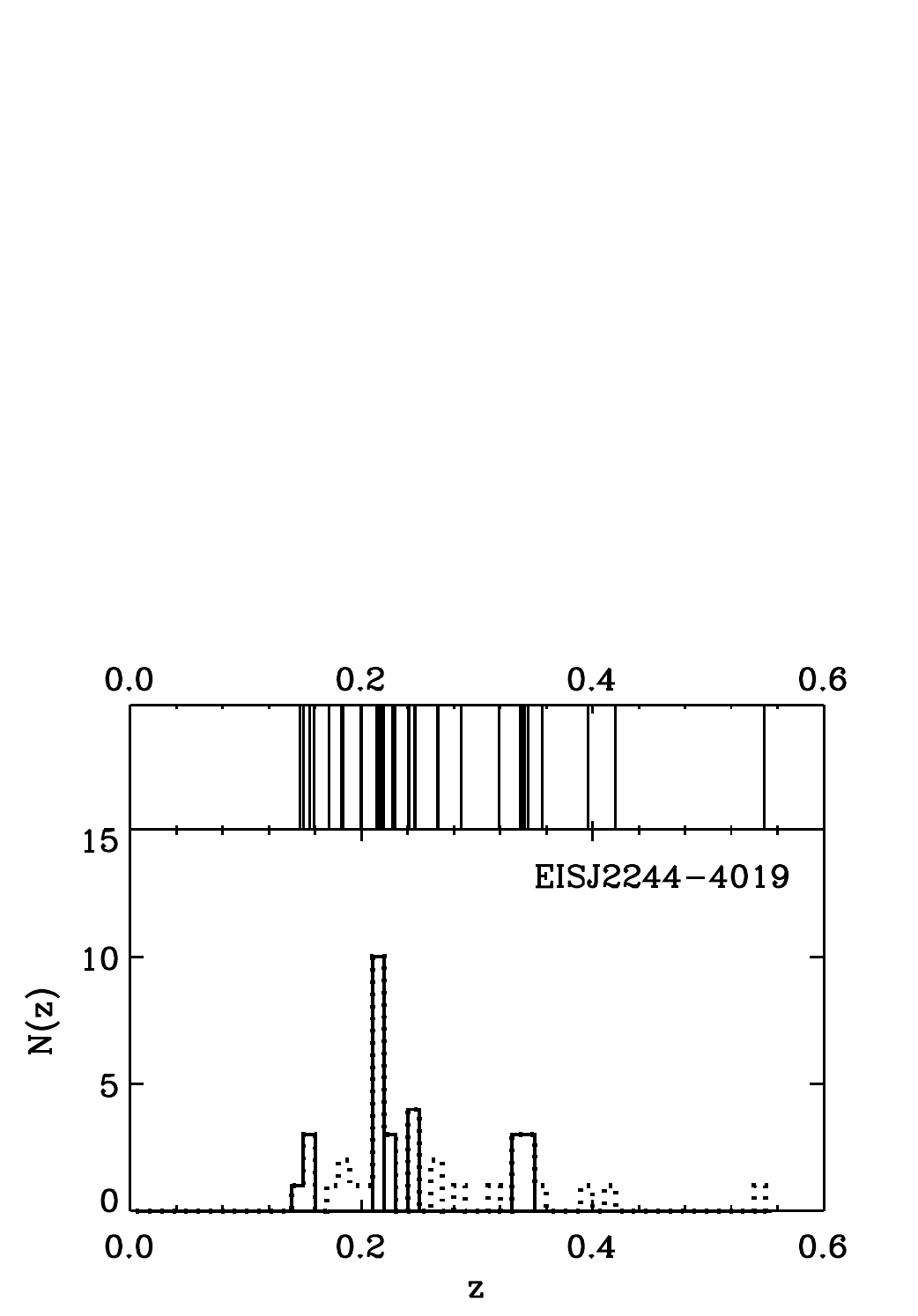}}
\resizebox{0.22\textwidth}{!}{\includegraphics[bb=0 0 283 226,clip]{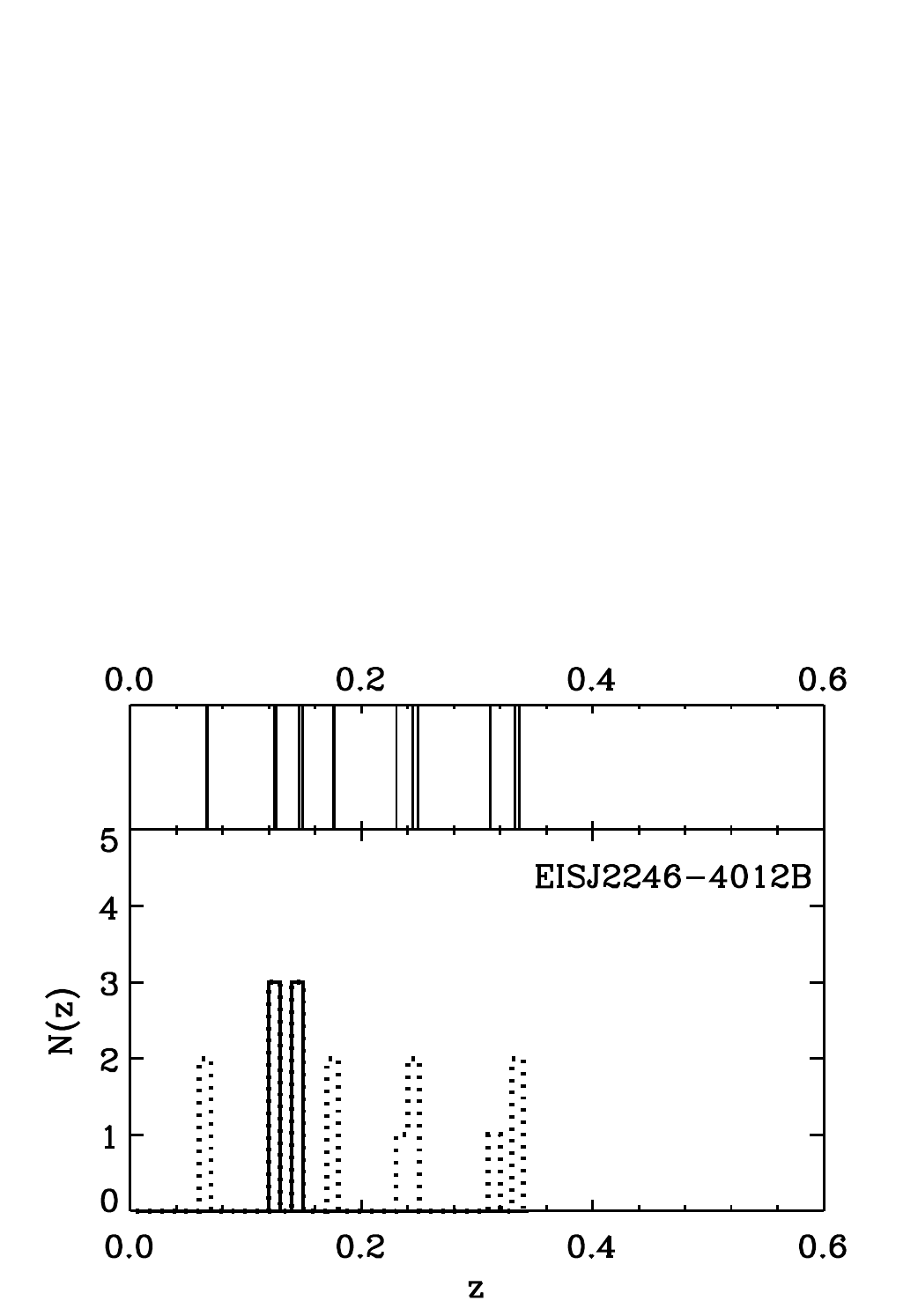}}
\resizebox{0.22\textwidth}{!}{\includegraphics[bb=0 0 283 226,clip]{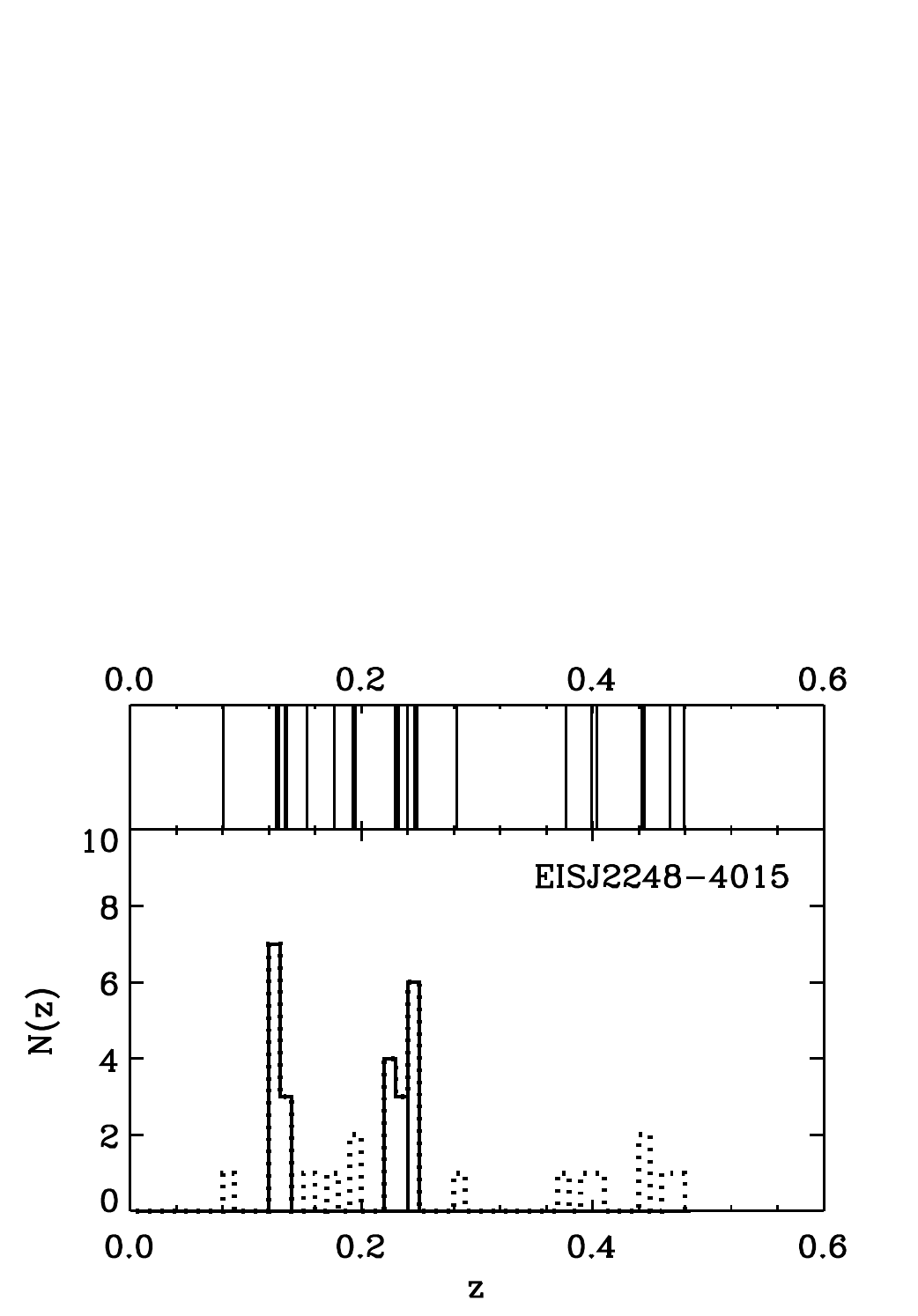}}
\caption{Redshift distributions for the observed cluster fields as
indicated in each panel. Note that the scale of the y-axis differs
between the panels. The upper panels show bar diagrams of the measured
redshifts, while the lower panels give the corresponding histograms of
the redshift distributions (dashed line). The solid lines mark the
detected groups.}
\label{fig:redshift_dists}
\end{center}
\end{figure*}

\subsection{Spectroscopic confirmation of systems}

\begin{table*}
\caption{Identified groups with a significance of at least 99\%. The
ones in bold face are the ones we interpret as related to the cluster
detection.}
\label{tab:EISgroups}
\begin{center}
\begin{minipage}{0.85\textwidth}
\begin{tabular}{llcccrrrr}
\hline\hline
ID & Cluster Field & Members & $\alpha$ (J2000) & $\delta$ (J2000) & z & $\sigma_v \mathrm{[km/s]}$\footnote{$\sigma_v=0$ reflects a measured velocity dispersion below the measurement error.} & $\sigma_1$ [\%] & Dist. [']\\
\hline
1 & EISJ0044-2950A\footnote{No significant groups were detected in this field.} & $-$ & $-$ & $-$ & $-$ & $-$ & $-$  & $-$ \\
\hline
2 & EISJ0045-2944 &  11 &   00 45 11.8  & -29 45 13.1 & 0.259 &   424 &   99.9 & 2.4\\
\hline
{\bf 3} & {\bf EISJ0047-2942} &   {\bf 6} & {\bf  00 47 17.2}  & {\bf -29 42 41.9} & {\bf 0.534} & {\bf   418} & {\bf  99.9} & {\bf 1.3}\\
\hline
4 & EISJ0048-2928\footnote{The candidate in this field could not be confirmed by the current data.} &   8 &  00 48 19.7  & -29 28 20.9 & 0.077 &   111 &  99.9 & 1.4\\
4 & EISJ0048-2928 &   9 &   00 48 40.3  &  -29 29 06.2 & 0.108 &    289 &   99.9 & 2.0\\
\hline
5 & EISJ0049-2920 &   7 &  00 49 26.8  & -29 22 04.9 & 0.108 &   773 &  99.9 & 1.8\\
{\bf 5} & {\bf EISJ0049-2920} & {\bf  10} &  {\bf 00 49 32.3}  & {\bf -29 20 34.7} & {\bf 0.187} &  {\bf  781} & {\bf  99.9} & {\bf 0.2}\\
\hline
6 & EISJ2236-3935 &   5 &  22 36 20.3 & -39 35 30.6 & 0.065 &   161 &  99.9 & 3.3\\
6 & EISJ2236-3935 &   3 &  22 35 51.8 & -39 36 03.5 & 0.074 &     0 &  99.9 & 2.2\\
{\bf 6} & {\bf EISJ2236-3935} &  {\bf  9} & {\bf 22 36 07.7} & {\bf -39 36 08.7} & {\bf 0.158} &  {\bf  915} &  {\bf 99.9} & {\bf 1.1}\\
6 & EISJ2236-3935 &   8 &  22 35 56.2 & -39 35 08.0 & 0.269 &   125 &  99.9 & 1.4\\
\hline
{\bf 7} & {\bf EISJ2236-4026} & {\bf   9} &  {\bf 22 36 38.1} & {\bf -40 26 50.8}  & {\bf 0.184} &  {\bf  151} & {\bf 99.9} &{\bf 1.9}\\
\hline
8 & EISJ2236-4014$^c$ &   3 &  22 36 50.9 & -40 17 24.6 & 0.053 &   295 &  99.9 & 2.5\\
\hline
9 & EISJ2237-4000 &   7 &  22 37 19.8 & -39 59 38.5 & 0.169 &   239 &  99.9 & 1.7\\
{\bf 9} & {\bf EISJ2237-4000} &  {\bf  9} & {\bf  22 37 08.1} & {\bf -40 00 32.9} & {\bf 0.196} &  {\bf 604} &  {\bf 99.9} &{\bf 0.7}\\
\hline
10 & EISJ2238-3934 &   3 &  22 38 16.8 & -39 35 57.0 & 0.065 &   489 &  99.9 & 2.8\\
{\bf 10} & {\bf EISJ2238-3934} &  {\bf 25} & {\bf  22 38 01.6} & {\bf -39 34 49.7} & {\bf 0.243} &  {\bf 1103} & {\bf 99.9} & {\bf 0.3}\bf \\
\hline
11 & EISJ2239-3954 &   4 &  22 39 22.5 & -39 57 38.2 & 0.115 &     0 &  99.2 & 3.2\\
{\bf 11} & {\bf EISJ2239-3954} &  {\bf 21} & {\bf 22 39 17.8} & {\bf -39 55 32.8} & {\bf 0.195} &  {\bf  420} &  {\bf 99.9}& {\bf 1.0}\\
\hline
12 & EISJ2240-4021 &   5 &  22 40 12.6 & -40 22 18.5 & 0.195 &   545 &  99.0 & 1.5\\
{\bf 12} & {\bf EISJ2240-4021} &  {\bf 8} & {\bf  22 40 08.5} & {\bf -40 21 19.5} & {\bf 0.247} & {\bf   199} & {\bf  99.6} & {\bf 0.2}\\
\hline
13 & EISJ2241-4006$^b$ & $-$ & $-$ & $-$ & $-$ & $-$ & $-$  & $-$ \\
\hline
14 & EISJ2241-3932$^c$ &   3 &  22 41 39.2 & -39 32 55.2 & 0.074 &   102 &  99.9 & 1.7\\
\hline
15 & EISJ2243-4010A$^{b,}$$^c$ & $-$ & $-$ & $-$ & $-$ & $-$ & $-$  & $-$ \\
\hline
16 & EISJ2243-3952 &   8 &  22 43 14.6 & -39 54 56.7 & 0.062 &   723 &  99.9 & 2.4\\
16 & EISJ2243-3952 &   8 &  22 43 31.1 & -39 54 11.8 & 0.285 &   33 &  99.2 & 2.7\\
\hline
17 & EISJ2243-3959 &   5 &  22 43 33.4 & -39 57 46.8 & 0.061 &   680 &  99.9 & 1.9\\
{\bf 17} & {\bf EISJ2243-3959} & {\bf  10} & {\bf  22 43 29.4} & {\bf -39 59 27.8} & {\bf 0.285} & {\bf   117} & {\bf 99.8} & {\bf 0.1}\\
\hline
18 & EISJ2243-4010B &   8 &  22 43 31.8 &  -40 09 21.3 &  0.215 &    457 &  99.9 & 2.4\\
\hline
19 & EISJ2243-3947 &   3 &  22 43 46.3 & -39 48 00.3 & 0.062 &   677 &  99.9 & 1.9\\
19 & EISJ2243-3947 &    7 &  22 44 15.2 & -39 48 42.5 &  0.126 & 176 &   99.9 & 3.9\\
19 & EISJ2243-3947 &   4 &  22 43 53.8 & -39 48 15.3 & 0.138 &    53 &  99.3 & 0.9\\
\hline
20 & EISJ2244-4008 &   3 &  22 44 07.7 & -40 09 08.1 & 0.102 &     0 &  99.8 & 2.8\\
20 & EISJ2244-4008 &   5 &  22 44 33.2 & -40 08 05.7 & 0.471 &    63 &  99.6 & 2.2\\
\hline
21 & EISJ2244-4019 &   4 &  22 44 16.3 & -40 19 53.7 & 0.153 &  1210 &  99.0 & 2.3\\
{\bf 21} & {\bf EISJ2244-4019} & {\bf  10} & {\bf  22 44 25.5} & {\bf -40 19 42.2} & {\bf 0.216} &  {\bf  325} &  {\bf 99.9} & {\bf 0.6}\\
\hline
22 & EISJ2246-4012B$^{b,}$$^c$ & $-$ & $-$ & $-$ & $-$ & $-$ & $-$  & $-$ \\
\hline
23 & EISJ2248-4015 &  10 &  22 48 54.6 & -40 15 39.2 & 0.129 &   863 &  99.9 & 0.3\\ 
23 & EISJ2248-4015 &   7 &  22 48 55.3 & -40 14 44.4 & 0.230 &   176 &  99.2 & 0.6\\
\hline
\end{tabular}
\end{minipage}
\end{center}
\end{table*}

The measured redshifts were used to search for significant groups in
redshift space to identify the physical systems associated to the
detections by the matched-filter algorithm.  As described in previous
papers (Paper~I, II and III) we identify groups in redshift space by
the ``gap''-technique of \cite{katgert96} with a gap-size of $\Delta
z=0.005(1+z)$ corresponding to $1500\mathrm{km/s}$ in the
restframe. The identified groups are shown as the solid
histograms. For assessing the significance of the identified groups we
use the CNOC2 0223+00 catalogue \citep{yee00}.  As detailed in
Paper~II, the significance, $\sigma_1$, is determined from the
probability of finding a group with the same number of objects or more
at the same redshift imposing the same magnitude limit as for the
survey. This procedure is a simple way to take into account the
selection function of the present survey and the clustering of field
galaxies. More advanced applications of such approaches have been
presented by for example \cite{holden99,ramella00,gilbank2004}.

To give an overview of the redshifts and their clustering,
Fig.~\ref{fig:redshift_dists} shows the redshift distribution for each
field. The upper parts show the bar diagram of the redshifts while the
lower parts give the redshift histogram with a bin size of $\Delta
z=0.01$. We find a total of 81 groups in redshift space with a least 3
members.

\begin{figure*}
\begin{center}
\resizebox{0.24\textwidth}{!}{\includegraphics{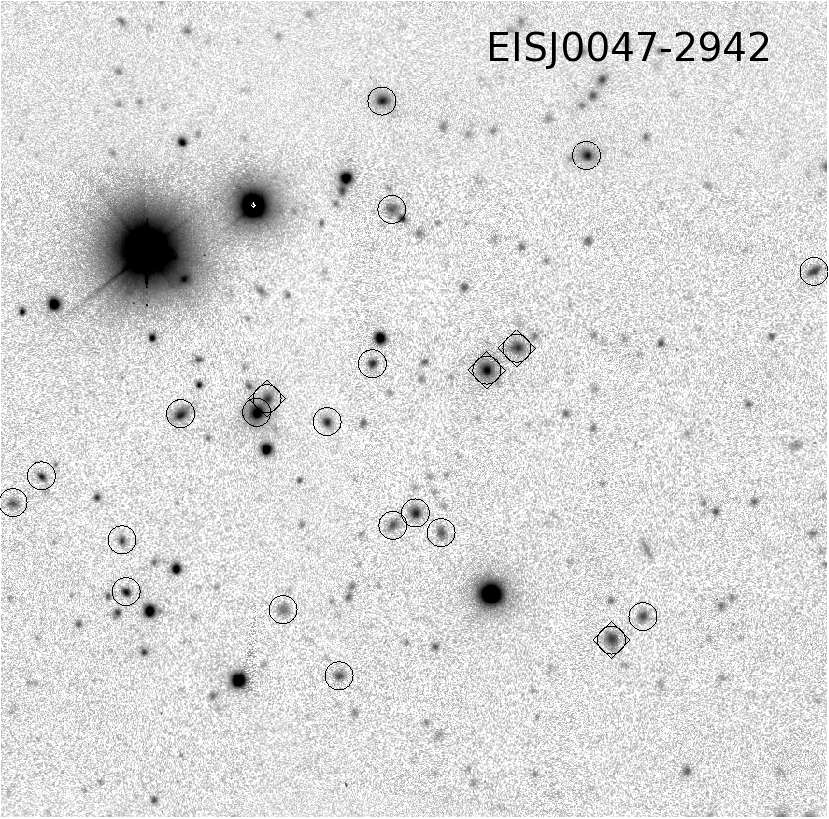}}
\resizebox{0.24\textwidth}{!}{\includegraphics{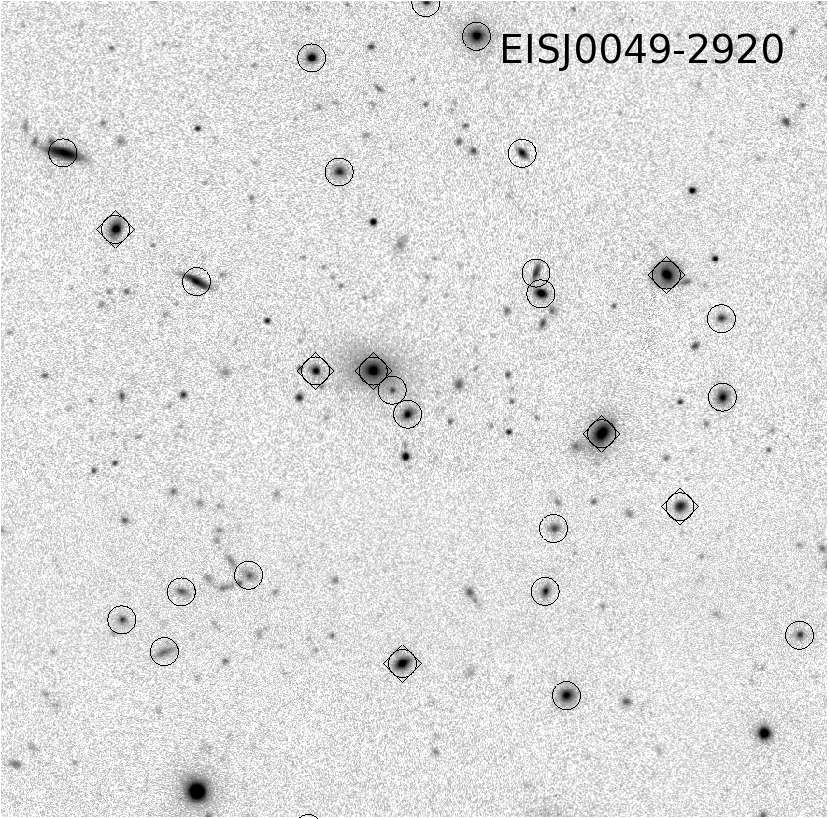}}
\resizebox{0.24\textwidth}{!}{\includegraphics{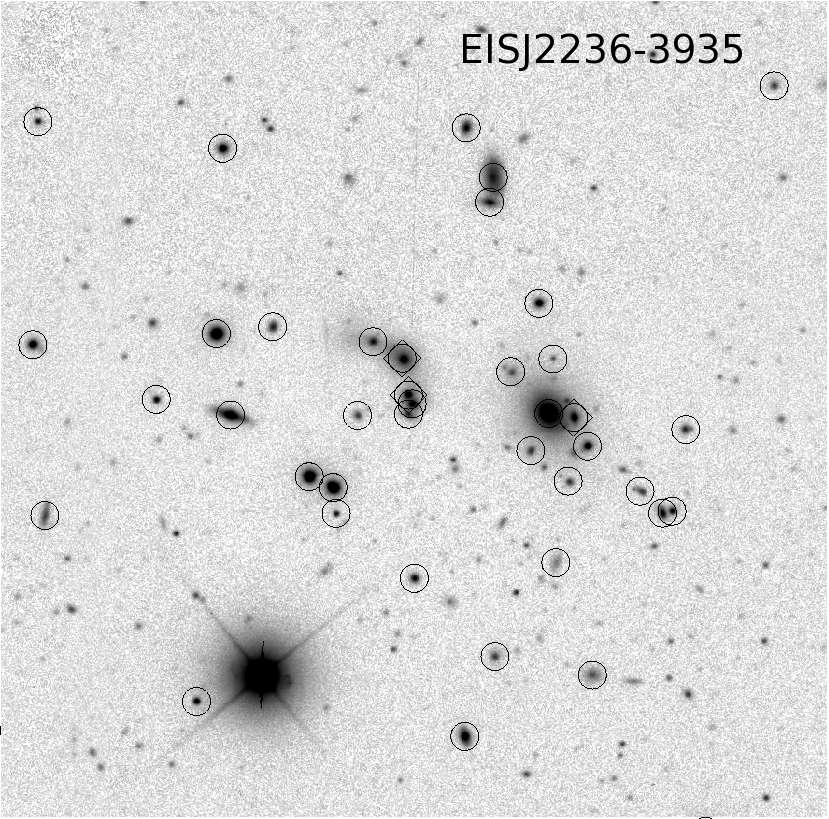}}
\resizebox{0.24\textwidth}{!}{\includegraphics{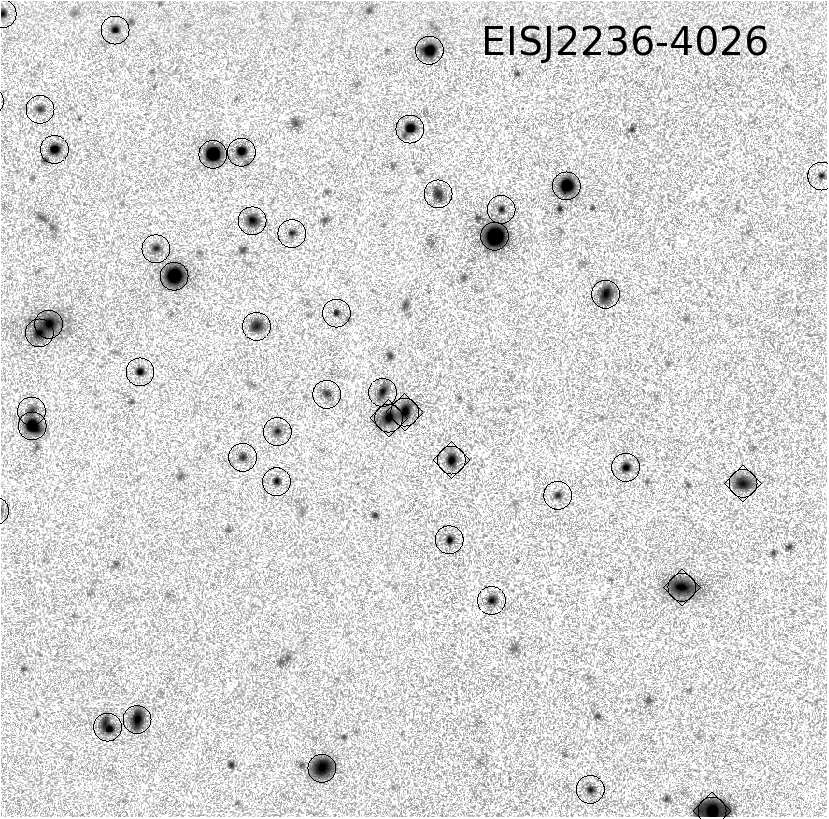}} 
\resizebox{0.24\textwidth}{!}{\includegraphics[bb=0 0 226 226,clip]{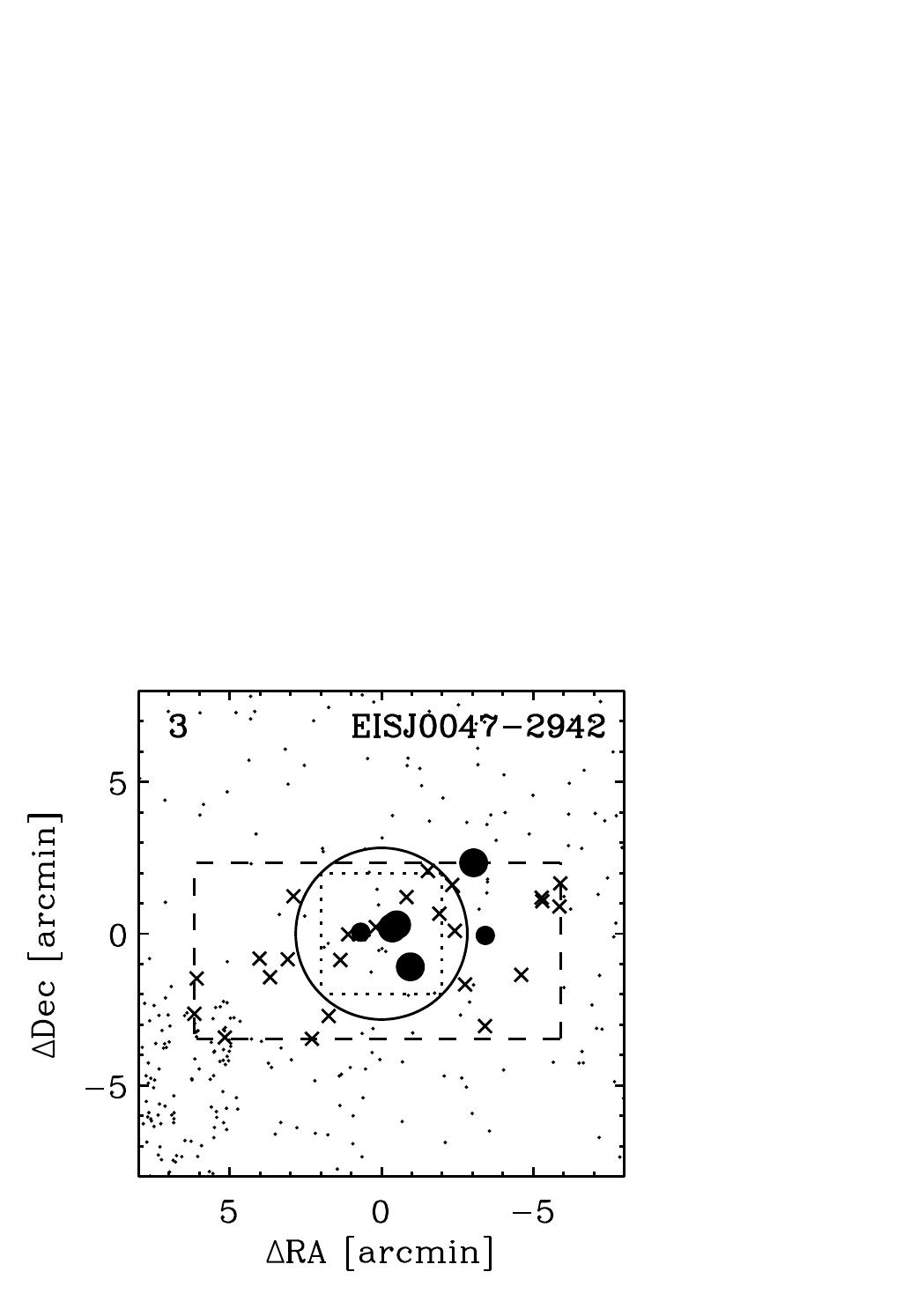}}
\resizebox{0.24\textwidth}{!}{\includegraphics[bb=0 0 226 226,clip]{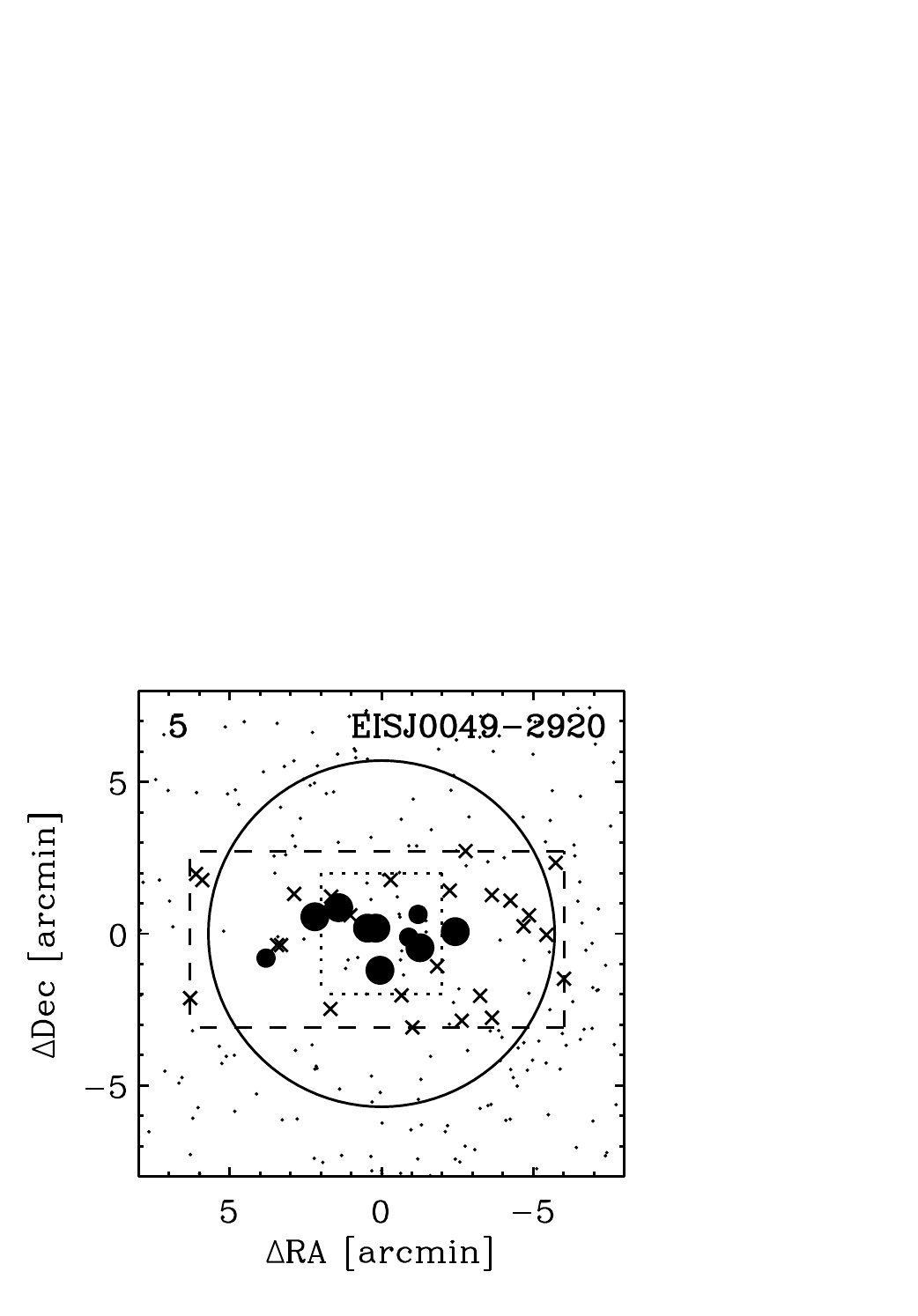}}
\resizebox{0.24\textwidth}{!}{\includegraphics[bb=0 0 226 226,clip]{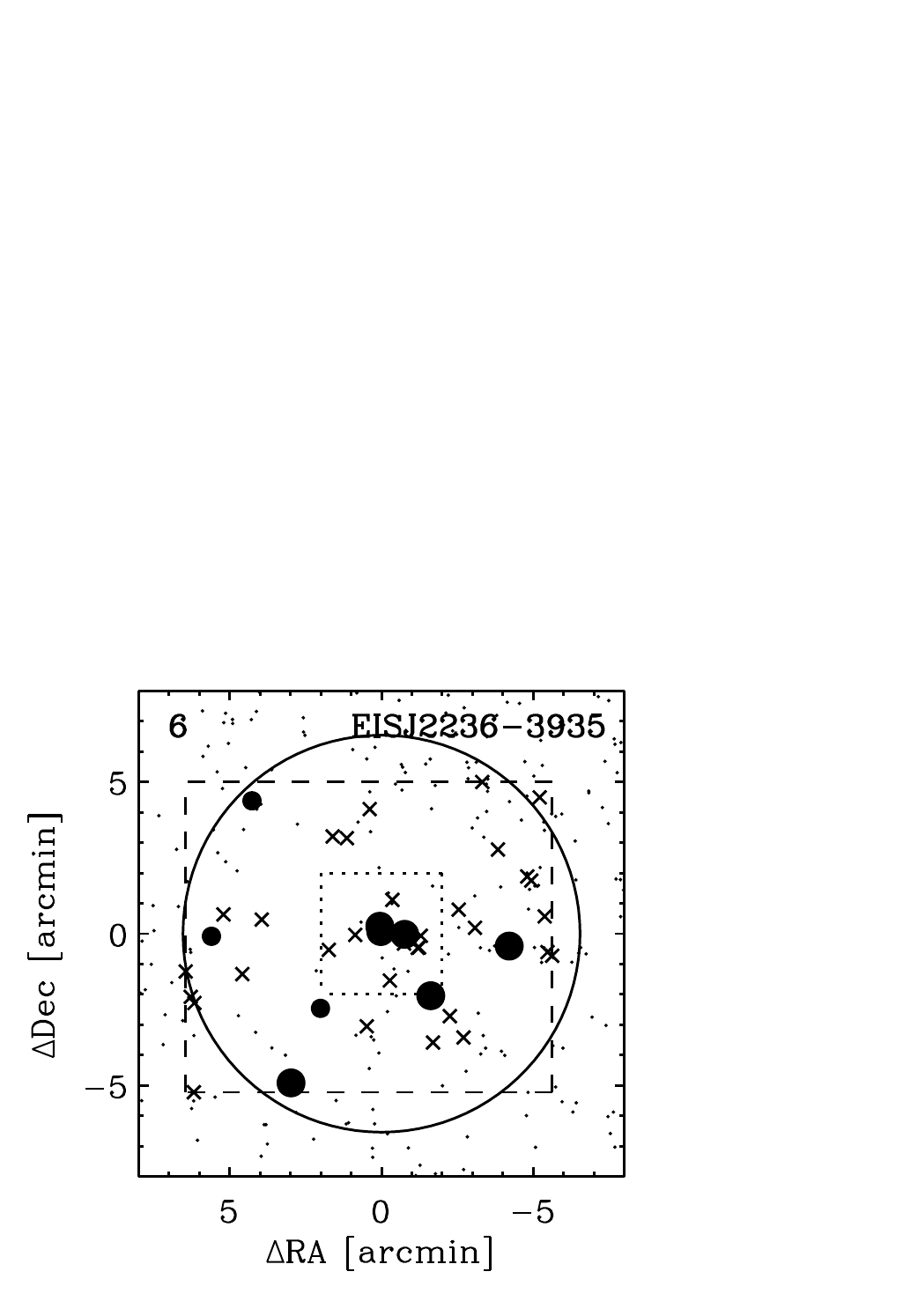}}
\resizebox{0.24\textwidth}{!}{\includegraphics[bb=0 0 226 226,clip]{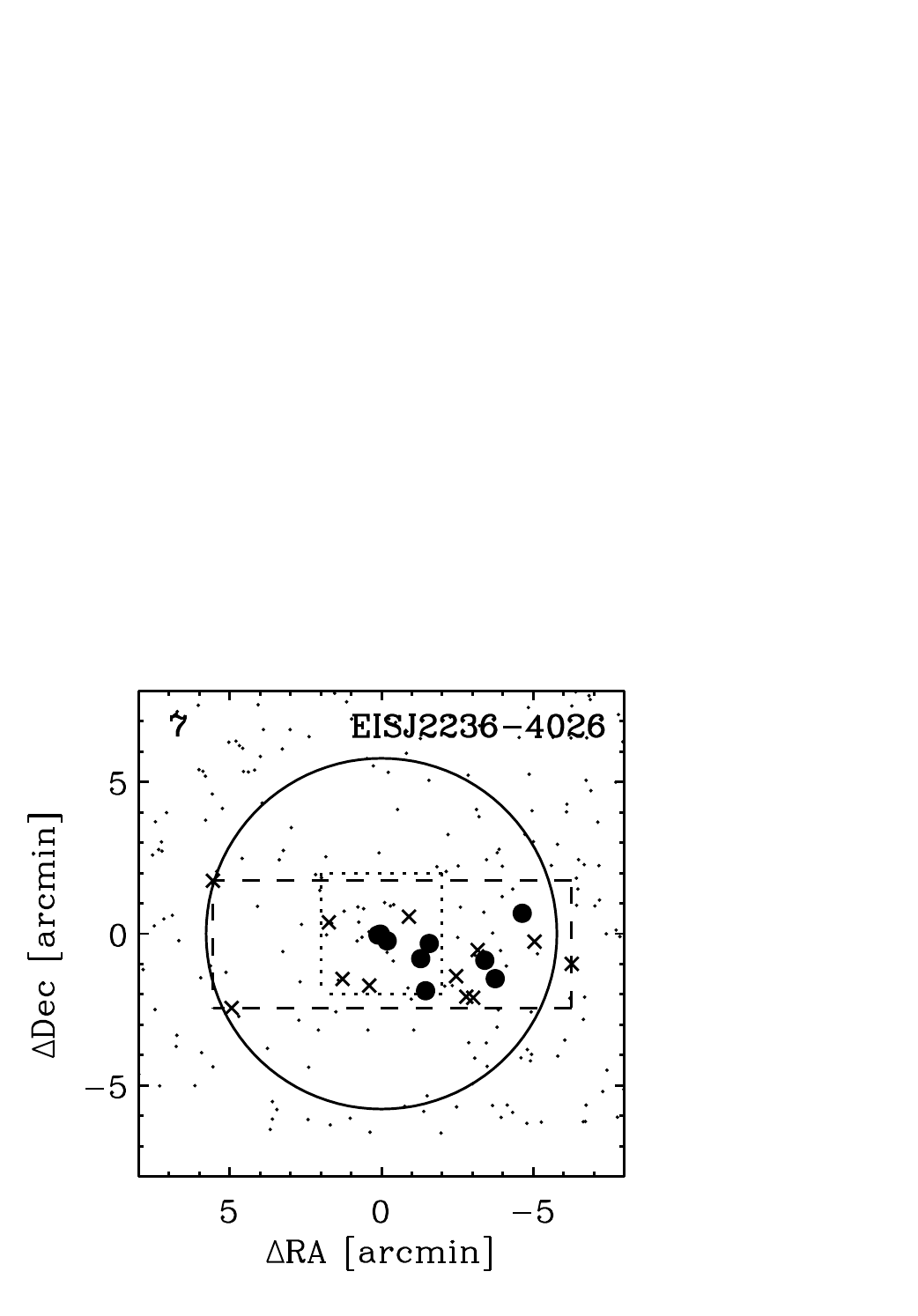}}
\resizebox{0.24\textwidth}{!}{\includegraphics{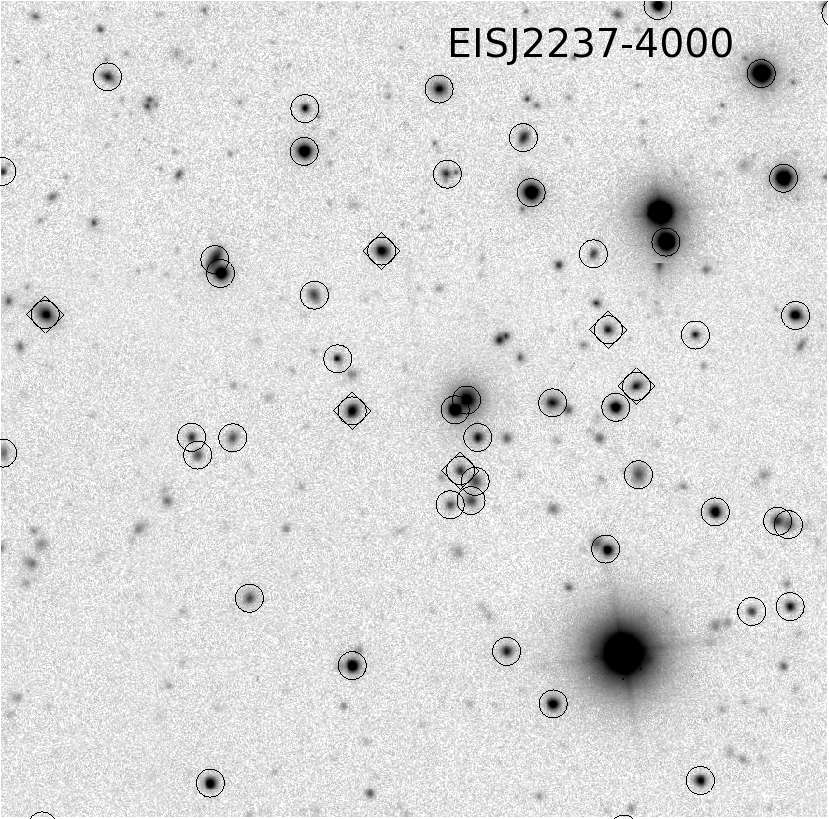}}
\resizebox{0.24\textwidth}{!}{\includegraphics{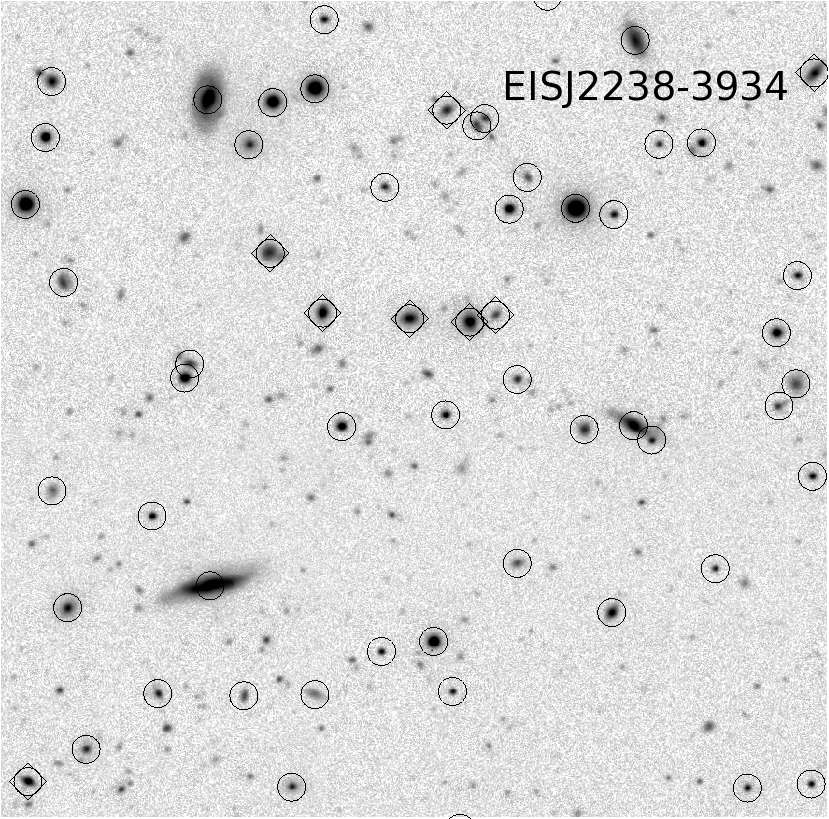}}
\resizebox{0.24\textwidth}{!}{\includegraphics{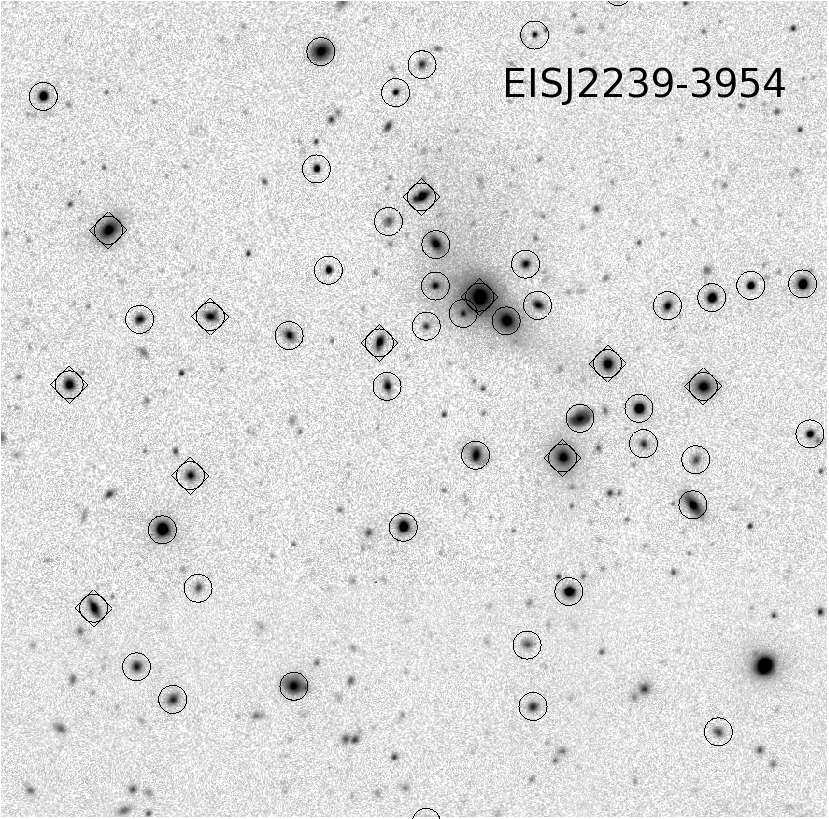}}
\resizebox{0.24\textwidth}{!}{\includegraphics{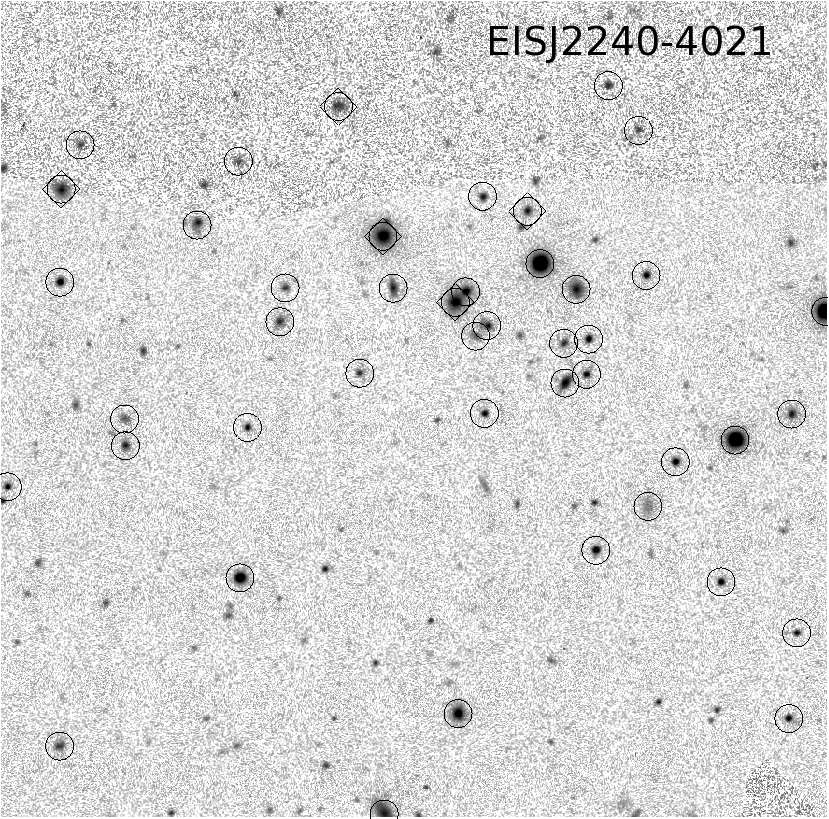}}
\resizebox{0.24\textwidth}{!}{\includegraphics[bb=0 0 226 226,clip]{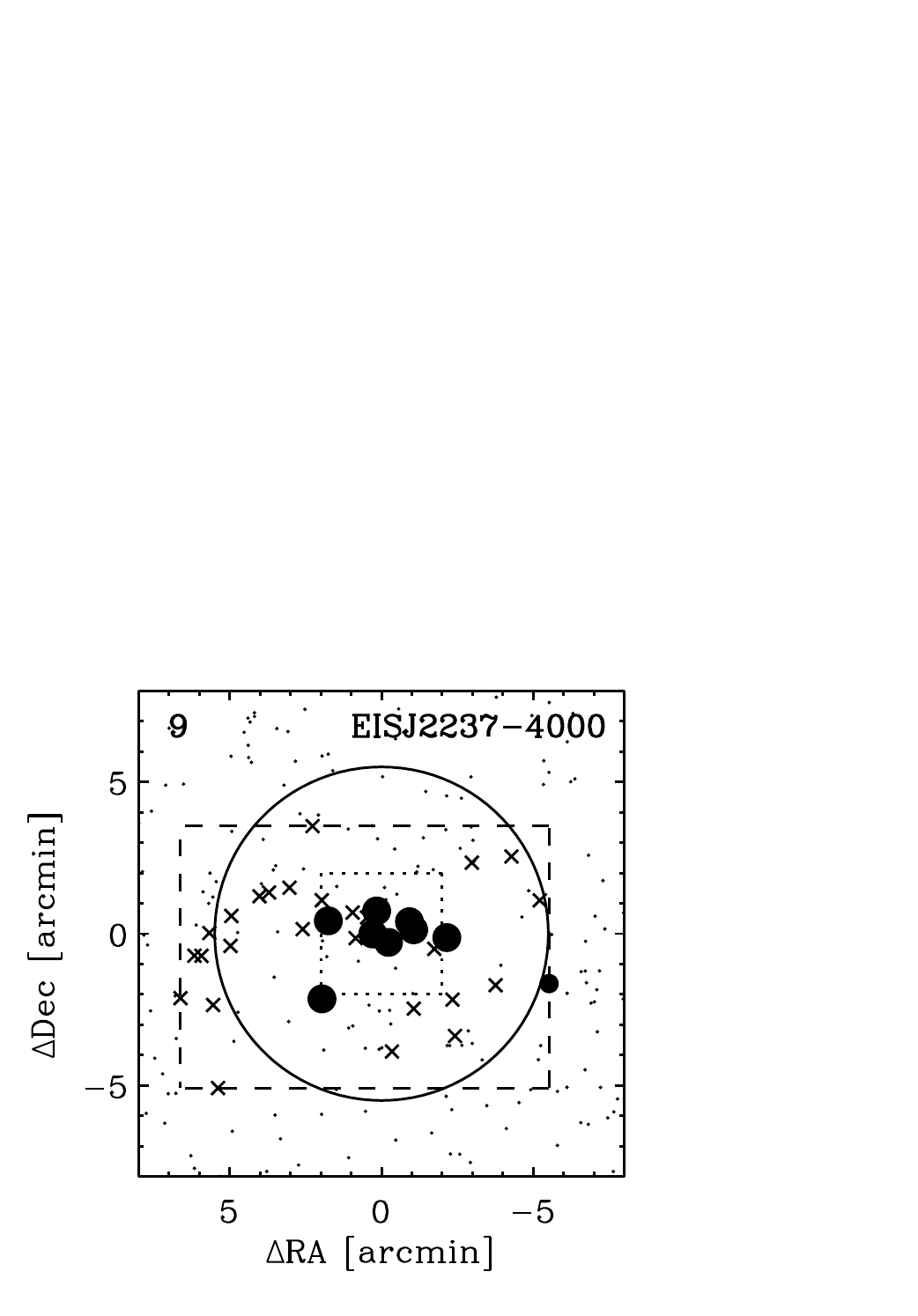}}
\resizebox{0.24\textwidth}{!}{\includegraphics[bb=0 0 226 226,clip]{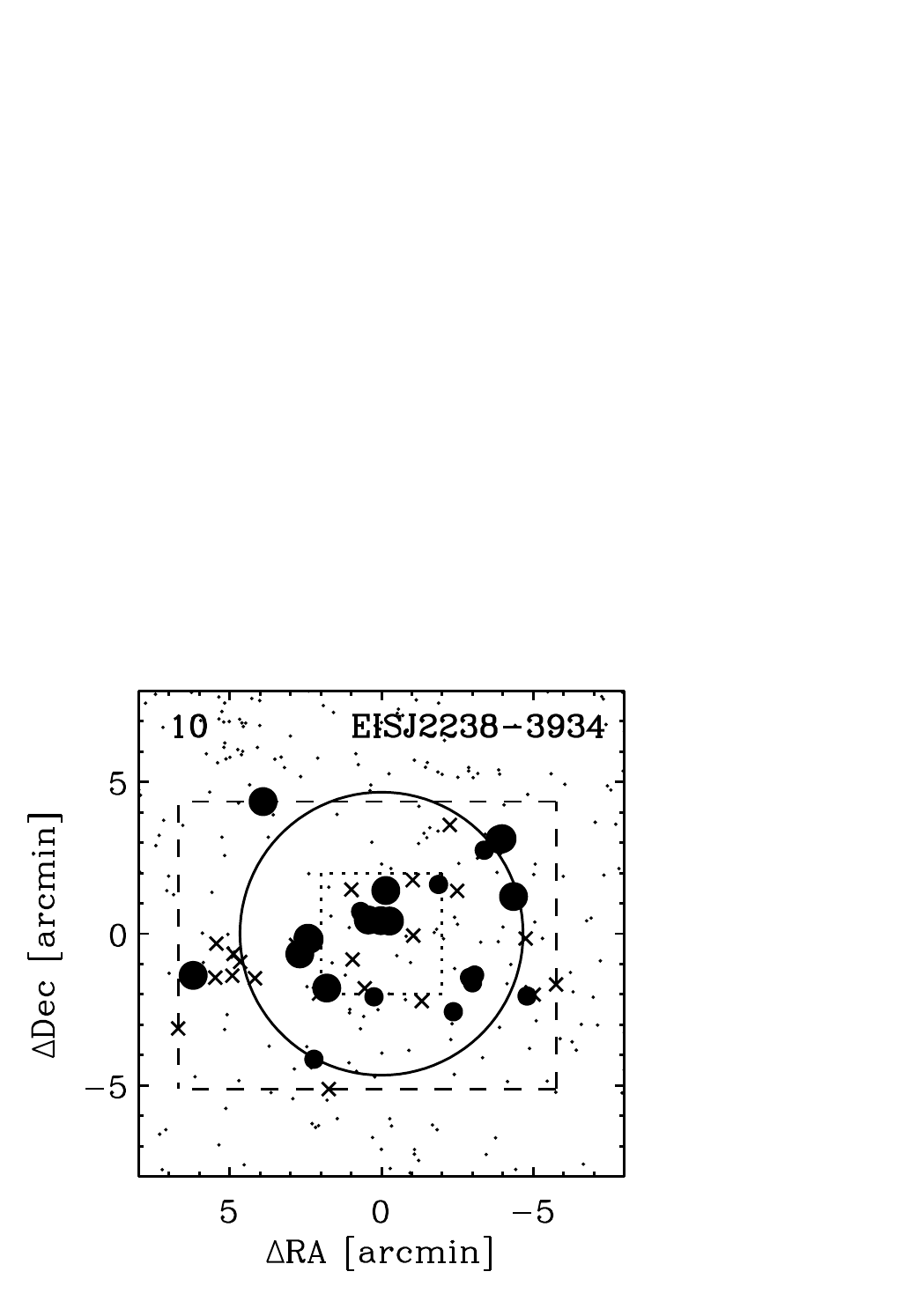}}
\resizebox{0.24\textwidth}{!}{\includegraphics[bb=0 0 226 226,clip]{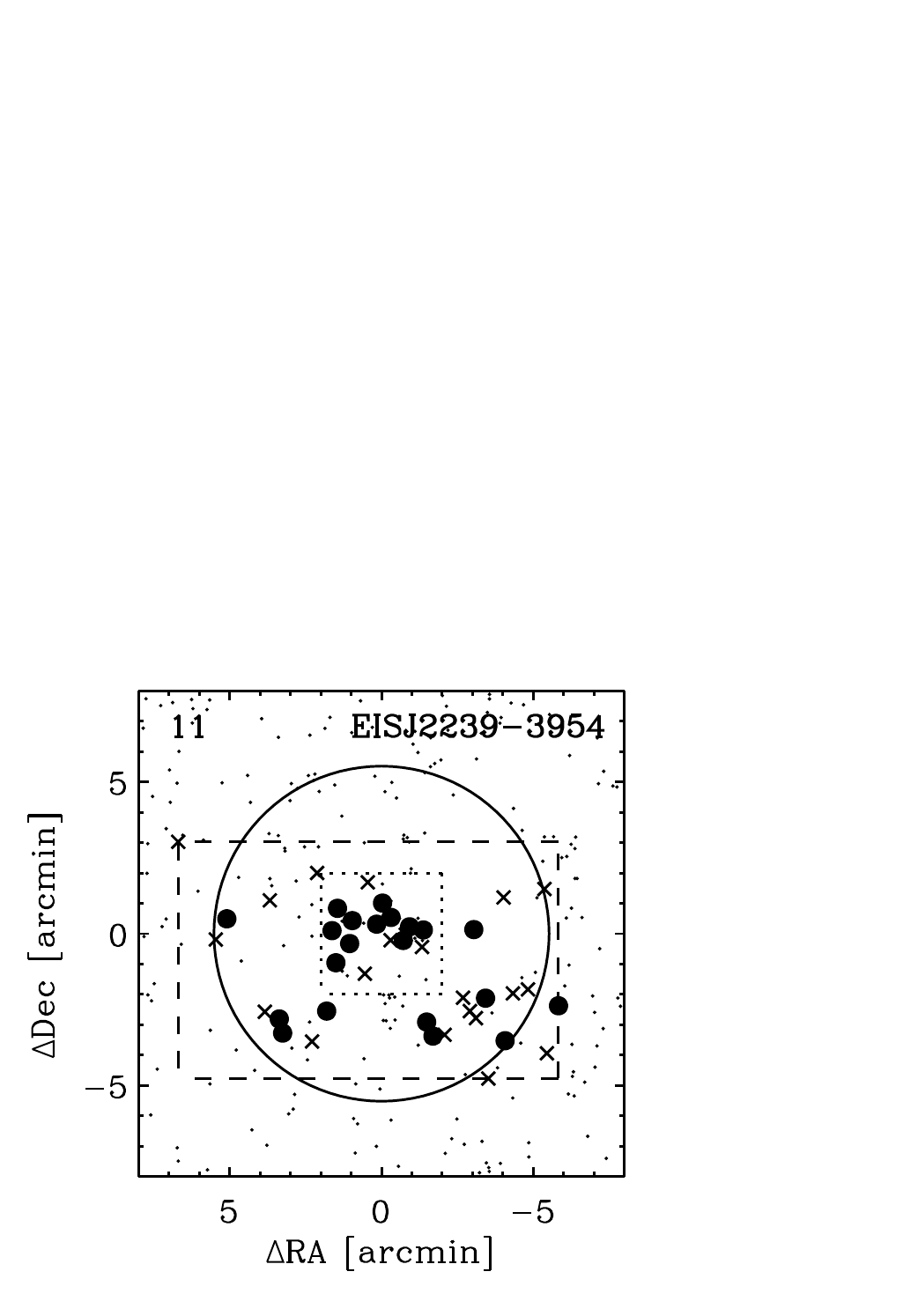}}
\resizebox{0.24\textwidth}{!}{\includegraphics[bb=0 0 226 226,clip]{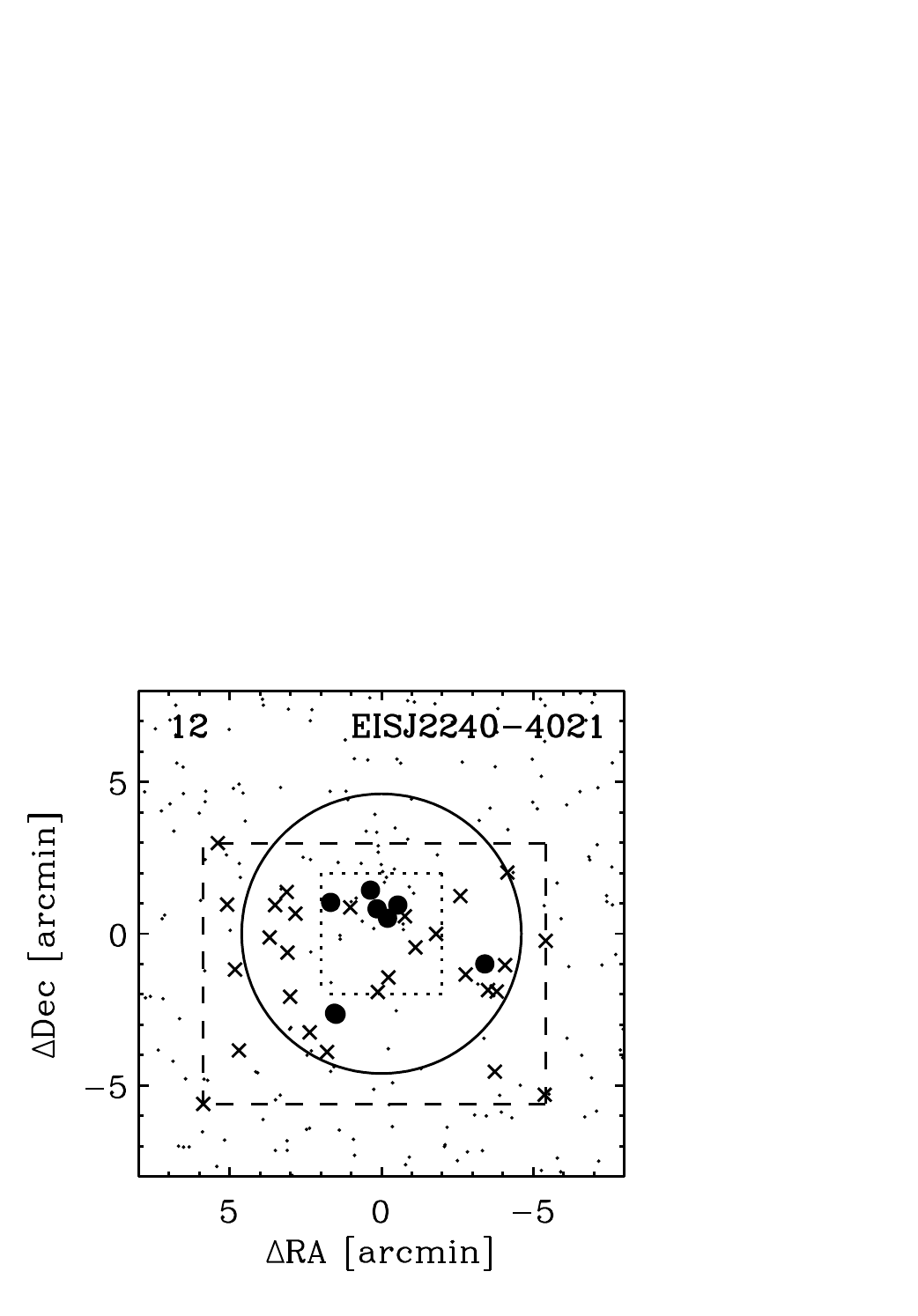}}
\resizebox{0.24\textwidth}{!}{\includegraphics{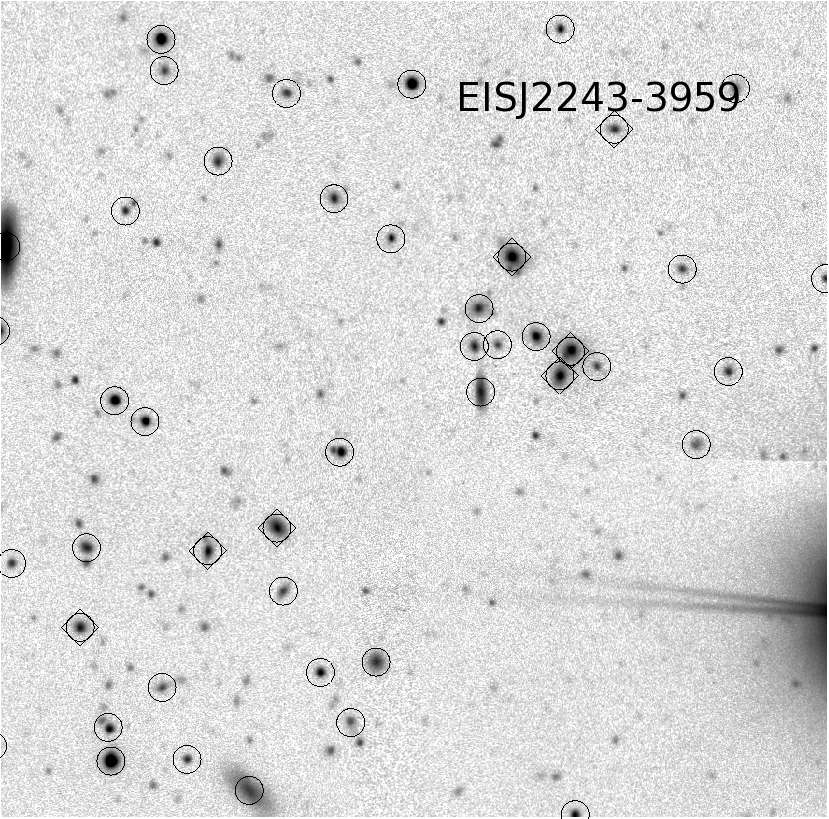}}
\resizebox{0.24\textwidth}{!}{\includegraphics[bb=0 0 226 226,clip]{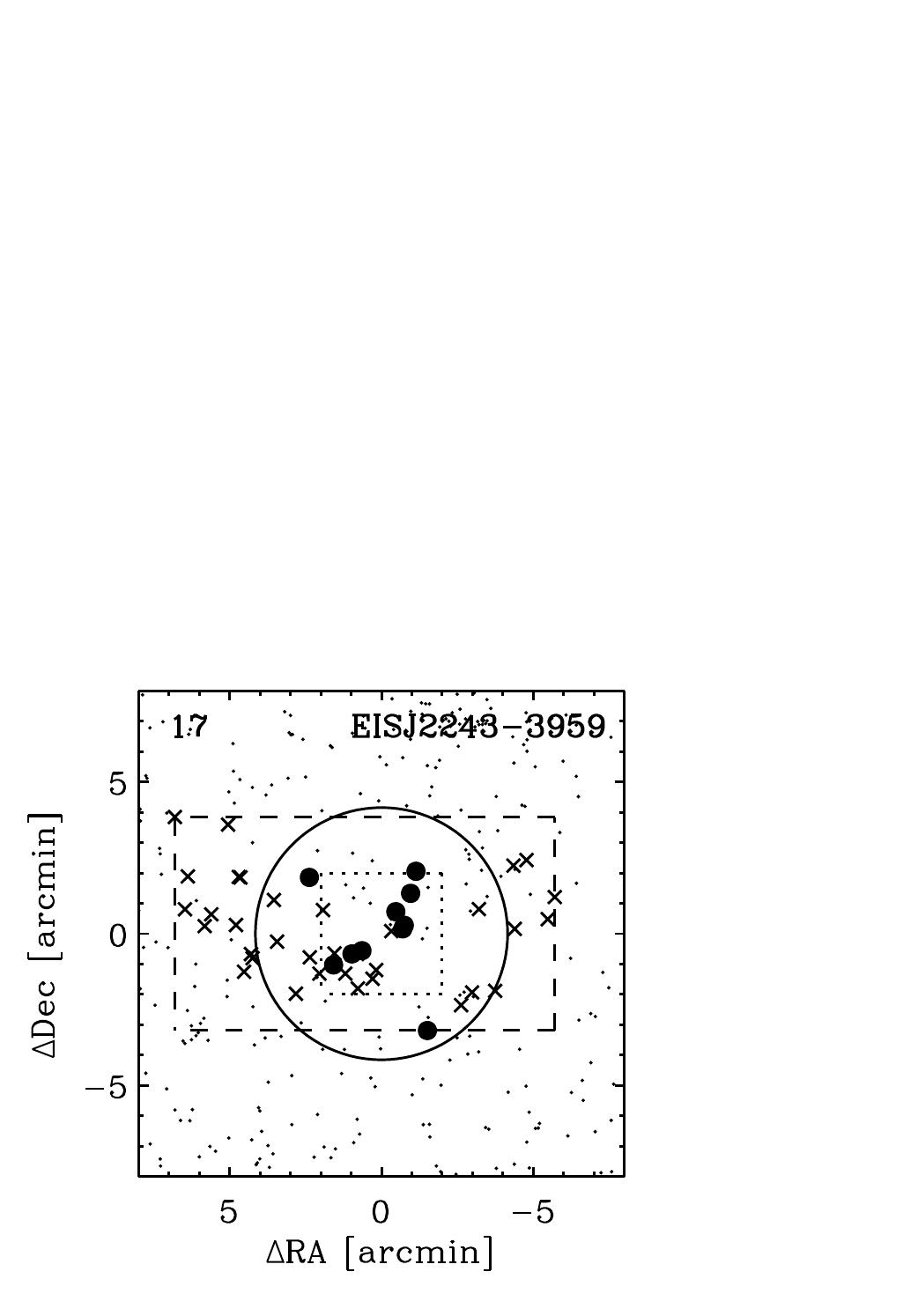}}
\resizebox{0.24\textwidth}{!}{\includegraphics{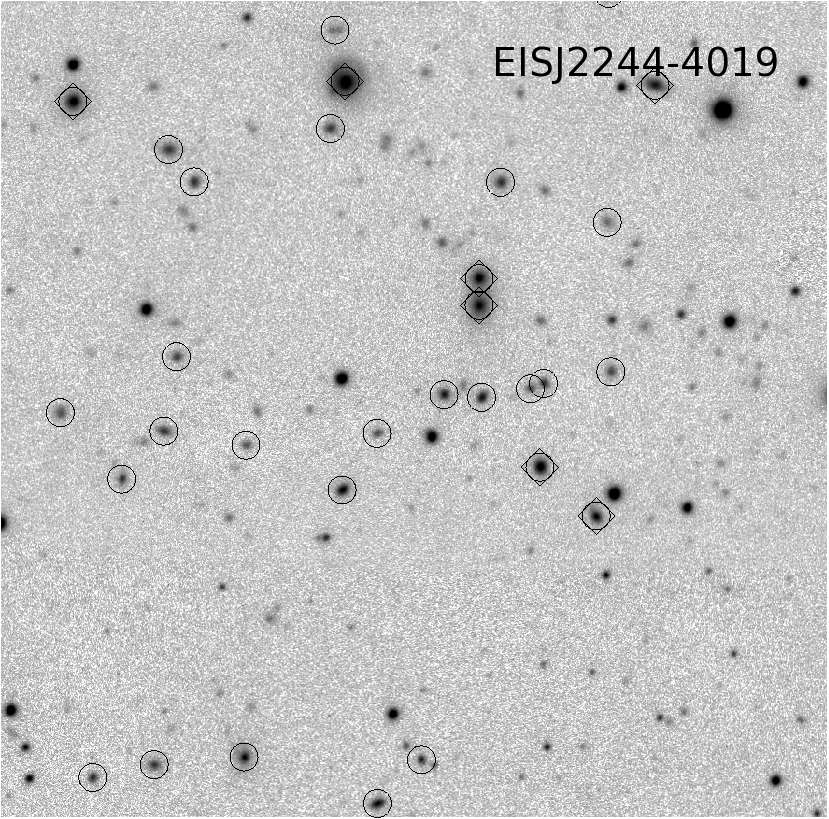}}
\resizebox{0.24\textwidth}{!}{\includegraphics[bb=0 0 226 226,clip]{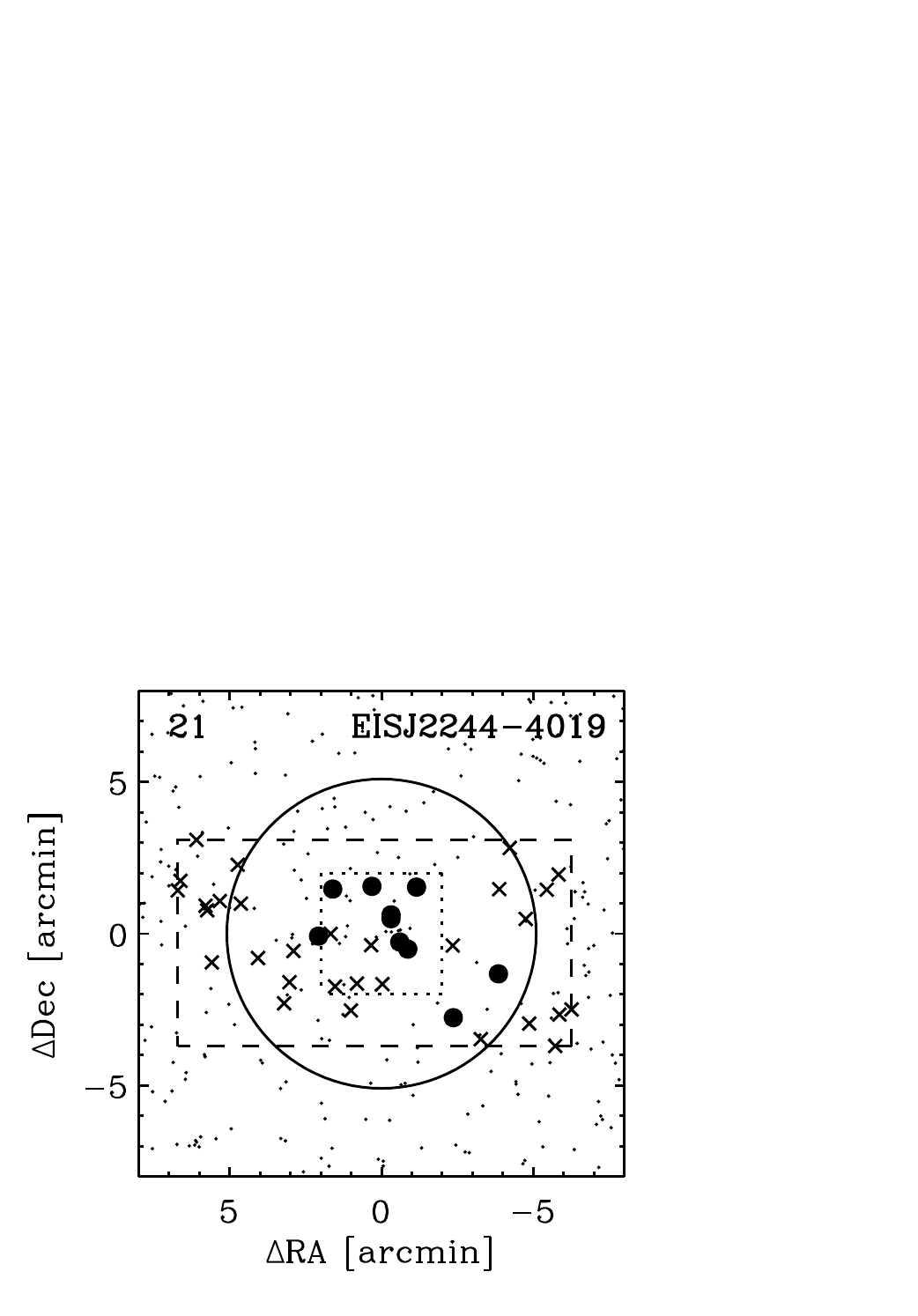}}
\end{center}
\caption{Image cutouts, centred at the original MF position and with
a size of $\sim4'$, and position plots for the ten fields with groups
assigned to the EIS detection. In the image cutouts galaxies with
$I<21$ are marked by circles and galaxies with redshifts and belonging
to the group are indicated by diamonds. In the projected distributions
all galaxies with $I\leq20$ are indicated by the small dots, crosses
mark galaxies with redshift and the member galaxies are marked by
filled circles. The large circles centred on the MF position indicate
a radius of $0.5h_{75}^{-1}$Mpc. In cases where there are symbols of
different size it relates to substructures in the redshift
distribution.  The dashed rectangle marks the region of the
redshift survey and the dotted square outlines the area of the image
cutout.}
\label{fig:conf}
\end{figure*}

\begin{figure*}
\begin{center}
\resizebox{0.24\textwidth}{!}{\includegraphics{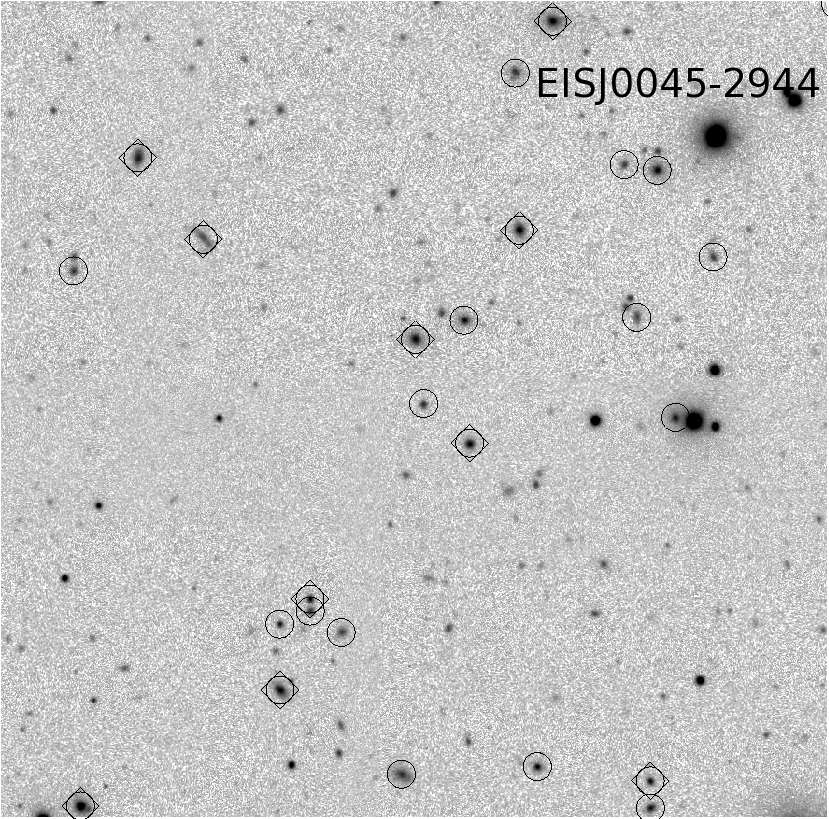}}
\resizebox{0.24\textwidth}{!}{\includegraphics{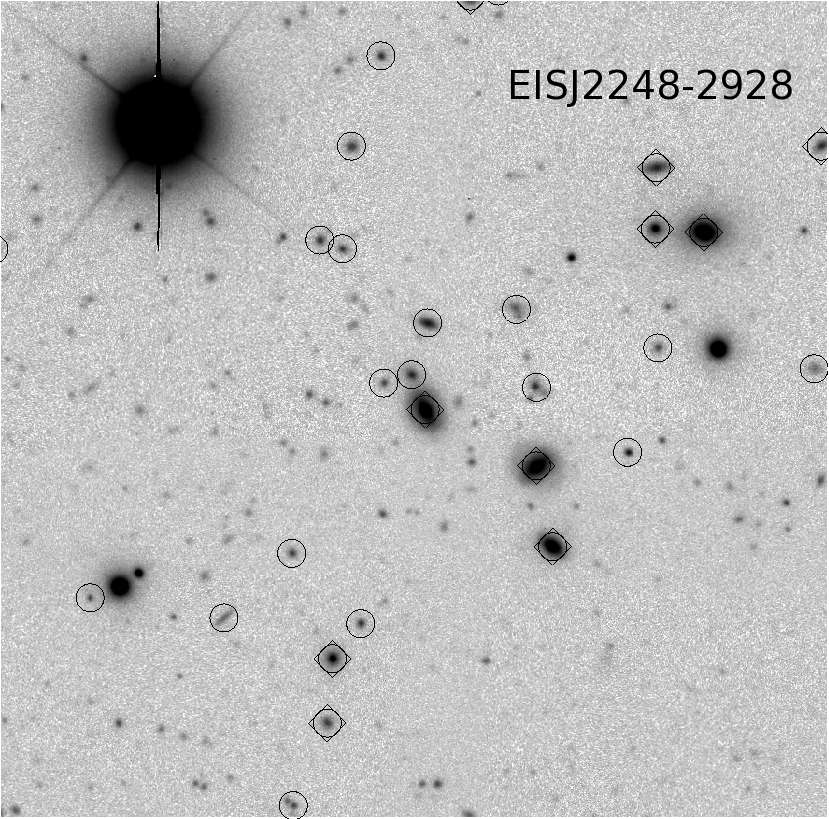}}
\resizebox{0.24\textwidth}{!}{\includegraphics{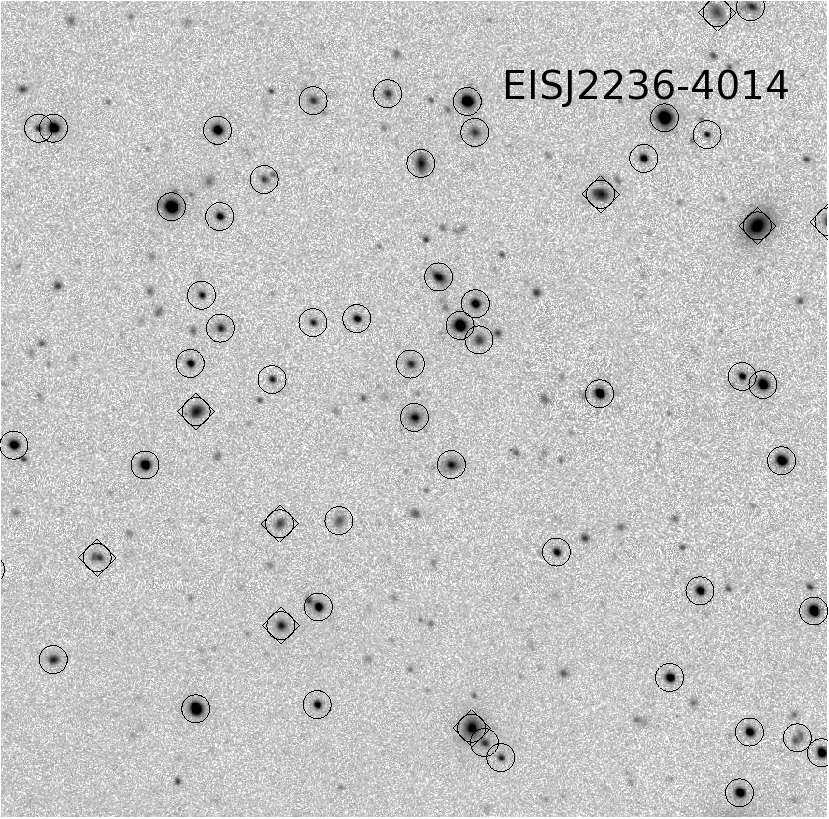}}
\resizebox{0.24\textwidth}{!}{\includegraphics{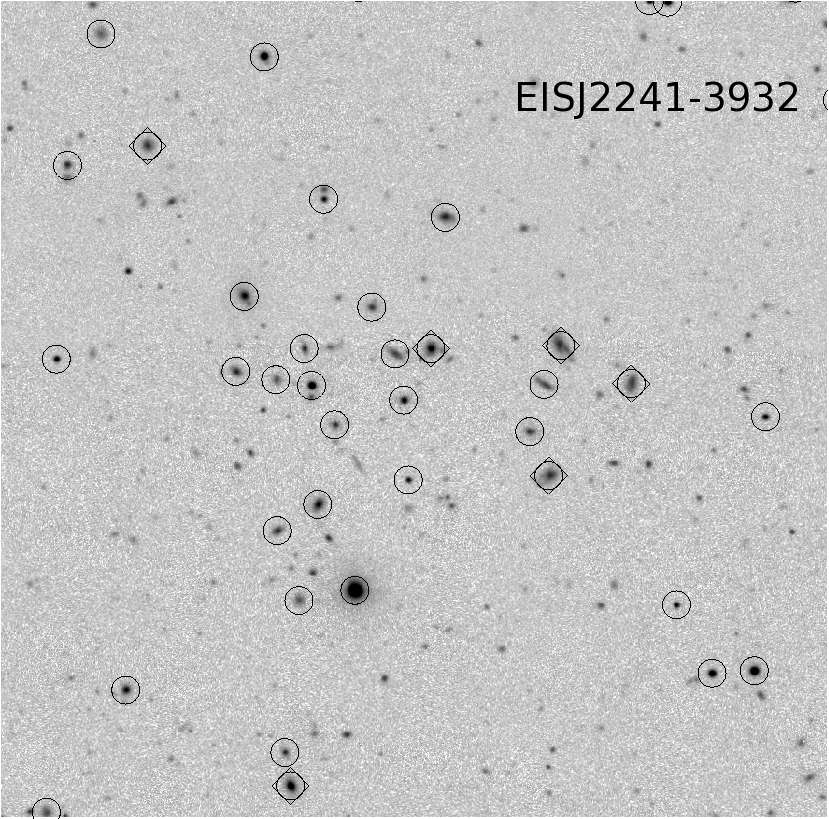}}
\resizebox{0.24\textwidth}{!}{\includegraphics[bb=0 0 226 226,clip]{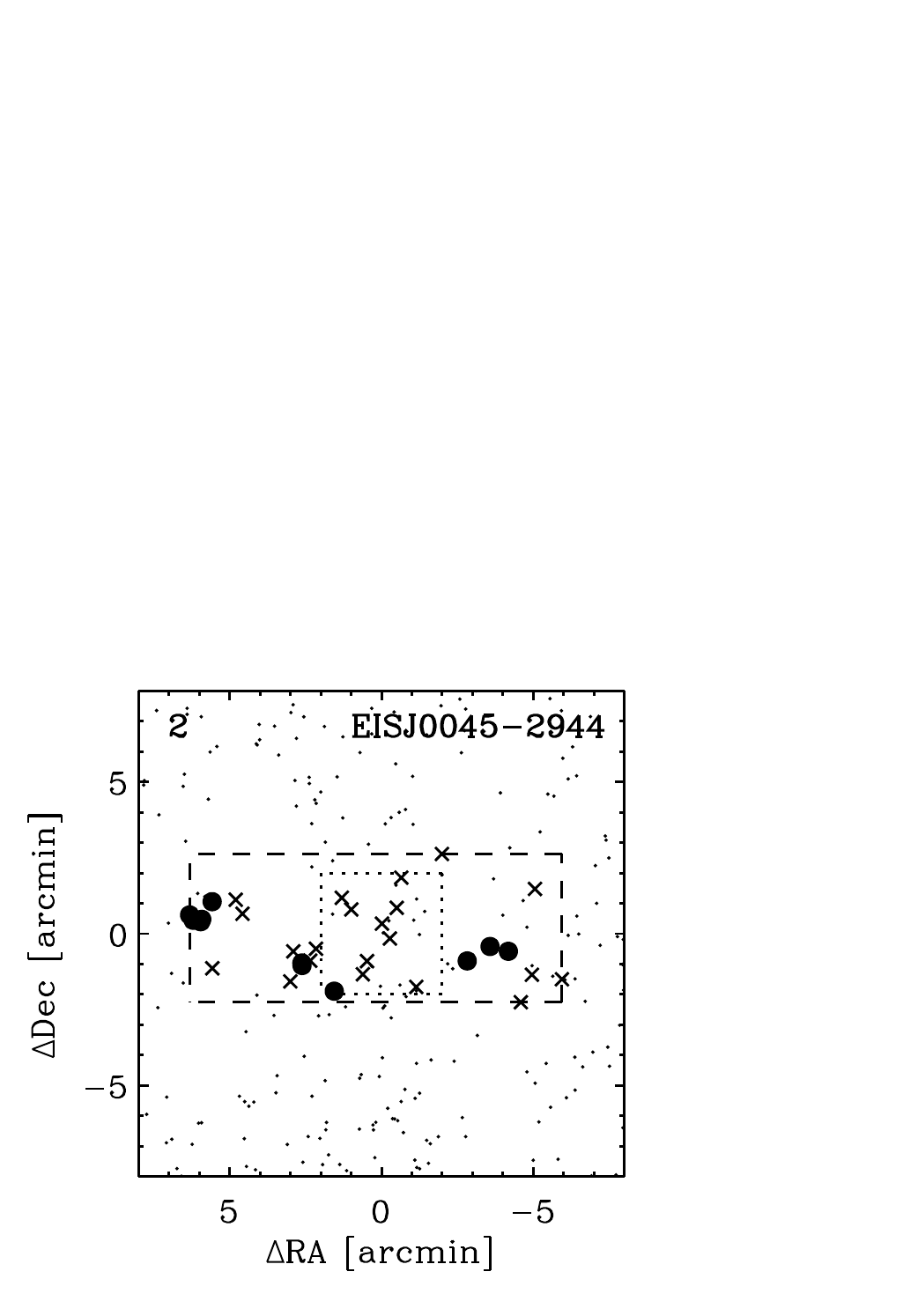}}
\resizebox{0.24\textwidth}{!}{\includegraphics[bb=0 0 226 226,clip]{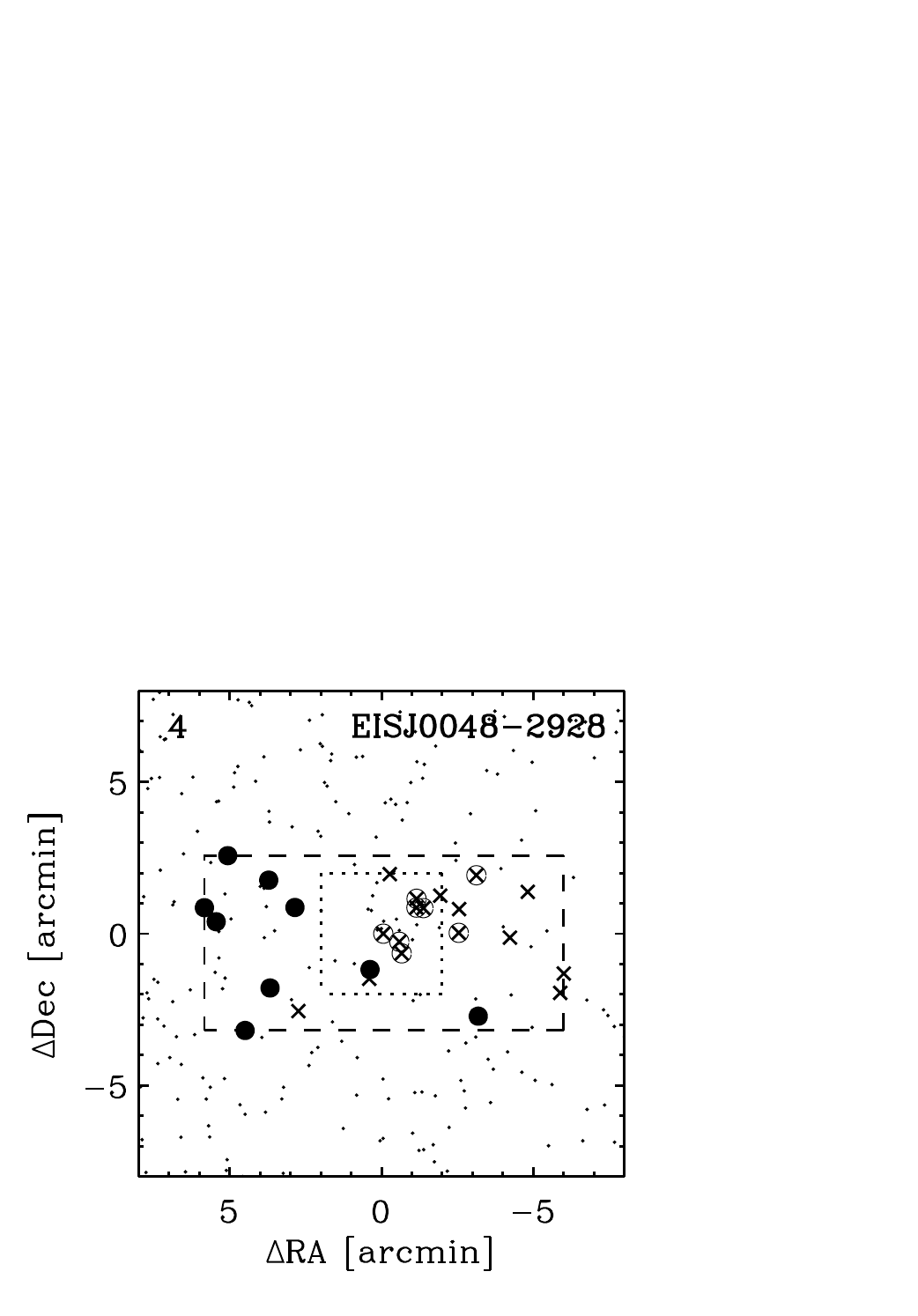}}
\resizebox{0.24\textwidth}{!}{\includegraphics[bb=0 0 226 226,clip]{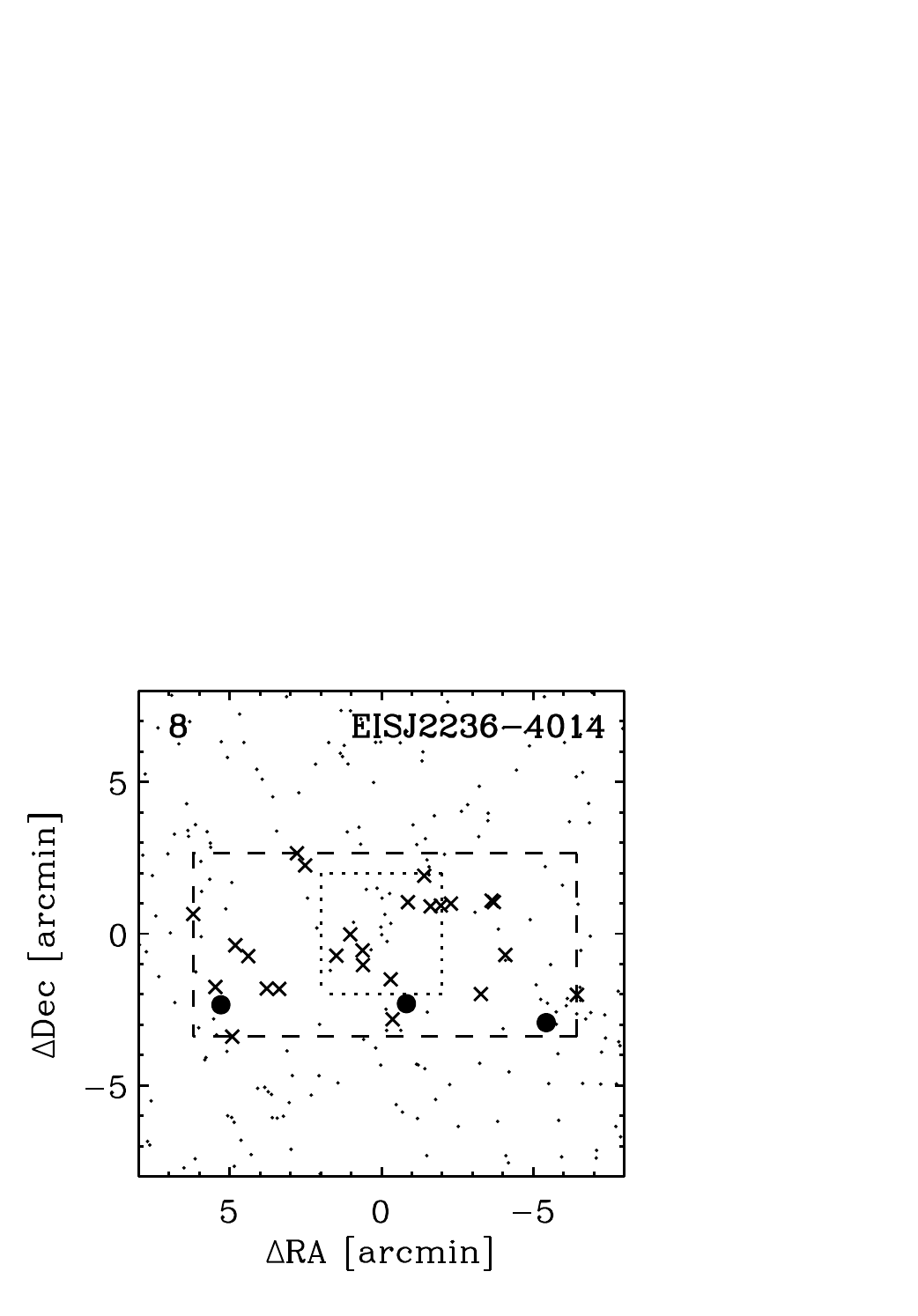}}
\resizebox{0.24\textwidth}{!}{\includegraphics[bb=0 0 226 226,clip]{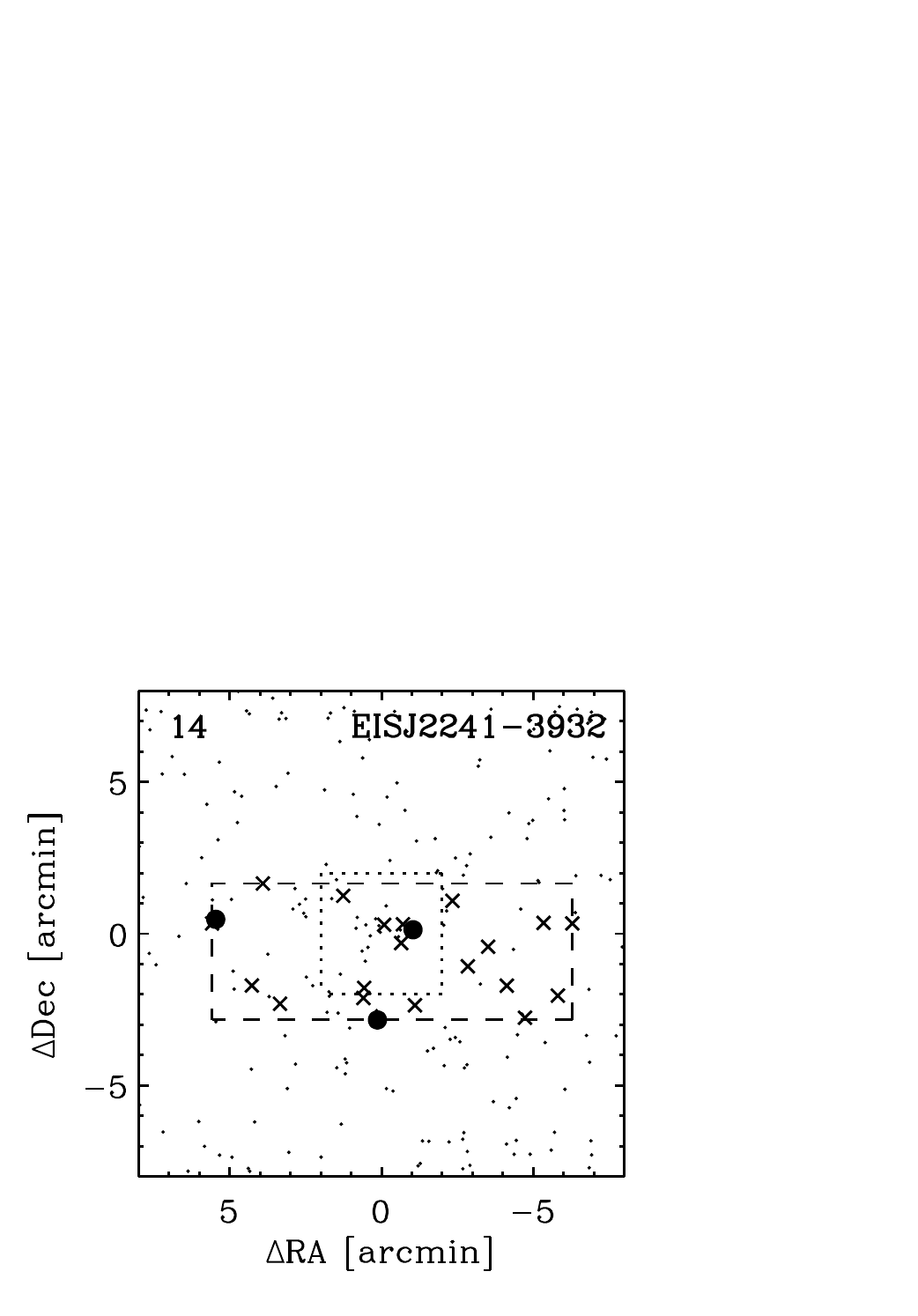}}
\resizebox{0.24\textwidth}{!}{\includegraphics{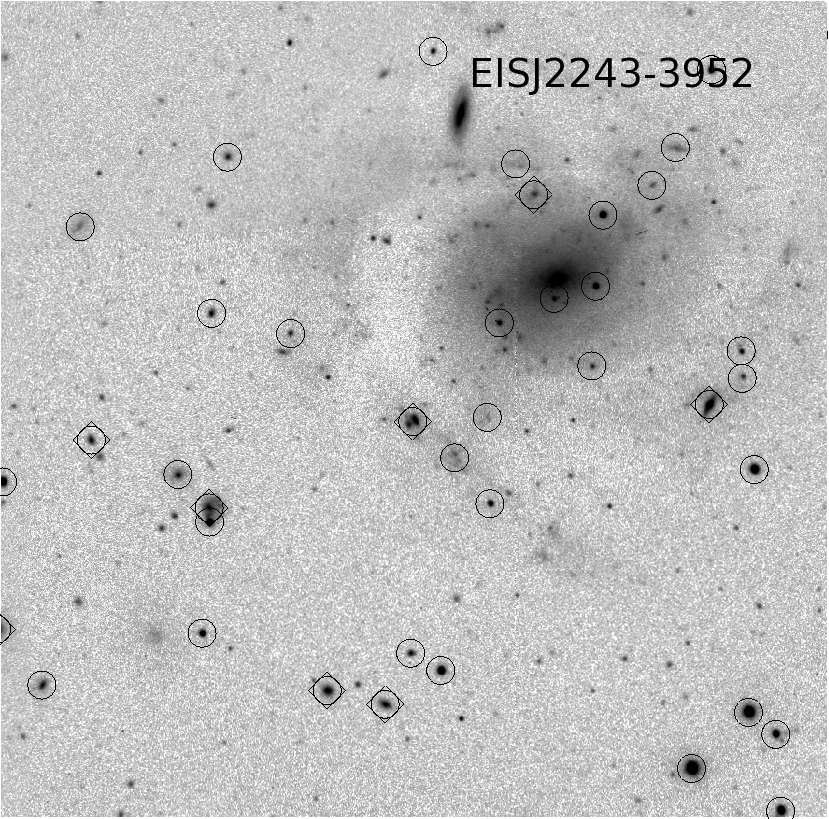}}
\resizebox{0.24\textwidth}{!}{\includegraphics{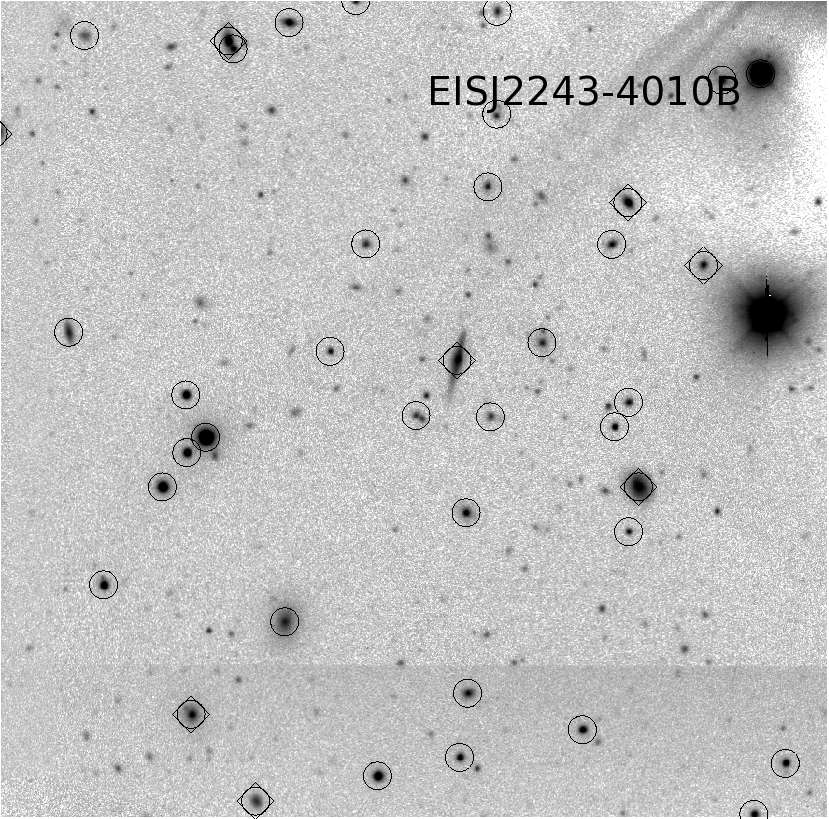}}
\resizebox{0.24\textwidth}{!}{\includegraphics{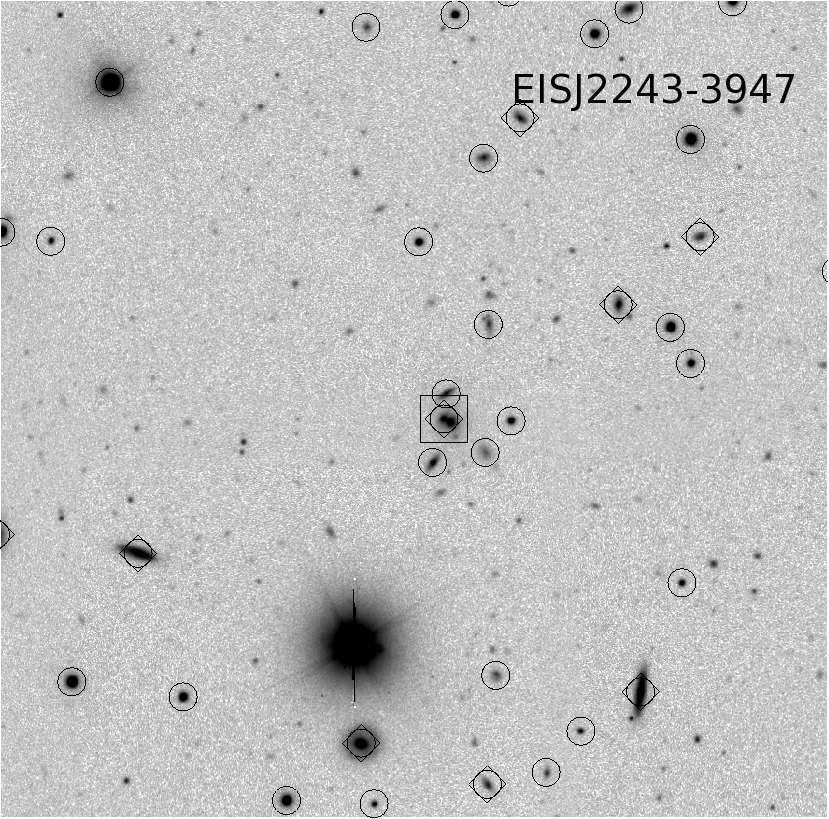}}
\resizebox{0.24\textwidth}{!}{\includegraphics{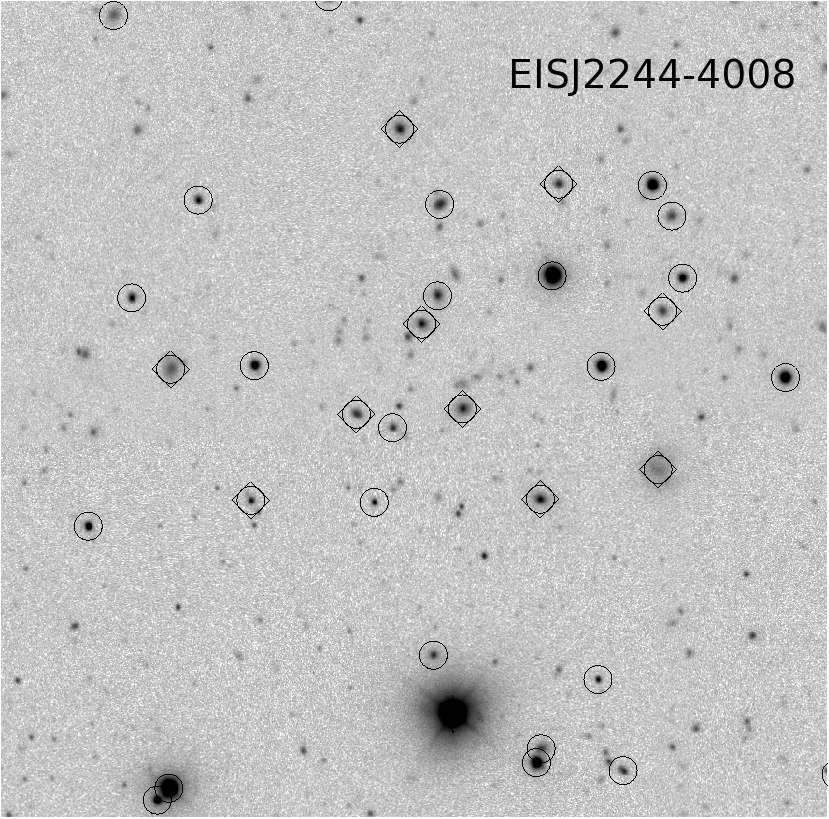}}
\resizebox{0.24\textwidth}{!}{\includegraphics[bb=0 0 226 226,clip]{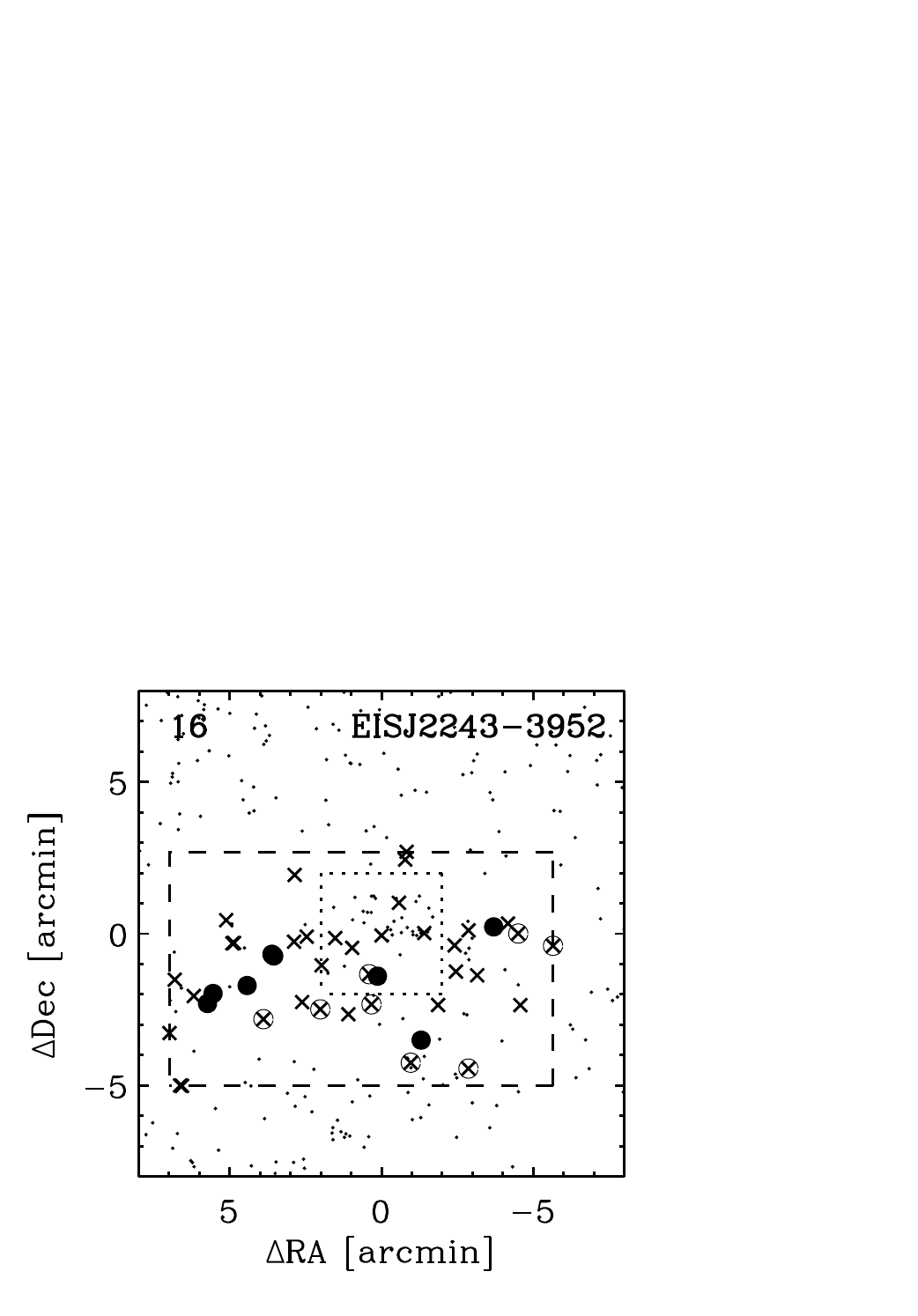}}
\resizebox{0.24\textwidth}{!}{\includegraphics[bb=0 0 226 226,clip]{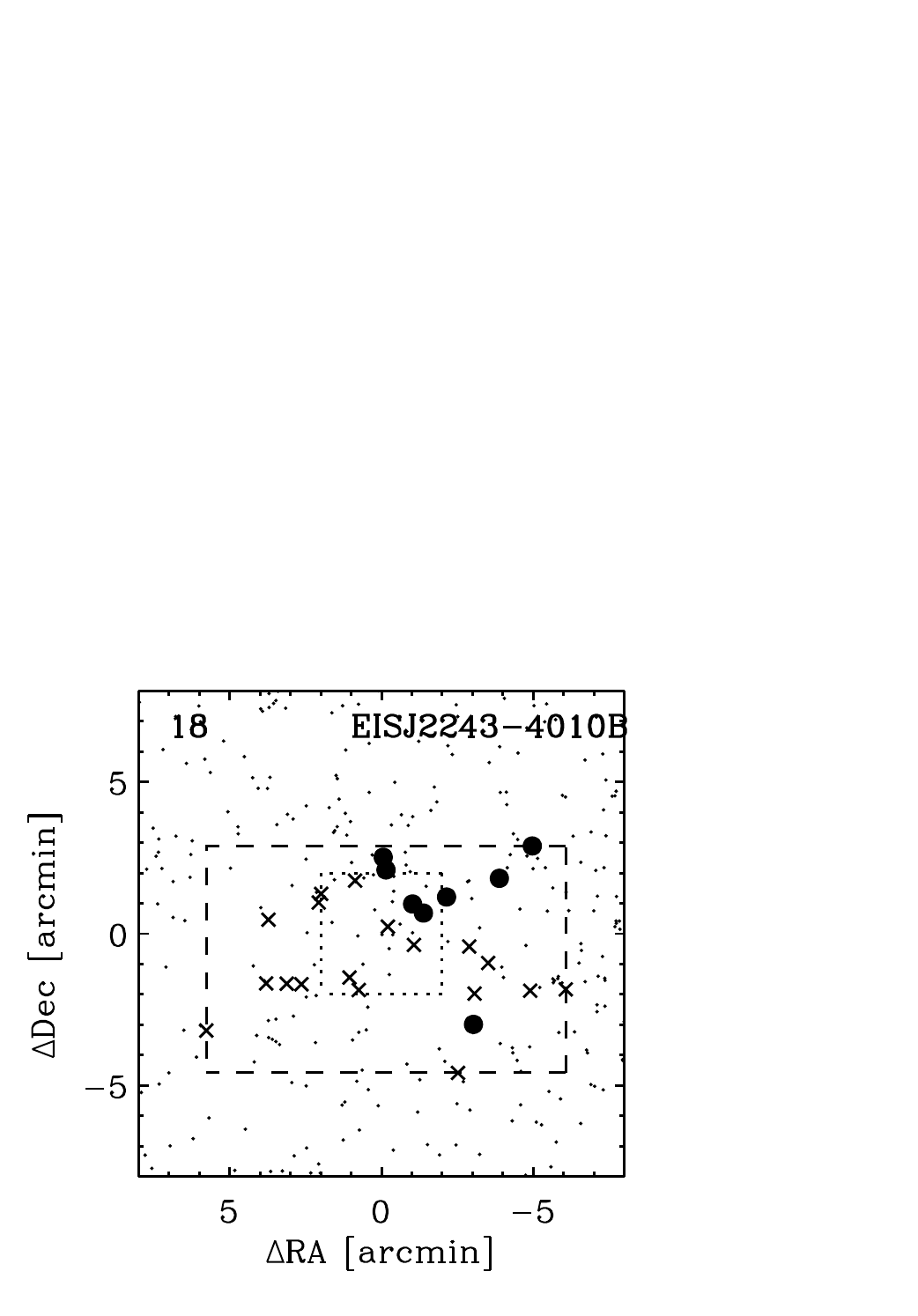}}
\resizebox{0.24\textwidth}{!}{\includegraphics[bb=0 0 226 226,clip]{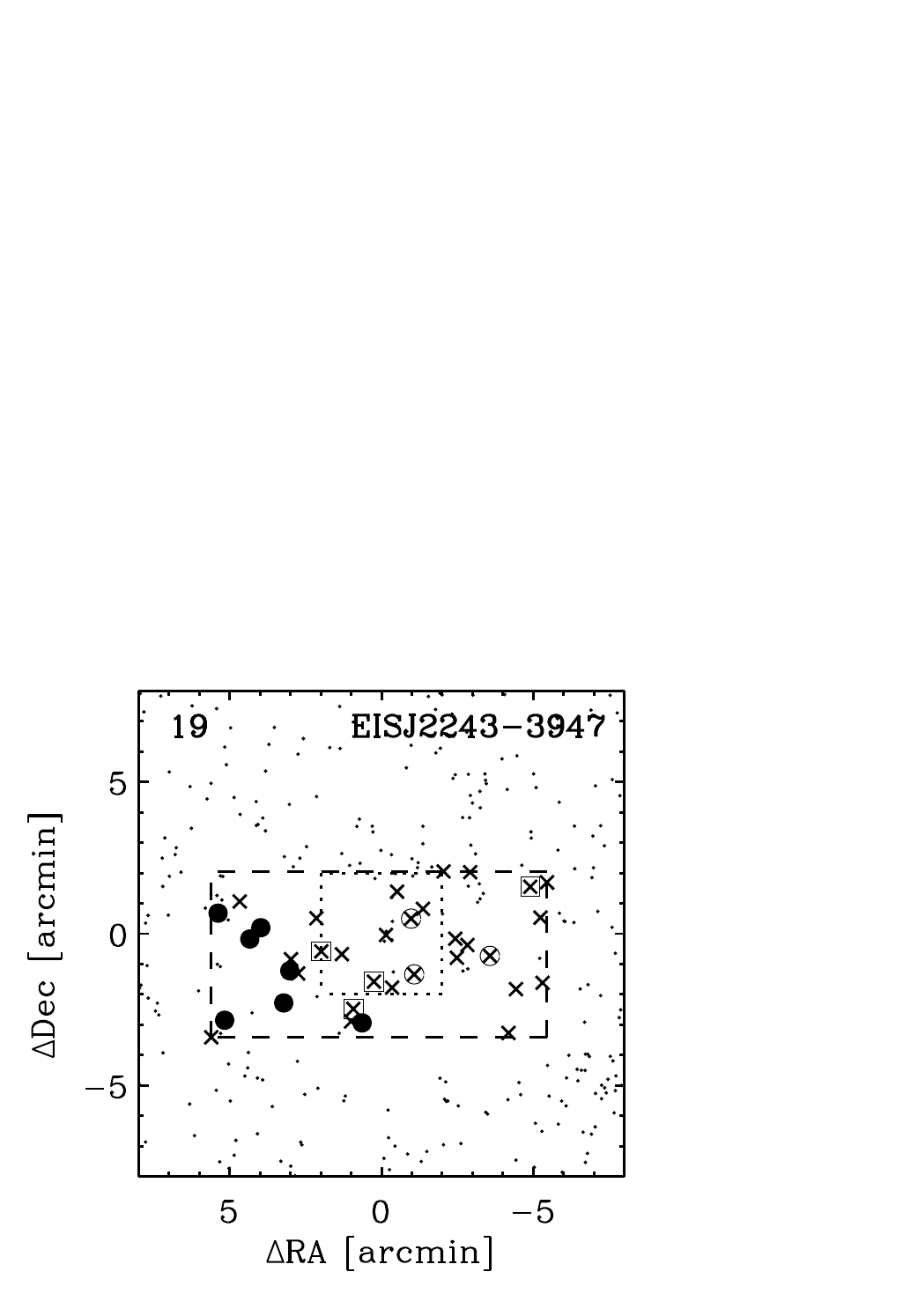}}
\resizebox{0.24\textwidth}{!}{\includegraphics[bb=0 0 226 226,clip]{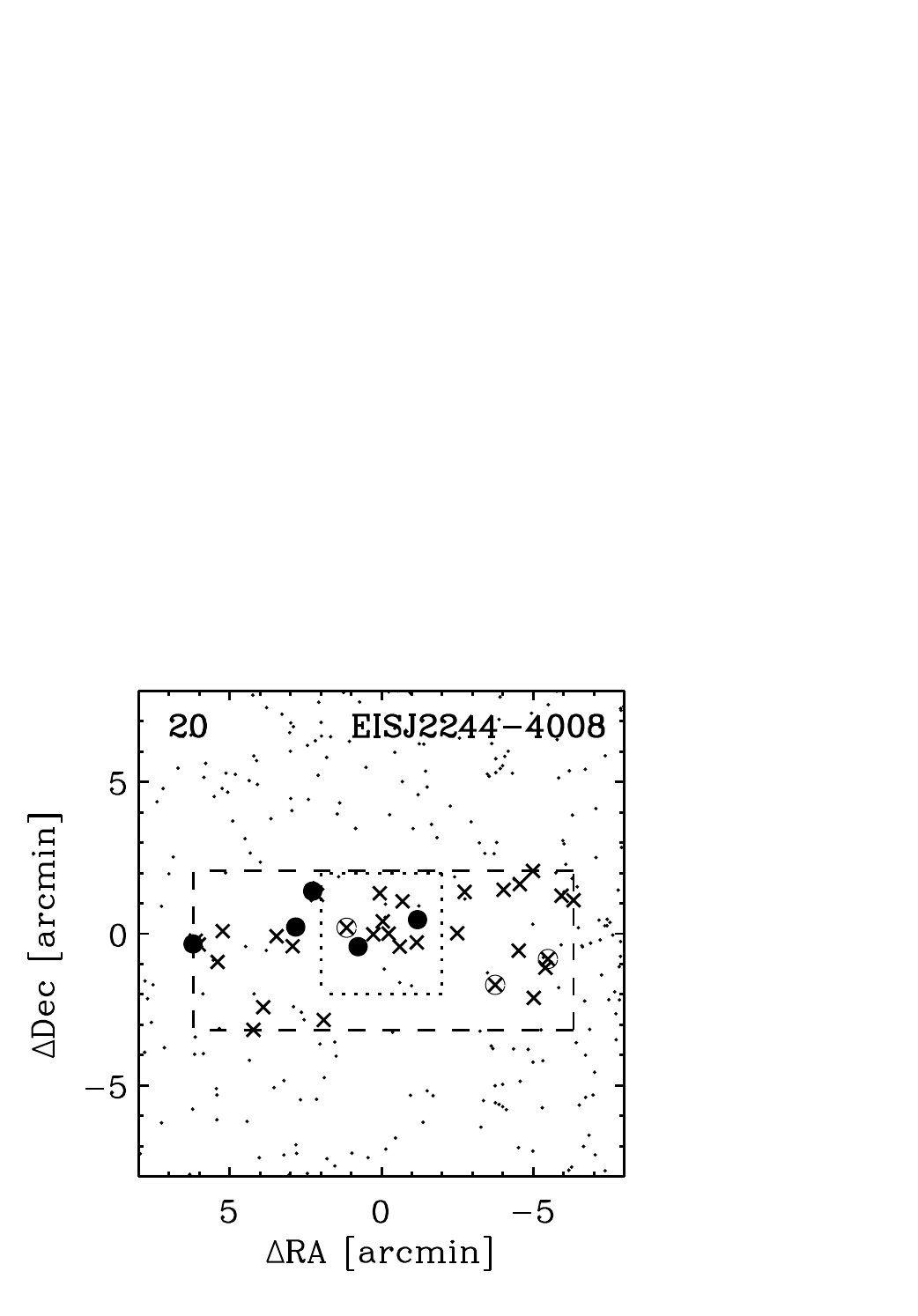}}
\resizebox{0.24\textwidth}{!}{\includegraphics{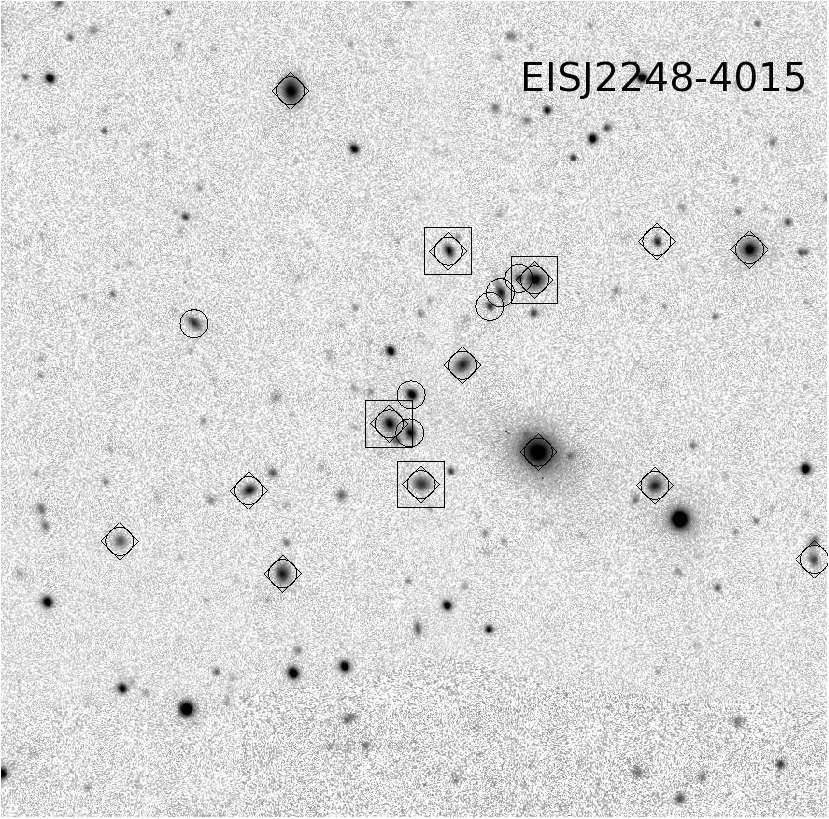}}
\resizebox{0.24\textwidth}{!}{\includegraphics[bb=0 0 226 226,clip]{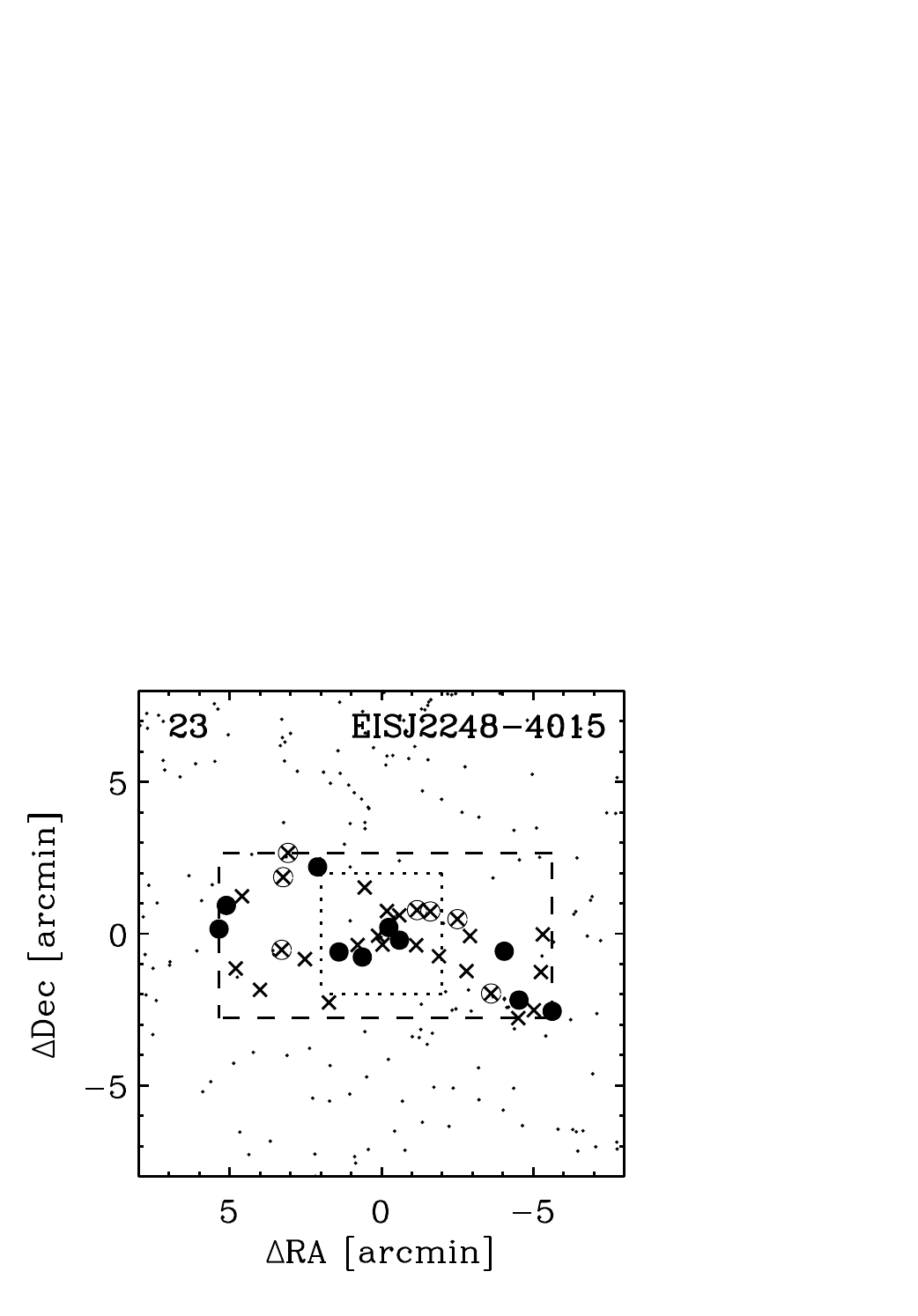}}
\end{center}
\caption{Image cutouts, centred at the original MF position and with
a size of $\sim4'$, and position plots for the nine fields with
significant groups detected but not assigned to the EIS cluster
candidate.  In the image cutouts galaxies with $I<21$ are marked by
circles and galaxies with redshifts are indicated by diamonds. In the
field of EISJ2243-3942 the galaxy marked by the big square has a
redshift of 0.2301. For EISJ2248-4015 the big squares mark four
galaxies with redshifts in the range 0.2460-0.2484. In the projected
distributions all galaxies with $I\leq20$ are indicated by the small
dots, crosses mark galaxies with redshift and each significant group
is denoted by its own symbol. The dashed rectangle marks the region of
the redshift survey and the dotted square outlines the area of the
image cutout.}
\label{fig:nonconf}
\end{figure*}

In Table~\ref{tab:EISgroups} we list groups with at least three
members and a significance $\sigma_1 \geq 99\%$ identified in each
cluster field. The remaining less significant detections are
considered marginal and given in Table~\ref{tab:marginal}. The tables
list in Cols.~1 and 2 the field identifier and the cluster field name;
in Col.~3 the number of spectroscopic members of the group; in Cols.~4
and 5 the mean position in J2000; in Col.~6 the mean redshift of the
group members; in Col.~7 the velocity dispersion corrected for our
measurement accuracy. In cases where the measured velocity dispersion
is smaller than the measurement error we list the value of
$\sigma_v=0$; in Col.~8 the significance as defined above and in
Col.~9 the distance in arcmin between the group and the original MF
position. 

\begin{figure*}
\begin{center}
\resizebox{0.24\textwidth}{!}{\includegraphics{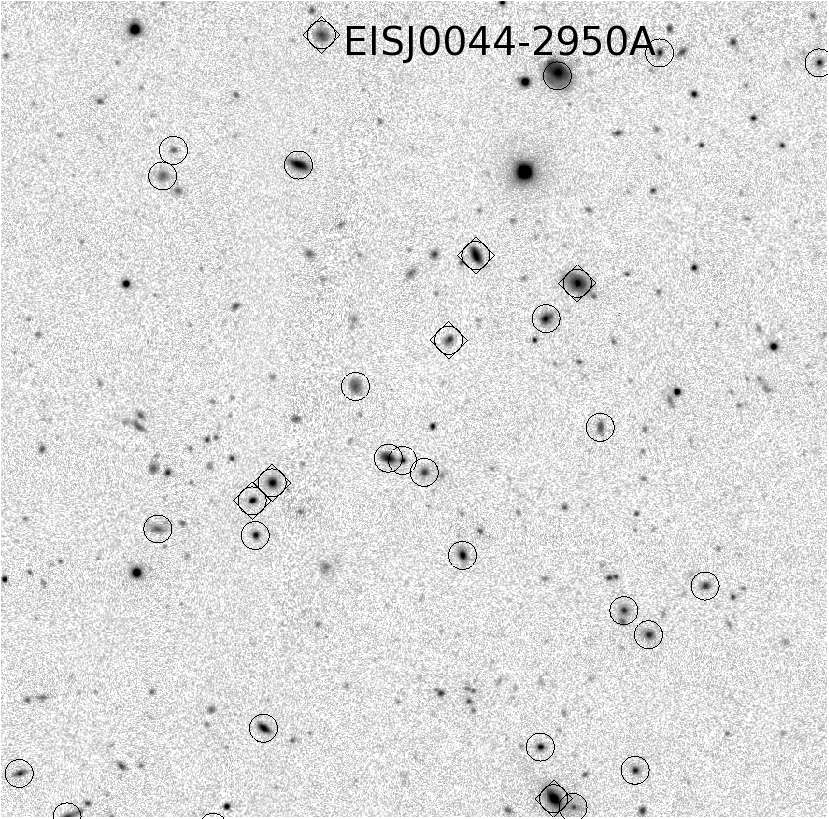}}
\resizebox{0.24\textwidth}{!}{\includegraphics{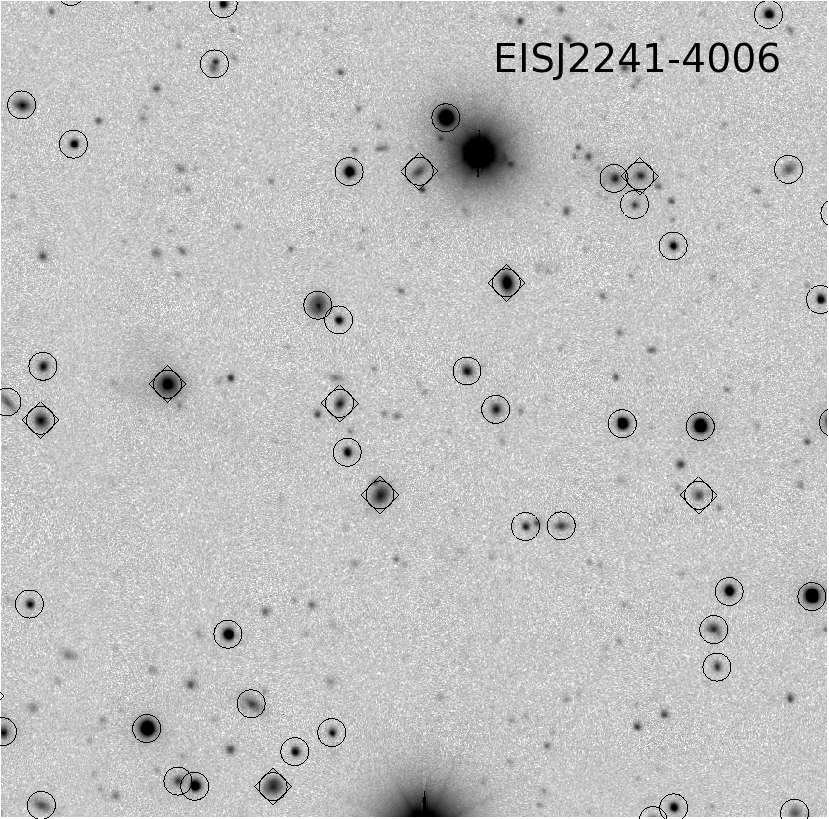}}
\resizebox{0.24\textwidth}{!}{\includegraphics{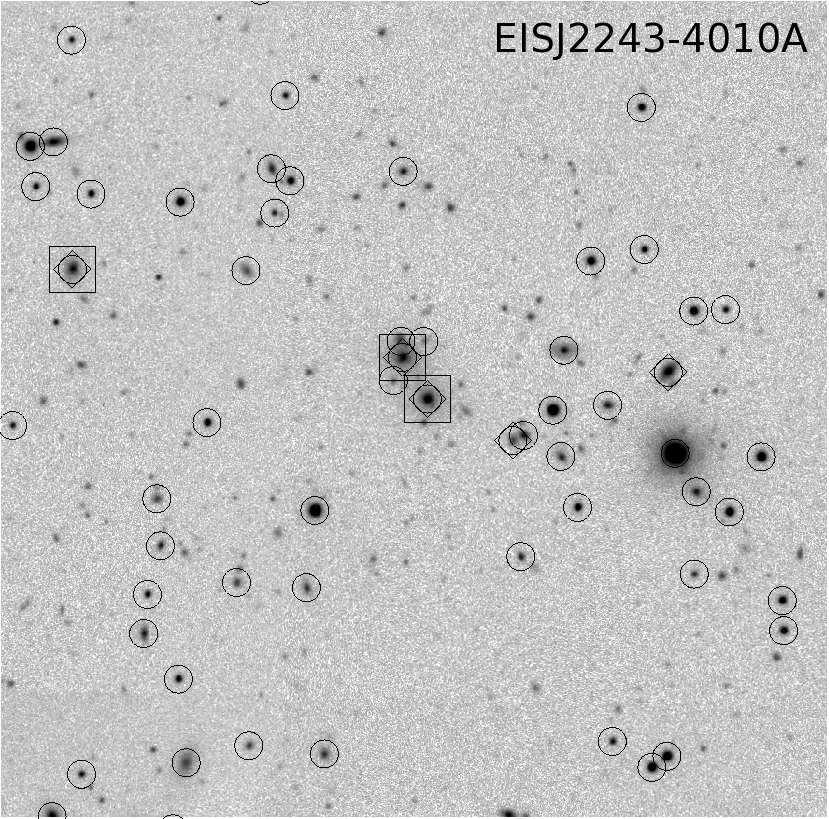}}
\resizebox{0.24\textwidth}{!}{\includegraphics{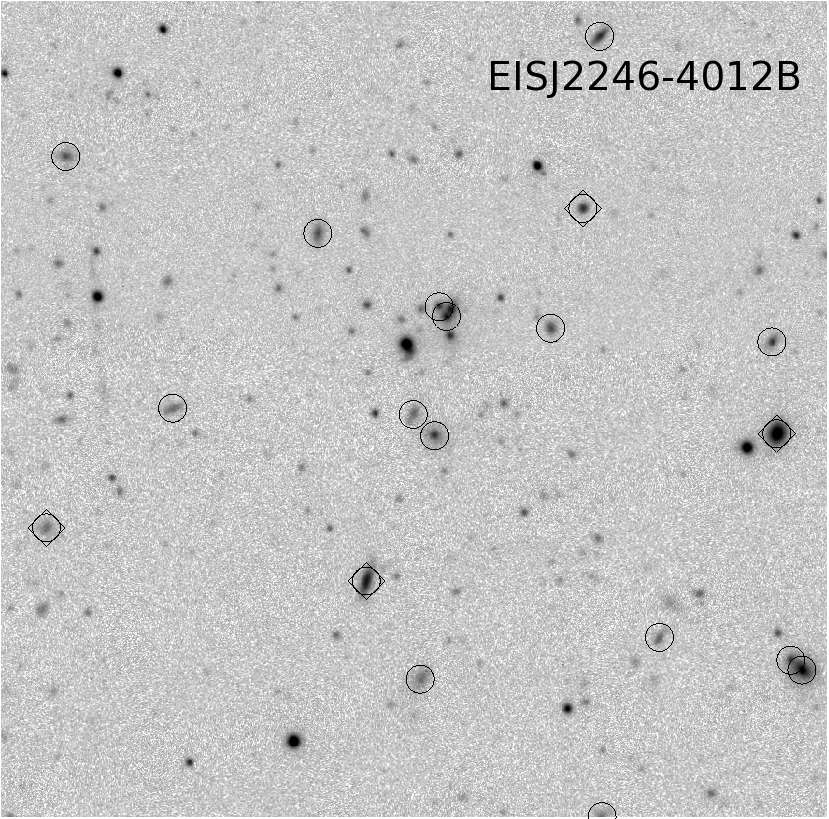}}
\end{center}
\caption{Image cutouts for the four fields without significant groups
detected. From left to right the panels show the field of
EISJ0044-2950A, EISJ2241-4006, EISJ2243-4010A and EISJ2246-4012B. The
cutouts are centred at the original position and have a size of
$\sim4\arcmin$. Galaxies with $I<21$ are marked by circles, additional
diamonds denote galaxies with measured redshifts. In the field of
EISJ2243-4010A the galaxies marked by the big squares have redshifts
in the range 0.3431-0.3471.}
\label{fig:nogroups}
\end{figure*}

The table lists 35 significant groups, ranging between zero and four
groups per cluster field, having from three to 25 members. In four
fields no significant groups were identified. As in Paper~III the
group associated with a matched filter detection is chosen as follows:
1) The richest group in the field, if it has a significantly larger
number of members than the other groups; 2) The one closest to the EIS
position, if two groups have roughly the same number of members; 3)
The most concentrated group, if two groups are close to the EIS
position and have almost the same number of members. To help in
the analysis the images and the spatial distribution of the galaxies
are shown in Figs.~\ref{fig:conf} and \ref{fig:nonconf}. The figures
show the central image cutout with $\sim4\arcmin$ on the side and the
spatial distributions of the galaxies within the surveyed area. The
projected distribution includes all galaxies with $I\leq20.0$ in the
cluster regions. In the figures symbols indicate galaxies with
measured redshift associated (solid symbols) or not (crosses) to a
group. In Fig.~\ref{fig:conf} only the group associated to the EIS
detection is marked. Note that small and large symbols are shown in
four cases. This is because examination of the redshift distribution
of these groups suggests the presence of sub-structures. The circle
has a radius of $0.5h_{75}^{-1}\mathrm{Mpc}$ at the redshift of the
confirmed group and is centred at the original position of the matched
filter detection.

From the analysis we find that in 19 of the 23 fields there are
between one and four significant overdensities in redshift space. In
10 fields (43\% of the entire set) we assigned one of the significant
detections to the EIS cluster. These detections are marked in bold
face in Table~\ref{tab:EISgroups}. In each field the assigned group
turned out to be the one closest to the original position with
distances ranging between $0.1\arcmin$ and $1.9\arcmin$. It can be
seen that most of the assigned groups are found within $1\arcmin$
corresponding to a few core radii at the redshifts discussed here. For
the other groups the distance to the original position is in general
much larger than this.  All, except one, of the 10 confirming
systems have at least six members located within a radius of
$0.5h^{-1}_{75}$Mpc. Whether these systems are bound or not can only
be properly assessed by much denser spectroscopic
observations. However, \cite{gal2008} found that with this search
radius, it is common to find un-bound systems with up to four members,
but not six. Therefore, we consider it likely, but not conclusive,
that the assigned groups represent bound systems.  As discussed
below, we should point out that close inspection of the redshift
distribution of one of the fields (EISJ2238-3934, \#10) led us to
split this system into two.  Another system worth mentioning is
EISJ2236-4026 (\#7) that was associated to a lower redshift group
despite the poor sampling of the luminosity function as indicated by
the ratio of faint to bright galaxies, and the lack of measured
redshifts at its originally estimated redshift.  This association was
based on the visual inspection of the image, which does not show sign
of a concentration of fainter galaxies.  Finally, note that due to the
poor sampling of the luminosity function at $z\geq0.4$ only one system
was confirmed at this redshift from the present data, demonstrating
the intrinsic bias of our results.
  
For nine fields, significant groups are identified but not assigned to
an EIS detection, even though in some cases the redshift is comparable
to the matched-filter estimates. This is because they fail in some
other respect. For instance, in some cases, we discard a group as the
true confirmation due to poor coverage of the central regions or
strong evidence of a concentration of fainter galaxies at the
matched-filter candidate position. The analysis shows that for one
cluster (EISJ2248-4015, \#23) a marginal detection with a
significance of 98.7\% is likely the origin of the detection. A
detailed account for each group is presented in
Appendix~\ref{app:ind_cases}.

Finally, for the remaining four fields, no significant group was
detected. The image cutouts of these fields are shown in
Fig.~\ref{fig:nogroups}. From the images it can be seen that for the
clusters EISJ0044-2950 (\#1) and EISJ2241-4006 (\#13) there is little
indication of a concentration of galaxies. The origin of the matched
filter detection is in these cases unclear. In the field of
EISJ2243-4010A a compact group of galaxies dominated by two bright
galaxies is found. The two bright galaxies have very similar redshifts
consistent with that originally estimated for the cluster
($z_{MF}=0.3$). This is suggestive of a group of galaxies but in
redshift space the group only has a significance of 73.4\% (listed in
Table~\ref{tab:marginal}), thus is not part of the sample considered
significant. Additional spectroscopy of the region may confirm the
existence of a group of galaxies. In the last field, EISJ2246-4012B
(\# 22) essentially no redshifts were obtained in the central region
of the field. This field was only covered by one slit mask due to the
lack of time. The weak concentration of galaxies in the centre may
represent a group of galaxies. The confirmation of this requires
additional spectroscopic data.

To summarise, we find 10 candidates confirmed by the present
spectroscopic data. For the remaining 13 fields the available data are
not sufficient to assign a significant group to the EIS
detection. However, careful examination of the available data showed
that in two cases a marginally significant detection in redshift space
is likely the origin of the matched-filter candidate. Four systems
($\sim20\%$) do not show any signs of a concentration of galaxies and
the origin of the matched filter detection remains unclear. For the
remaining seven cases the available data are insufficient to draw a
final conclusion. Of the original sample 14 were at $z_{MF}=0.3$ of
which we confirm 8 ($\sim57\%$). Of the 9 candidates with $z_{Mf}=0.4$
we confirm only 2 ($\sim22\%$). 

Below we investigate in more detail the properties of the confirmed
systems.  Table~\ref{tab:colour} summarises these results and will be
used throughout the rest of the paper. The table gives in Col.~1 the
field identifier; in Col.~2 the name of the cluster field; in Col.~3
the number of spectroscopic members; in Col.~4 the matched filter
redshift estimate; in Col.~5 the spectroscopic redshift; in Col.~6 the
velocity dispersion with 68\% bootstrap errors; in Col. 7 updated
$\Lambda_{cl, new}$-richness as described below; in Col.~8 and 9 the
colour of the identified photometric red sequence and the confidence
level of its detection as described in Sect.~\ref{sec:colours}; in
Col.~10 and 11 the colour of the red sequence of the spectroscopic
members and its significance (Sect.~\ref{sec:colours}); and in Col.~12
the measured colour scatter for the spectroscopic members.  The five
systems without colour information are indicated by N.A. in the table.

\begin{table*}
\caption{Properties of the confirmed EIS clusters and groups.}
\label{tab:colour}
\begin{center}
\begin{tabular}{rlrrrrlrrrrrr}
\hline\hline 

Id & Cluster & \#mem & $z_{MF}$ & $z_{spec}$ &
\multicolumn{2}{c}{$\sigma_v \mathrm[km/s]$ } & $\Lambda_{cl, new}$ &
$(V-I)_{ph}$ & $\sigma_{S/N}$ & $(V-I)_{sp}$ & $\sigma_{spec}$ &
Scatter\\

\hline

\vspace{1mm}

3 & EISJ0047-2942 &  6 & 0.4 & 0.534 &  $ 617$ & $^{+ 158}_{- 473}$  & 64.6& $-$ & $-$ & $-$   & $-$ &$-$\\ 

\vspace{1mm}

5 & EISJ0049-2920 &  10 & 0.3 & 0.187 &  $ 893$ & $^{+ 266}_{- 676}$ &23.9 & $-$ & $-$  & 1.500   & 99.8 & 0.151\\ 

\vspace{1mm}

6 & EISJ2236-3935 &  9 & 0.3 & 0.158 &  $1160$ & $^{+  60}_{-1004}$  & 13.3& N.A. & N.A.  & N.A.   & N.A. & N.A. \\ 

\vspace{1mm}

7 & EISJ2236-4026 &  9 & 0.4 & 0.184 & $ 187$ & $^{+  48}_{- 115}$ & 31.8& $-$ & $-$ &  1.575  & 99.6  & 0.104\\ 

\vspace{1mm}

9 & EISJ2237-4000 &  9 & 0.3 & 0.196 & $ 373$ & $^{+ 344}_{- 299}$   & 11.9& N.A.  & N.A.  &  N.A.  & N.A.  & N.A.\\ 

\vspace{1mm}

10 & EISJ2238-3934 &  25 & 0.3 & 0.243 &  $1433$ & $^{+ 158}_{- 167}$  & 34.9 & N.A.  & N.A.  &  N.A.  & N.A.  & N.A.\\ 

\vspace{1mm}

11 & EISJ2239-3954 &  21 & 0.3 & 0.195 &   $ 475$ & $^{+  93}_{- 117}$  & 30.9& N.A.  & N.A.  &  N.A.  & N.A.  & N.A.\\ 

\vspace{1mm}

12 & EISJ2240-4021 &  8 & 0.3 & 0.247 &   $ 268$ & $^{+  57}_{- 148}$  & 20.8& 1.800 & 94.8 & $-$   &$-$  & $-$\\ 

\vspace{1mm}

17 & EISJ2243-3959 &  10 & 0.3 & 0.285 &   $ 175$ & $^{+  41}_{- 115}$  & 38.1& N.A. & N.A. & N.A.   & N.A.  & N.A.\\ 

%\vspace{1mm}
%
%20 & EISJ2244-4008 & 5 & 0.4 & 0.471 & $ 127$ & $^{+  47}_{-  25}$ & 24.9& $-$  & $-$  &  $-$  &$-$  &$-$\\ 

\vspace{1mm}

21 & EISJ2244-4019 & 10 & 0.3 & 0.216 &  $ 393$ & $^{+  73}_{- 164}$ & 22.9 & 1.500  & 96.6 & 1.425   & $>99.9$ & 0.103 \\ 

\hline
\end{tabular}
\end{center}
\end{table*}

\begin{table*}
\caption{Updated properties of the groups showing substructure.}
\label{tab:colour_update}
\begin{center}
\begin{tabular}{rlrrrrrrlr}
\hline\hline 

Id & Cluster & $\alpha$ (J2000) & $\delta$ (J2000) & \#mem & $z_{MF}$ & $z_{spec}$ &
\multicolumn{2}{c}{$\sigma_v \mathrm[km/s]$ } \\

\hline

\vspace{1mm}

3 & EISJ0047-2942 &  00:47:17.4 & -29:42:33 & 4 & 0.4 & 0.532 &  $ 251$ & $^{+109 }_{-101 }$  \\

\vspace{1mm}

5 & EISJ0049-2920 &  00:49:31.7 & -29:20:33 & 7 & 0.3 & 0.189 &  $ 248$ & $^{+116}_{-112}$ \\

\vspace{1mm}

6 & EISJ2236-3935 &  22:35:59.9 & -39:36:45 & 6  & 0.3 & 0.161 &  $284$ & $^{+258}_{-169}$  \\

\vspace{1mm}

9 & EISJ2237-4000 &  22:37:11.3 & -40:00:23 & 8 & 0.3 & 0.195 & $ 350$ & $^{+225}_{-171}$   \\

%\vspace{1mm}

10 & EISJ2238-3934 &  22:38:06.3 & -39:34:03 & 13 & 0.3 & 0.240 &  $497$ & $^{+55}_{-148}$  \\
\vspace{1mm}

10 & EISJ2238-3934 &  22:37:54.0 & -39:35:51 & 10 & 0.3 & 0.249 &  $357$ & $^{+55}_{-99}$  \\

\hline
\end{tabular}
\end{center}
\end{table*}

\subsection{Reliability of the matched-filter redshifts}

\begin{figure}
\resizebox{0.9\columnwidth}{!}{\includegraphics[bb=0 0 283 198,clip]{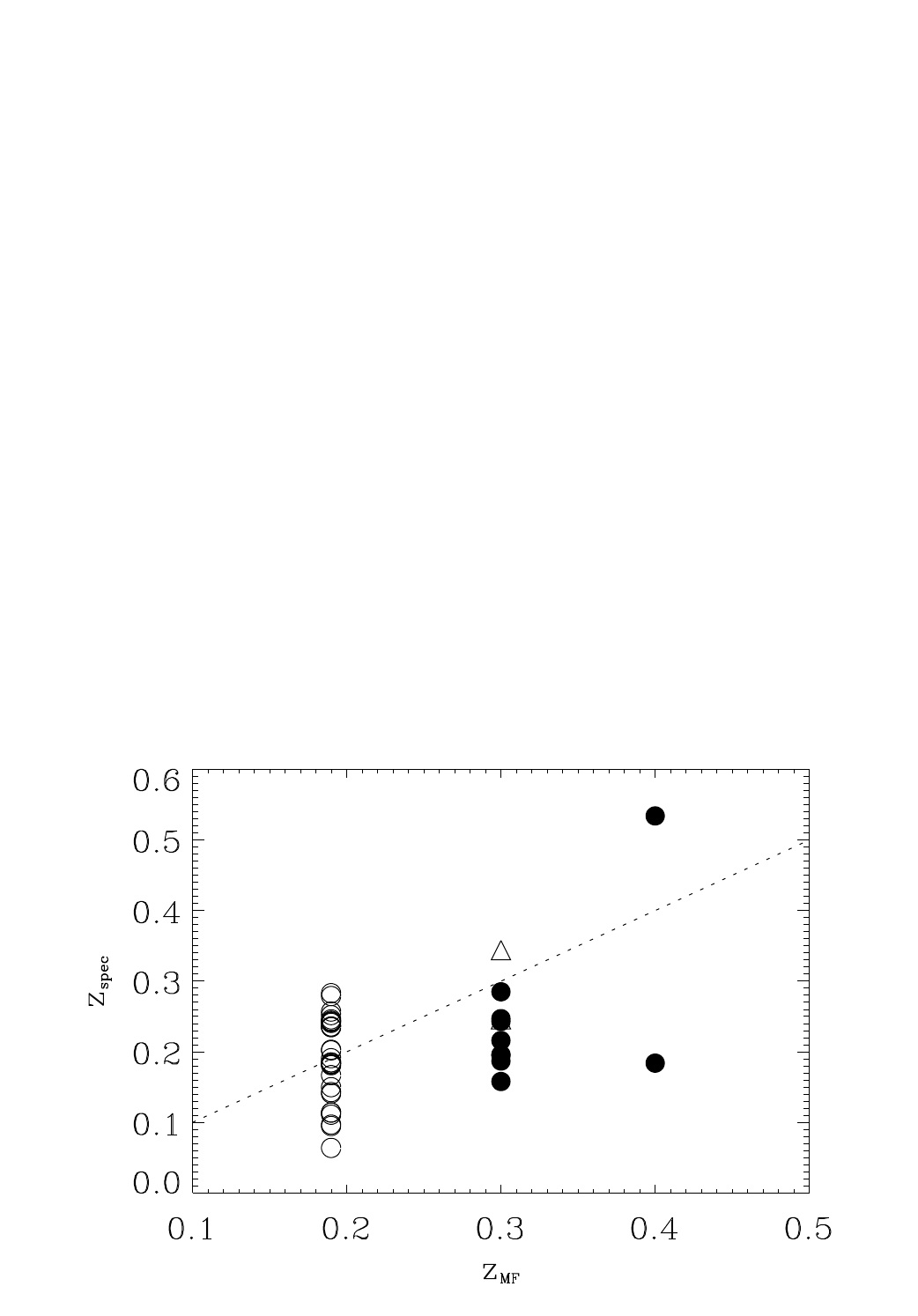}}
\resizebox{0.9\columnwidth}{!}{\includegraphics[bb=0 0 283 198,clip]{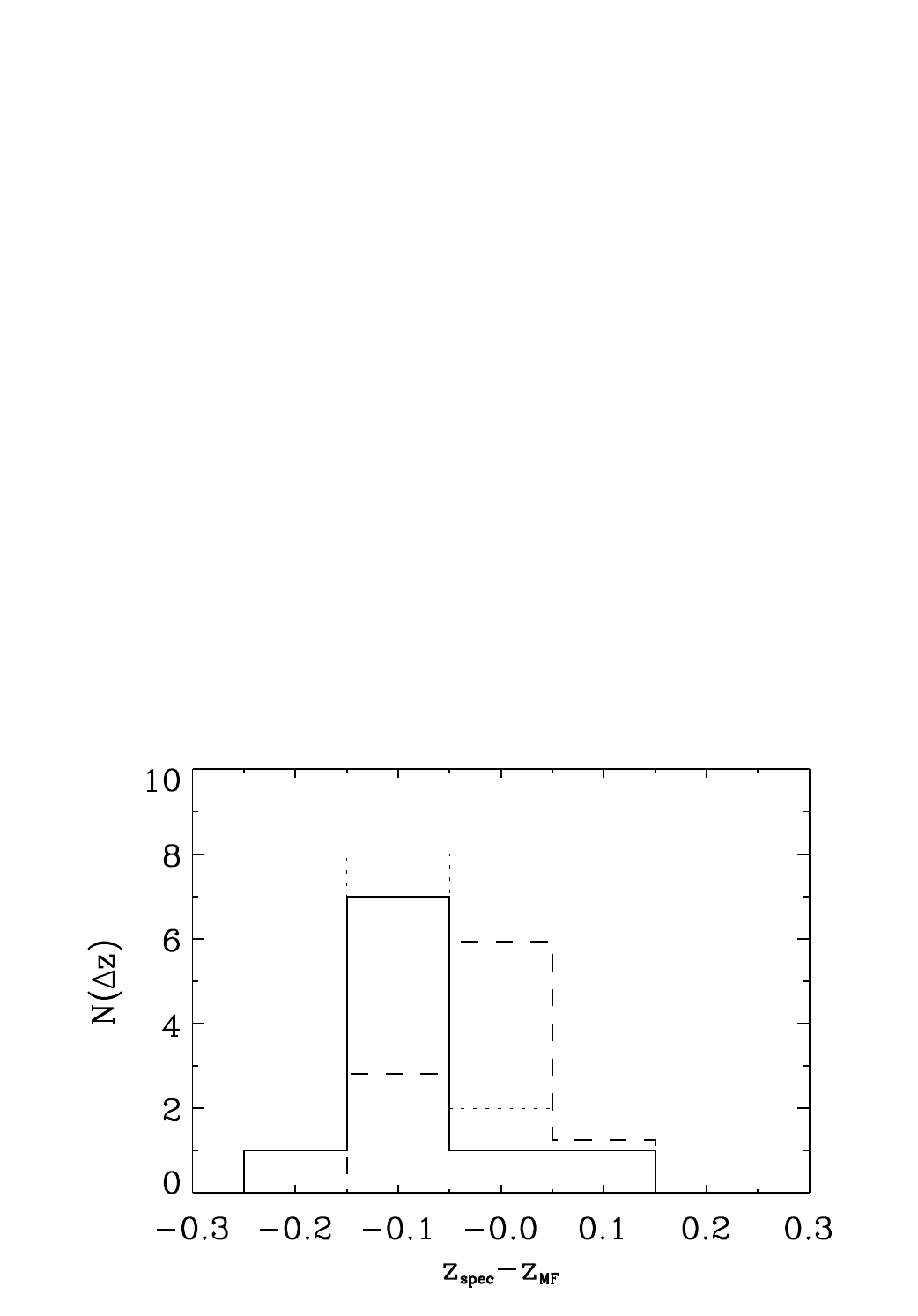}}
\caption{ Upper panel: The spectroscopic versus matched filter
redshifts. Solid circles mark the confirmed systems, open circles mark
systems from Paper~III and triangles mark those systems that have a
suggestive spectroscopic redshift. Lower panel: The distribution of
the offset between the spectroscopic and matched-filter estimated
redshifts of all the confirmed systems in this paper (solid line) and
in Paper~III (dashed line, scaled to match the same number of
objects). The dotted histogram includes the two systems with a
suggestive spectroscopic redshift.}
\label{fig:redshifts}
\end{figure}

For any catalogue of cluster candidates it is important to establish
the reliability of the estimated redshifts, since systematic
spectroscopic follow-up is, in general, not possible. Here we
investigate the reliability of the redshifts estimated by the
matched-filter algorithm. The redshift estimates, reported in
Table~\ref{tab:colour}, are obtained using the mean values for
consistency with previous papers. Due to the sample sizes we have
investigated the consistency of these mean values with the results of
the median and biweight estimators \citep[e.g.][]{beers90}. It was
established that the three measures agreed to within $\delta
z\sim0.002$ (Paper~III).

A summary of the comparison between the estimated and measured
redshifts is presented in Fig.~\ref{fig:redshifts}. In the upper panel
we show $z_{spec}$ versus $z_{MF}$ obtained for the sample of clusters
presented in this paper (filled circles). In the figure we also
include points from the $z_{MF}=0.2$ sample of paper III (open
circles) and for completeness the results obtained for the two
marginal detections discussed above (open triangles). In the lower
panel we show a normalised distribution of the redshift differences,
comparing that of the present sample (solid line) with that of
Paper~III (dashed line). The case of EISJ2236-4026 (\#7) stands
out as having a large discrepancy between the estimated and measured
redshift. However, its impact on the average values reported below is
only marginal.

From the upper panel it is seen that, except for one system, all have
measured redshifts smaller than 0.3, yielding a mean redshift of
$\langle z_{spec}\rangle =0.24\pm0.11$, only marginally higher than
the value of $\langle z_{spec}\rangle=0.18\pm0.06$ obtained for the
Paper~III sample, and definitely not as high as one would have
expected from our initial selection of systems with estimated
redshifts in the range $0.3\leq z_{MF}\leq0.4$. However, as mentioned
before this is most likely a consequence of the observational bias
created by the relatively bright limiting magnitude of the present
spectroscopic survey. As mentioned in the last section over half of
the systems in our sample have not been confirmed by the present data,
with some showing clear evidence for the presence of concentrations of
fainter galaxies. 

In this paper we find that a mean redshift difference of $\langle
\Delta z\rangle=\langle z_{spec}-z_{MF}\rangle= -0.08\pm0.1$. This is
consistent with the value of $\Delta z=-0.1$ estimated by
\cite{olsen00} from simulated data, but significantly larger than the
value of $\langle \Delta z\rangle=-0.01\pm 0.06$ obtained for the
$z_{MF}=0.2$ sample of Paper~III. Disregarding EISJ2236-4026 (\#7)
from consideration would change the average offset for the present
sample to $\langle \Delta z\rangle=-0.06\pm 0.08$, thus only a minor
difference.

Another potential source of bias that we have considered may arise
from the fact that the luminosity function of clusters of galaxies
varies with the mass of the system \citep{zandivarez2006} with poorer
systems having fainter magnitudes. If this is the case, conceivably
the difference between the matched-filter and spectroscopic redshifts
would vary depending on the cluster population being considered. If
this effect plays a significant role we would expect to find a
relation ship between the measured velocity dispersions and the
redshift offsets. However, such a relation is not found in our data,
so we conclude that this effect is not causing any significant bias in
this work.

\subsection{Velocity dispersions}

\begin{figure*}
\begin{center}
\resizebox{0.23\textwidth}{!}{\includegraphics[bb=0 0 283 198,clip]{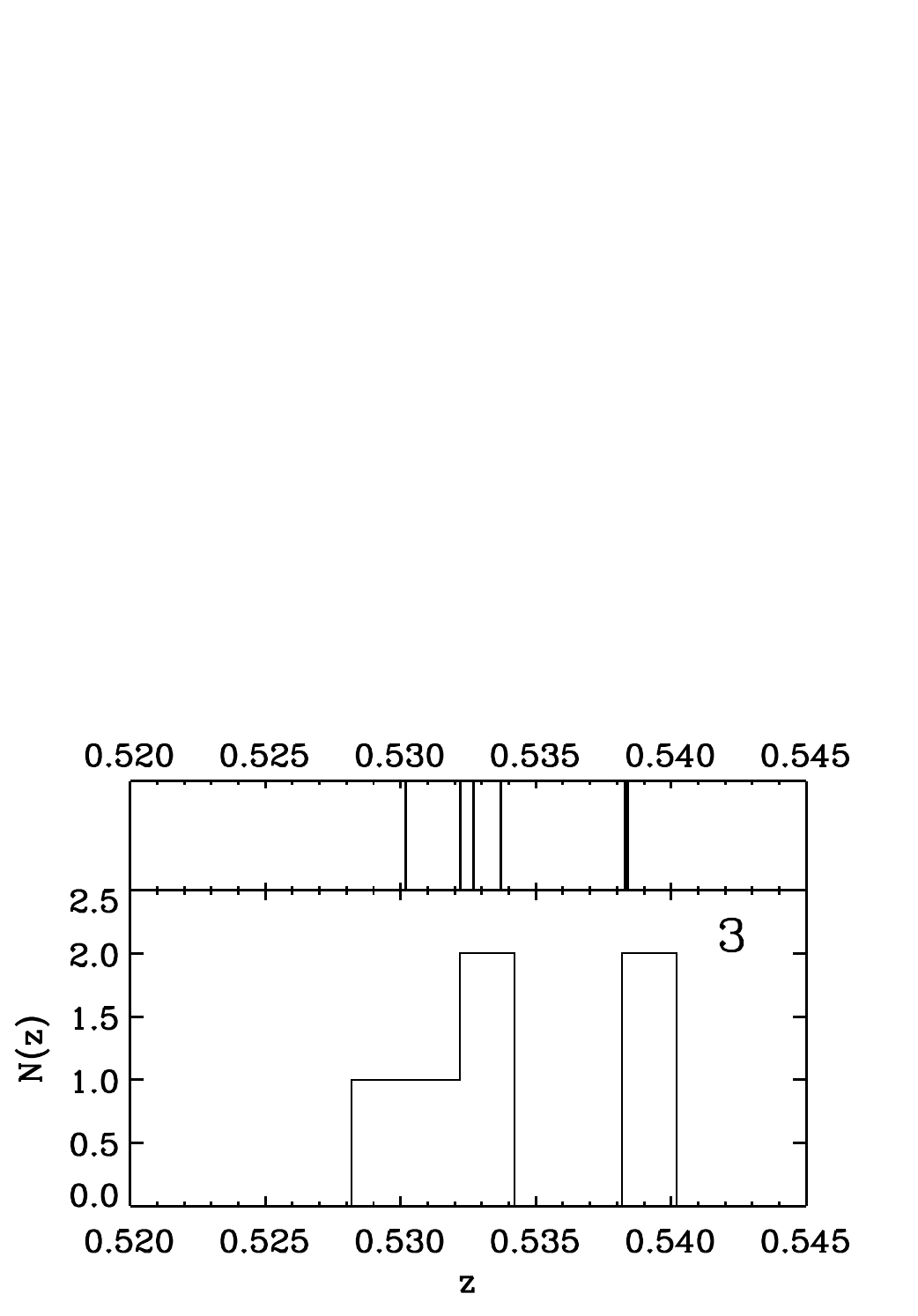}}
\resizebox{0.23\textwidth}{!}{\includegraphics[bb=0 0 283 198,clip]{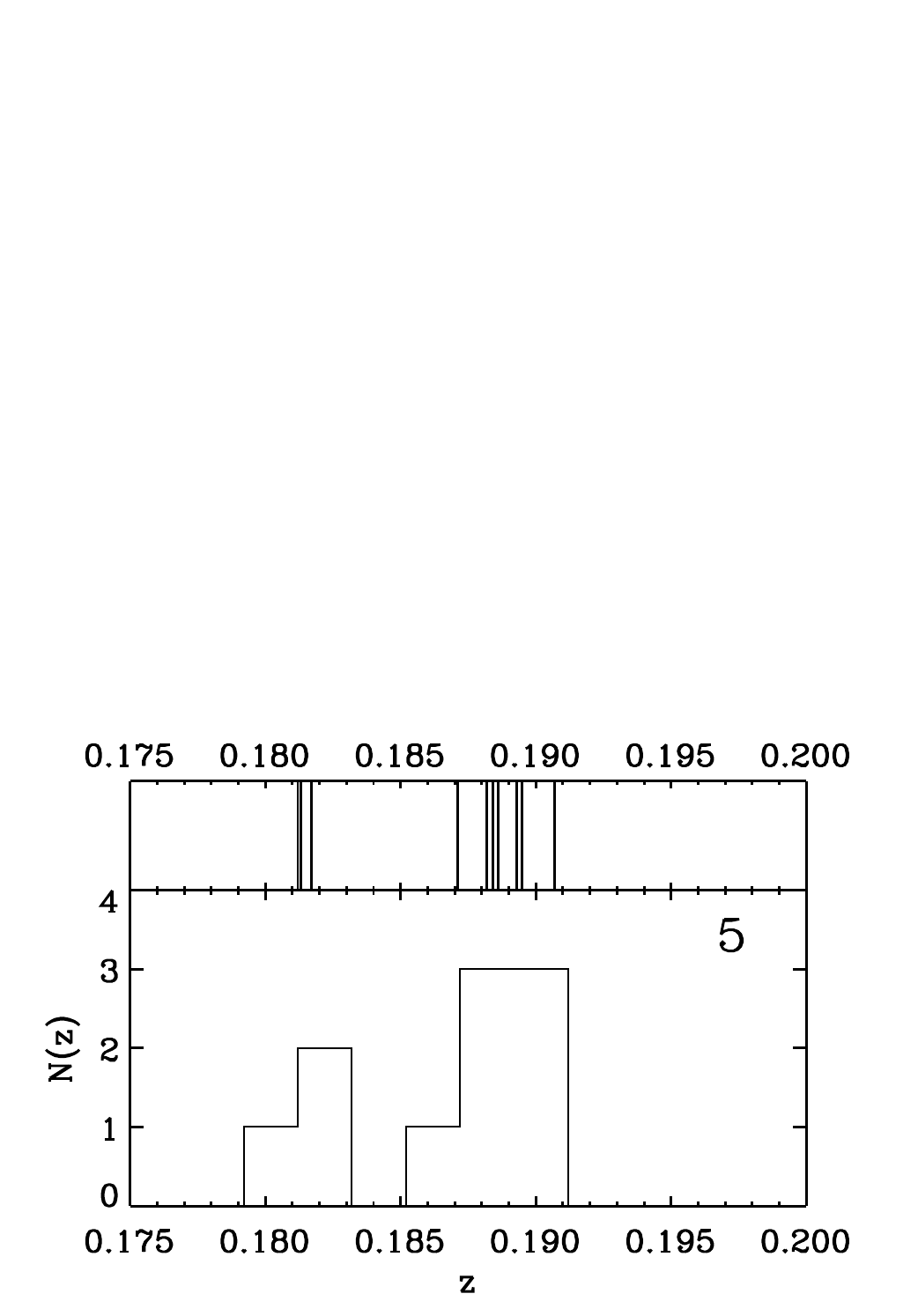}}
\resizebox{0.23\textwidth}{!}{\includegraphics[bb=0 0 283 198,clip]{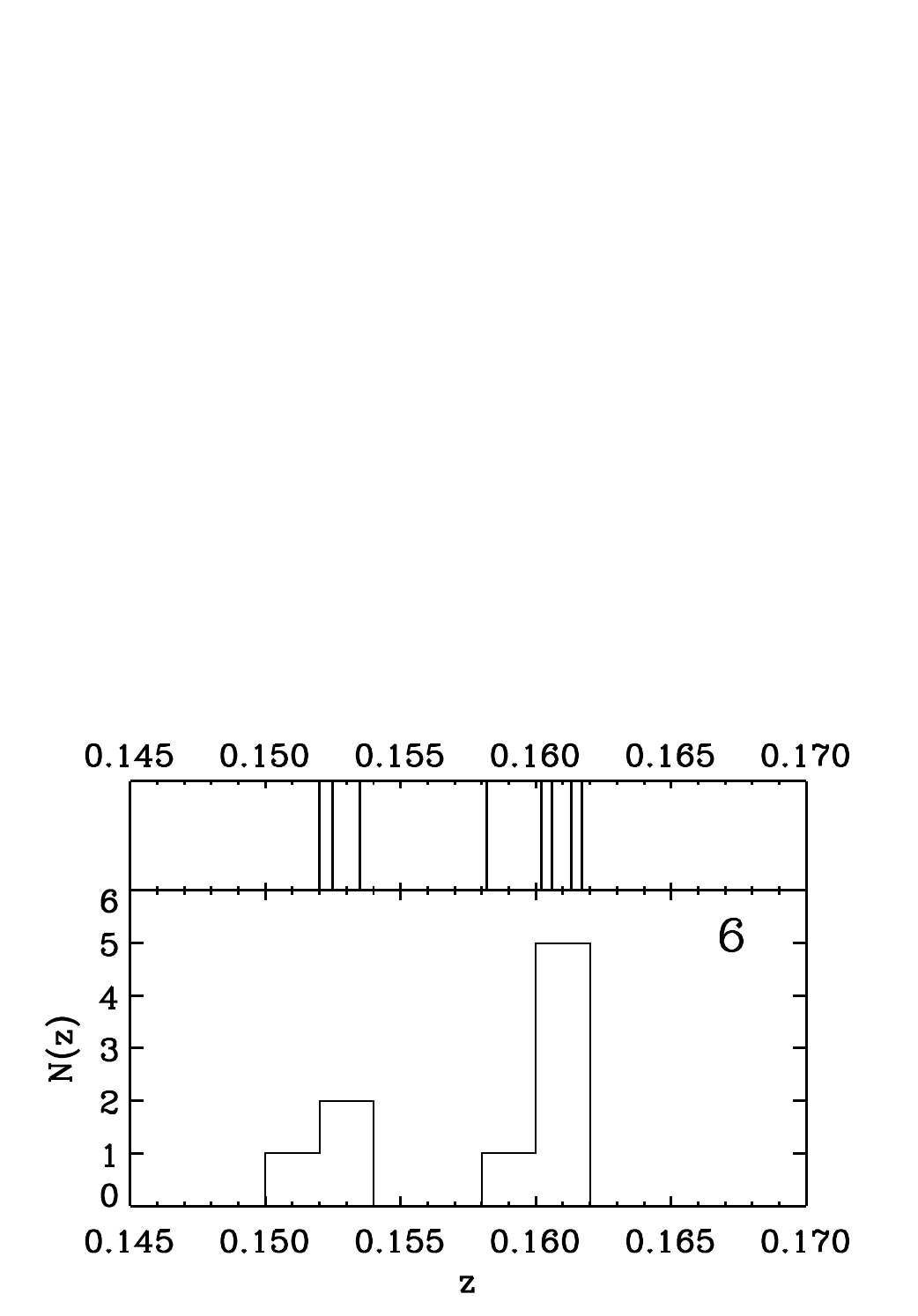}}
\resizebox{0.23\textwidth}{!}{\includegraphics[bb=0 0 283 198,clip]{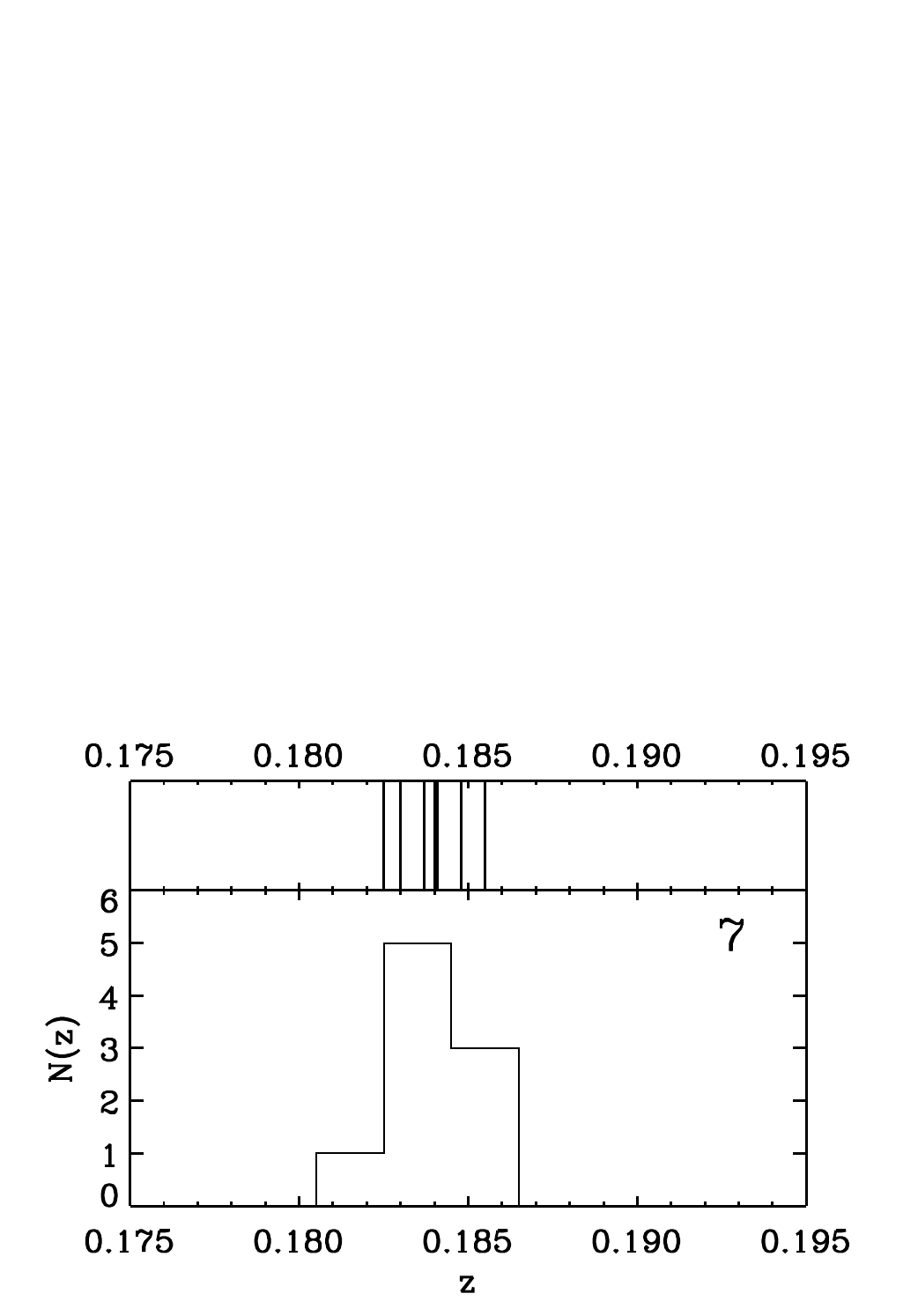}}
\resizebox{0.23\textwidth}{!}{\includegraphics[bb=0 0 283 198,clip]{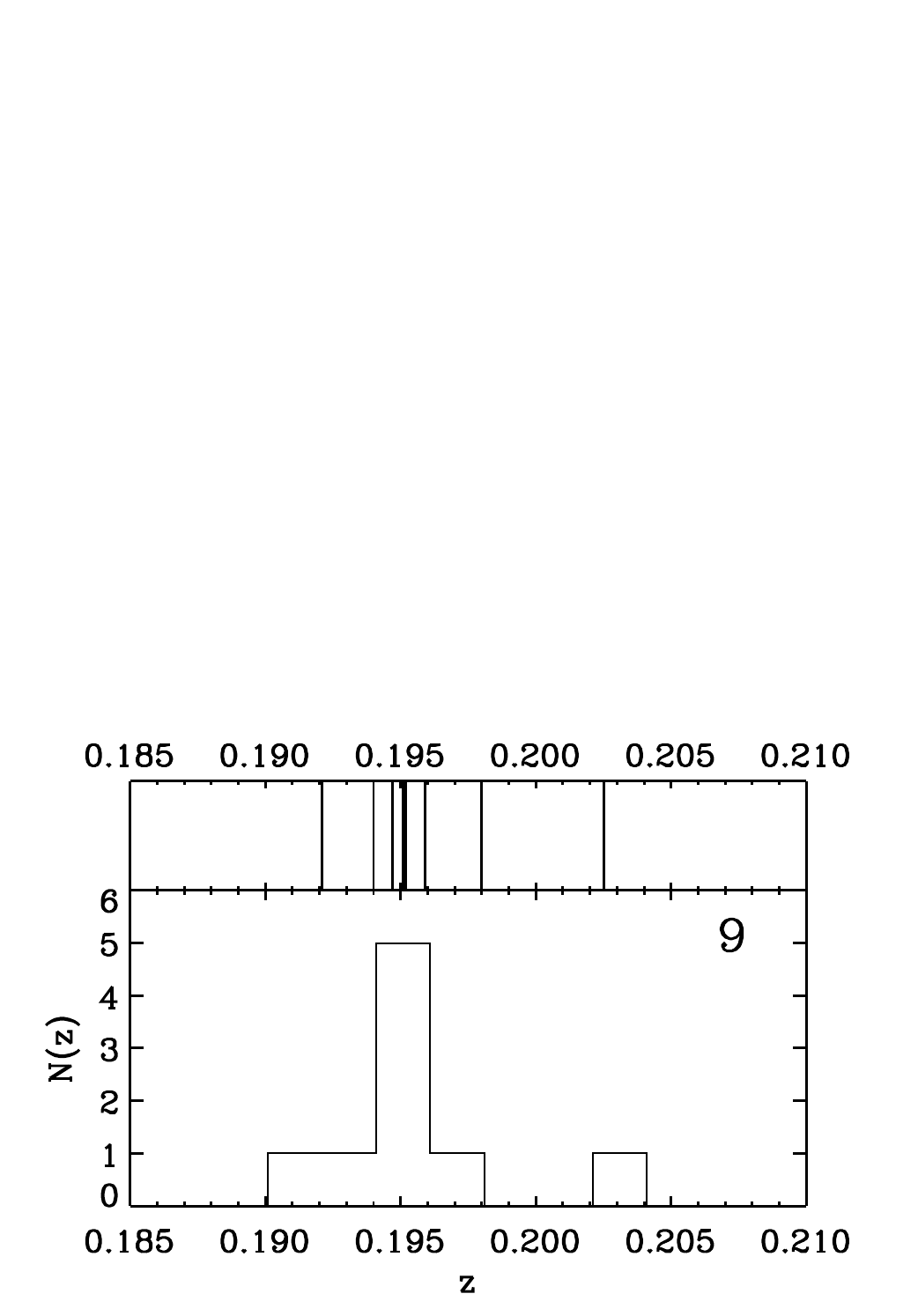}}
\resizebox{0.23\textwidth}{!}{\includegraphics[bb=0 0 283 198,clip]{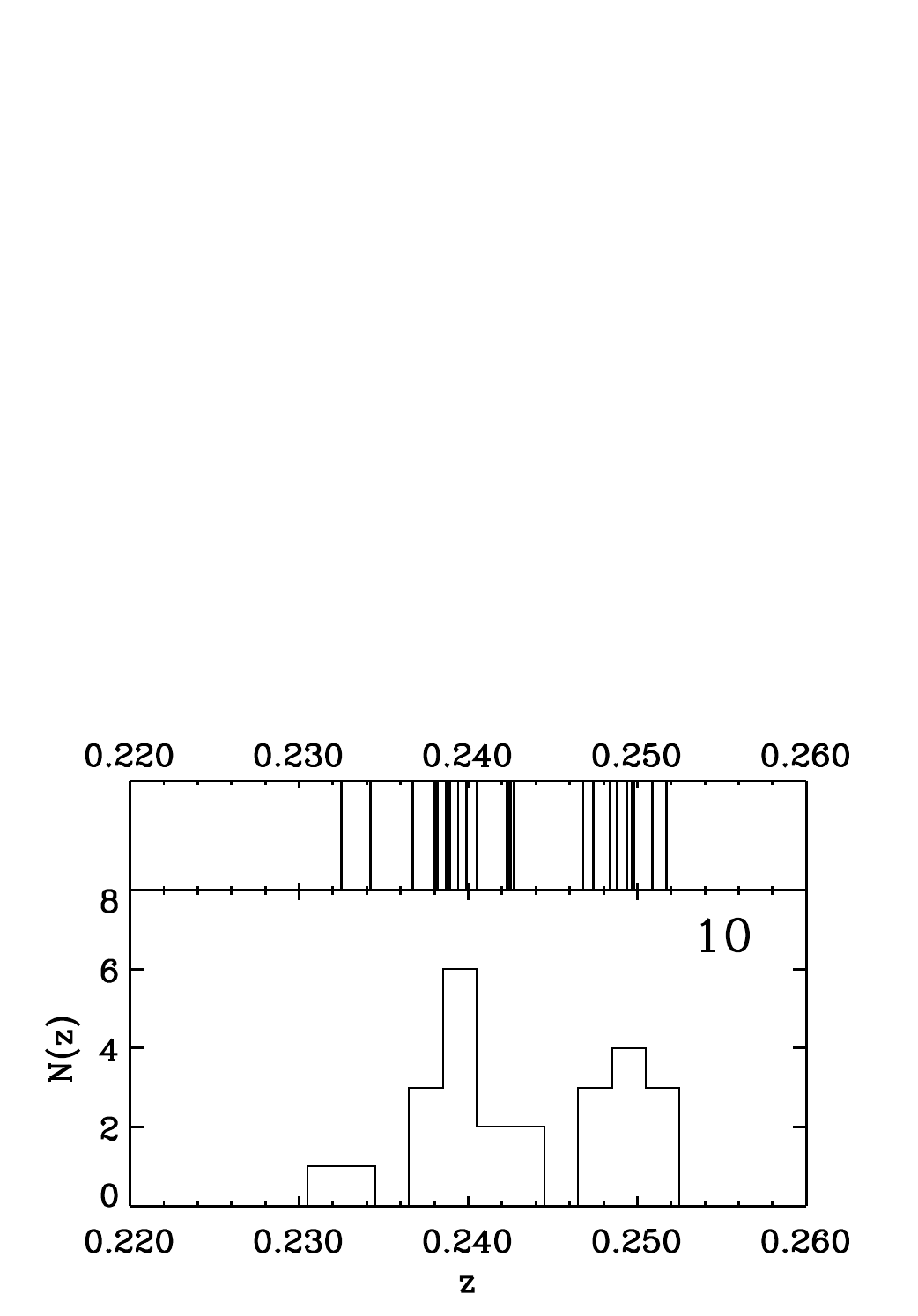}}
\resizebox{0.23\textwidth}{!}{\includegraphics[bb=0 0 283 198,clip]{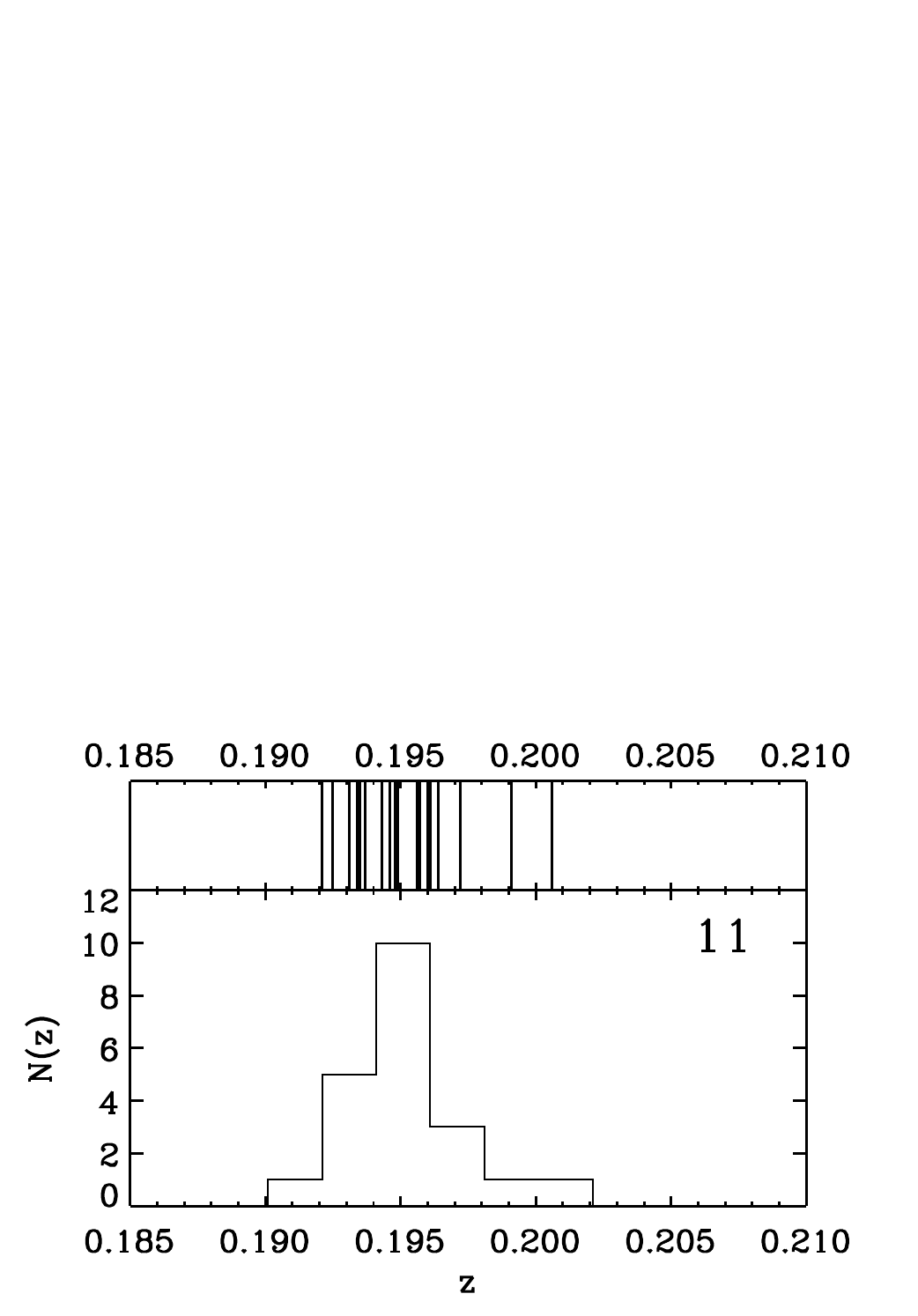}}
\resizebox{0.23\textwidth}{!}{\includegraphics[bb=0 0 283 198,clip]{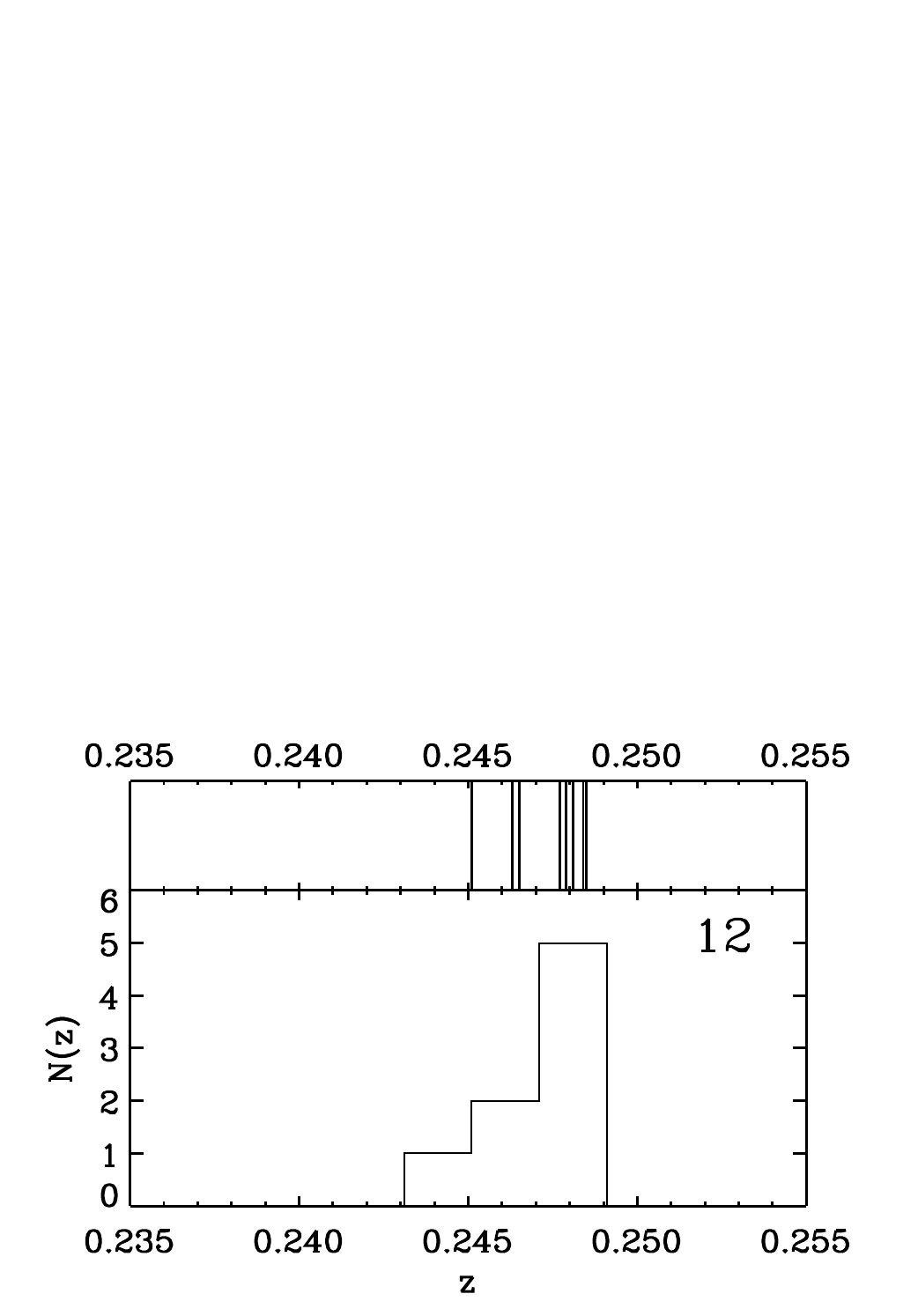}}
\resizebox{0.23\textwidth}{!}{\includegraphics[bb=0 0 283 198,clip]{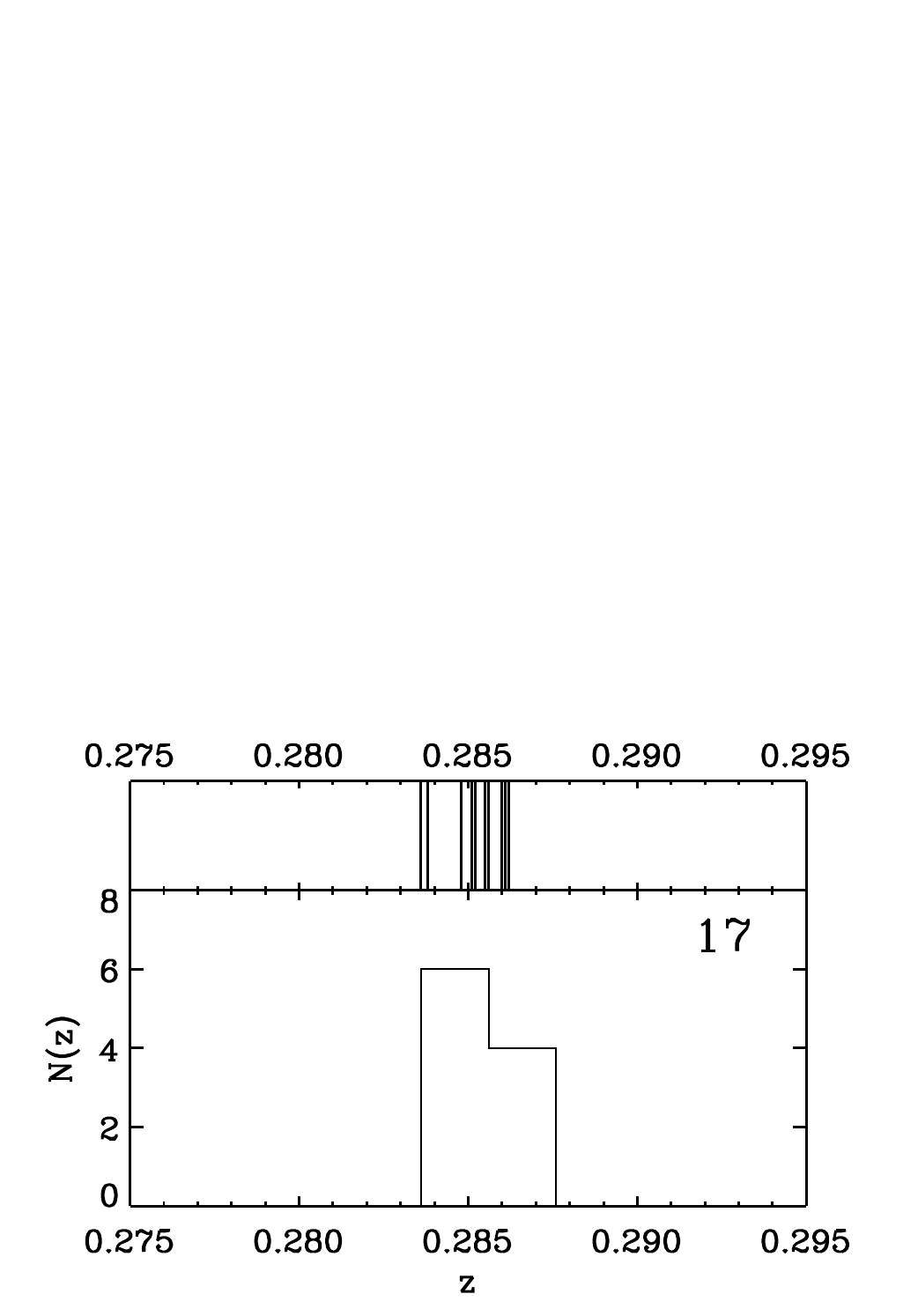}}
\resizebox{0.23\textwidth}{!}{\includegraphics[bb=0 0 283 198,clip]{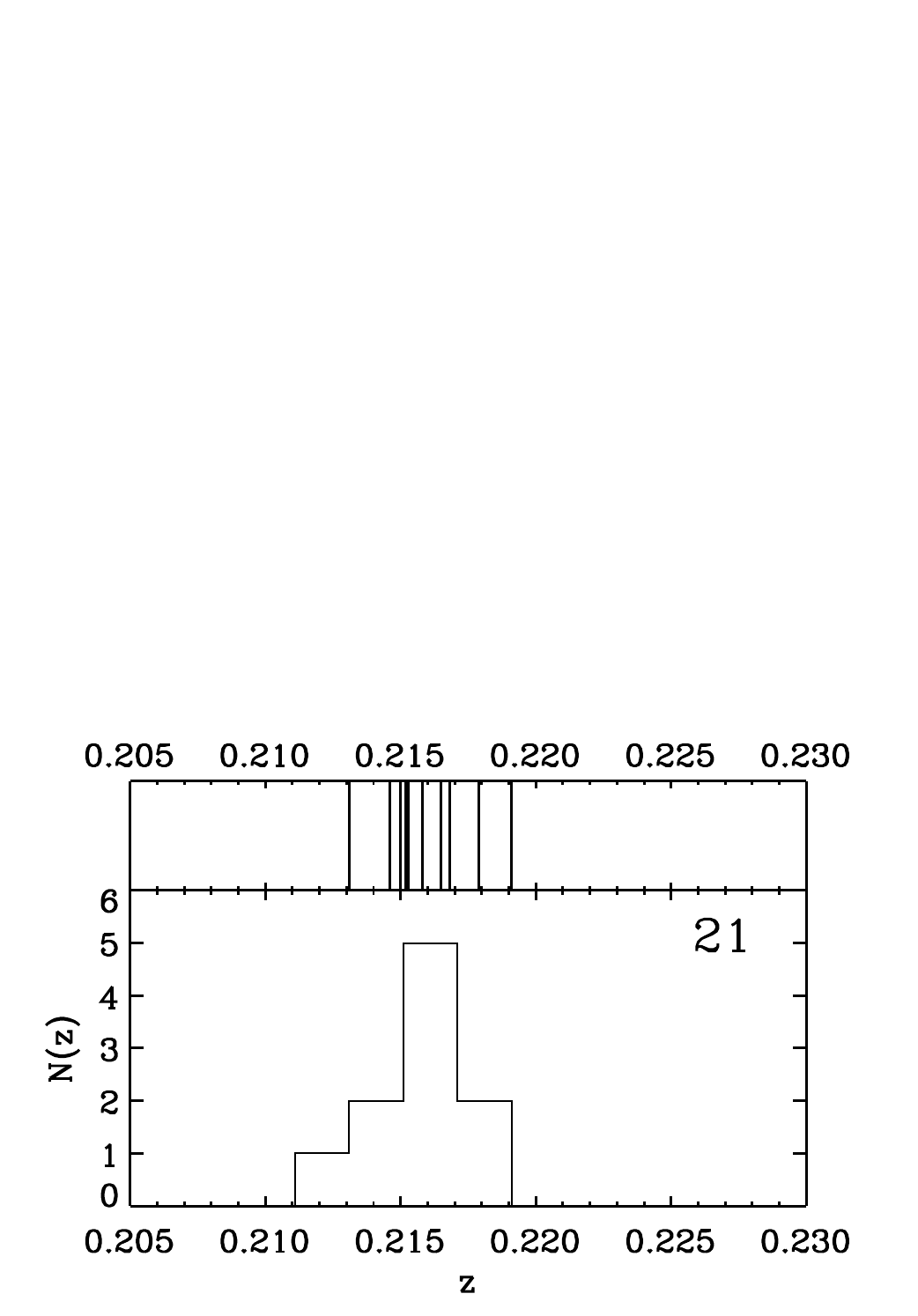}}
\end{center}
\caption{Detailed redshift distributions for all systems considered confirmed.}
\label{fig:zdist_detail}
\end{figure*}

The velocity dispersion of a galaxy cluster is a commonly used
estimator of its mass and thus its distribution is expected to reflect
the mass distribution of a cluster population.  As discussed in
Paper~III different estimators (gapper, biweight and standard
deviation) have been tested to determine how sensitive the values of
the velocity dispersion are when considering samples that have a small
number of members with measured redshifts and possible substructures
as observed in the redshift distributions shown in
Fig.~\ref{fig:zdist_detail}. As before, we found that the different
estimators yield consistent values, differing by no more than
$50\mathrm{km/s}$.  In Table~\ref{tab:colour} we use the biweight
estimator found to be more robust in the presence of outliers. The
velocity dispersions are rest-frame values corrected for our estimated
redshift accuracy. The estimates with 68\% bootstrap errors are given
in Table~\ref{tab:colour}. We note that the values listed in
Table~\ref{tab:colour} may differ from those in
Table~\ref{tab:EISgroups} because a different estimator was used.

\begin{figure}
\resizebox{0.9\columnwidth}{!}{\includegraphics[bb=0 0 283 198,clip]{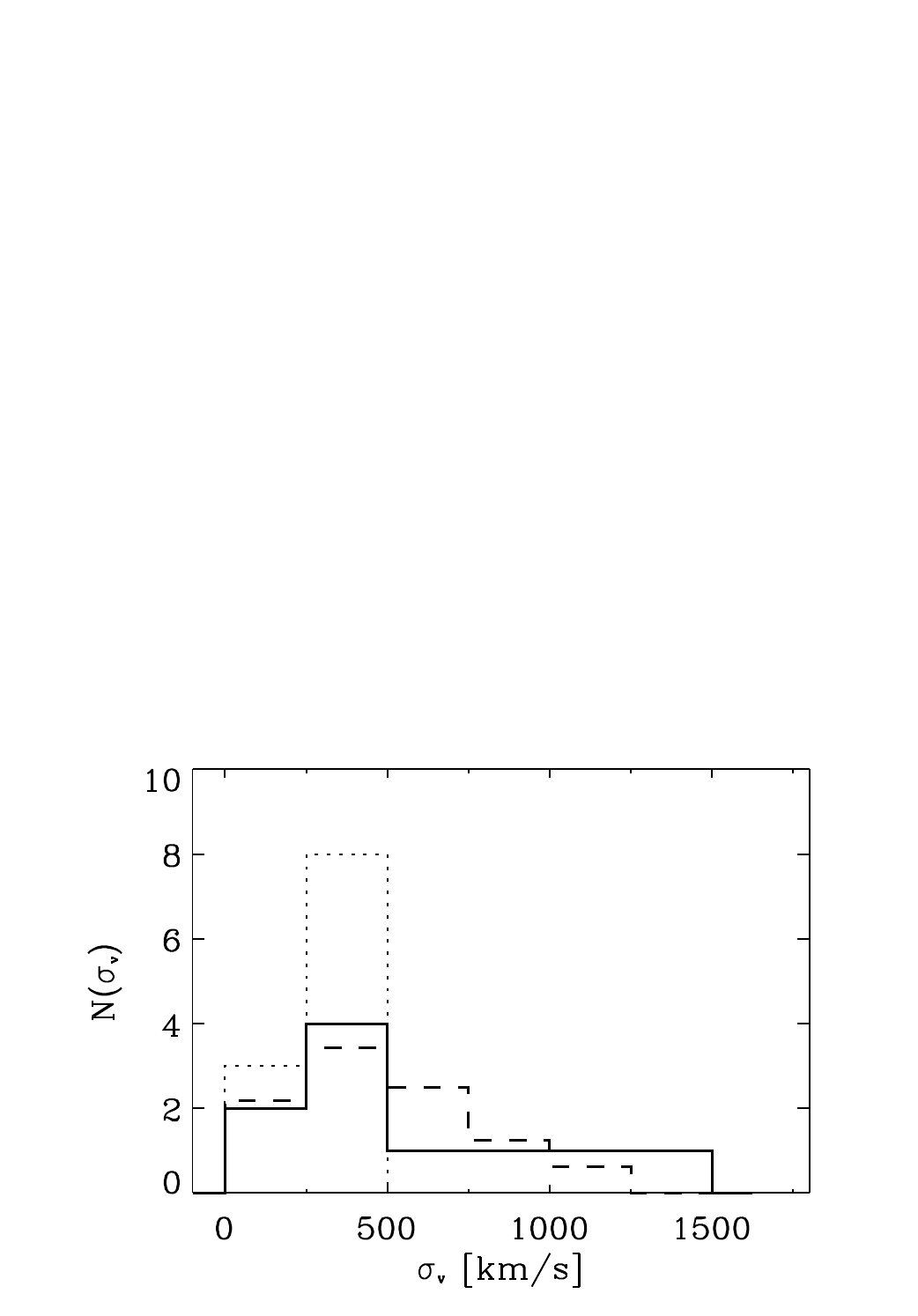}}
\caption{Distribution of velocity dispersions (solid line) compared
with that of the $z_{MF}=0.2$ sample scaled to the same number of
objects (dashed line, Paper~III). The dotted histogram denote the
distribution assuming that the four systems with substructure are
superpositions and only the richest group is included as the
confirmation.  }
\label{fig:veldisp_dist}
\end{figure}

The distribution of velocity dispersions is shown in
Fig.~\ref{fig:veldisp_dist} for the systems identified in this paper
(solid histograms). In the figure we also show the same distribution
for the $z_{MF}=0.2$ sample (dashed histogram) of Paper~III,
normalised to the same number of systems. Given the small number of
systems in the two samples the two distributions are very
similar. Moreover, as discussed in the previous paper, the observed
distributions are comparable to that of Abell clusters (Fig.~7 of
Paper~III) with, perhaps, a larger fraction of low velocity dispersion
systems. The measured velocity dispersions cover a broad range of
values, comparable to that recently reported for the SDSS cluster
sample \citep{becker2007}, varying from groups ($\sim$200~km/s) to
rich clusters ($\sim$800~km/s). There are, however, some systems with
large velocity dispersions such as EISJ2236-3935 (\#6) and
EISJ2238-3934 (\#10), with 1160 km/s and 1433 km/s suggestive of very
massive systems, unlikely to be detected within the small volumes
probed by our survey. Examination of Fig.~\ref{fig:zdist_detail}
suggests that in at least five cases (\#3, \#5, \#6, \#9, \#10), or
50\% of the confirmed systems, the presence of peaks in the redshift
distribution along the line of sight of these systems, separated by
$\sim$2000~km/s, are clearly visible and may influence our estimate of
the velocity dispersion. This may reflect the presence of interlopers
or possibly bound infalling systems frequently observed
\citep[e.g. ][]{buote1996}. The presence of these interlopers may bias
the velocity dispersion estimates, especially in the case of \#6 and
\#10. While the sample considered here is small, it is worth noting
the high frequency of such cases, even though comparable with other
recent findings \citep{jeltema2008}.

In order to investigate the nature of the interlopers, we have
examined the spatial distribution of these systems and the relative
position of the galaxies found in the different subgroups. In
Fig. ~\ref{fig:conf} galaxies belonging to different subgroups are
indicated by the size of the symbols. From a close examination of
Figs.~\ref{fig:zdist_detail} and ~\ref{fig:conf} we find that: 1)
EISJ0047-2942 (\#3) consists of one dominant group twice as large as
the secondary peak located in the background, which we consider to be
unrelated to the main cluster. Therefore we keep only the dominant
group; 2) EISJ0049-2920 (\#5) is similar to the previous case and we
keep only the dominant group which in this case is the most distant
one; 3) EISJ2236-3935(\#6) is similar to the previous case and we keep
only the group at the highest redshift, considerably decreasing our
measured velocity dispersion; 4) EISJ2237-4000 (\#9) is the easiest
case with a dominant group with a single interloper at a higher
redshift which can be neglected without significantly impacting the
value calculated for the velocity dispersion as can be seen by
comparing the two reported values; 5) EISJ2238-3934 (\#10) with the
largest velocity dispersion consists of two comparable clumps, one
with 13 members and the other with 10, and one foreground group
composed of only two galaxies, which we consider as field interlopers
and exclude from further analysis. Therefore, we consider these as two
independent clusters.

Based on the above considerations we recomputed the velocity
dispersion considering only the richest group in the systems \#3, \#5
and \#6, removing the interloper in \#9 and splitting the cluster \#10
into two. The results are summarised in Table~\ref{tab:colour_update}
which lists in Col.~1 the field identifier; in Col.~2 the name of the
cluster field; in Col.~3 the number of spectroscopic members; in
Col.~4 the matched filter redshift estimate; in Col.~5 the
spectroscopic redshift; and in Col.~6 the velocity dispersion with 68\%
bootstrap errors.  The
table shows that the five systems are now represented by six clusters
of which five have significantly smaller velocity dispersions than the
original detections, adding four new systems to the velocity
dispersions range 250km/s to 500km/s and one with velocity dispersion
$<$250km/s. This new distribution is also shown in
Fig.~\ref{fig:veldisp_dist} (dotted histogram) which shows a high
concentration of systems close to the peak value of $\sim$400 km/s
found for the much larger sample by \citet{becker2007}. Our sample is
arguably small making the above discussion very subjective. However,
only a much better sampling of each system would allow to fully
characterise the nature of the systems identified in redshift space.

%%%%%%%%%%%%%%%%%%%%%%%%%%%%%%%%%%%%%%%%%%%%%%%%%%%%%%%%%%%%%%%%%
\section{Photometric properties of the detected systems}
\label{sec:properties}
%%%%%%%%%%%%%%%%%%%%%%%%%%%%%%%%%%%%%%%%%%%%%%%%%%%%%%%%%%%%%%%%%

In addition to the spectroscopic results reported above, the
photometric properties of the confirmed systems are also important for
comparison with other cluster samples possibly having other selection
effects.  In this section we will further characterise the 10
confirmed clusters in terms of their richness and photometric
properties of their galaxy population as well as compare these
properties with those of the $z_{MF}=0.2$ sample discussed in
Paper~III.

\subsection{Richness}

The matched filter algorithm provides a measure of the richness
($\Lambda_{cl}$) for the cluster candidates based on the estimated
redshifts, where the $\Lambda_{cl}$-richness is the number of
$L^*$-galaxies required to match the total luminosity of the cluster.
Computation of the richness depends on the apparent Schechter
magnitude and angular extent of the cluster, which at these redshifts
vary rapidly. Therefore, even though the spectroscopic and estimated
redshifts are in reasonable agreement we have to recompute the cluster
richness using the assigned spectroscopic redshift and the centre
of the assigned groups as listed in Table~\ref{tab:EISgroups}.  The
new richness values ($\Lambda_{cl, new}$) are listed in
Table~\ref{tab:colour}. In general, since the matched-filter tends to
overestimate the redshifts it also overestimates the richness. Note,
however, that for EISJ0047-2942 (\#3) the opposite is true with the
corrected richness being much larger than the original estimate. This
is because the redshift was significantly underestimated.

A large scatter of the relation between the new and original
richness could already be expected based on results from simple
simulations. \citet[][ Fig. 8]{olsen07} show that even after
correcting for the redshift overestimate, the richness may be off by
typically up to 50\%. The scatter is of the order 25\%. This is slightly
worse than the value claimed by \cite{postman02}.

The final corrected values for the richness are in the range 12-65
comparable to that obtained in Paper~III and the range reported by
\citet{bahcall03} using a modified version of the matched filter in
their analysis of 400 square degrees of the SDSS survey. As in the
previous paper we find that the measured richness and velocity
dispersions follow the same relation as that determined by Bahcall et
al., with comparable scatter. This again demonstrates that the simple
matched-filter technique used by us does not measure richness very
accurately. Furthermore, while we can use the redshift distribution to
re-compute more accurate velocity dispersions in case of contamination
by nearby systems, this cannot be easily done for the richness, showing
that it is probably not a reliable proxy for the mass.  

\subsection{Galaxy population}
\label{sec:colours}

\begin{figure*}
\begin{center}

\resizebox{0.24\textwidth}{!}{\includegraphics[bb=0 0 283 198,clip]{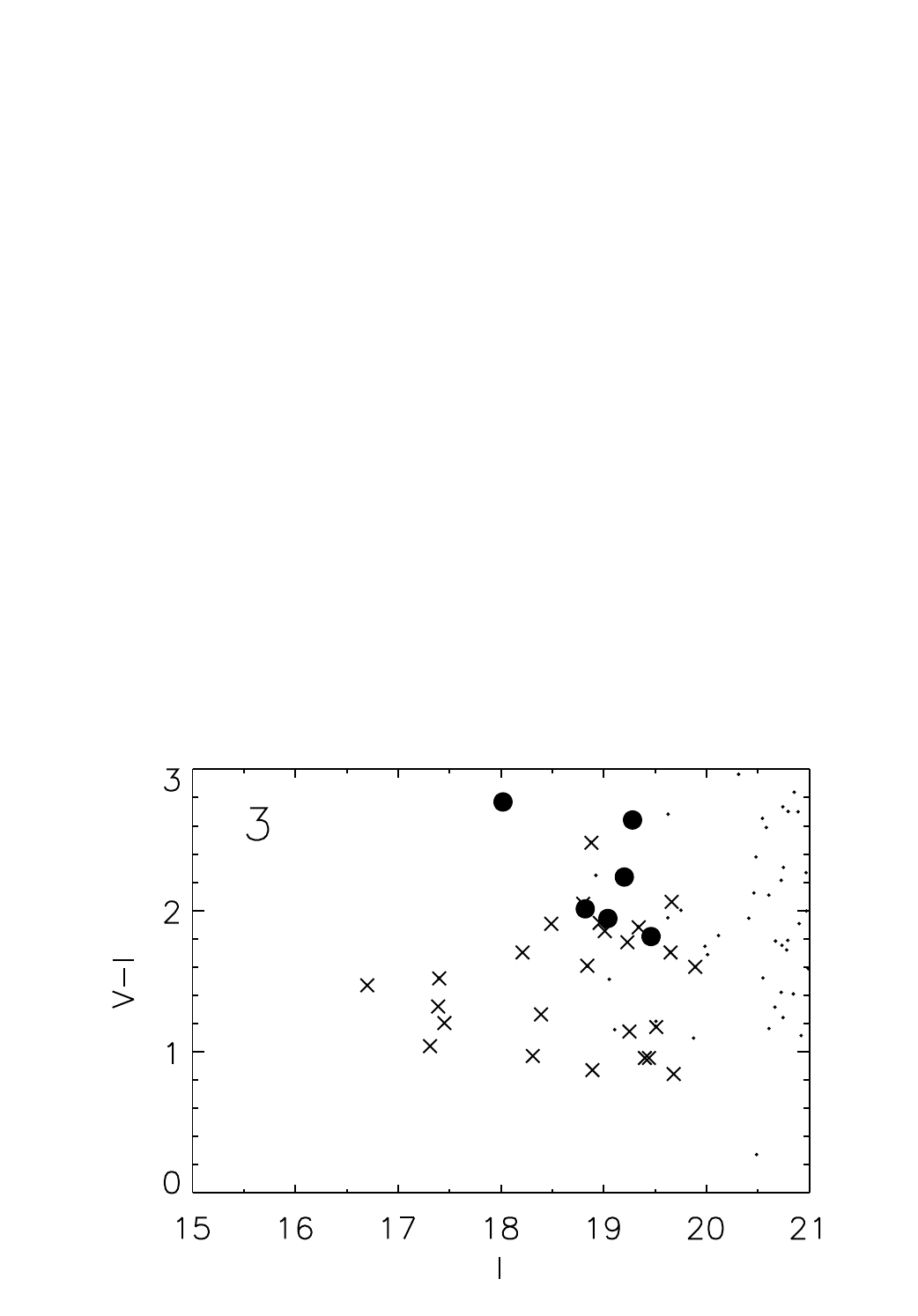}}
\resizebox{0.24\textwidth}{!}{\includegraphics[bb=0 0 283 198,clip]{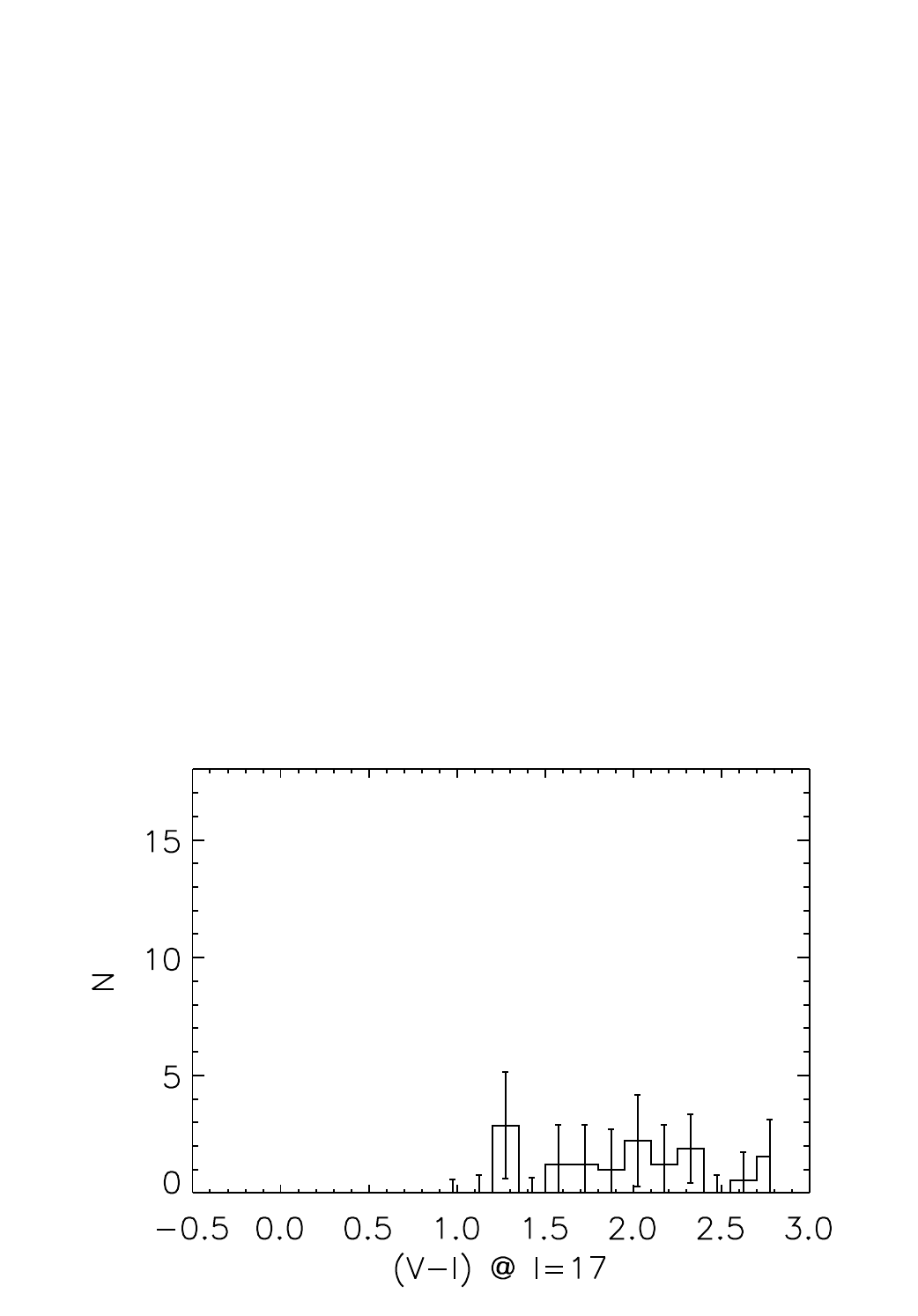}}
\resizebox{0.24\textwidth}{!}{\includegraphics[bb=0 0 283 198,clip]{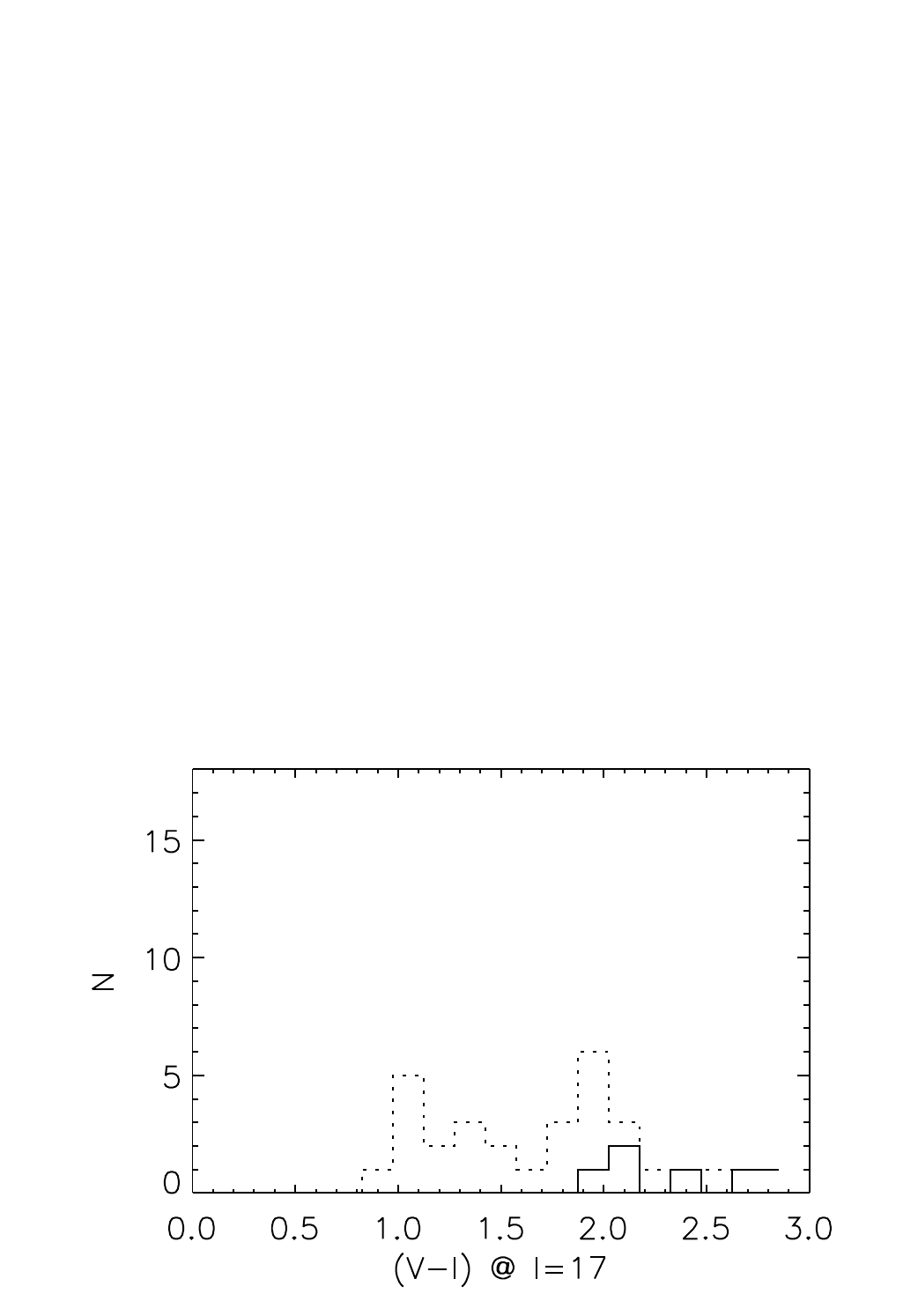}}

\resizebox{0.24\textwidth}{!}{\includegraphics[bb=0 0 283 198,clip]{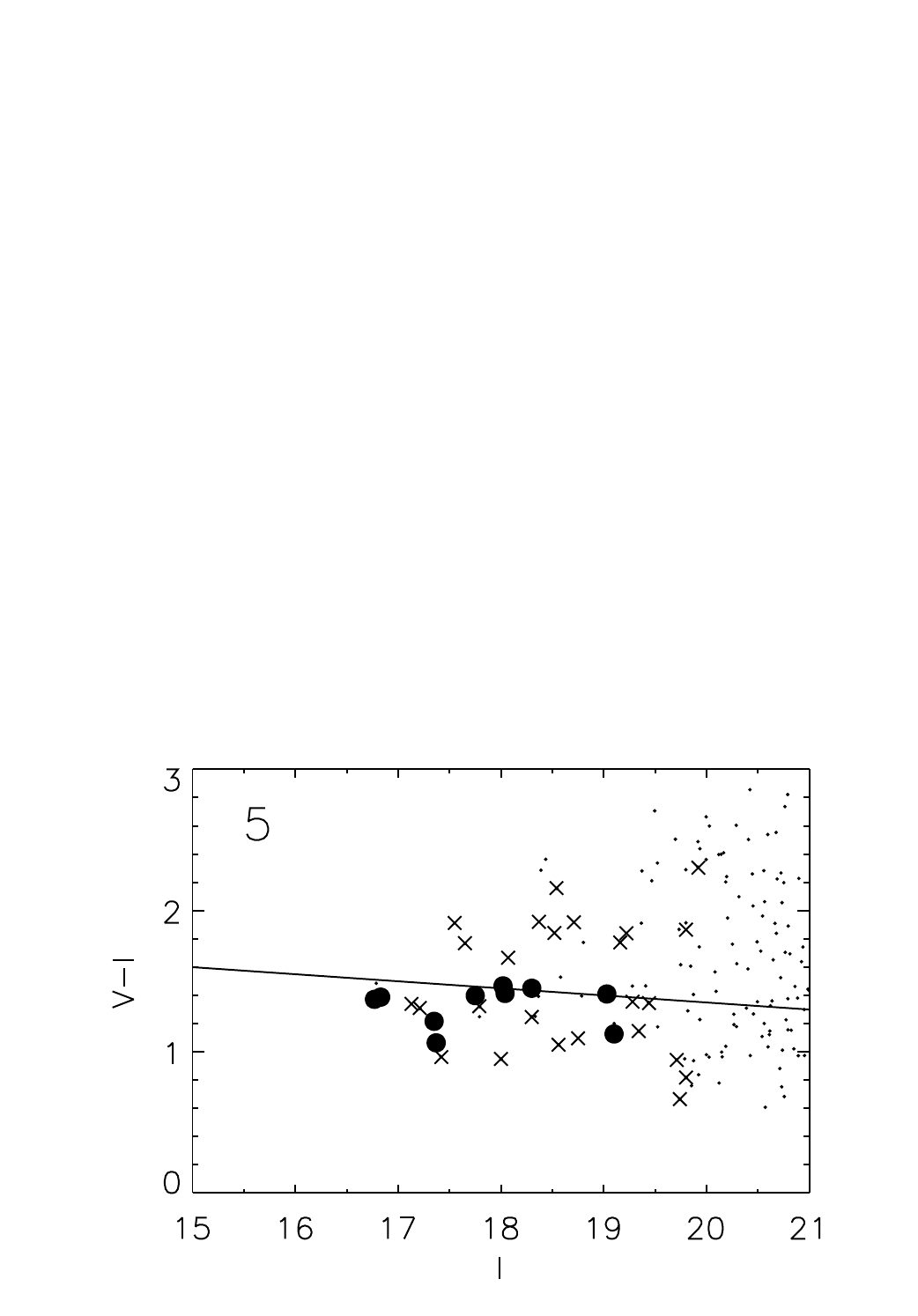}}
\resizebox{0.24\textwidth}{!}{\includegraphics[bb=0 0 283 198,clip]{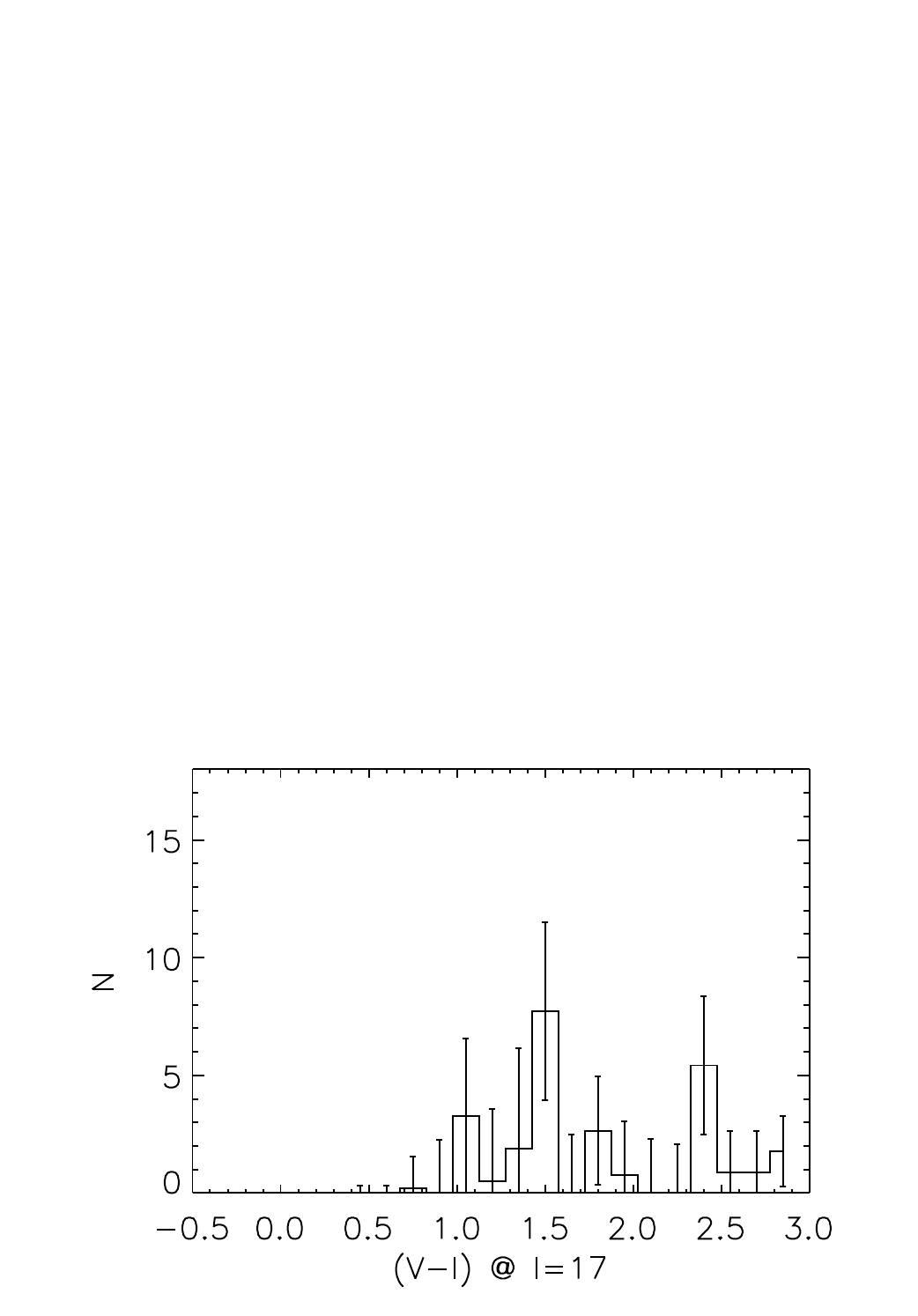}}
\resizebox{0.24\textwidth}{!}{\includegraphics[bb=0 0 283 198,clip]{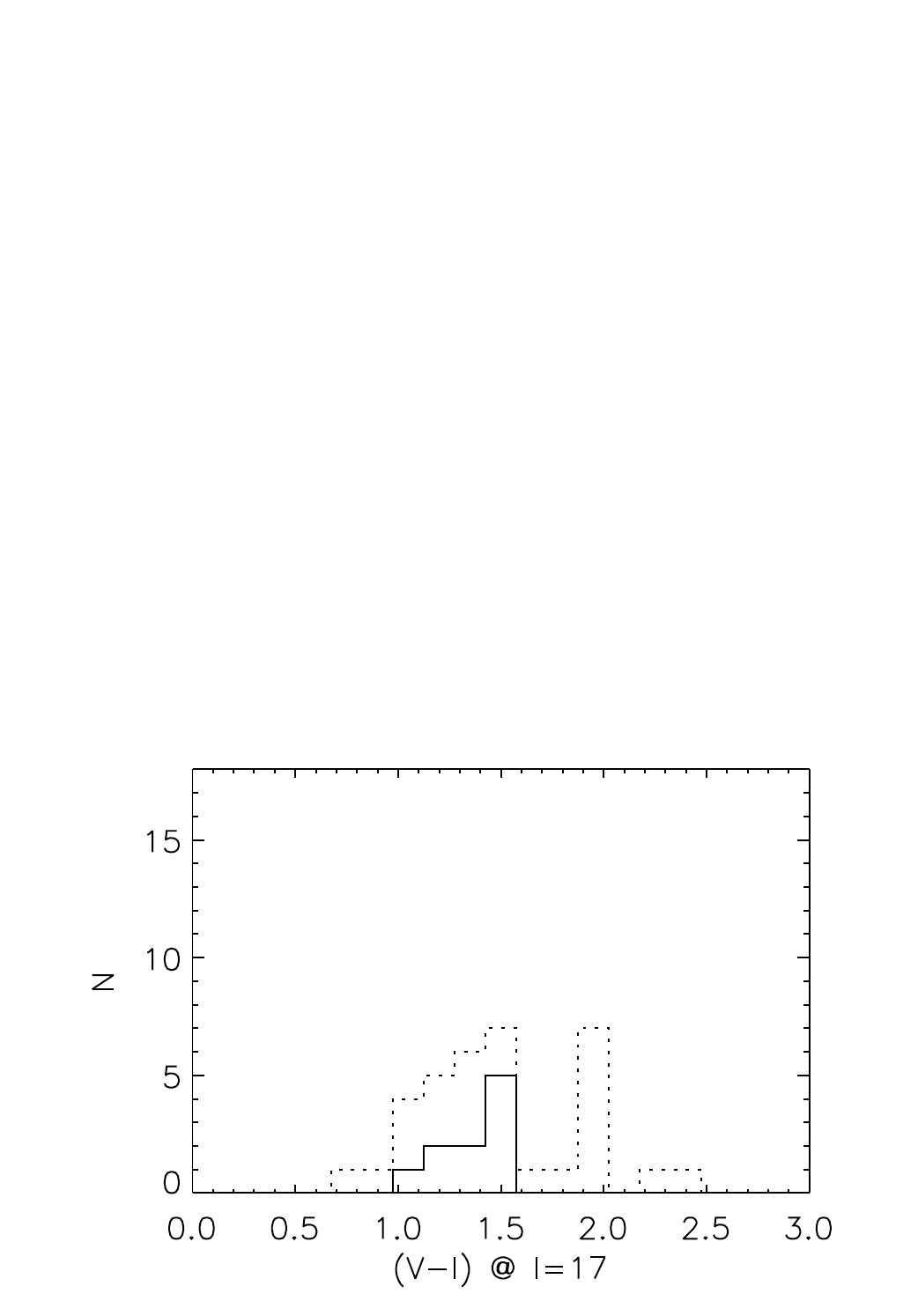}}

\resizebox{0.24\textwidth}{!}{\includegraphics[bb=0 0 283 198,clip]{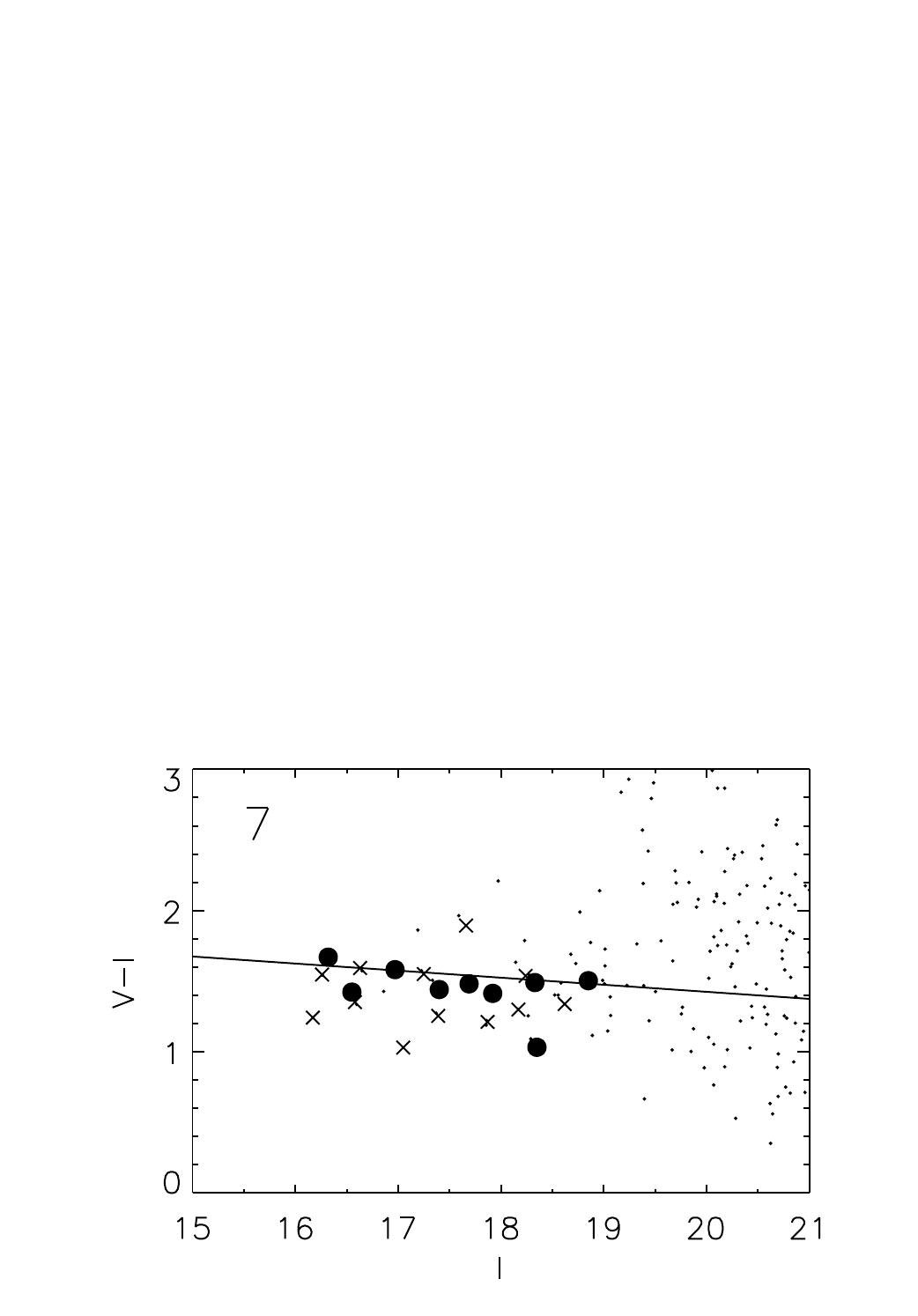}}
\resizebox{0.24\textwidth}{!}{\includegraphics[bb=0 0 283 198,clip]{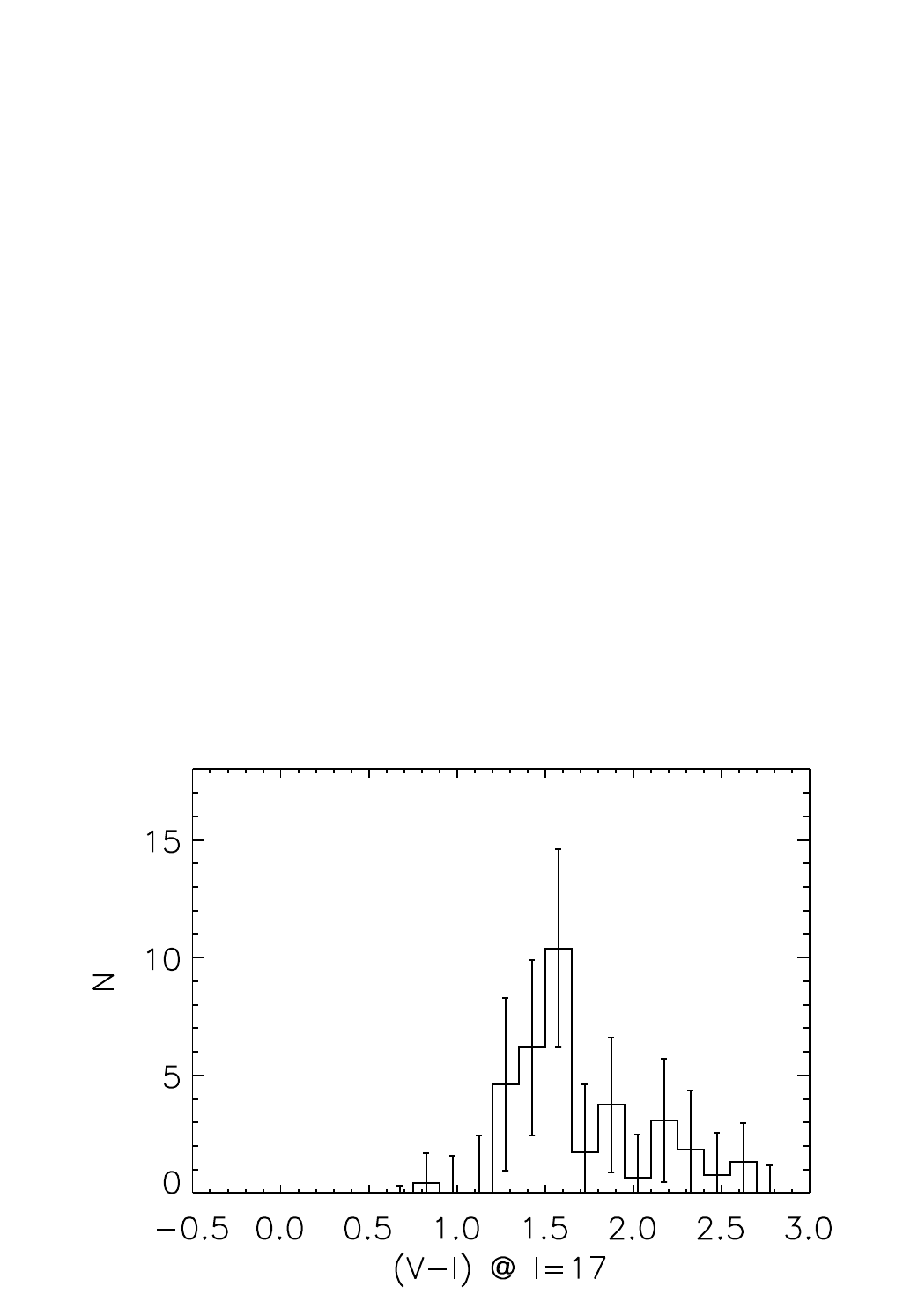}}
\resizebox{0.24\textwidth}{!}{\includegraphics[bb=0 0 283 198,clip]{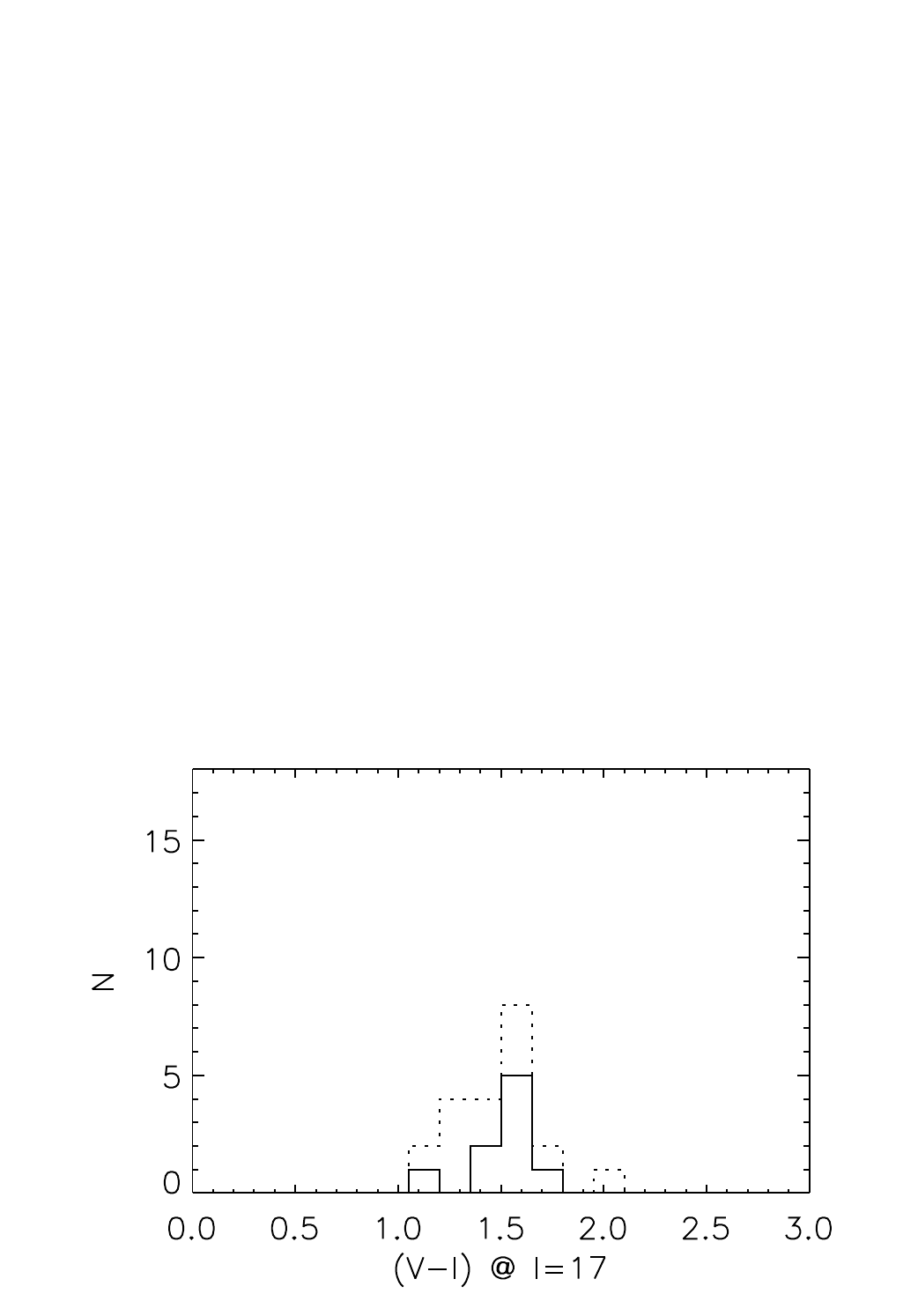}}

\resizebox{0.24\textwidth}{!}{\includegraphics[bb=0 0 283 198,clip]{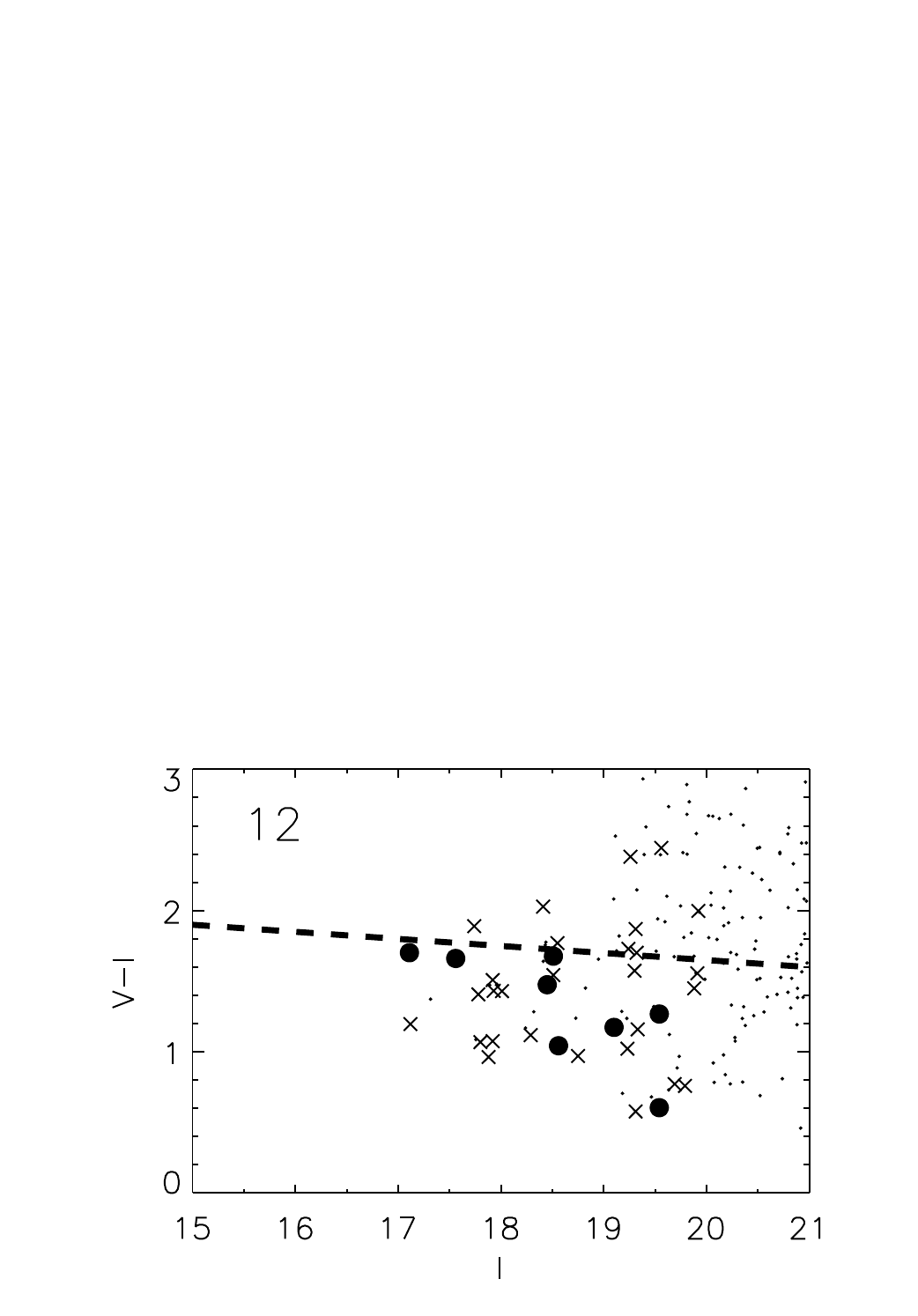}}
\resizebox{0.24\textwidth}{!}{\includegraphics[bb=0 0 283 198,clip]{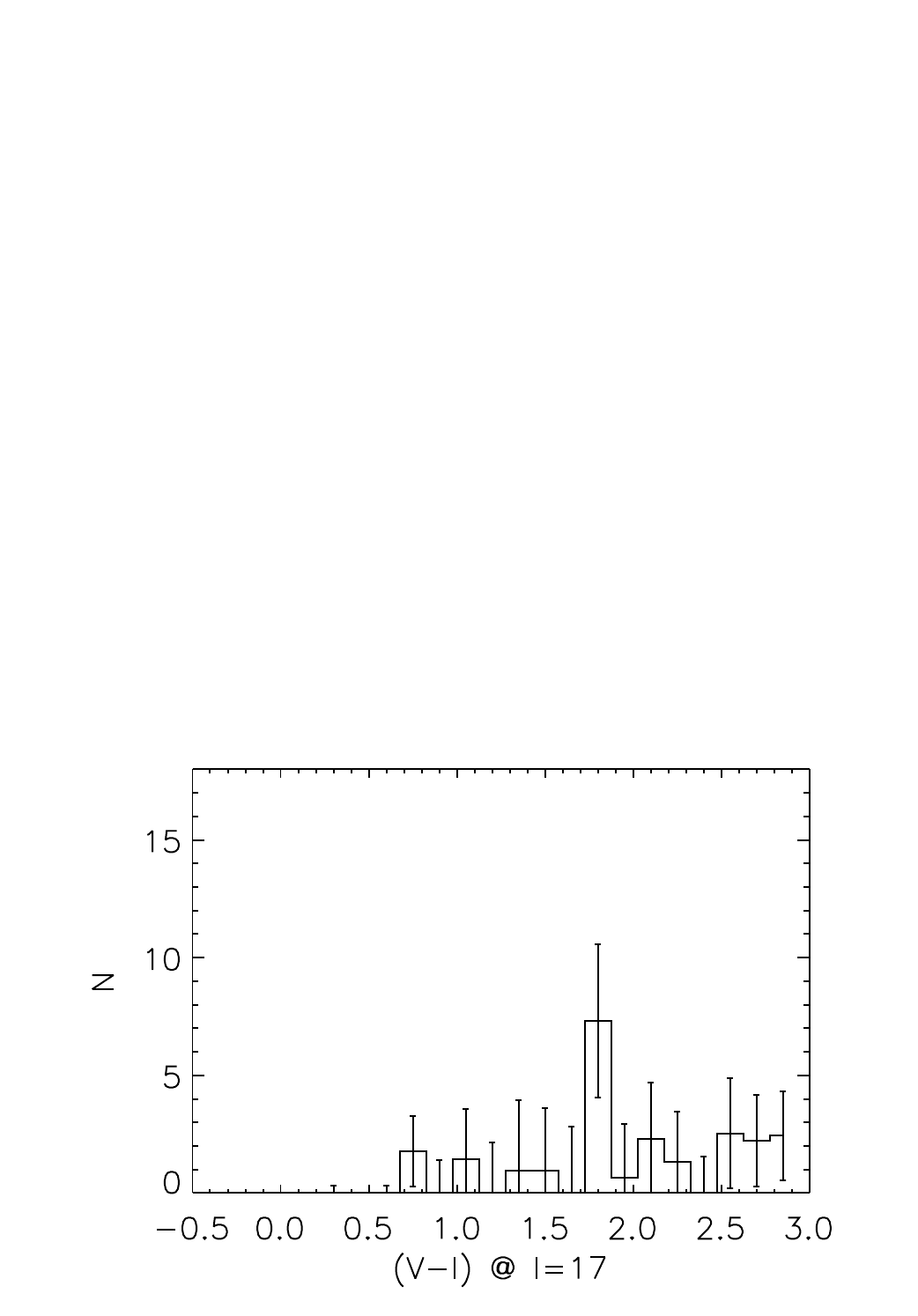}}
\resizebox{0.24\textwidth}{!}{\includegraphics[bb=0 0 283 198,clip]{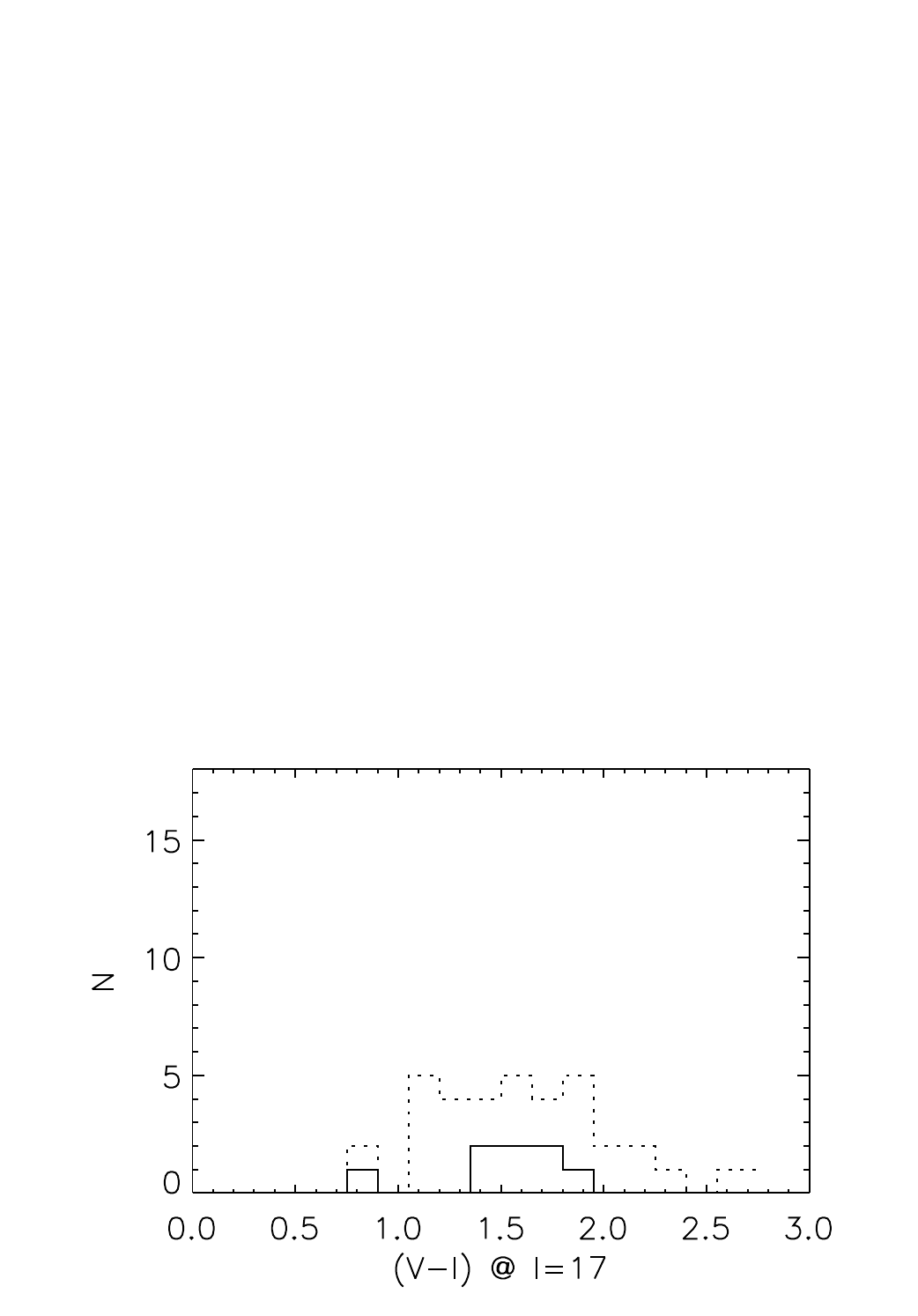}}

\resizebox{0.24\textwidth}{!}{\includegraphics[bb=0 0 283 198,clip]{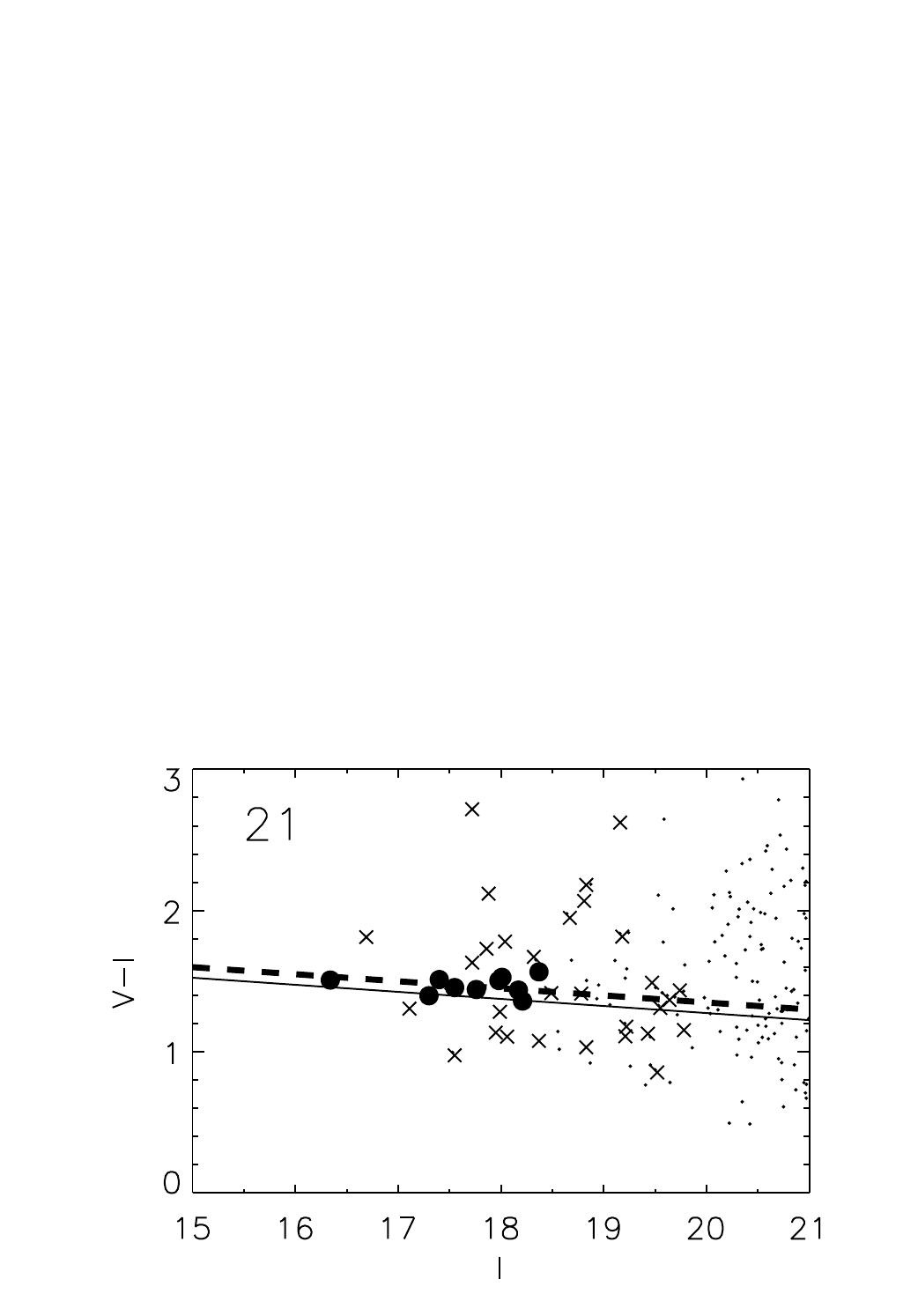}}
\resizebox{0.24\textwidth}{!}{\includegraphics[bb=0 0 283 198,clip]{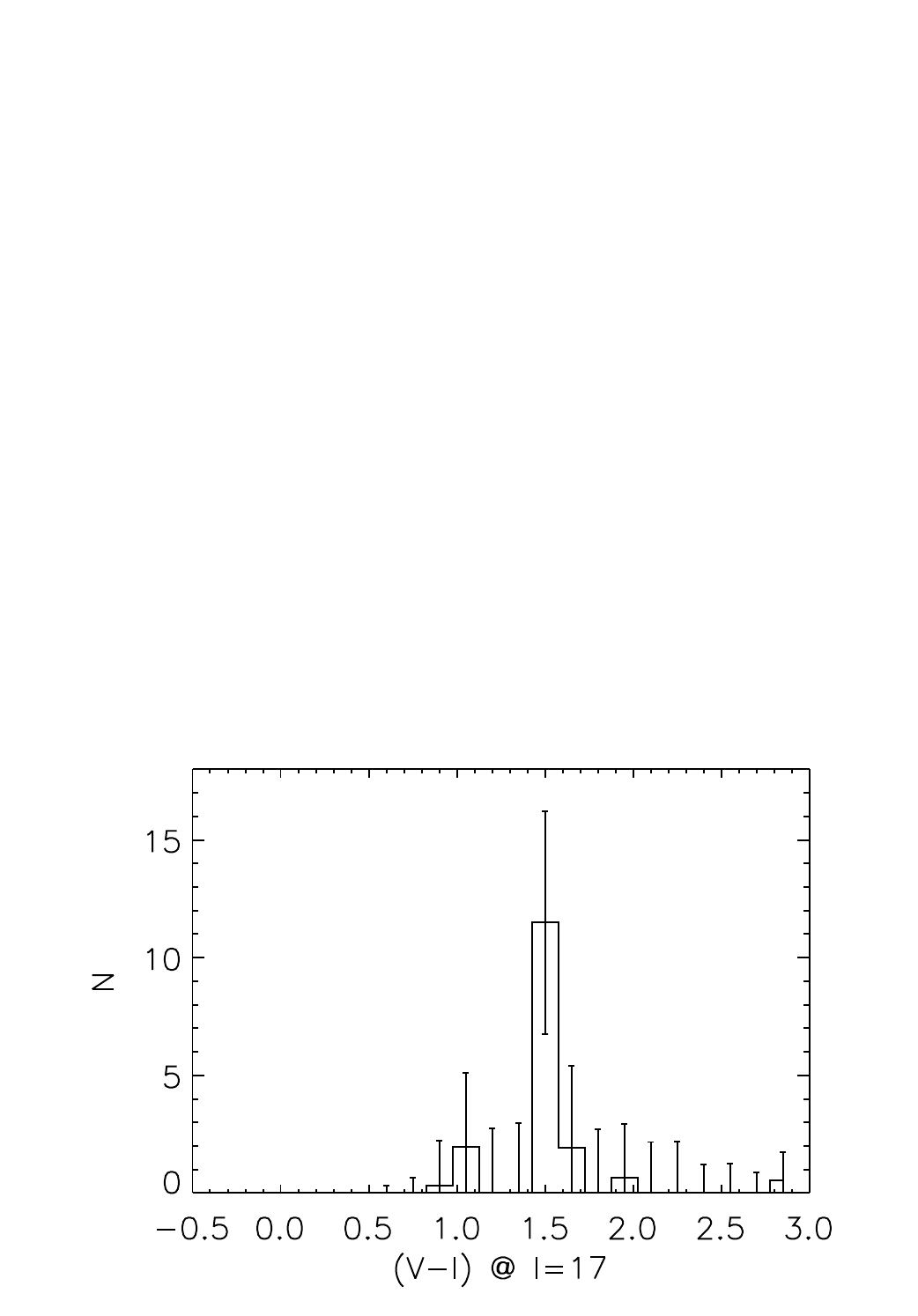}}
\resizebox{0.24\textwidth}{!}{\includegraphics[bb=0 0 283 198,clip]{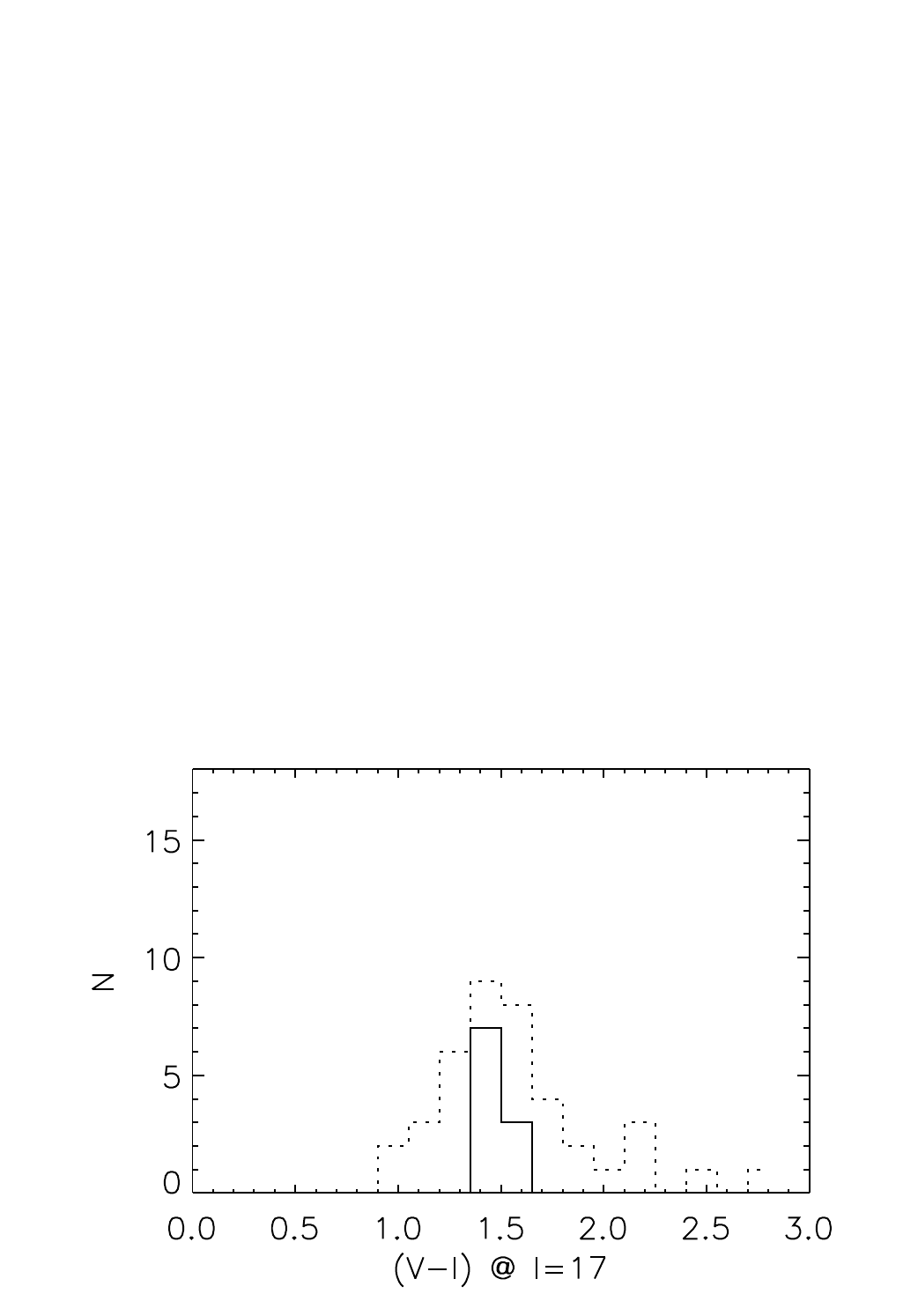}}

\end{center}
\caption{For each cluster we show three diagrams with the cluster
identification number indicated in the first one. The first diagram is
the colour-magnitude diagram for all galaxies within
$0.75h_{75}^{-1}\mathrm{Mpc}$ from the cluster centre (dots). On top
of that we mark by solid circles the spectroscopic members of the
confirmed group and by crosses the remaining galaxies with
redshifts. The solid line is the locus of the red sequence detected
from the spectroscopic members and the dashed line is the one detected
in the photometric analysis. In both cases we only show the line if we
consider the sequence significant (see the text for details). The
second plot is the ``tilted colour histogram'' for the galaxies
with $I\leq20$ in the same region statistically corrected for the
background contribution. The last panel is the ``tilted colour
histogram'' for the spectroscopic members (solid line) and for all
galaxies with a redshift (dotted line).}
\label{fig:cmd}
\end{figure*}

Finally, we characterise the galaxy population of the confirmed
clusters by studying the colour properties of the member galaxies. In
particular, the colour-magnitude diagram of cluster members normally
reveals the presence of a narrow sequence of bright, early-type
galaxies known as the ``red sequence'' \citep[e.g. ][]{gladders98,
stanford98, holden04, lopez-cruz04}, which has been used to identify
clusters \citep[e.g. ][]{gladders00}. The presence of this
colour-magnitude (CM) relation serves as unambiguous evidence for the
presence of a real physical system. In Paper~III we used this fact as
the basis for an objective method for detecting red sequences for the
cluster candidates identified by the use of the matched-filter method
applied to single passband photometric data. While most of the
previous studies focused on relatively rich systems, the presence of
red sequences in very poor clusters and groups has also been reported
\citep[e.g.][]{andreon03}.

For five of the ten systems analysed here, $V$-band images are
available, thus allowing us to construct and investigate the CM
diagram of the ``cluster'' members.  The $(V-I)\times I$ CM diagrams
were constructed considering galaxies within a radius of
$0.75h_{75}^{-1}\mathrm{Mpc}$ (the ``cluster region'').  The
identification method is based on ``tilted colour histograms'',
counting galaxies within slices of a given width and characterised by
a slope taken to be comparable to that typically observed for the CM
relations of nearby clusters \citep[e.g. ][]{lopez-cruz04}. For each
cluster we construct two histograms shifted in colour by half a bin
width to assure good sensitivity at any colour.  For each histogram we
identify the most significant peak based on the probability of finding
a similar colour overdensity at a randomly chosen position. Readers
are referred to Paper~III for the details of the red sequence
detection method.

The red sequence identification is applied separately to two samples.
First, we consider the sample of all galaxies brighter than $I=20$
with and without redshifts. This has the advantage of good statistics
but is susceptible to projection effects possibly leading to
contamination by non-cluster members and thus dilution of the red
sequence, if one exists. Second, we restrict the analysis to those
galaxies for which we have redshifts.  While these are not affected by
projection effects, the statistics are usually poor in particular
considering the sparse sampling of our survey.  

The results of the analysis of the cluster CM diagrams are shown in
Fig.~\ref{fig:cmd}.  The left panels of this figure show the
colour-magnitude diagrams for galaxies (dots), including all galaxies
brighter than $I=21$ within the cluster region (defined above).
Filled circles indicate spectroscopically confirmed members and
crosses other (field) galaxies with measured redshifts. Also shown in
this panel are the best-estimated loci characterising the red sequence
as determined from the photometric data alone (dashed line) and that
obtained considering only the confirmed spectroscopic members (solid
line).  The middle panels show the background-corrected colour
distributions of galaxies brighter than $I=20.0$ within the cluster
region. The right panels show the colour distributions of
spectroscopic members (solid histograms) and that of all galaxies with
measured redshifts (dotted histograms).

Using the photometric sample we find only two systems (EISJ2240-4021,
\#12 and EISJ2244-4019, \#21) showing signs of a red sequence, while
we identify three systems (indicated by dashed lines in the CM
diagrams of Fig.~\ref{fig:cmd}) when the spectroscopic sample is used.
In one case (EISJ2244-4019, \#21) the red sequence is detected using
both the photometric and spectroscopic samples.  For EISJ2240-4021
(\#12) a red sequence is only detected from the photometric data and
not from the confirmed spectroscopic members, showing the difficulties
in identifying red sequences from photometric data alone.

Even though the total number of systems analysed is arguably small,
the fraction of systems with a red sequence identified using the
spectroscopic data is $\sim$60\%, similar to the fraction of 59\%
reported in Paper~III. However, only one system shows both a
photometrically and spectroscopically identified sequence, which is
significantly fewer than expected from Paper~III.

In summary, a total of four systems show evidence of having a red
sequence. We find two systems from the photometric data alone and
three from the spectroscopic analysis with one system being in
common. Both the colour and scatter of the detected red sequences are
consistent with our previous work.  Moreover, the colours of the
galaxies along the red sequences are consistent with the passive
evolution scenario for galaxy evolution. Based on our current
discussion and that of Paper~III we find that up to 40\% of the
clusters identified by the matched-filter method and confirmed
spectroscopically do not have a detectable red sequence.

There are three possible explanations for the apparent lack of a red
sequence in some of our systems: 1) That the systems detected by the
matched-filter and tentatively confirmed by the spectroscopic data are
just density enhancements in redshift space but do not form bound
systems; 2) That spectroscopic results are biased either in the
selection of targets or in the measurement of the redshift leading to
a sample consisting of predominantly spiral galaxies; 3) That there
are systems formed predominantly by a population of blue
galaxies. Unfortunately, with the present data it is impossible to
disentangle these various options, and must await much larger samples.

\begin{figure*}
\begin{center}
\resizebox{0.3\textwidth}{!}{\includegraphics[bb=0 0 283 198,clip]{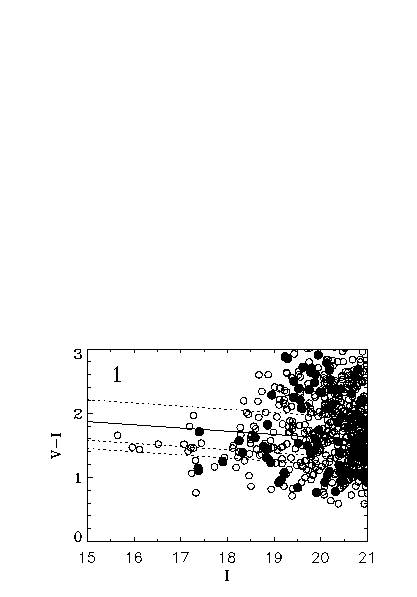}}
\resizebox{0.3\textwidth}{!}{\includegraphics[bb=0 0 283 198,clip]{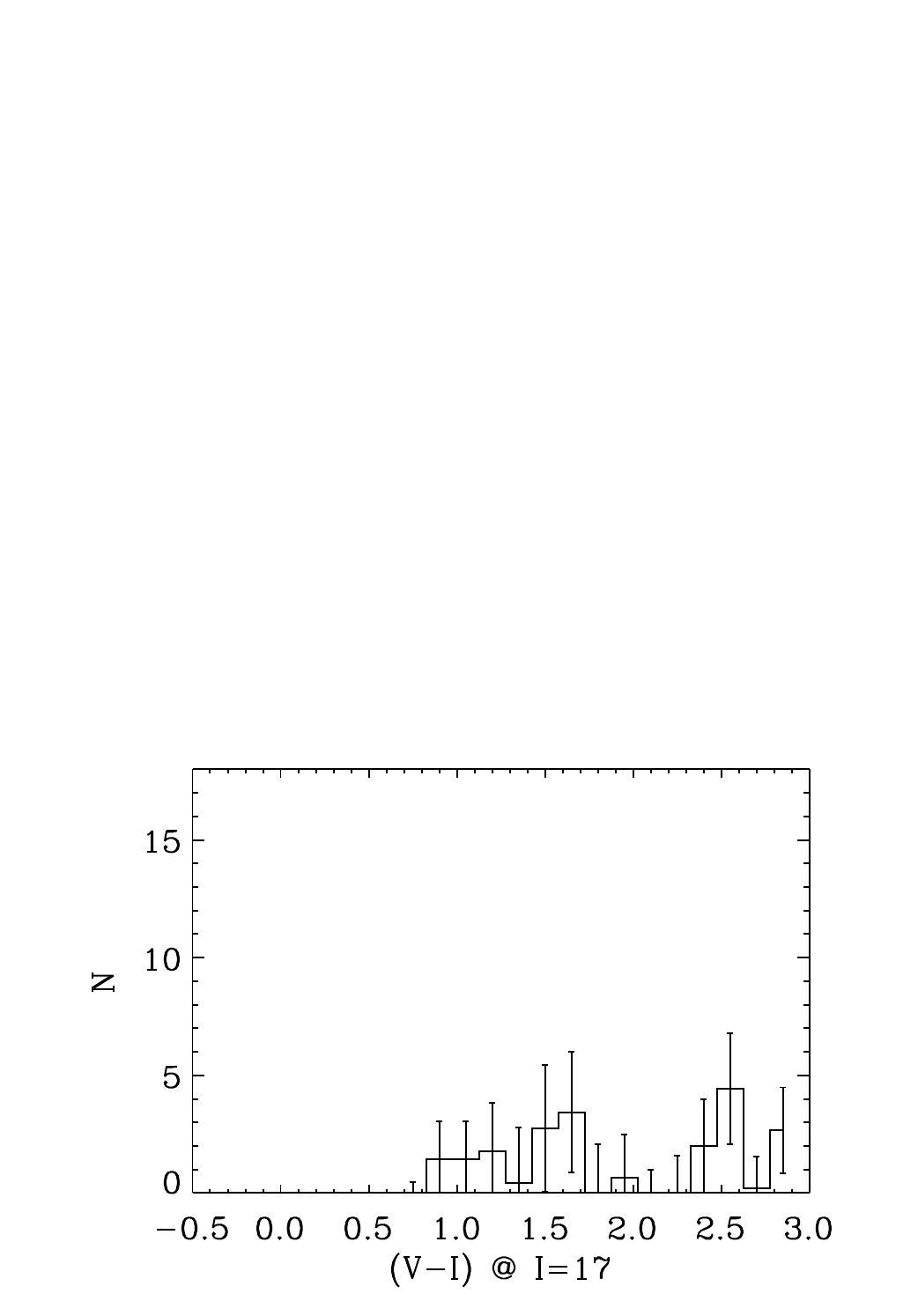}}
\resizebox{0.3\textwidth}{!}{\includegraphics[bb=0 0 283 198,clip]{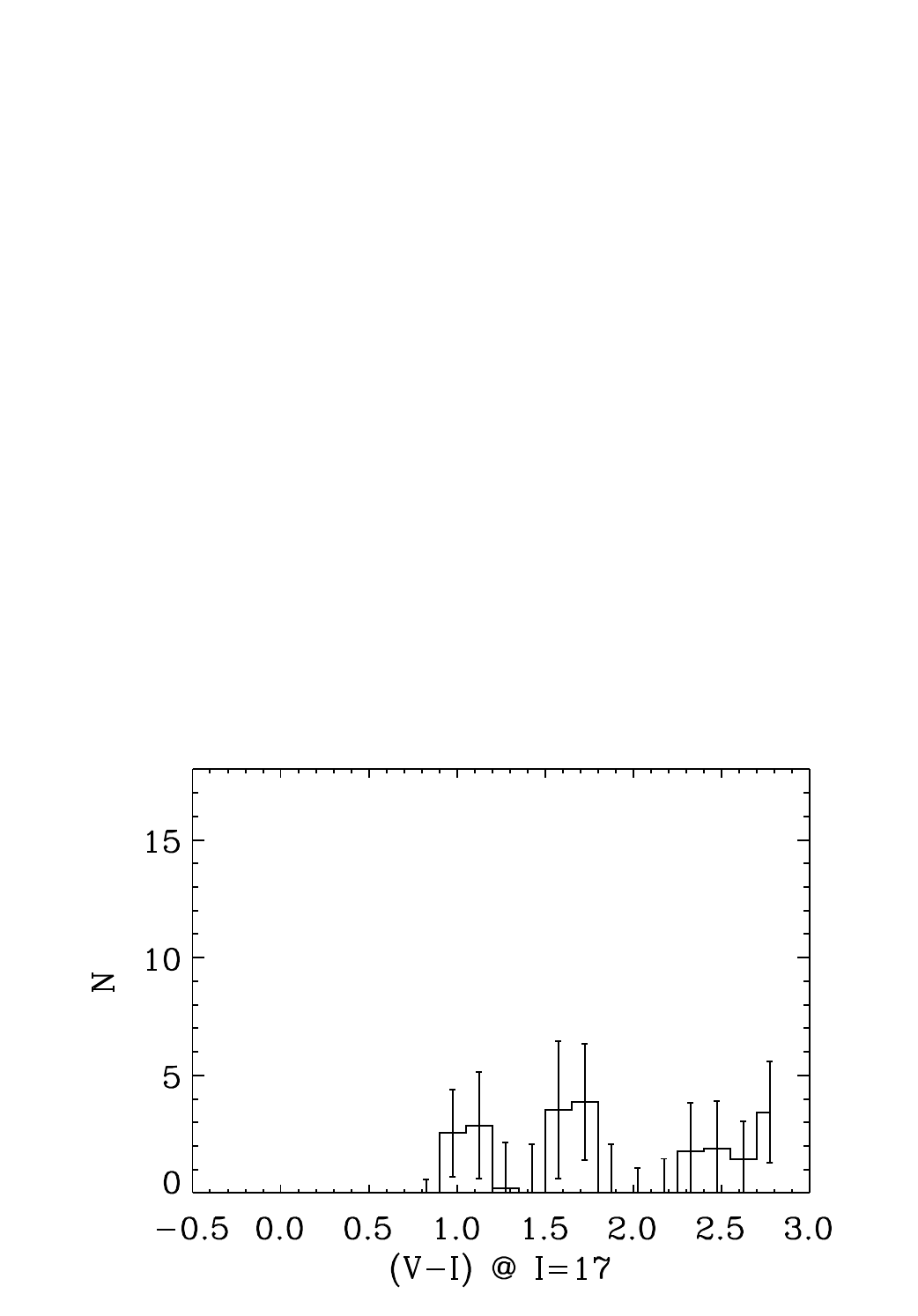}}
\resizebox{0.3\textwidth}{!}{\includegraphics[bb=0 0 283 198,clip]{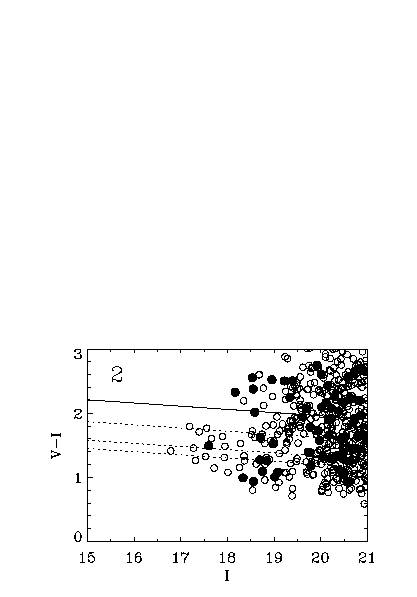}}
\resizebox{0.3\textwidth}{!}{\includegraphics[bb=0 0 283 198,clip]{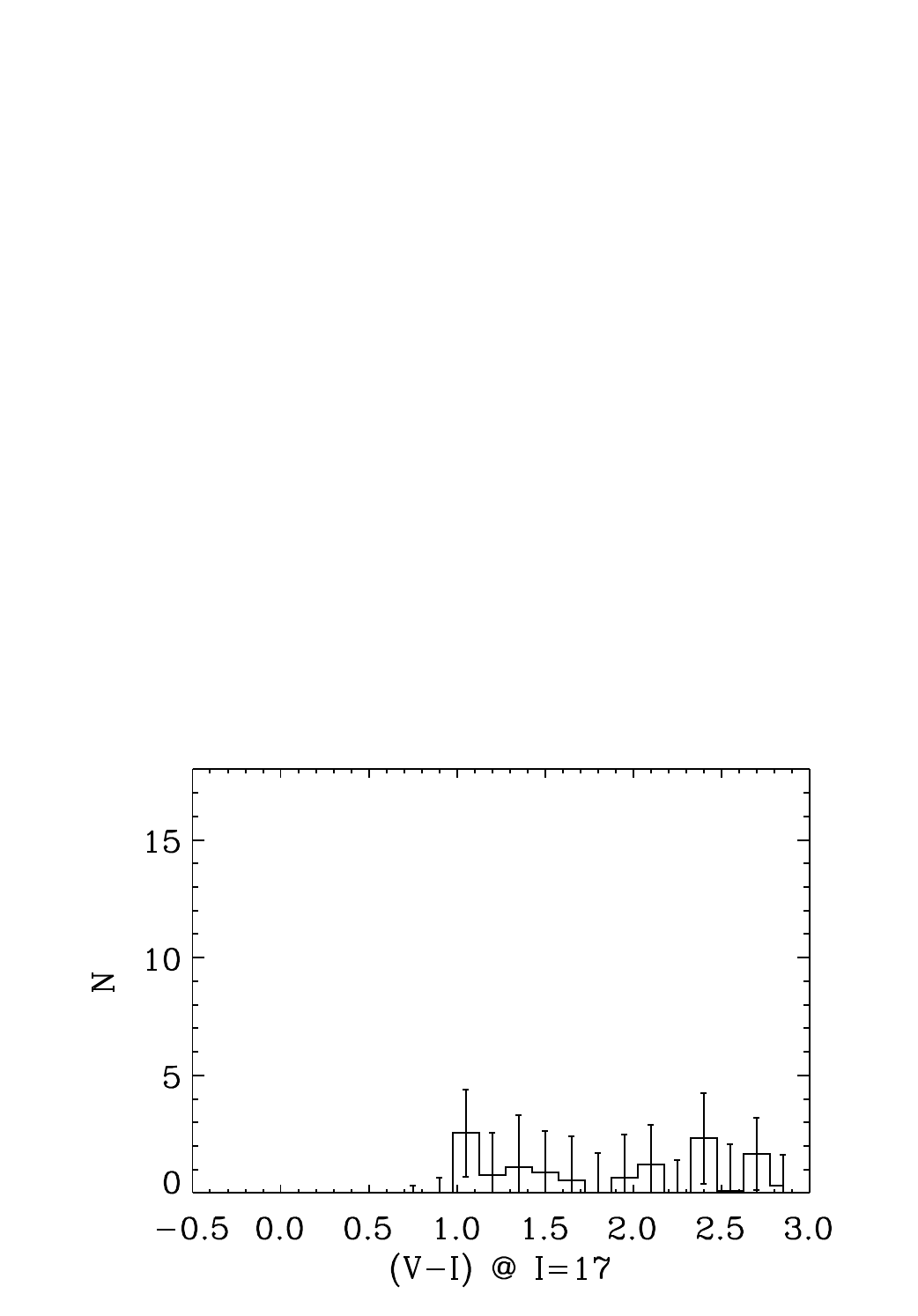}}
\resizebox{0.3\textwidth}{!}{\includegraphics[bb=0 0 283 198,clip]{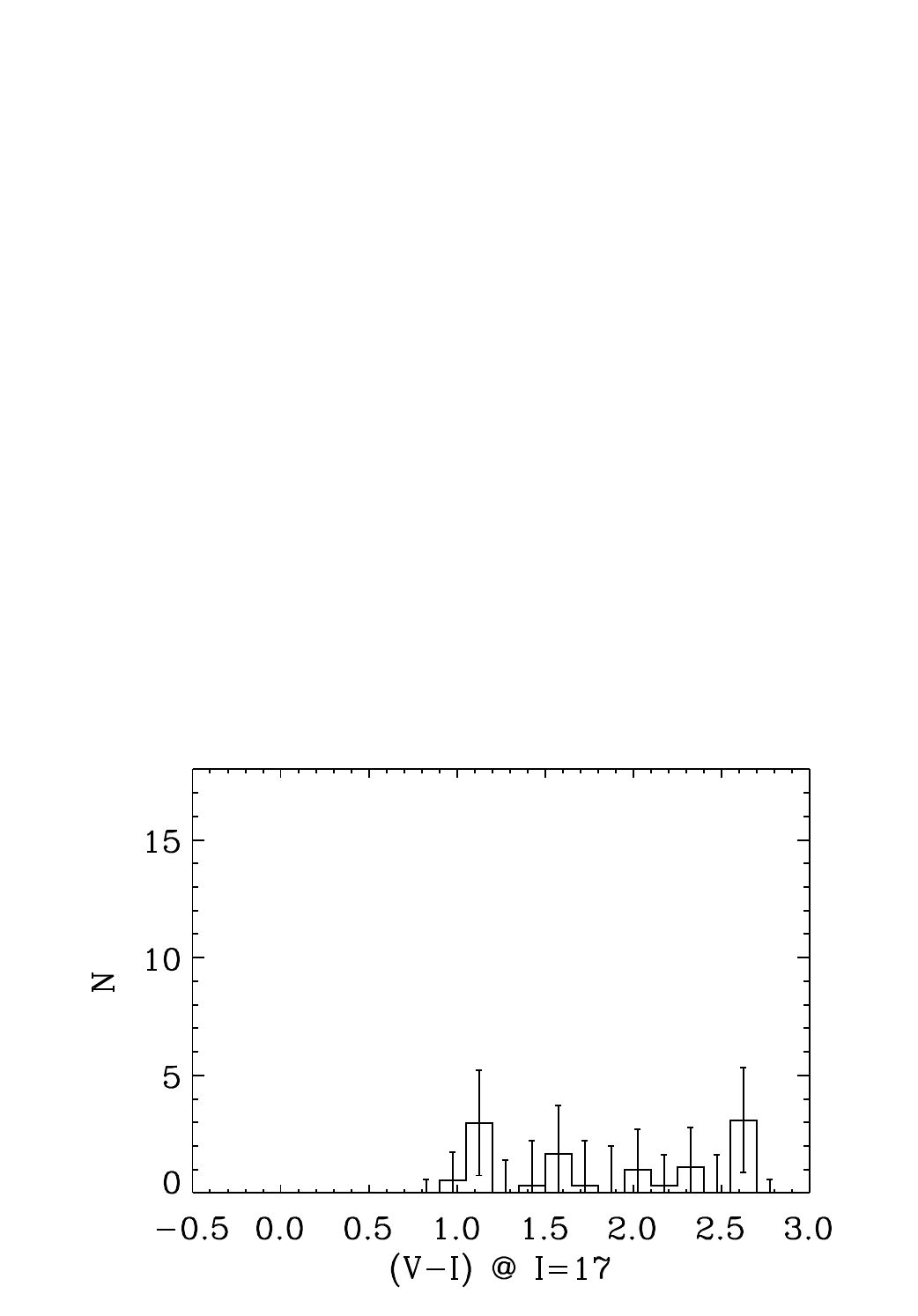}}
\resizebox{0.3\textwidth}{!}{\includegraphics[bb=0 0 283 198,clip]{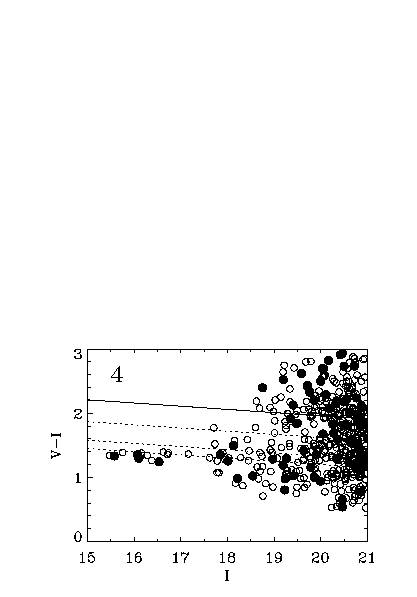}}
\resizebox{0.3\textwidth}{!}{\includegraphics[bb=0 0 283 198,clip]{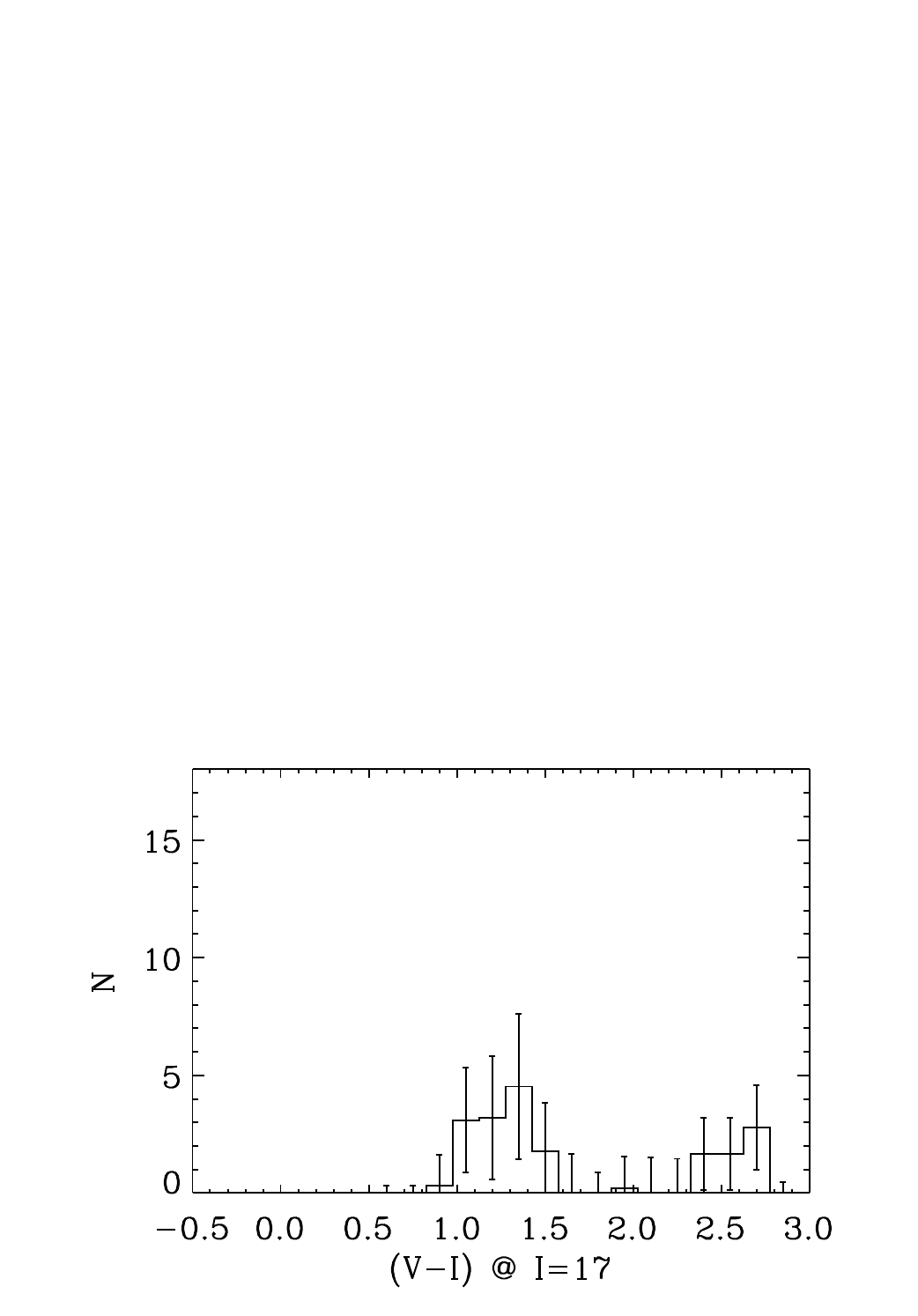}}
\resizebox{0.3\textwidth}{!}{\includegraphics[bb=0 0 283 198,clip]{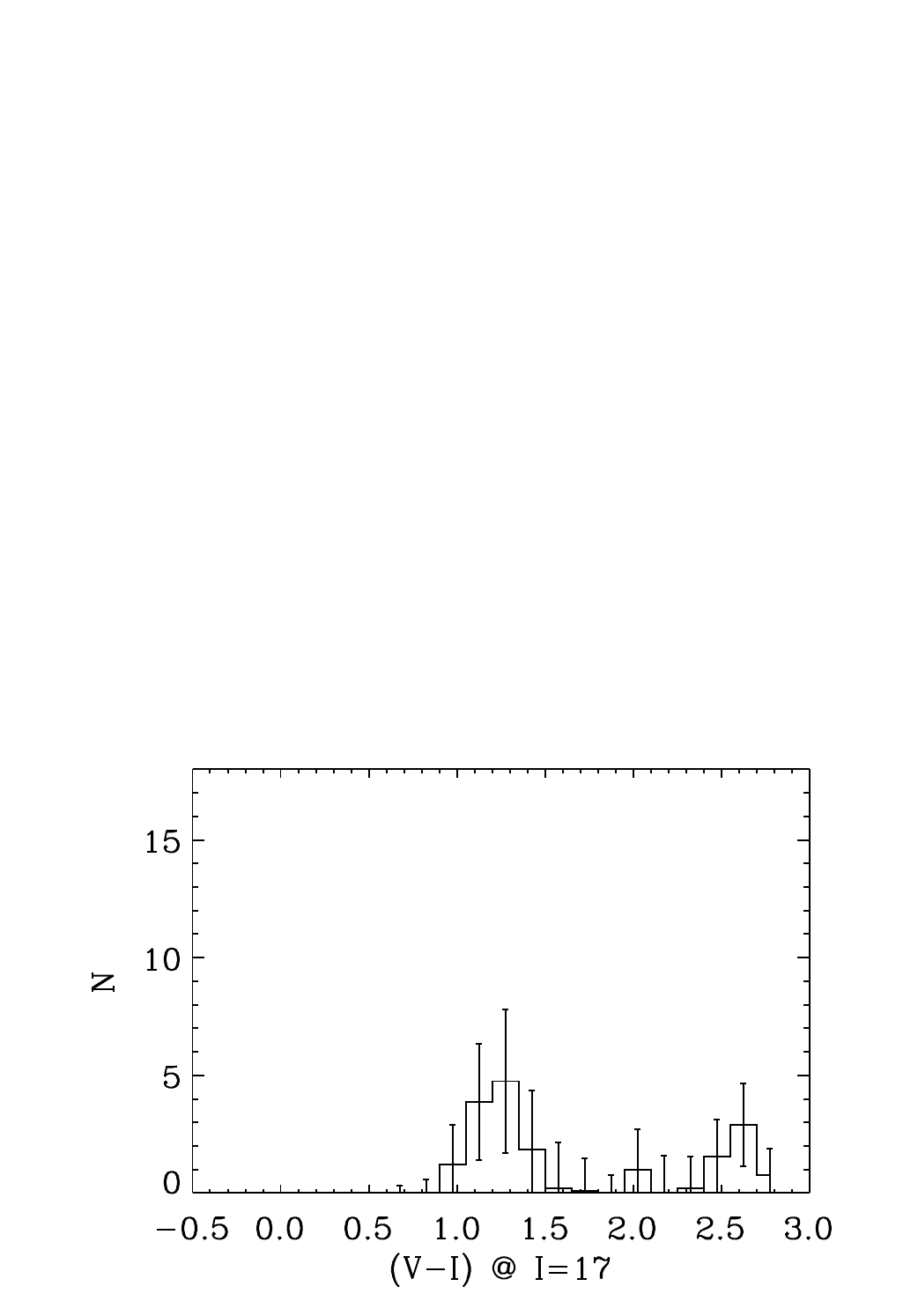}}
\resizebox{0.3\textwidth}{!}{\includegraphics[bb=0 0 283 198,clip]{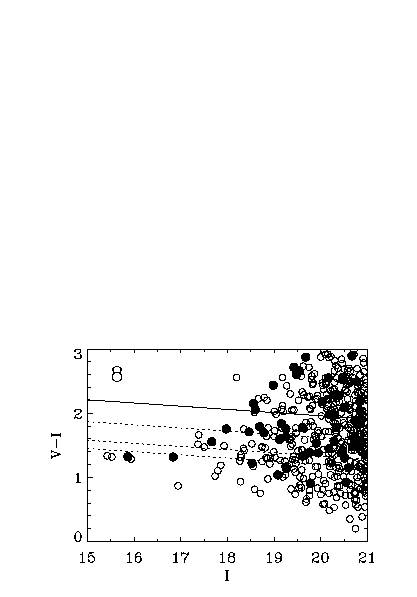}}
\resizebox{0.3\textwidth}{!}{\includegraphics[bb=0 0 283 198,clip]{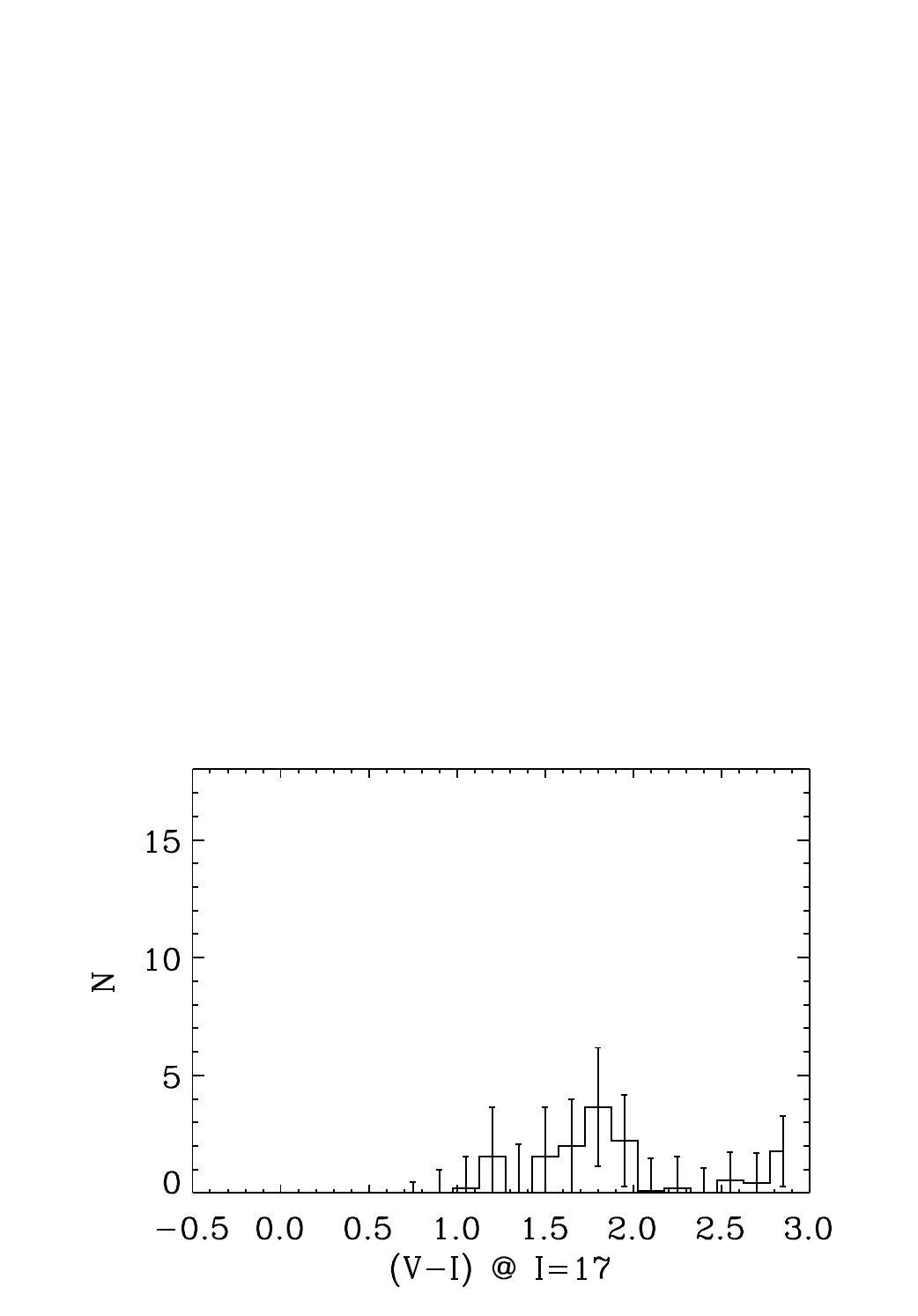}}
\resizebox{0.3\textwidth}{!}{\includegraphics[bb=0 0 283 198,clip]{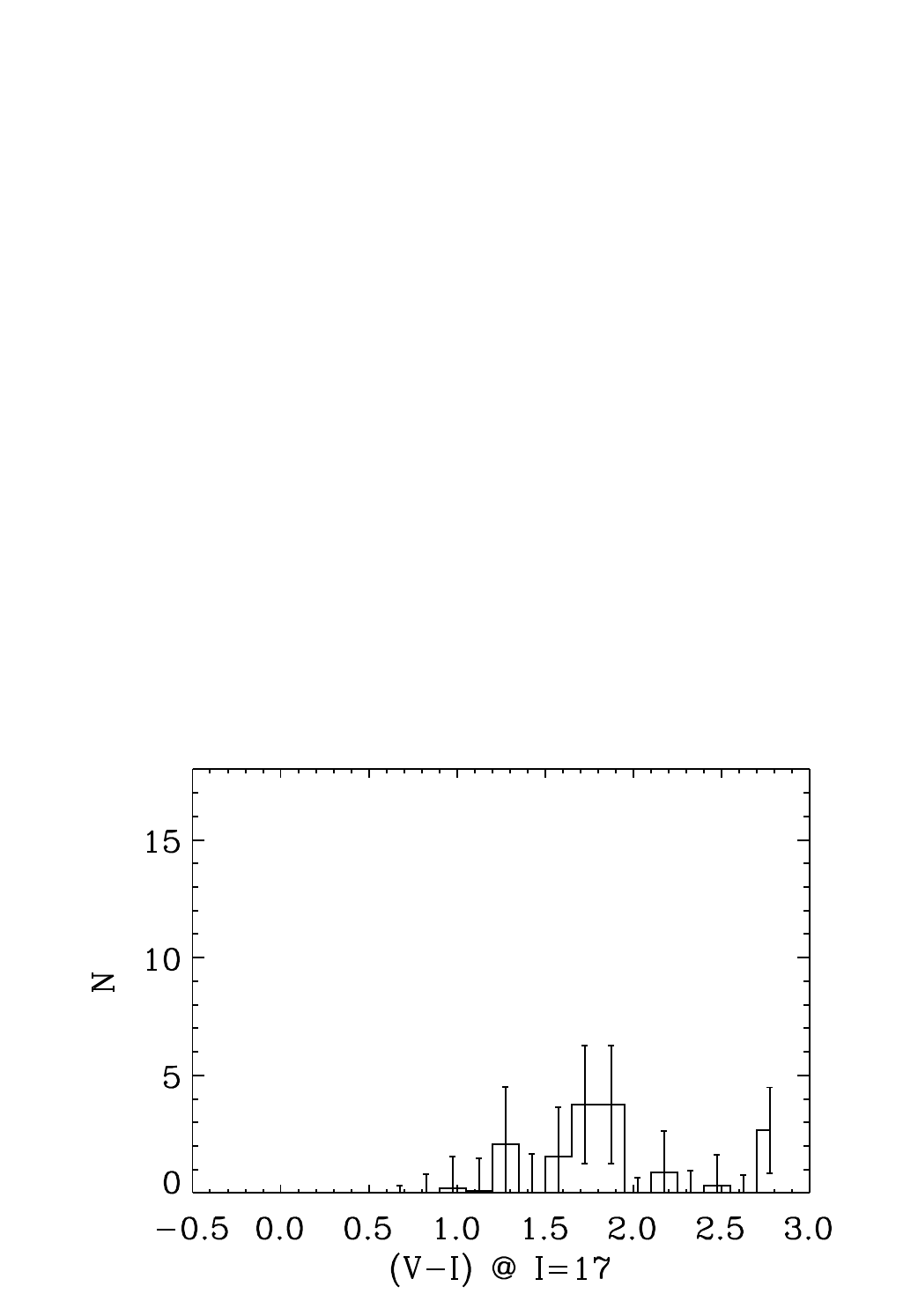}}
\resizebox{0.3\textwidth}{!}{\includegraphics[bb=0 0 283 198,clip]{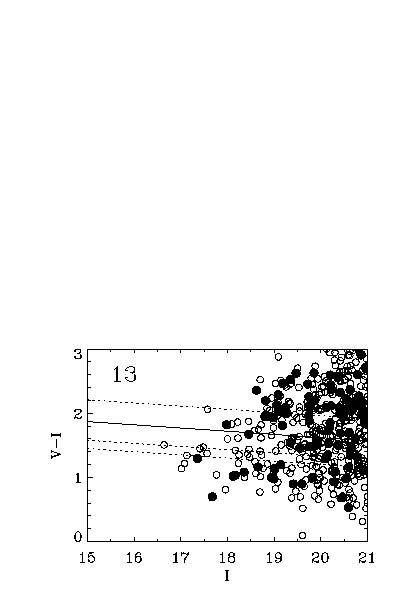}}
\resizebox{0.3\textwidth}{!}{\includegraphics[bb=0 0 283 198,clip]{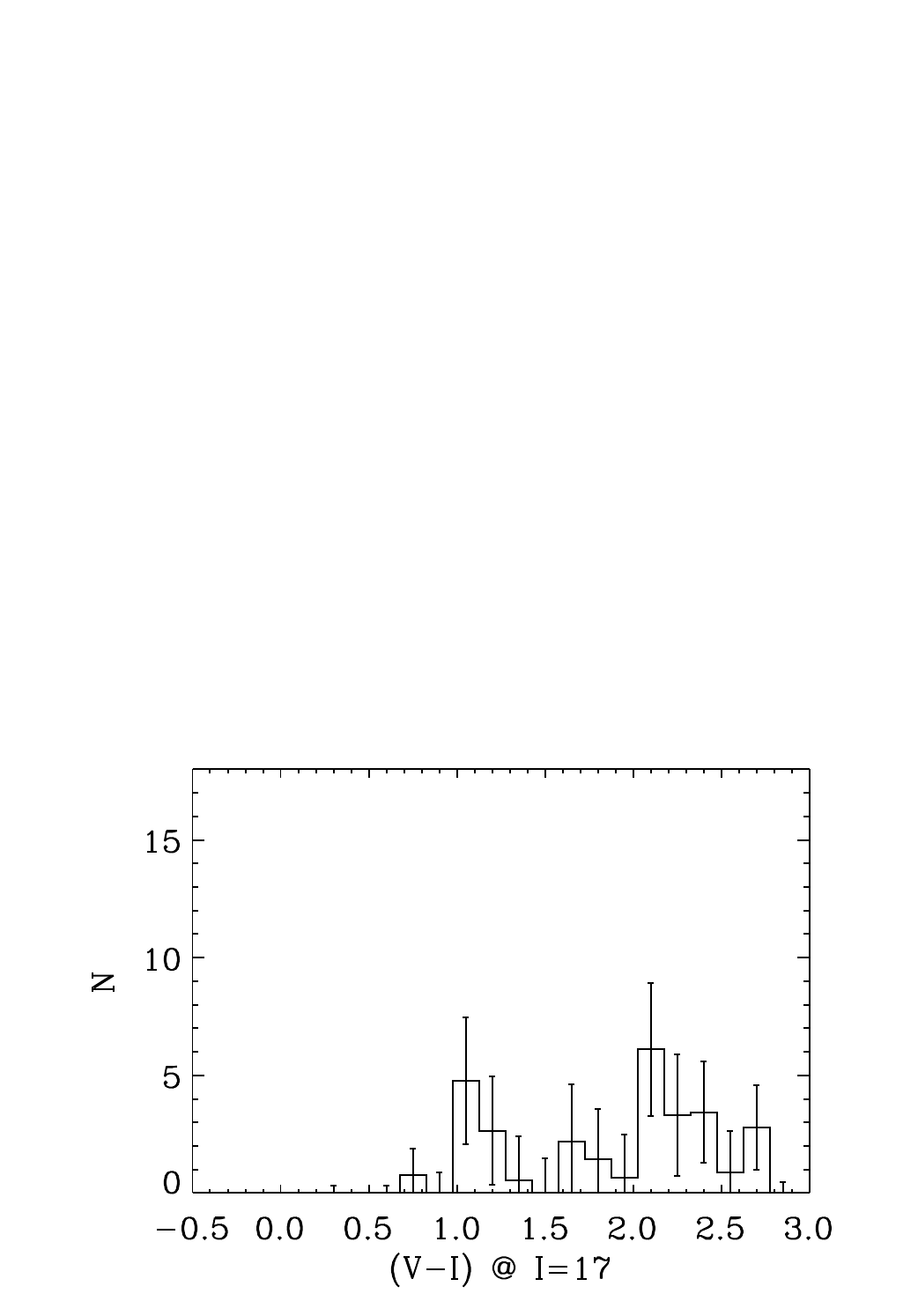}}
\resizebox{0.3\textwidth}{!}{\includegraphics[bb=0 0 283 198,clip]{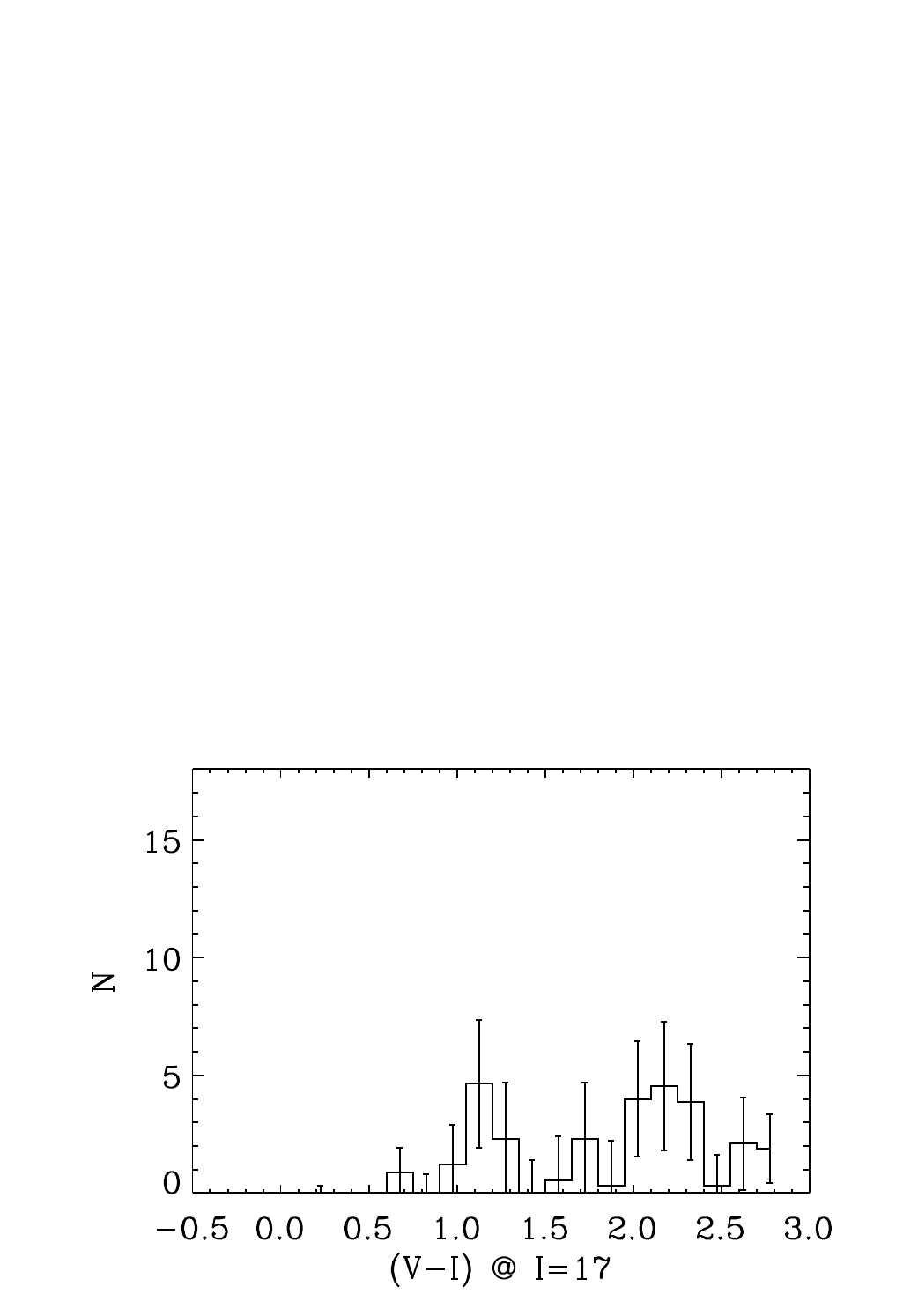}}
\resizebox{0.3\textwidth}{!}{\includegraphics[bb=0 0 283 198,clip]{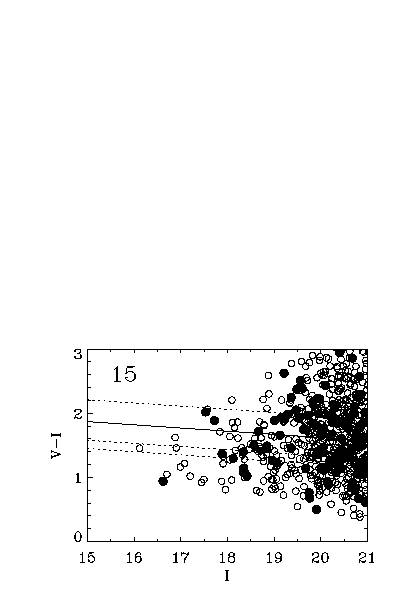}}
\resizebox{0.3\textwidth}{!}{\includegraphics[bb=0 0 283 198,clip]{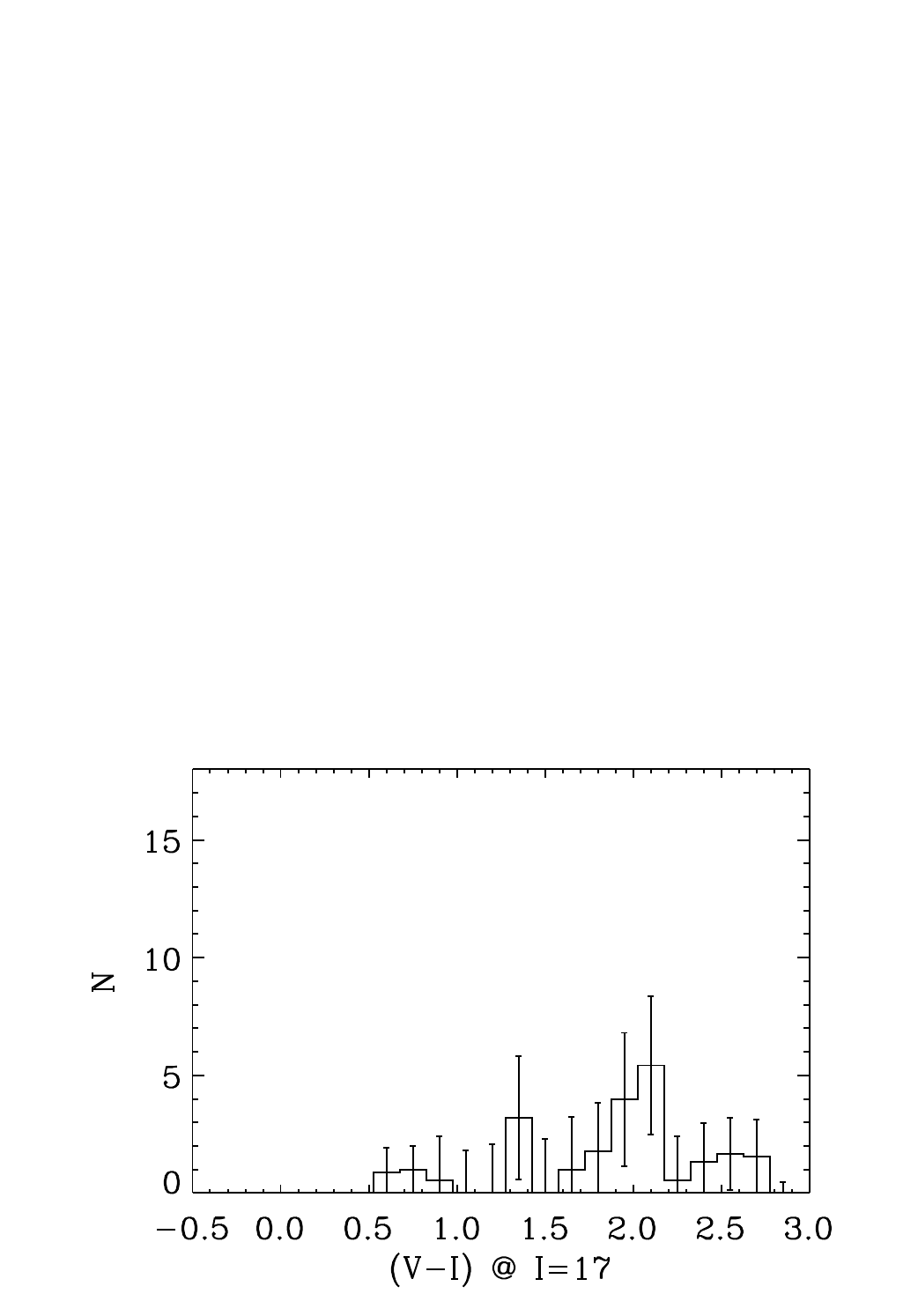}}
\resizebox{0.3\textwidth}{!}{\includegraphics[bb=0 0 283 198,clip]{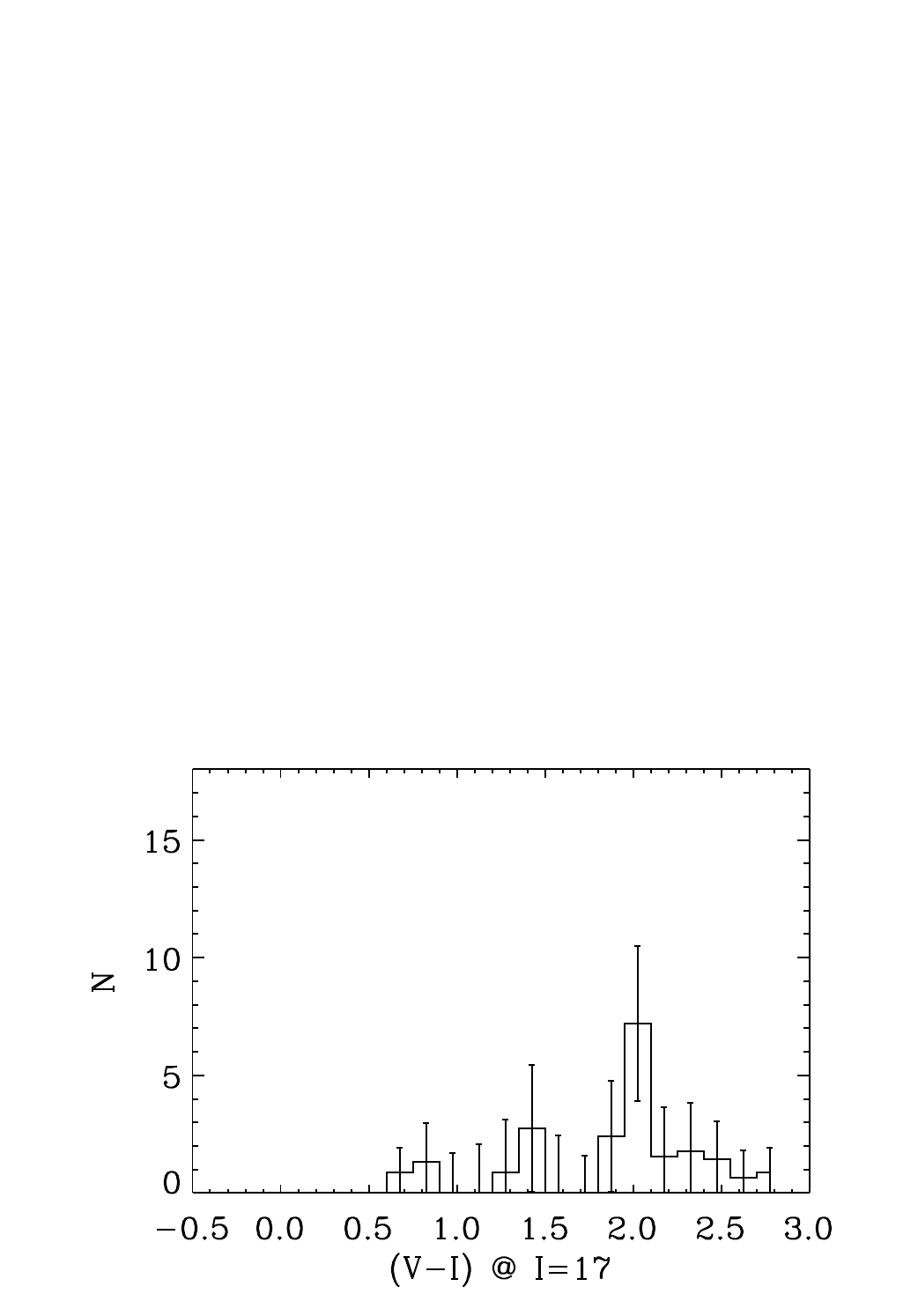}}
\end{center}
\caption{ Colour-magnitude diagrams for the ten non-confirmed
systems with $V-$ and $I-$band data available. The filled circles mark
galaxies inside the cluster region, while open circles mark objects in
the background region. The lines mark the expected position of the red
sequence at redshifts 0.1, 0.2, 0.3 and 0.4, with the solid line
marking the one expected from the matched-filter estimated
redshift. The second and third columns the two shifted, tilted
histograms similar to the ones of Fig.~\ref{fig:cmd} in the middle
column. A magnitude cut of $I\leq20$ is applied for constructing the
tilted histograms.}
\label{fig:cmd_nonconf}
\end{figure*}

\addtocounter{figure}{-1}

\begin{figure*}
\begin{center}
\resizebox{0.3\textwidth}{!}{\includegraphics[bb=0 0 283 198,clip]{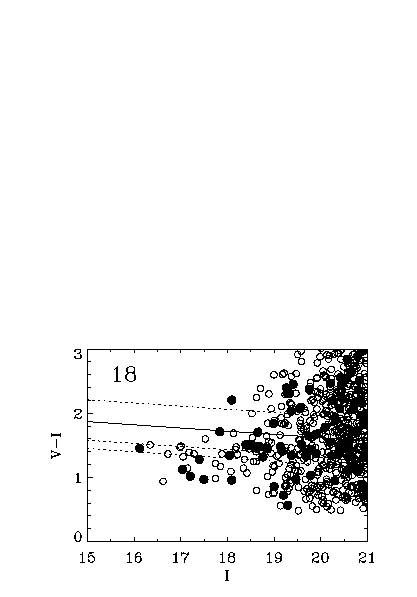}}
\resizebox{0.3\textwidth}{!}{\includegraphics[bb=0 0 283 198,clip]{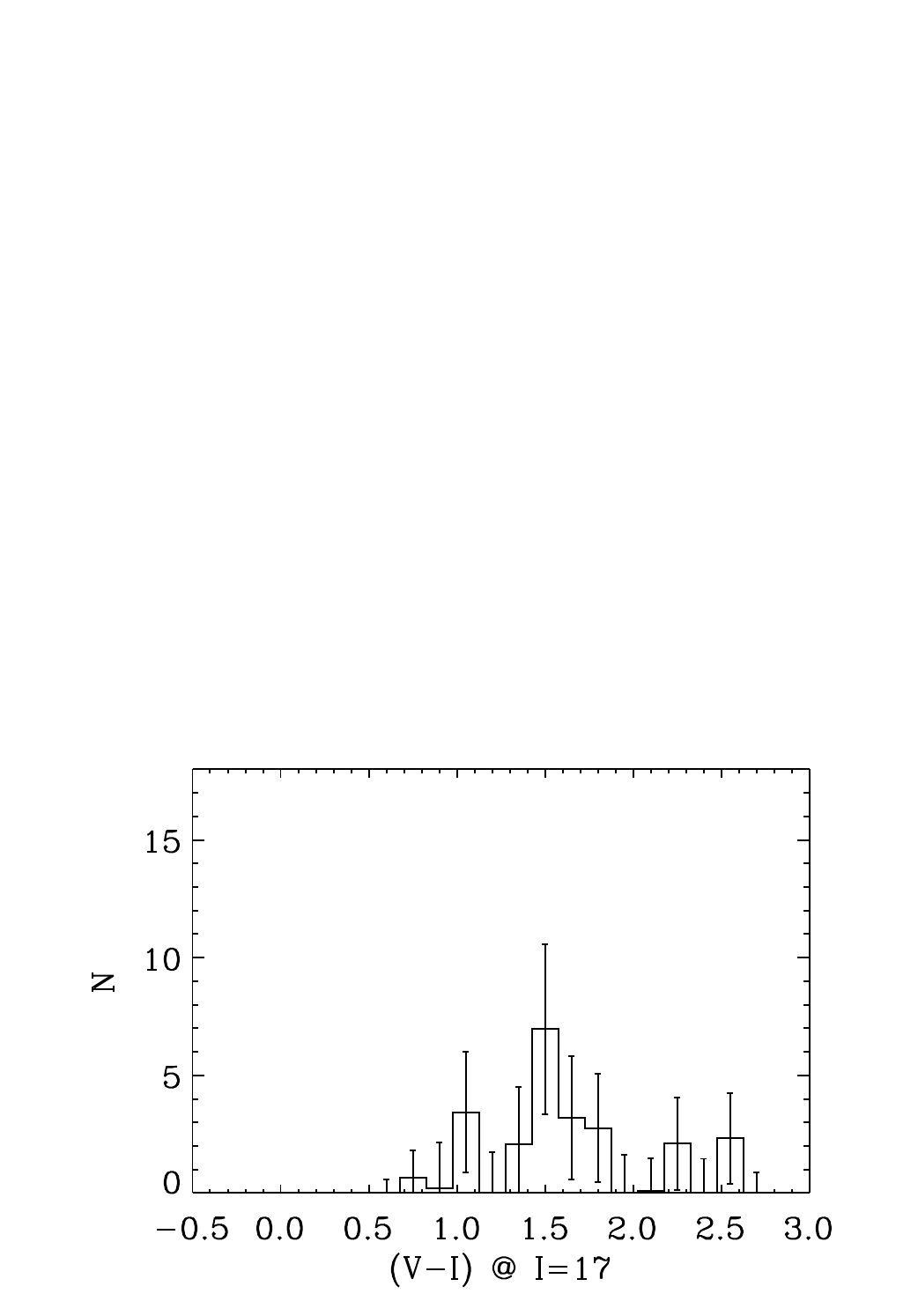}}
\resizebox{0.3\textwidth}{!}{\includegraphics[bb=0 0 283 198,clip]{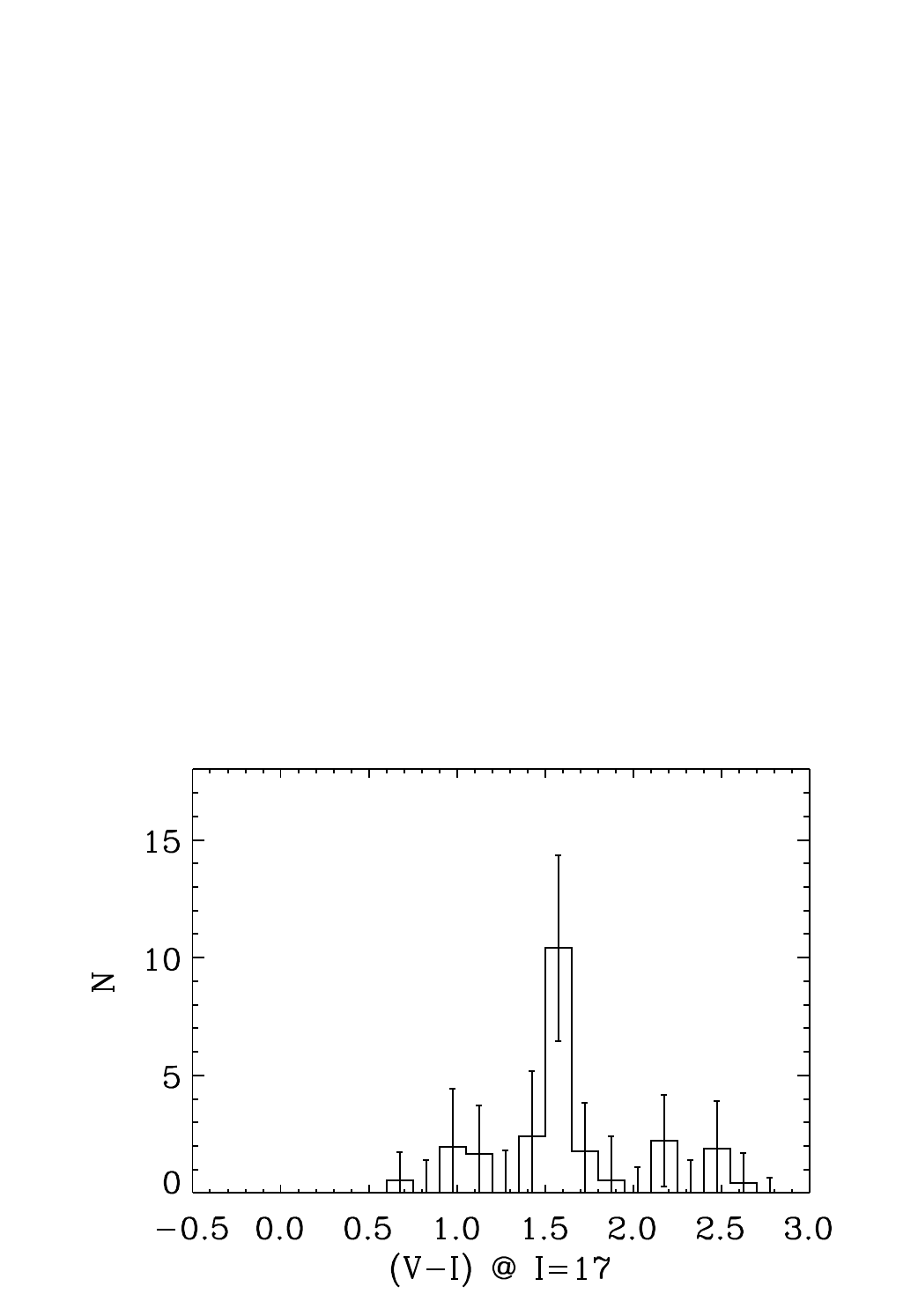}}
\resizebox{0.3\textwidth}{!}{\includegraphics[bb=0 0 283 198,clip]{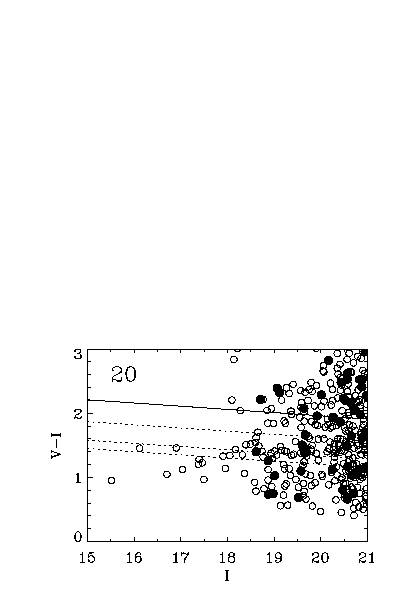}}
\resizebox{0.3\textwidth}{!}{\includegraphics[bb=0 0 283 198,clip]{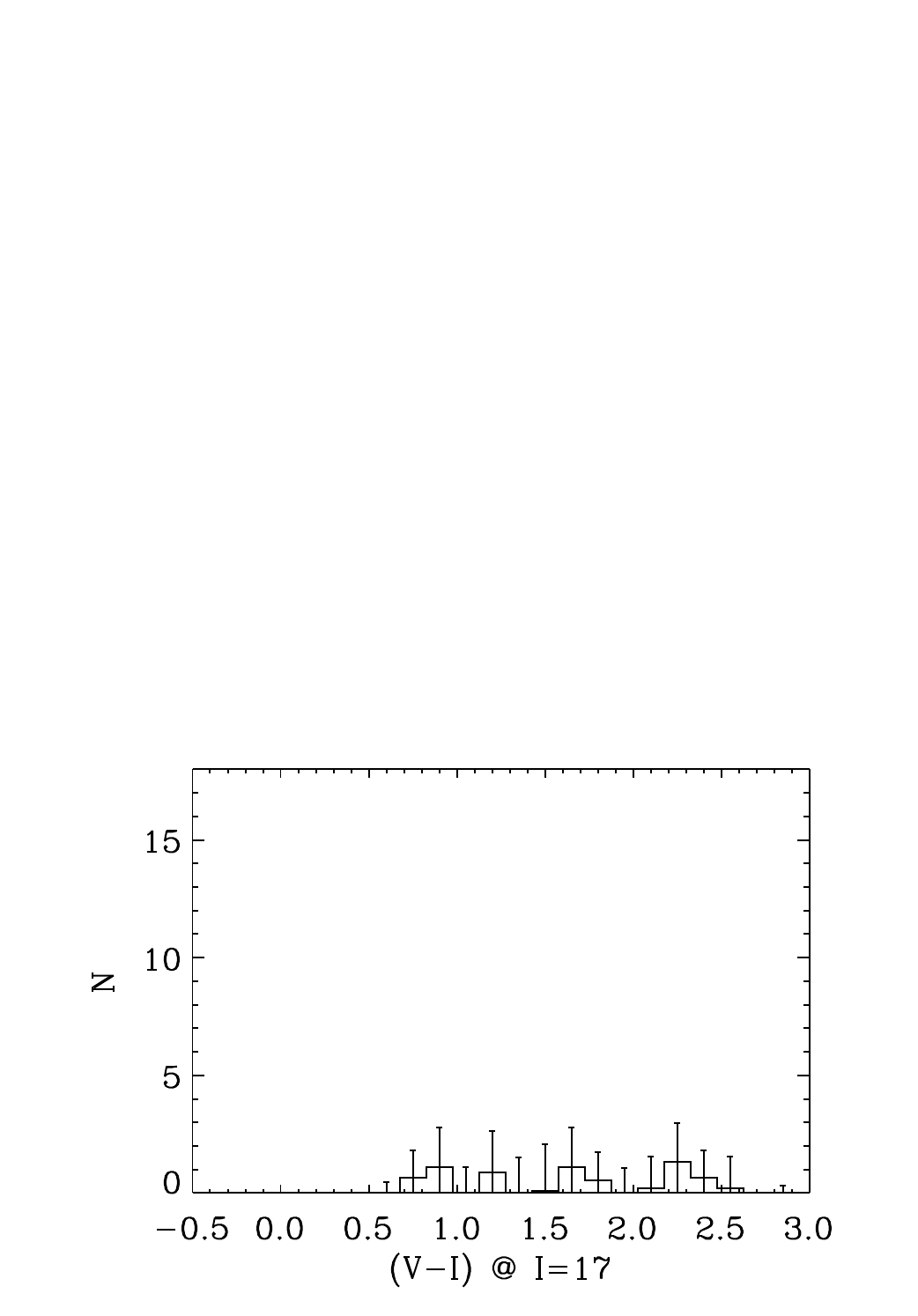}}
\resizebox{0.3\textwidth}{!}{\includegraphics[bb=0 0 283 198,clip]{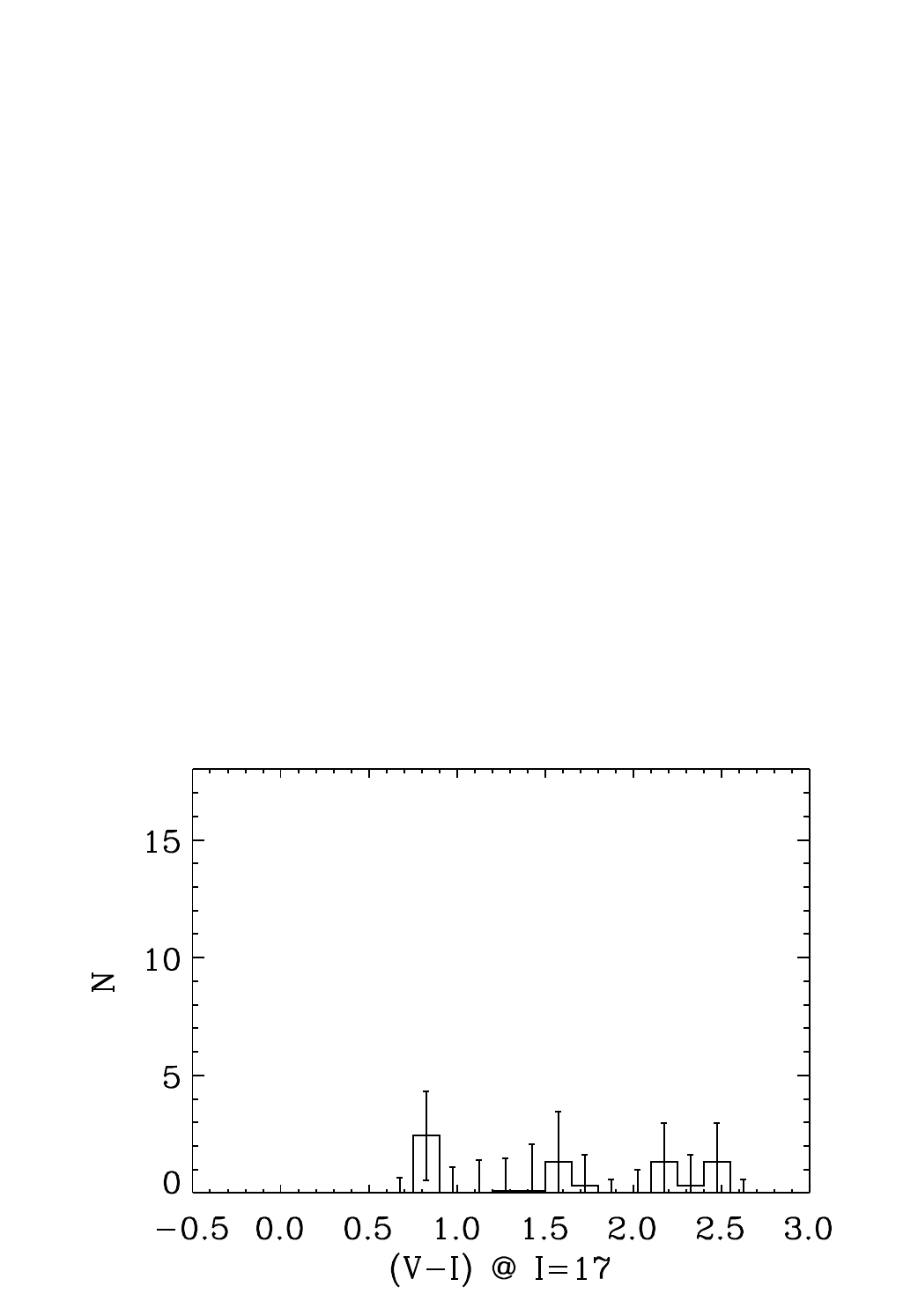}}
\resizebox{0.3\textwidth}{!}{\includegraphics[bb=0 0 283 198,clip]{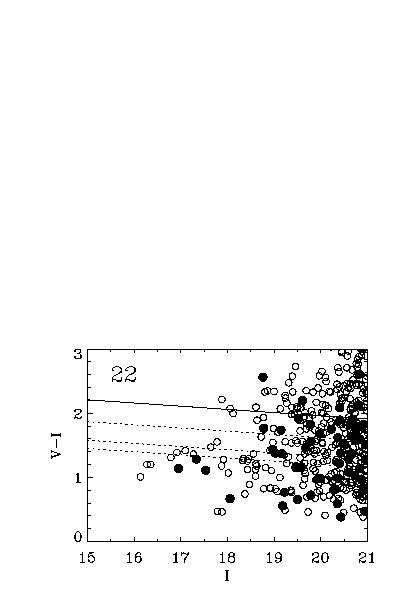}}
\resizebox{0.3\textwidth}{!}{\includegraphics[bb=0 0 283 198,clip]{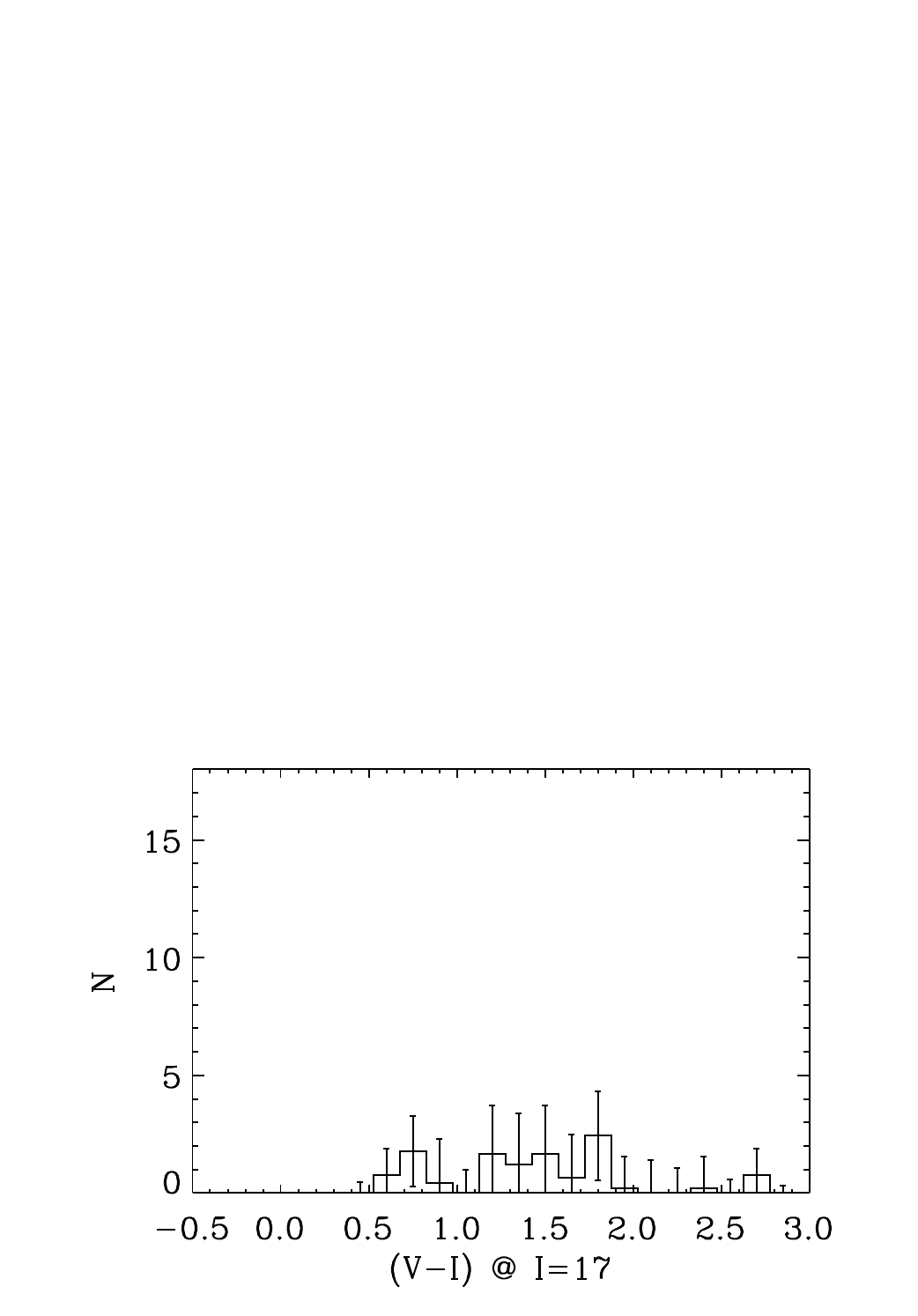}}
\resizebox{0.3\textwidth}{!}{\includegraphics[bb=0 0 283 198,clip]{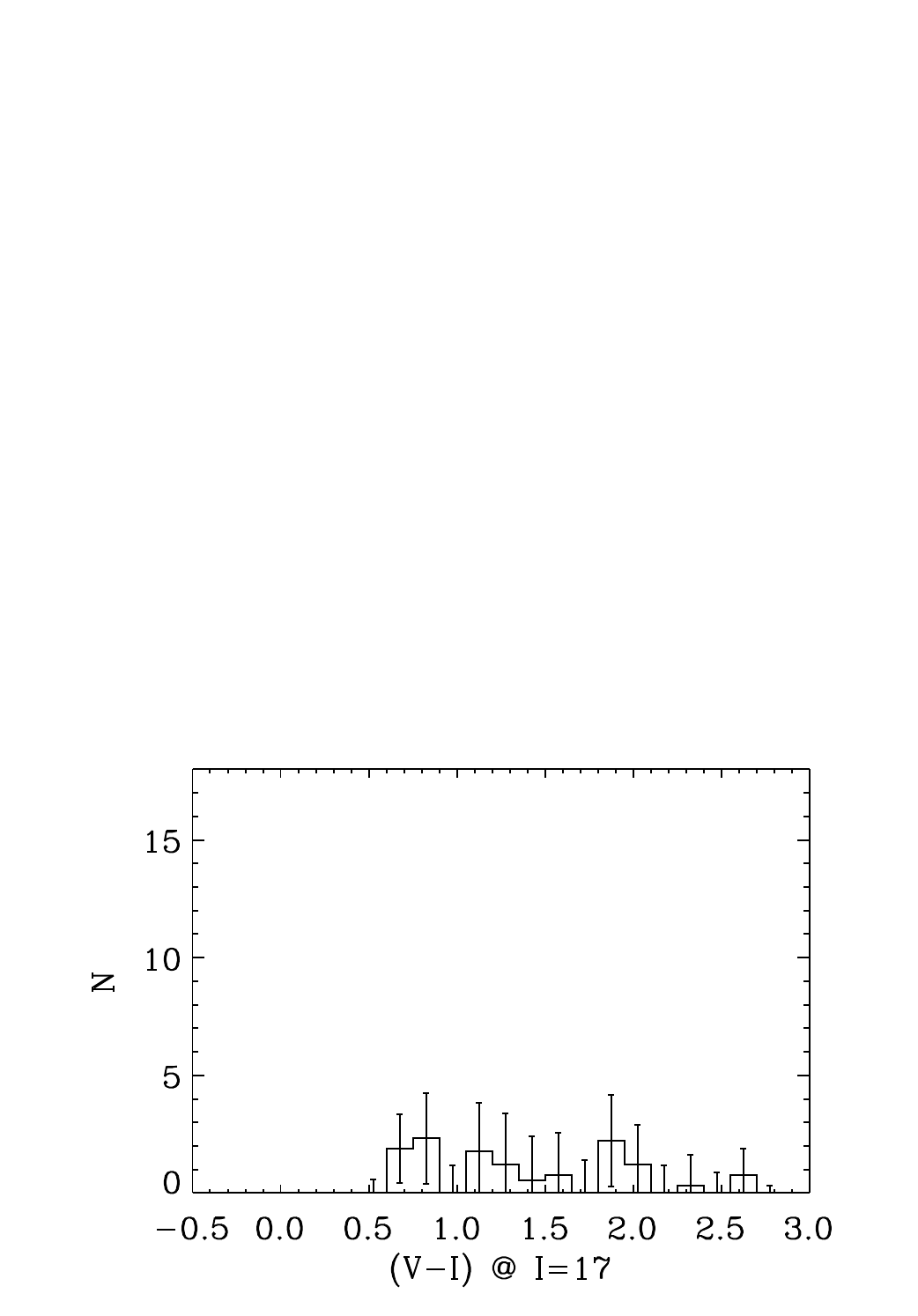}}
\resizebox{0.3\textwidth}{!}{\includegraphics[bb=0 0 283 198,clip]{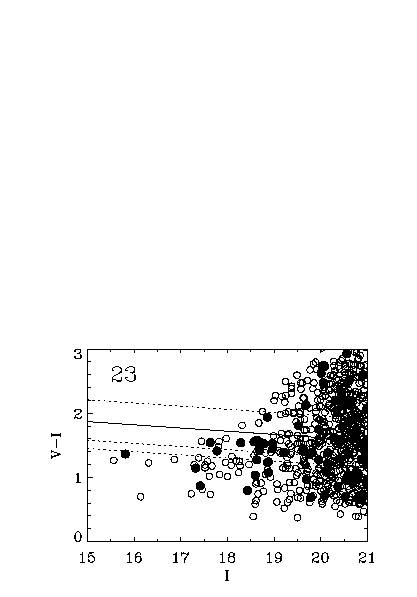}}
\resizebox{0.3\textwidth}{!}{\includegraphics[bb=0 0 283 198,clip]{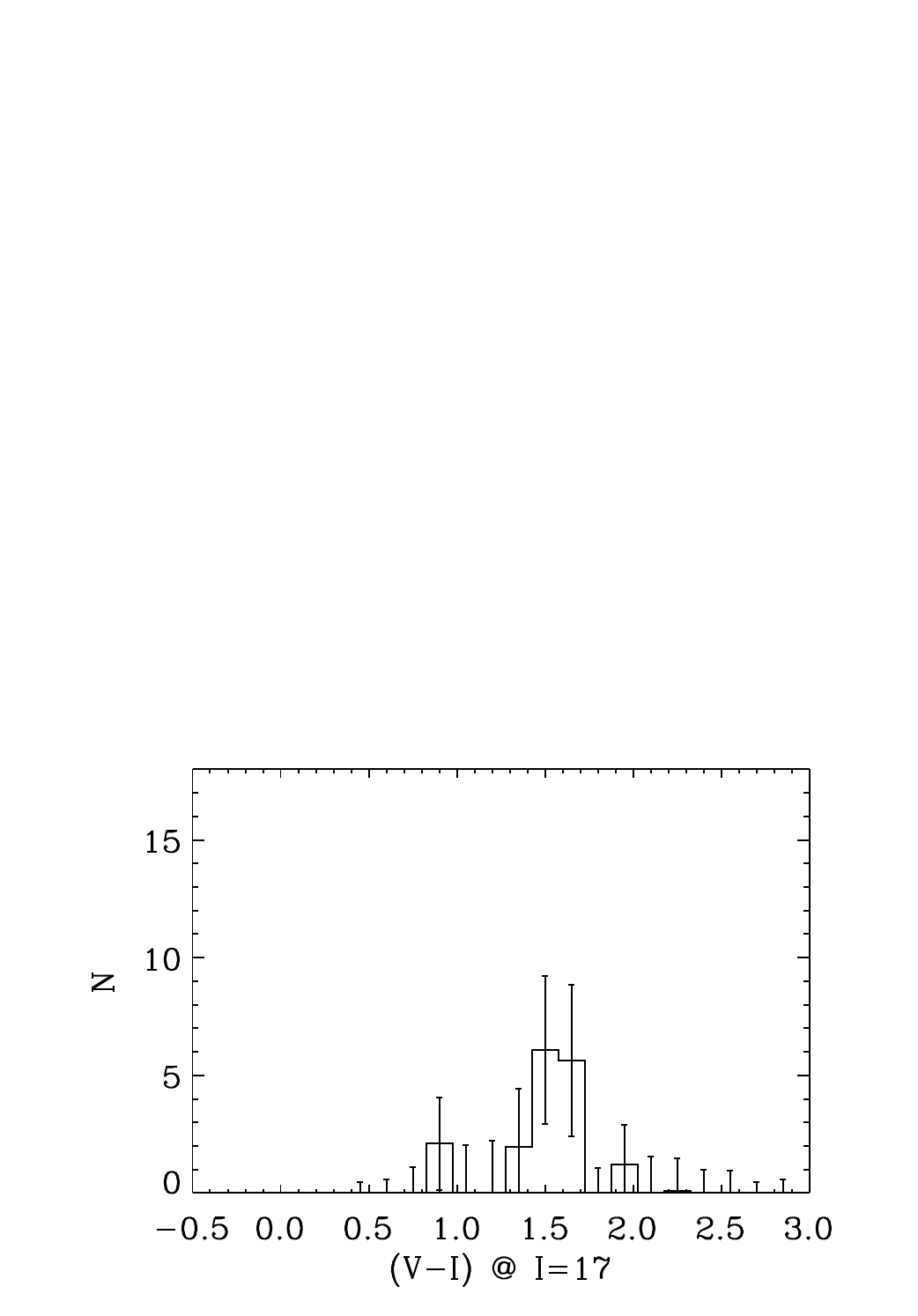}}
\resizebox{0.3\textwidth}{!}{\includegraphics[bb=0 0 283 198,clip]{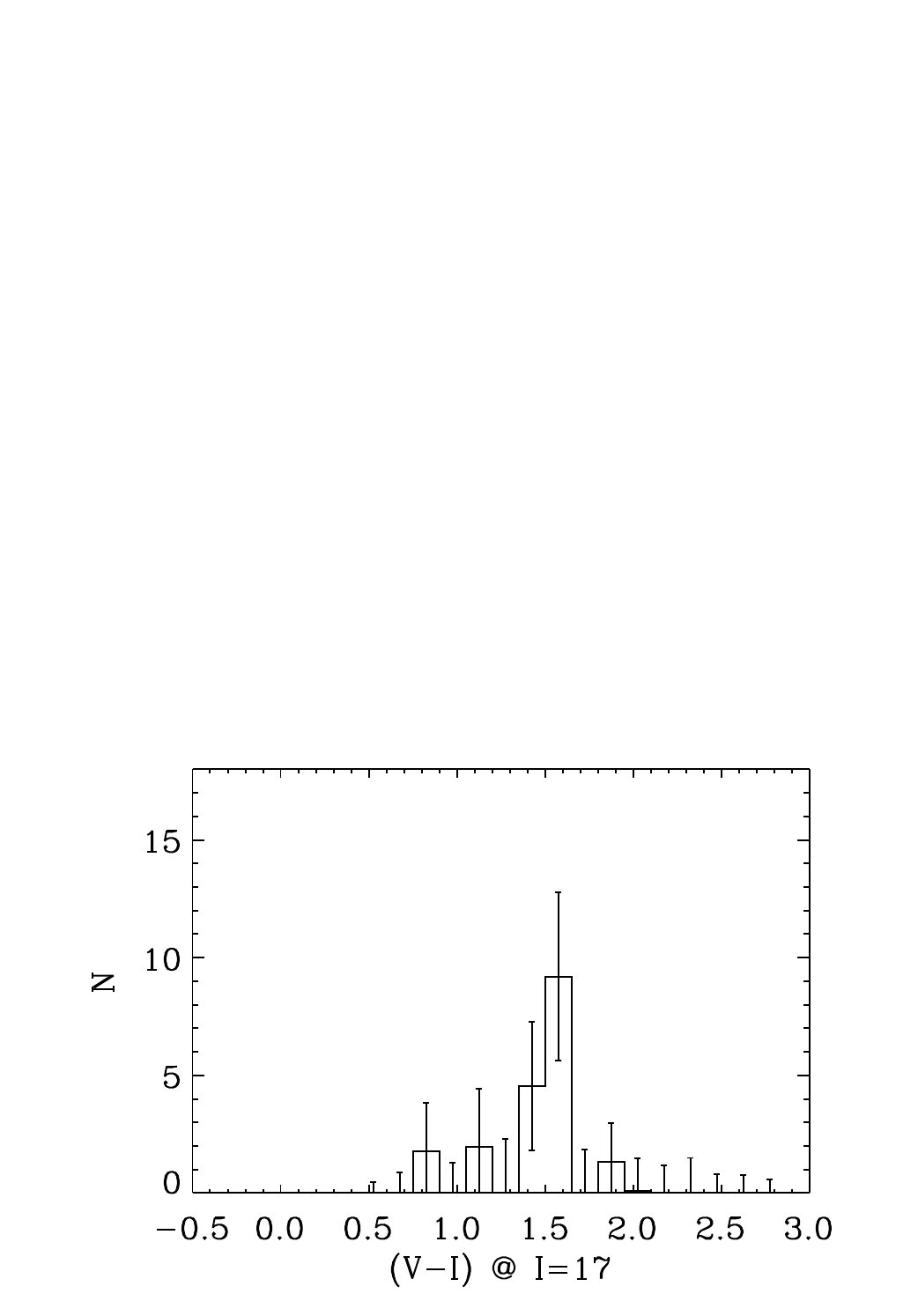}}
\end{center}
\caption{\it -- Continued.}
\end{figure*}

\subsection{CMDs for non-confirmed candidates}
\label{sec:nonconf}

We have also used the available colour information, to explore in more
detail the fields within which we were unable to identify significant
groups from the spectroscopic data possibly associated with
matched-filter detections. From the 13 fields without a detection,
there are 10 fields for which we have colour information. The CM
diagrams for these 10 fields are shown in
Fig.~\ref{fig:cmd_nonconf}. The filled symbols represent galaxies
within a radius of $0.75h^{-1}_{75}\mathrm{Mpc}$ as defined by the
redshift estimated for the matched-filter detection. The lines mark
the expected position of the red sequence for redshifts $z=0.1, 0.2,
0.3, 0.4$. The solid lines indicate the relation for the
matched-filter estimated redshifts.  Also shown are tilted histograms
also used in the analysis of the galaxy populations of the confirmed
systems. The histograms are constructed considering galaxies brighter
than $I\leq20$ and shifted by half the bin width in colour. These
histograms are used to identify the most significant peaks and their
S/N ratios with the results summarised in
Table~\ref{tab:nonconf_color}. The table gives: in Cols.~1 and ~2 the
field identifier and cluster name, in Cols.~3 and 5 the colour of the
peaks for each of the tilted histograms; and in Cols.~4 and 6 the
related S/N values.  From the figures and the values of S/N ratio
listed in Table~\ref{tab:nonconf_color}, it is clear that there are no
obvious red sequences. The marginal detections of red sequences with
S/N$>2$ are compared with the marginal detections of groups in
redshift space listed in Table~\ref{tab:marginal}. There are four such
cases that are discussed below:

\begin{itemize}

\item For EISJ2241-4006 (\#13) the colour is found to be $V-I\sim2.1$,
this roughly corresponds to a redshift of $z\sim0.4$. In the redshift
distribution in Fig.~\ref{fig:redshift_dists} we find a group with six
members at $z_{spec}=0.344$ with a significance of 94.5\%.

\item For EISJ2243-4010A (\#15) the colour is found to be 2.025, which
corresponds to a redshift slightly smaller than $z=0.4$. In the
redshift distribution this could correspond to the group at $z=0.345$
with three members and a significance of 73.4\%.

\item For EISJ2243-4010B (\#18) the colour is found to be 1.575
roughly corresponding to $z=0.25$. This could correspond to both
groups found in the field. However, the richest group has its centre
very far from the MF position and is thus unlikely to give rise to
the red sequence at the MF position. The other group with three
members at $z\sim0.285$ and a significance of 80.5\% is very spread
over the field and from these data we cannot conclude that it is
related to the MF detection.

\item For EISJ2248-4015 (\#23) the colour is also 1.575 corresponding
to $z\sim0.25$. This is consistent with the group at $z=0.246$ with a
significance of 98.7\%, which was also mentioned above as the likely
origin of the MF detection.

\end{itemize}

Even though these results are by themselves inconclusive, they provide
support that the matched-filter detections are, at least for the four
cases listed above, indeed related to a cluster. However, a better
spectroscopic coverage is needed to firmly establish the possible
presence both of a cluster and a red sequence.

\begin{table}
\caption{The colours and S/N ratio for the detected peaks in each of the
tilted histograms for the non-confirmed candidates.}
\label{tab:nonconf_color}
\begin{tabular}{rlrrrr}
\hline\hline
ID & Field & $(V-I)_1$ & $(S/N)_1$& $(V-I)_2$ & $(S/N)_2$ \\
\hline
1  & EISJ0044-2950A & 2.550 & 1.9 & 1.725 & 1.6\\
2  & EISJ0045-2944  & 1.050 & 1.4 & 2.625 & 1.4\\
4  & EISJ0048-2928  & 1.350 & 1.5 & 1.275 & 1.6\\
8  & EISJ2236-4014  & 1.800 & 1.4 & 1.725 & 1.5\\
13 & EISJ2241-4006  & 2.100 & 2.2 & 1.125 & 1.7\\
15 & EISJ2243-4010A & 2.100 & 1.9 & 2.025 & 2.2\\
18 & EISJ2243-4010B & 1.500 & 1.9 & 1.575 & 2.6\\
20 & EISJ2244-4008  & 2.250 & 0.8 & 1.575 & 0.6\\ 
22 & EISJ2246-4012B & 1.800 & 1.3 & 1.875 & 1.1\\
23 & EISJ2248-4015  & 1.500 & 1.9 & 1.575 & 2.6\\
\hline
\hline
\end{tabular}
\end{table}

%__________________________________________________________________

\section{Discussion}
\label{sec:conclusions}
%__________________________________________________________________

In this paper we present new redshifts for 747 galaxies in the fields
of 23 EIS cluster candidates with redshifts $z_{MF}=0.3-0.4$.  In one
of the fields two candidates were found, but the spectroscopic
coverage was only sufficient for investigating the main target.  The
main aim of this spectroscopic survey was to identify overdensities in
redshift space corresponding to the cluster candidates detected by the
matched-filter search technique in order to evaluate the reliability
of the results of this algorithm as well as investigate the
characteristics of the detected systems. The combination of
spectroscopic and photometric data sets available for these cluster
candidates was used to characterise the confirmed systems.

For 10 of the 23 ($\sim43\%$) targeted candidates we confirm the
presence of an overdensity in redshift space. This confirmation rate
is significantly smaller than that of Papers~I--III, where $\sim95\%$
of the systems with $z_{MF}=0.2$ were confirmed.  This lower rate is
most likely due to the brighter magnitude limit of the spectroscopic
survey, which also implied probing primarily the low end of the
redshift range being considered.  It is worth noting, that in five of
the fields observed even though no system was confirmed there was
clear evidence, based on the visual inspection of the images of these
fields, for the presence of a concentration of fainter galaxies at the
position of the cluster candidate. Unfortunately, these fainter
galaxies did not have a measured redshift. A final word on the nature
of these candidates will require further observations of these fields.
Furthermore, four out of five confirmed candidates with colour
information show evidence for a red sequence lending further support
to the interpretation that these systems are bound and have early-type
galaxies in their core.

From the inspection of the colour-magnitude diagrams for 10
non-confirmed candidates suggest that four of these are possibly
associated with a bound system based on the presence of a weak signal
in the colour histograms matching the redshift of a  marginal
detection in redshift space.

Comparison of the overall properties of the sample discussed in the
present paper ($z_{MF}=0.3-0.4$) and that of our previous
Papers~I--III ($z_{MF}=0.2$) leads to the following conclusions:

\begin{itemize}

\item The mean redshift of the confirmed sample is 0.24 higher than
0.18 measured in Paper~III. This indicates that we have successfully
identified systems at higher redshift, even though at a lower rate,
which as indicated throughout the paper was mostly due to a bright
limiting magnitude of the spectroscopic data.

\item Consistent with our findings in Paper~III, $\sim$60\% of the
identified overdensities in redshift space with photometric data in
two passbands show evidence for a red sequence among their
spectroscopic members with a colour consistent with their measured
redshift. This lends credence to the interpretation that these
overdensities are indeed bound systems with a well-defined population
of early-type galaxies. On the other hand, the remaining cases may not
be totally discarded as they may still be spiral-rich bound systems. A
final conclusion will require a considerable improvement in sampling,
a hard task considering the angular extent of these systems.

\item The distribution of velocity dispersions is consistent with
that obtained by the much larger SDSS survey.

\item The range of richness of our confirmed systems is in good
agreement with those of \citet{bahcall03} analysing a subset of the
SDSS data using a similar implementation of the matched-filter
algorithm. As previous authors we find that richness as estimated by
the matched filter is not an adequate proxy for the mass of the
system.

\end{itemize}

\section{Summary}
\label{sec:summary}

In this series of four papers we have measured 1954 galaxy redshifts
in the range 0.0065--0.6706 over 58 fields around EIS candidate
clusters identified using the matched-filter algorithm and with
estimated redshifts below $z_{MF}=0.4$. For a total of 42 cases we
were able to associate the candidate with density enhancements in
redshift space with mean redshifts between 0.095 and 0.534.  This
represents a yield of $\sim$75\% for the matched-filter technique,
which is consistent with the original estimates by \cite{olsen99a}
based on simulations and also comparable to the estimates
reported by \citet{kim02}. The method tends to overestimate the
system redshift reaching an offset of 0.1 at $z_{MF}\lesssim0.4$. The
number of galaxies with concordant redshifts range from four to
35. The one-dimensional velocity dispersions of the identified density
enhancements, which we have referred to as systems, vary from very low
values up to $\sim$1400~km/s, with the equivalent number of $L^*$
galaxies varying from 11 to 65. Due to the undersampling of the
systems by our spectroscopic observations, these values, in general,
have large errors and cannot by themselves distinguish between chance
enhancements, small groups and rich clusters. Based on the
results of this series we find that the identified clusters consist of
substructures in more than 35\% of the cases, impacting the
matched-filter richness estimates.

For 34 of the confirmed groups and clusters colour information is
available and was used to search for a red sequence. For 20 systems,
corresponding to $\sim$60\%, we were able to detect one using the
spectroscopically confirmed members.  While having a red sequence is
not a necessary condition for a bound system, its existence definitely
lends further credence to the detection. On the other hand, the
matched-filter technique may detect systems with a large fraction of
late-type galaxies or where the early-type galaxies are affected by
recent starbursts that may occur during merging of groups or clusters.
The confirmation of this idea will require more extensive
spectroscopic surveys than the one carried out in the present program.

This series complements the work carried out by
\cite{ramella00,benoist02,olsen05b} who carried out spectroscopic
observations of EIS candidate clusters at intermediate and high
redshifts, obtaining confirmation yields, redshift offsets relative to
the matched-filter estimates and velocity dispersions similar to those
reported here.

Combining the results of this series with those of the previous
papers, we have obtained roughly 2500 redshifts for a total of 74
cluster candidate fields out of 302 in the original candidate list,
spanning the redshift range from 0.2 to 1.3, being perhaps one of the
most extensive spectroscopic follow-up works of its kind. Taken
altogether, these results show that the matched-filter technique leads
to the detection of real density enhancements in redshift space. It
should be used in conjunction with other cluster search techniques for
the optimal identification of clusters of galaxies from optical and
near-infrared imaging data. This is extremely important considering
that large surveys such as UKIDSS, Dark Energy Survey and VISTA are
either on-going or are envisioned to start in the next few years
providing unprecedented samples of candidate clusters. Understanding
the differences and possible biases of different detection techniques
is particularly important for determining the completeness of the
observed samples and be able to use, for instance, number counts of
clusters of galaxies as a tool to constrain cosmological parameters as
proposed by several new dark energy projects.

On-going and future surveys will provide considerable more
information enabling the application of other detection algorithms
based on for instance colours or photometric redshifts. However,
regardless of the detection algorithm optical cluster surveys will
always have to rely on extensive spectroscopic follow-up both for
confirmation, membership assignment, subclustering and for fully
characterising the systems.  Some of these issues are addressed in the
present work.

%__________________________________________________________________

\begin{acknowledgements}
We thank the anonymous referee for useful comments which greatly helped
improving the paper.  We thank John Pritchard, Lisa Germany and Ivo
Saviane for making the pre-imaging observations of the fields. We
would also like to thank the 2p2 team, La Silla, for their support at
any time during the observations.  We are also in debt to Morten
Liborius Jensen for preparing the slit masks. This research has made
use of the NASA/IPAC Extragalactic Database (NED) which is operated by
the Jet Propulsion Laboratory, California Institute of Technology,
under contract with the National Aeronautics and Space Administration.
This work has been supported by The Danish Board for Astronomical
Research.  LFG acknowledges financial support from the Carlsberg
Foundation and the Danish Natural Science Research Council.  The Dark
Cosmology Centre is funded by the Danish National Research Foundation.

\end{acknowledgements}

\bibliographystyle{aa}
\bibliography{/home/lisbeth/tex/lisbeth_ref}

\begin{appendix}

\section{Marginal detections of groups in redshift}
\label{app:marginal}

This appendix presents the 46 groups identified in redshift space but
which do not fulfil our significance criterion. These groups are
listed here to help the discussion of Sect.~\ref{sec:nonconf}.
Table~\ref{tab:marginal} gives in Cols.~1 and 2 the field identifier
and the cluster field name; in Col.~3 the number of spectroscopic
members of the group; in Cols.~4 and 5 the mean position in J2000; in
Col.~6 the mean redshift of the group members; in Col.~7 the velocity
dispersion corrected for our measurement accuracy. In cases where the
measured velocity dispersion is smaller than the measurement error we
list the value of $\sigma_v=0$; in Col.~8 the significance as defined
above and in Col.~9 the distance in arcmin between the group and the
original MF position.

\begin{table*}
\caption{Identified groups with a significance less than 99\%.}
\label{tab:marginal}
\begin{center}
\begin{minipage}{0.85\textwidth}
\begin{tabular}{rlrcccrrrr}
\hline\hline
ID & Cluster Field\footnote{ Fields where no additional marginal
detections were found are not included in the table.} & Members & $\alpha$ (J2000) & $\delta$ (J2000) & z & $\sigma_v \mathrm{[km/s]}$\footnote{$\sigma_v=0$ reflects a measured velocity dispersion below the measurement error.} & $\sigma_1$ [\%] & Dist. [']\\
\hline
1 & EISJ0044-2950A &   3 &  00 44 49.9 &  -29 50 04.0 & 0.173 &     0 &  98.0 & 2.0\\
1 & EISJ0044-2950A &   3 &  00 44 45.5 & -29 50 38.8 & 0.259 &   444 &  91.3 & 2.9\\
\hline
2 & EISJ0045-2944 &   4 &  00 45 07.1 & -29 45 36.3 & 0.186 &   627 &  97.9 & 1.5\\
2 & EISJ0045-2944 &   3 &  00 45 09.0 & -29 44 49.6 & 0.226 &     0 &  84.9 & 1.8\\
2 & EISJ0045-2944 &   4 &  00 45 07.6 & -29 45 31.9 & 0.373 &   398 &  80.2 & 1.6\\
\hline
3 & EISJ0047-2942 &   3 &  00 47 14.1 & -29 43 13.3 & 0.217 &     0 &  82.3 & 1.9\\
3 & EISJ0047-2942 &   5 &  00 47 29.4 & -29 43 06.1 & 0.264 &   747 &  98.9 & 1.4\\
3 & EISJ0047-2942 &   4 &  00 47 35.4 & -29 44 46.5 & 0.335 &   386 &  90.7 & 3.2\\
\hline
5 & EISJ0049-2920 &   3 &  00 49 34.9 & -29 20 15.7 & 0.293 &   187 &  81.8 & 0.8\\
5 & EISJ0049-2920 &   5 &  00 49 24.3 & -29 20 27.5 & 0.325 &   163 &  97.0 & 1.5\\
\hline
7 & EISJ2236-4026 &   3 &  22 36 38.7 & -40 28 01.8 & 0.153 &    37 &  98.3 & 2.4\\
7 & EISJ2236-4026 &   3 &  22 36 26.1 & -40 27 29.6 & 0.194 &   263 &  92.4 & 4.2\\
\hline
8 & EISJ2236-4014 &   5 &  22 36 49.1 & -40 15 22.8 & 0.274 &   579 &  98.7 & 0.8\\
8 & EISJ2236-4014 &   3 &  22 36 41.6 & -40 15 47.3 & 0.373 &   510 &  67.0 & 2.3\\
\hline
9 & EISJ2237-4000 &   4 &  22 37 19.7 & -39 58 06.6 & 0.242 &   147 &  93.8 & 2.7\\
9 & EISJ2237-4000 &   5 &  22 37 22.1 & -40 00 50.6 & 0.270 &   662 &  98.8 & 2.1\\
\hline
10 & EISJ2238-3934 &   3 &  22 38 17.3&  -39 35 34.4 & 0.145 &   564 &  96.6 & 2.8\\
\hline
11 & EISJ2239-3954 &   4 &  22 39 26.3&  -39 53 27.8 & 0.233 &   468 &  92.8 & 1.9\\
\hline
12 & EISJ2240-4021 &   3 &  22 40 11.7 & -40 25 09.6 & 0.115 &   135 &  97.7 & 4.1\\
12 & EISJ2240-4021 &   3 &  22 40 07.7 & -40 22 59.2 & 0.291 &   593 &  79.0 & 1.9\\
\hline
13 & EISJ2241-4006 &   3 &  22 41 15.4 & -40 07 49.3 & 0.196 &   503 &  90.0 & 2.6\\
13 & EISJ2241-4006 &   5 &  22 41 31.6 & -40 07 58.2 & 0.246 &   165 &  96.5 & 1.8\\
13 & EISJ2241-4006 &   3 &  22 41 28.7 & -40 06 30.2 & 0.294 &   939 &  83.2 & 0.4\\
13 & EISJ2241-4006 &   6 &  22 41 40.4 & -40 08 27.6 & 0.344 &   147 &  94.4 & 3.3\\
13 & EISJ2241-4006 &   3 &  22 41 13.5 & -40 06 26.7 & 0.531 &   146 &  98.5 & 2.5\\
\hline
15 & EISJ2243-4010A &   3 &  22 42 48.4 & -40 10 52.0 & 0.150 &   142 &  98.4 & 2.6\\
15 & EISJ2243-4010A &   3 &  22 43 14.6 & -40 09 03.0 & 0.216 &   398 &  83.9 & 2.8 \\
15 & EISJ2243-4010A &   3 &  22 43 04.7 & -40 10 05.4 & 0.345 &   313 &  73.4 & 0.6\\
\hline
16 & EISJ2243-3952 &   3 &  22 43 12.4 & -39 53 48.7 & 0.171 &    70 &  97.5 & 1.7\\
16 & EISJ2243-3952 &   5 &  22 43 08.2 & -39 53 13.2 & 0.245 &    70 &  95.0 & 2.2\\
\hline
17 & EISJ2243-3959 &   5 &  22 43 40.2 & -40 00 30.9 & 0.214 &   357 &  94.8 & 2.3\\
17 & EISJ2243-3959 &   4 &  22 43 45.4 & -39 59 58.3 & 0.259 &   796 &  94.1 & 3.1\\
17 & EISJ2243-3959 &   3 &  22 43 30.5 & -40 01 03.8 & 0.347 &   568 &  81.6 & 1.5\\
\hline
18 & EISJ2243-4010B &   3 &  22 43 44.4 & -40 12 00.5 & 0.285 &     0 &  80.5 & 1.5\\
\hline
19 & EISJ2243-3947 &   5 &  22 43 55.6 & -39 47 42.7 & 0.196 &   545 &  98.5 & 0.2\\
19 & EISJ2243-3947 &   3 &  22 43 34.5 & -39 48 26.1 & 0.341 &   421 &  79.3 & 4.2\\
\hline
20 & EISJ2244-4008 &   3 &  22 44 29.8 & -40 08 44.2 & 0.178 &   501 &  96.2 & 1.6\\
20 & EISJ2244-4008 &   3 &  22 44 52.4 & -40 08 51.9 & 0.199 &     0 &  88.5 & 5.9\\
20 & EISJ2244-4008 &   5 &  22 44 10.8 & -40 07 56.5 & 0.213 &   383 &  96.9 & 2.1\\
20 & EISJ2244-4008 &   3 &  22 44 11.2 & -40 07 19.6 & 0.359 &   681 &  78.5 & 2.3\\
\hline
21 & EISJ2244-4019 &   3 &  22 44 32.6 & -40 21 00.4 & 0.228 &   182 &  83.9 & 1.5\\
21 & EISJ2244-4019 &   4 &  22 44 57.2 & -40 17 56.2 & 0.244 &   632 &  92.3 & 5.8\\
21 & EISJ2244-4019 &   6 &  22 44 18.4 & -40 20 40.1 & 0.340 &   431 &  93.9 & 2.1\\
\hline
22 & EISJ2246-4012B &   3 &  22 46 40.6 & -40 13 11.6 & 0.126 &   138 &  98.4 & 1.6\\
22 & EISJ2246-4012B &   3 &  22 47 05.2 & -40 14 29.3 & 0.148 &   405 &  98.6 & 3.6\\
\hline
23 & EISJ2248-4015 &   6 &  22 48 57.9 & -40 15 40.5 & 0.246 &   567 &  98.7 & 0.7\\
\hline
\end{tabular}
\end{minipage}
\end{center}
\end{table*}

\section{Non-confirmed candidates with significant redshift detections.}
\label{app:ind_cases}

This appendix presents detailed descriptions for the nine fields
within which a significant group was identified but not associated to
the EIS detection.

\begin{itemize}

\item EISJ0045-2944 (\#2): the identified group in redshift space
covers a large area of the field and shows no concentration at the
expected position. Therefore, it was discarded as a confirmation. From
the image it is likely that the four galaxies of similar brightness
located at the centre of the field are responsible for the
signal. Note that the two redshifts measured in this group do not
agree.

\item EISJ0048-2928 (\#4): in this field two groups were found, both
at quite low redshift with the nearest group being located almost at
the position of the matched-filter detection. However, inspecting the
image one finds in the same region a handful of galaxies with
magnitudes $I\sim19$ matching the Schechter magnitude at $z=0.4$. We
believe that the latter group is the one responsible for the
detection. A firm conclusion about this candidate will require
considerable more data.

\item EISJ2236-4014 (\#8): there is only one significant group, but it
is poor, spread over the field and far away from the centre. From the
image it can be seen that there are no redshifts obtained for the
galaxies close to the position of the matched-filter detection. There
are a number of galaxies with magnitudes matching those expected for a
group at the cluster candidate estimated redshift.

\item EISJ2241-3932 (\#14): in this field a significant group with
three members is found to the east of the matched-filter
position. From the image a promising concentration of galaxies is
found at the matched-filter position. We conclude that there could be
a cluster but more spectroscopic data are needed for a firm
conclusion.

\item EISJ2243-3947 (\#19): three significant groups are identified,
all at quite low redshift. The richest group with seven members is
located to the east of the cluster position. The two other groups have
three and four members with the smallest one also being displaced to
the west of the centre of the matched-filter detection. Inspecting the
image we find a compact group of galaxies most likely causing the
matched filter detection. Only one of these galaxies have a redshift
measured at $z=0.2301$, thus not belonging to any of the identified
groups in redshift space. It is thus inconclusive whether this is the
true redshift of the concentration or whether it is caused by chance
alignment.

\item EISJ2243-3952 (\#16): two significant groups are found, both of
them spread over the entire field with no clear concentration. From
the image it can be seen that this detection is close to the position
of a nearby galaxy. We believe that spurious objects found in the
vicinity of this bright galaxy has caused the MF detection. We
consider this detection totally spurious and an artifact caused by
problems in the original galaxy catalogue.

\item EISJ2243-4010B (\#18): only one significant group is found with
a centre very far from the original matched filter position. From the
image there is a sparse group of galaxies at the position of the
matched filter detection, however only a foreground galaxy has a
redshift measured.

\item EISJ2244-4008 (\#20): we find two significant groups. Both of
these are wide spread and displaced from the original matched filter
position. The concentration of galaxies at the matched filter position
all match the expected magnitudes but the obtained redshifts vary a
lot. We therefore conclude that there is likely no cluster at this
position.

\item EISJ2248-4015 (\#23): in this field three groups were detected -
two are significant and listed in Table~\ref{tab:EISgroups} and one
less significant listed in Table~\ref{tab:marginal}. All of them have
redshifts ranging from 0.13 to 0.25 making the association with the
matched filter detection difficult. We do not believe that the
significant groups are associated with the matched filter detection as
they cover the entire field of view. On the other hand, from the image
we find that the less significant group (at 98.6\%) forms a compact
system very close to the matched filter position, with six galaxies
having concordant redshifts near $z=0.246$ . We, therefore, believe
that this system is responsible for the detection. This conclusion is
supported by the analysis of the colour-magnitude diagram in
Sect.~\ref{sec:nonconf}.

\end{itemize}

\end{appendix}

\end{document}